%% file: diss.tex
\documentclass[12pt,a4paper,headcount,headline,nocenter,bold,noupper,twoside]{thesis}

\usepackage{commands}
\usepackage{exscale}

\newcommand{\emptypage}{\newpage\thispagestyle{empty}\begin{center}\end{center}\newpage}
\newcommand{\stand}{}%{\bf Stand: \today}}
\begin{document}

\include{titel}

%\listoffigures
\pagestyle{empty}
\include{abstract}
\emptypage
\include{danke}

\emptypage
\pagestyle{headings}
\setcounter{page}{1}
\pagenumbering{roman}
\tableofcontents
\clearpage
\emptypage
\pagenumbering{arabic}
\setcounter{page}{1}
\include{kapitel1}

\clearpage
\thispagestyle{empty}
\include{kapitel2}

\clearpage
\thispagestyle{empty}
\include{kapitel3}

\clearpage
\thispagestyle{empty}
\include{kapitel4}

\clearpage
\thispagestyle{empty}
\include{kapitel5}

\clearpage
\thispagestyle{empty}
\include{kapitel6}

\clearpage
\thispagestyle{empty}
\include{kapitel7}

\clearpage
\thispagestyle{empty}
\include{kapitel8}

\clearpage
\thispagestyle{empty}
\chapapp{Chapter}
\appendix
\include{anhanga}

\clearpage
\thispagestyle{empty}
\include{anhangb}

\clearpage
\thispagestyle{empty}
\include{anhangc}

\clearpage
\thispagestyle{empty}
\include{anhangd}

\clearpage
\thispagestyle{empty}
\addcontentsline{toc}{chapter}{References}

\include{zitate}
\end{document}

%% file: titel.tex
\begin{titlepage}
\begin{center}
  \vspace{1.5cm}

  %{\LARGE \bf Universality in Quantum Chaos from   \vspace{0.2cm}

  %Classical Periodic Orbits}

  {\LARGE \bf Periodic-Orbit Approach to \vspace{0.2cm}

  Universality in Quantum Chaos}

\begin{center}
\vspace{2cm}
  \includegraphics[scale=0.8]{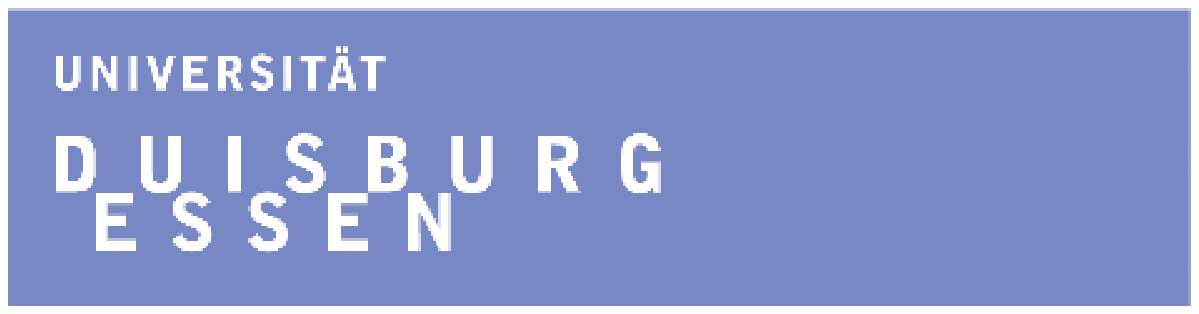}
\vspace{2cm}
\end{center}

  {\sf
    Dissertation\\
    zur Erlangung des Grades\\
    Doktor der Naturwissenschaften\\
    (Dr.~rer.~nat.)\\
    vorgelegt
    am Fachbereich Physik der\\ 
    Universit\"at Duisburg-Essen\\\vspace{1.2cm} von\\\vspace{1.2cm}
    {\Large Sebastian M\"uller}\\\vspace{0.8cm}
    aus Essen
    \vfill
    eingereicht: Essen, September 2005\\
    %{\bf Entwurf}}%
    %{\bf Stand: \today}
    }
\end{center}
\end{titlepage}
\thispagestyle{empty}
\normalsize
\mbox{}
\vfill

\noindent
\begin{tabbing}
Koreeferent:~~~\=~Prof.~Dr.~Robert~Graham\kill
Referent:\>~Prof.~Dr.~Fritz~Haake\\
Korreferent:\>~Prof.~Dr.~Robert~Graham\\[.3cm]
%Disputation:\>~Essen, am~20.~Juni~2000
\end{tabbing}

%% file: abstract.tex
%\vspace{-1cm}
%\abstract

\vspace{\afterchaptervspace}

{\noindent\LARGE\bfseries Abstract}

\vspace{\afterchaptervspace}

\stand We show that in the semiclassical limit, classically chaotic systems have
universal spectral statistics.
%..., full classical chaos is paralleled by universal spectral statistics.
Concentrating on short-time statistics,
we identify the pairs of classical periodic orbits
determining the small-$\tau$ behavior of the spectral form factor $K(\tau)$ of fully
chaotic systems. The two orbits within each pair differ only
by their connections
inside close self-encounters in phase space. The frequency of occurrence of these self-encounters
is determined by ergodicity. Permutation theory is used to systematically
sum over all topologically different families of such orbit pairs.
The resulting expansions of the form factor in powers of $\tau$ coincide with the
predictions of random-matrix theory, both for systems
with and without time-reversal invariance, and to all orders
in $\tau$. Our results are closely related to the zero-dimensional nonlinear $\sigma$ model of quantum field
theory. The relevant families of orbit pairs are
in one-to-one correspondence to Feynman diagrams
appearing in the perturbative treatment of the $\sigma$ model.

\vspace{\afterchaptervspace}

{\noindent\LARGE\bfseries Kurzfassung}

\vspace{\afterchaptervspace}

\begin{otherlanguage}{german}

Wir zeigen, dass klassisch chaotische Systeme sich im semiklassischen Limes durch
die universelle Statistik ihrer Quantenspektren auszeichnen. Dabei konzentrieren
wir uns auf das Kurzzeitverhalten des spektralen Formfaktors $K(\tau)$. Wir
weisen nach, dass f{\"u}r dieses Verhalten Paare periodischer Bahnen verantwortlich
sind, die sich voneinander nur durch ihre Verbindungen innerhalb naher Selbstbegegnungen
im Phasenraum unterscheiden. Die H{\"a}ufigkeit solcher Selbstbegegnungen wird
durch die Ergodizit{\"a}t der klassischen Dynamik bestimmt.
Wir verwenden Methoden der Permutationstheorie, um {\"u}ber alle topologisch verschiedenen Familien
solcher Bahnpaare zu summieren. Die resultierenden
Entwicklungen des
Formfaktors in Potenzen von $\tau$ stimmen
in allen Ordnungen mit den Vorhersagen der Zufallsmatrixtheorie
{\"u}berein,
sowohl f\"ur zeitumkehrinvariante Systeme als auch f\"ur Systeme
ohne Zeitumkehrinvarianz.
Unsere Ergebnisse haben einen engen Bezug zum nulldimensionalen
nichtlinearen $\sigma$-Modell
der Quantenfeldtheorie. Die betrachteten Familien von Bahnpaaren entsprechen
Feynman-Diagrammen, die bei der perturbativen Behandlung des $\sigma$-Modells
auftreten.

\end{otherlanguage}

\enlargethispage{0.5cm}%!!!!

%% file: danke.tex
\vspace{\afterchaptervspace}

{\noindent\LARGE\bfseries Acknowledgement}

\vspace{\afterchaptervspace}

\stand

I am extremely grateful to my thesis advisor, Prof.  Fritz Haake, for
giving me the opportunity to work on this interesting topic, and for
his continuous interest and support.

I also wish to thank Prof. Robert Graham for agreeing to co-referee this thesis.

Most of this thesis is part of a very close and enjoyable cooperation
with Prof. Petr Braun, Prof. Fritz Haake, and Stefan Heusler.  Later,
Prof. Alexander Altland joined the team and provided crucial insights
into field-theoretical aspects.  It is a special pleasure to thank all
of them for this great experience and for many important
contributions.  Furthermore, I am indebted to
Petr Braun, Fritz Haake, and Stefan Heusler
for their comments on
the manuscript.

I also want to thank Marko Turek, Dominique Spehner, and Prof. Klaus
Richter for the joint work on our publication \cite{Higherdim},
containing results on multi-dimensional systems included in this
thesis.

I am grateful to my colleagues Sven Biermann, Julia Ernst, Thorsten
Feldmann, Gregor Hackenbroich, Birgit Hein, Konstantin Krutitsky,
Philipp Kuhn, Andrea Lambert, Christopher Manderfeld, Bernhard Mieck,
Axel Pelster, Wieland Ronalter, Dima Savin, Holger Schaefers, Urs
Schreiber, Lionel Sittler, Prof. Hans-J{\"u}rgen Sommers, Prof. Stefan
Thomae, Matthias Timmer, and Carlos Viviescas for the pleasant working
atmosphere and numerous interesting discussions.

During conferences, seminars, and visits, I enjoyed helpful
discussions with Arnd B{\"a}cker, Prof. Gregory Berkolaiko, Prof. Bruno
Eckhardt, Sven Gnutzmann, Prof. Gerhard Knieper, Jan M{\"u}ller, Jens
Marklof, Prof. J{\"u}rgen M{\"u}ller, Prof. Taro Nagao, Holger Schanz, Henning
Schomerus, Martin Sieber, Prof. Ben Simons, Prof. Uzy Smilansky,
Wen-ge Wang, and Prof. Martin Zirnbauer.

I am grateful to Barbara Sacha for help with organizational problems,
and to R{\"u}diger Oberhage who maintained our computer system and was
always available for questions.

This work was supported by the Sonderforschungsbereich TR/12
``Symmetries and Universality in Mesoscopic Systems'' of the Deutsche
Forschungsgemeinschaft.

Finally, I am grateful to my parents for all they have done for me.

%% file: kapitel1.tex
\chapter{Introduction}

\stand

Chaotic quantum systems display universal behavior
\cite{Stoeckmann,Haake}. Their energy eigenvalues have universal
statistics, and show a tendency to repel each other.  The
conductance and shot noise of chaotic quantum dots are of universal
form, as well as the fluctuations of cross sections in chaotic
scattering systems.  Many more examples for universality can be found
in diverse ares ranging from quantum chromodynamics and molecular
spectroscopy to the study of normal-metal/superconductor
heterostructures.

A quantitative prediction of such universal features is possible if,
rather than considering an individual system, we {\it average} over
all Hamiltonians (represented by large matrices) sharing the same
symmetry properties. This approach, termed random-matrix theory, was
pioneered by Wigner and Dyson in the 1950s, for the level statistics
of atomic nuclei \cite{Wigner,Mehta}.  Surprisingly, the predictions
of random matrix theory, even though derived using {\it ensembles} of
matrices, are typically respected even by {\it individual} chaotic
dynamics.  Growing evidence for the success of random-matrix theory
outside of its initial realm of nuclear physics (see \cite{PreBGS} for
preceding works) led Bohigas, Giannoni, and Schmit \cite{BGS} to
conjecture that {\it the statistics of energy levels of individual
  classically chaotic systems is faithful to random-matrix theory}.

One of the fundamental problems of quantum chaos is to find a
first-principles proof for this conjecture. The reasons (and
conditions) for universality should be related to the {\it classical}
signatures of chaos \cite{Gaspard} displayed by the same type of
systems.  The classical time-evolution of chaotic dynamics {\it
  depends sensitively on the initial conditions}, a property known as
hyperbolicity. Typically, a small initial separation between two
trajectories will grow exponentially in time. Only separations along
certain ``stable'' phase-space directions rather {\it shrink}
exponentially.  Moreover, chaotic dynamics are ergodic: Long
trajectories {\it uniformly fill the energy shell}.

How can we relate universality to these classical manifestations of
chaos?  To answer this question, several approaches have been
suggested, such as parametric level dynamics \cite{Haake} or
field-theoretical methods
\cite{BallisticSigma,Aleiner,Disorder1,Disorder2}.  Following
pioneering work in \cite{Berry,Argaman,SR,Sieber}, our starting point
will be Gutzwiller's trace formula\cite{Gutzwiller}. In the
semiclassical limit, the level density of a (bounded) chaotic {\it
  quantum} system can be written as a sum over its {\it classical}
periodic orbits.  By Fourier transforming Feynman's path integral for
the (trace of) the time-evolution operator, we express the level
density as an integral over classical trajectories closed in
configuration space, with an integrand ${\rm e}^{{{\rm i}} S/\hbar}$ involving
the classical action $S$.  A stationary-phase approximation brings
about a sum over trajectories extremizing the action -- which are
nothing but classical periodic orbits solving the equations of motion.
Each orbit $\gamma$ then comes with a phase factor $\rm{e}^{{{\rm i}} S_\gamma/\hbar}$.

As a prominent example for universality, we will consider the
two-point correlator of the level density, and its Fourier transform,
the so-called spectral form factor $K(\tau)$.  The form factor is
naturally expressed as a function of a time variable $\tau$, conjugate
to the energy difference; this time is made dimensionless by referral
to the Heisenberg time $T_H\propto \hbar^{1-f}$ ($f$ denoting the number of
degrees of freedom), the time scale associated to the mean level
spacing.  The prediction of random-matrix theory depends only on
whether the system in question is time-reversal invariant or not.  In
absence of time-reversal invariance, we have to average over Hermitian
matrices (the Gaussian Unitary Ensemble, GUE) and obtain $K(\tau)=\tau$.
Time-reversal invariant systems require an average over real symmetric
matrices (the Gaussian Orthogonal Ensemble, GOE), yielding $K(\tau)=
2\tau-\tau\ln(1+2\tau)= 2\tau-2\tau^2+2\tau^3-\ldots$.  Here, we momentarily excluded
complications due to spin or geometric symmetries and, most
importantly, restricted ourselves to $\tau<1$.  Broad experimental and
numerical support suggests that {\it individual} chaotic systems are
faithful to the above predictions if we (i) consider highly excited
states (justifying the semiclassical limit $\hbar\to 0$), and (ii) perform
two averages, usually over small intervals of energy and time, to turn
$K(\tau)$ into a smooth and plottable function.

The form factor, involving a product of level densities, leads to a
double sum over long periodic orbits $\gamma$, $\gamma'$ with periods close to
$\tau T_H$, and thus of the order of the Heisenberg time; the associated
phase factor ${\rm e}^{{{\rm i}} (S_\gamma-S_{\gamma'})/\hbar}$ depends on the
difference between their classical actions.  This double sum relates
correlations in {\it quantum spectra} to correlations among the {\it
  actions} of {\it classical orbits} \cite{Argaman}.  In the
semiclassical limit, it may draw systematic contributions only from
pairs of orbits with a small action difference, at most of the order
of Planck's constant. The contributions of all other orbit pairs will
interfere destructively and effectively vanish after performing the
two averages indicated above.

Following Berry \cite{Berry}, we first consider pairs of orbits with
{\it identical} action. In absence of time-reversal invariance (or
other degeneracies of the periodic-orbits spectrum), this leaves only
pairs of identical orbits $\gamma=\gamma'$.  The resulting ``diagonal''
contribution $\tau$ coincides with the random-matrix result. In presence
of time-reversal invariance, mutually time-reversed orbits must also
be taken into account.  The number of relevant orbit pairs is thus
doubled. The overall contribution, $2\tau$, reproduces only the leading
term of the GOE form factor.

Can this approach be extended to all orders of $K(\tau)$?
As first seen numerically by Argaman et al. \cite{Argaman},
off-diagonal orbit pairs are capable of producing additional contributions
to the form factor. We know by now that one must account not
only for pairs of orbits {\it identical} up to time reversal, but also
for orbits composed of parts
{\it similar} up to time reversal.  Within each family of such orbit
pairs, the action difference can be steered to zero, by varying
parameters defining the family.

\begin{figure}
\begin{center}
  \includegraphics[scale=0.57]{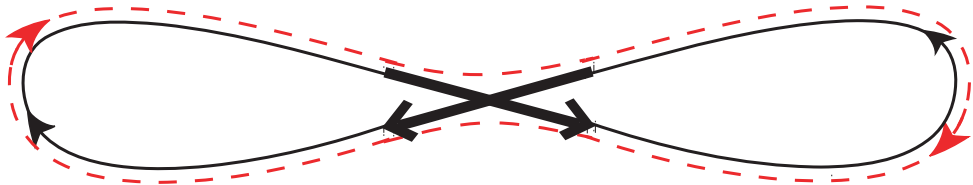}
\end{center}
\caption{Sketch of a Sieber/Richter pair in configuration space:
  The partner orbits, depicted by solid and dashed lines, differ
  noticeably only inside an encounter of two orbit stretches (marked
  by opposite arrows, indicating the direction of motion). The sketch
  greatly exaggerates the difference between the two partner orbits in
  the loops outside the encounter and depicts the loops
  disproportionally short; similar remarks apply to all subsequent
  sketches of orbit pairs.}
\label{fig:SR}
\end{figure}

In a paradigmatic breakthrough, the first such family was identified
by Sieber and Richter \cite{SR,Sieber}, based on intuition drawn from
quantum field theory \cite{Aleiner,Disorder1,Disorder2}.  Their
original formulation is based on small-angle self-crossings in
configuration space.  Inside each Sieber/Richter pair ($\gamma$, $\gamma'$),
one orbit differs from its partner by narrowly avoiding one of its
many self-crossings; see Fig.  \ref{fig:SR}.

In phase-space language, both $\gamma$ and $\gamma'$ contain a region (a
so-called ``encounter'') where two stretches of {\it the same orbit}
are almost mutually time-reversed.  Between the two encounter
stretches, each orbit goes through two ``loops''.  The two orbits
noticeably differ from each other only by their connections inside the
encounter. In contrast, they practically coincide in one loop, and are
mutually time-reversed in the other loop.  The closer the stretches
are, the smaller will be the resulting action difference
$S_\gamma-S_{\gamma'}$.  Sieber/Richter pairs exist only for time-reversal
invariant systems, where they give rise to the quadratic contribution
to the form factor, $-2\tau^2$.

\begin{figure}
\begin{center}
  \includegraphics[scale=0.32]{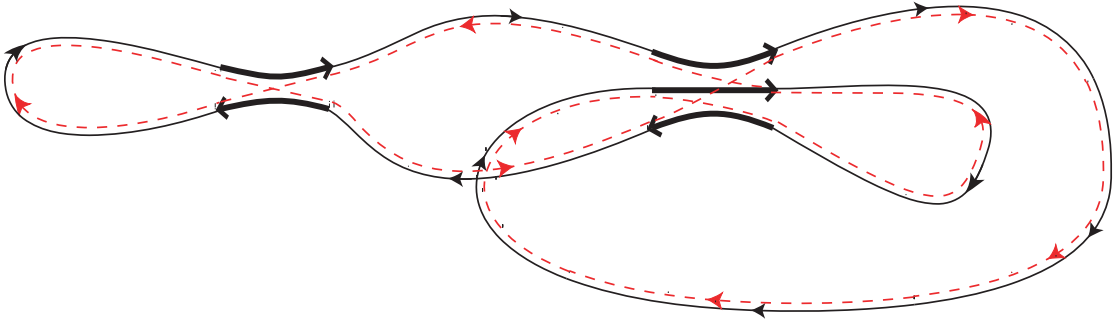}
\end{center}
\caption{Periodic orbit $\gamma$  with one 2- and one 3-encounter
  highlighted, and a partner $\gamma'$ obtained by reconnection.}
\label{fig:23connected}
\end{figure}

In this thesis, we will extend Sieber's and Richter's reasoning (first
formulated for surfaces of constant negative curvature) to general
chaotic dynamics and, moreover, extract all remaining terms in the
power series of $K(\tau)$. To go beyond the quadratic approximation, we
have to consider pairs of orbits differing in {\it any number of
  encounters}, with {\it arbitrarily many stretches}.  For instance,
the cubic contribution to the spectral form factor arises from pairs
of orbits differing in either two close encounters each involving two
stretches, or in one encounter of three stretches. See Fig.
\ref{fig:23connected} for an example of an two partner orbits
differing in one encounter of two and one encounter of three
stretches, contributing to the order $\tau^4$.

We shall see how the classical signatures of chaos -- hyperbolicity
and ergodicity -- determine the universal contributions of these orbit
pairs to the spectral form factor.  It is only due to {\it
  hyperbolicity} that we can obtain partner orbits via {\it local}
reconnections inside encounters, leaving the intervening loops almost
unaffected. Two different encounter stretches, respectively belonging
to $\gamma$ and $\gamma'$, may lead to approximately coinciding loops if their
phase-space difference is close to the stable direction and thus
shrinks exponentially in time. Hyperbolicity also determines the
duration of an encounter.  The stretches stay close as long as their
exponential divergence permits. Hence, the duration of an encounter
will be a logarithmic function of the smallest phase-space separations
involved. Since the encounters relevant for spectral universality
involve a separation of the order of a Planck cell, encounter
durations will be of the order of Ehrenfest time
$T_E\propto\ln\frac{\text{const.}}{\hbar}$ -- divergent in the semiclassical
limit, but much smaller than the orbit periods.

Furthermore, {\it ergodicity} determines the frequency of occurrence
of close self-encounters inside long periodic orbits. We thus see that
all system-specific properties fade away, leaving us indeed with {\it
  universal} contributions to the form factor.  (The conditions
mentioned here will be refined to guarantee that all classical time
scales remain finite and thus negligible compared to $T_H$ and $T_E$.)

A second problem, decoupled from the phase-space considerations
sketched above, is to systematically account for {\it all} families of
orbit pairs differing in encounters. These families will be
characterized not only by the number of related encounters and of the
stretches involved, but also on the order in which stretches and loops
are passed. To deal with this problem, we shall define the notion of
``structures'' of orbit pairs, and relate these structures to
permutations. Roughly speaking, each structure corresponds to one way
of reordering the loops $1,2,\ldots,L$ of $\gamma$ ($L$ denoting the total
number of loops) to a new sequence in $\gamma'$ -- possibly changing the
sense of traversal for dynamics with time-reversal invariance.  Using
the theory of permutations, we can determine by recursion the number
of structures contributing to each order in $\tau$, and hence the Taylor
coefficients of $K(\tau)$.  The resulting series fully coincide with the
predictions of random-matrix theory, both for systems with and without
time-reversal invariance.

Our approach is closely related to the non-linear $\sigma$ model of
quantum field theory. In its zero-dimensional version, that model
provides an efficient way of implementing random-matrix averages of
spectral correlators, or impurity averages for disordered rather than
chaotic systems.  An alternative approach to universality was even
aimed at deriving a ``ballistic'' $\sigma$ model for individual chaotic
systems, by investigating the underlying classical dynamics
\cite{BallisticSigma}.

A perturbative implementation of the $\sigma$ model yields a diagrammatic
expansion mirroring our expansion in terms of (families of) orbit
pairs, with vertices corresponding to encounters and propagator lines
corresponding to orbit loops.  The present orbit pairs thus give a
classical interpretation for Feynman diagrams.  That said, it is no
surprise that the importance of close self-encounters in periodic
orbits was first realized in a field-theoretical context
\cite{Aleiner,Disorder1,Disorder2}.

This thesis is organized as follows. To set up the stage we first want
to review the necessary concepts from classical and quantum chaos.

We shall then move on to discuss the orbit pairs responsible for the
two lowest orders of the spectral form factor.  The treatment of
Sieber/Richter pairs in Chapter \ref{sec:tau2} will exemplify the
phase-space geometry of encounters.  The separations between encounter
stretches will be measured in a coordinate system spanned by the
stable and unstable directions;  Sieber's and Richter's original
treatment in terms of crossing angles rather than phase-space
coordinates, and our generalizations of this approach, will be the
topic of a separate Appendix.  The statistics of phase
space-separation will be derived using (i) ergodicity and (ii) the
necessity of having encounter stretches separated by non-vanishing
loops. We shall see that indeed the leading off-diagonal contribution
to the form factor, $-2\tau^2$, comes about.

In the fourth Chapter, we shall classify the orbit pairs responsible
for the cubic contribution to $K(\tau)$, either involving two encounters
of two, or one encounter of three orbit stretches. We will see that
reconnections inside encounters may give rise to either one connected
or several disjoint orbits; only connected partner orbits contribute
to the form factor. We will summarize the lessons learned for more
complicated orbit pairs, and precisely define the notion of
``structures'' of orbit pairs.

Thus prepared, we will attack the general problem in Chapters
\ref{sec:geometry} and \ref{sec:combinatorics}. In Chapter
\ref{sec:geometry}, we will investigate the phase-space geometry of
encounters with arbitrarily many stretches.  For each structure of
orbit pairs, we will determine the corresponding density of
phase-space separations.  Integration leads to a simple formula for
the contribution to the form factor arising from each structure.

Structures will be counted in Chapter \ref{sec:combinatorics} with the
help of permutation theory, leading to a recursion for the Taylor
coefficients of $K(\tau)$.  In absence of time-reversal invariance, the
contributions of all orbit pairs differing in encounters sum up to
zero.  For time-reversal invariant dynamics, we indeed recover the
logarithmic behavior predicted by the GOE.  This completes our
semiclassical derivation of the random-matrix form factor.

In Chapter \ref{sec:sigma} we give a brief introduction to the bosonic
replica variant of the nonlinear $\sigma$ model, and show that its
perturbative implementation directly parallels our semiclassical
treatment.

Conclusions will be presented in Chapter \ref{sec:conclusions}. We
shall discuss open problems (such as the large-time behavior of
$K(\tau)$), and point to interesting applications in mesoscopic physics.

Appendices will provide important details and extensions, e.g., an
alternative treatment in terms of crossing angles, and generalizations
to multidimensional and ``inhomogeneously hyperbolic'' systems.
Moreover, we shall show that contributions to the form factor arise
only from those encounters which have all their stretches separated by
intervening loops.

Parts of the results of this thesis have been published in
\cite{ThePaper,Mueller,Tau3,Letter,Higherdim,LongPaper}.

%% file: kapitel2.tex
\chapter{Classical and quantum chaos}

\stand

\section{Classical chaos}

\label{sec:classical}

In this thesis, we will be concerned with {\it quantum} properties of
{\it classically chaotic} Hamiltonian systems. In particular, we need
to introduce two notions of classical chaos: hyperbolicity and
ergodicity; see \cite{Gaspard} for a more detailed account.

\subsection{Hyperbolicity}

\label{sec:hyperbolicity}

Roughly speaking, a system is hyperbolic if its {\it time evolution
  depends sensitively on the initial conditions}.  To prepare for a
more precise definition we introduce, for each phase-space point ${\bf
  x}=({\bf q}, {\bf p})$, a {\it Poincar{\'e} surface of section} ${\cal
  P}$ orthogonal to the trajectory passing through ${\bf x}$.
Assuming a Cartesian configuration space (and thus a Cartesian
momentum space), ${\cal P}$ consists of all points ${\bf x}+\delta{\bf
  x}=({\bf q}+ \delta{\bf q}, {\bf p}+\delta{\bf p})$ in the same energy shell
as ${\bf x}$ whose configuration-space displacement $\delta{\bf q}$ is
orthogonal to ${\bf p}$.  For systems with $f$ degrees of freedom,
${\cal P}$ is a $(2f-2)$-dimensional surface within the
$2f-1$-dimensional energy shell.  As long as two trajectories
respectively passing through the points ${\bf x}$ and ${\bf x}+\delta{\bf
  x}$ in ${\cal P}$ remain sufficiently close, we may follow their
separation by linearizing the equations of motion around one
trajectory,
\begin{equation}
 \delta{\bf x}(t)=M({\bf
  x}_0,t)\delta{\bf x}(0).
\end{equation}
Here $\delta{\bf x}(t)$ denotes the phase-space separation in a {\it
  co-moving} Poincar{\'e} section, transversal to ${\bf x}(t)$, the image
of ${\bf x}(0)={\bf x}_0$ under time evolution over time $t$.  The
so-called {\it stability matrix} $M$ with $M_{ij}({\bf
  x}_0,t)=\frac{\partial(\delta x_i(t))} {\partial(\delta x_j(0))}$ defines a linear map
from the Poincar{\'e} section at ${\bf x}(0)$ to that at ${\bf x}(t)$.

We can now define hyperbolicity, first for systems with $f=2$ degrees
of freedom.  Given hyperbolicity, the (two-dimensional) Poincar{\'e}
section ${\cal P}$ is spanned by one {\it stable direction} ${\bf e}^s({\bf
  x})$ and one {\it unstable direction} ${\bf e}^u({\bf x})$ \cite{Gaspard}.
We may thus decompose $\delta{\bf x}$ as
\begin{equation}
  \label{decompose_general}
  \delta{\bf x}={s}{\bf e}^s({\bf x})+{u}{\bf e}^u
({\bf x})\,.
\end{equation}
The linearized equations of motion now read
\begin{eqnarray}
\label{linearize}
s(t)&=&\Lambda(\x_0,t)^{-1}s(0)\nonumber\\
u(t)&=&\Lambda(\x_0,t)u(0)\,.
\end{eqnarray}
Here, ${s}(t)$ and ${u}(t)$ denote stable and unstable components in a
co-moving Poincar{\'e} section at ${\bf x}(t)$.  In the long-time limit,
the fate of the {\it stretching factor} $\Lambda({\bf x}_0,t)$ and thus of
the stable and unstable components is governed by the (local) {\it
  Lyapunov exponent} $\lambda({\bf x}_0)>0$
\begin{equation}
\label{asymptotics}
\Lambda({\bf x}_0,t)\sim{\rm e}^{\lambda({\bf x}_0)t}\,;
\end{equation}
the stable components shrink exponentially, whereas the unstable ones
grow exponentially.  Indeed, this behavior leads to a sensitive
dependence on initial conditions: Two trajectories whose initial
difference has a non-vanishing unstable component will diverge
exponentially fast.

Equation (\ref{linearize}) implies that the stable and unstable
coordinates change with velocities
\begin{eqnarray}
\label{stretching}
\frac{d s}{dt}&=&-\chi(\x(t))s\nonumber\\
\frac{d u}{dt}&=&\chi(\x(t))u\,,
\end{eqnarray}
depending on the so-called {\it local stretching rate}
$\chi(\x(t))=\frac{d\ln|\Lambda(\x_0,t)|}{dt}$.
Incidentally, the Lyapunov exponent
$\lambda(\x_0)$ coincides with the average of the local stretching rate
$\chi(\x(t))$ over an infinite trajectory starting at $\x_0$, given that
\begin{equation}
\lambda(\x_0)\stackrel{(\ref{asymptotics})}{=}\lim_{T\to\infty}\frac{1}{T}
\ln|\Lambda(\x_0,T)|=\lim_{T\to\infty}\frac{1}{T}\int_0^T dt\,\chi(\x(t))\,.
\end{equation}

For so so-called {\it homogeneously hyperbolic} systems $\lambda$, $\Lambda$, and
$\chi$ are independent of the point on the energy shell, i.e.,
$\chi(\x)=\lambda(\x)=\lambda$, $\Lambda(\x,t)={\rm e}^{\lambda t}$ for all $\x$.  The $\x$
dependence of $\lambda$, $\Lambda$, and $\chi$ will be relevant only in Appendix
\ref{sec:maths}, when dealing with certain subtle issues related to
inhomogeneous hyperbolicity; until then, we may think of these
quantities as constants.

As in \cite{Spehner,Turek,Higherdim}, the directions $\vs({\bf x})$
and $\vu({\bf x})$ are mutually normalized by fixing their symplectic
product as
\begin{equation}
\label{norm}
\vu({\bf x})\land\vs({\bf x})={\bf e}^u({\bf x})^T
\left(\begin{array}{cc}0&1\\-1&0\end{array}\right)
{\bf e}^s({\bf
  x})=1\,.
\end{equation}
We note that (\ref{norm}) does not determine uniquely the norms of
$\vs({\bf x})$, $\vu({\bf x})$ for a given dynamics. However, the
following results are valid for any normalization respecting
(\ref{norm}).

In hyperbolic systems with an {\it arbitrary number $f$ of degrees of
  freedom}, the Poincar{\'e} section ${\cal P}$ at point ${\bf x}$ is
spanned by $f-1$ pairs of stable and unstable directions $\vs_m({\bf
  x})$, $\vu_m({\bf x})$ ($m=1,2,\ldots,f-1$).  A displacement $\delta {\bf
  x}$ inside ${\cal P}$ may thus be decomposed as
\begin{equation}
\label{decompose_multi}
  \delta{\bf x}=\sum_{m=1}^{f-1}\left(s_m\vs_m({\bf x})+u_m\vu_m({\bf x})\right),
\end{equation}
compare (\ref{decompose_general}).  Each pair of directions comes with
a separate stretching factor $\Lambda_m({\bf x},t)$, Lyapunov exponent
$\lambda_m({\bf x})$, and stretching rate $\chi_m({\bf x})$.  In extension of
Eq. (\ref{norm}), the directions will be mutually normalized as
\cite{Higherdim}
\begin{eqnarray}
\label{norm_multi}
&\vu_m({\bf x})\land\vs_n({\bf x})=\delta_{mn}&\nonumber\\
&\vu_m({\bf x})\land\vu_n({\bf x})=\vs_m({\bf x})\land\vs_n({\bf x})=0,&
\end{eqnarray}
where $\vu_m({\bf x})\land\vs_m({\bf x})=1$ is a useful convention,
whereas all other relations immediately follow from hyperbolicity.

\subsection{Ergodicity}

\label{sec:ergodicity}

In {\it ergodic} systems, almost all trajectories fill the
corresponding energy shell uniformly. The time average
$\lim_{T\to\infty}\frac{1}{T}\int_0^T F({\bf x}(t))$ of any observable
$F(\x)$ along such a trajectory coincides with an energy-shell average
$\overline{F}=\int\frac{d\mu(\x)}{\Omega} F({\bf x})$ in the Liouville
measure $\frac{d\mu(\x)}{\Omega}$.  Here $d\mu(\x)$ is defined by $d\mu(\x)\equiv
d^{2f}x\,\delta(H({\bf x})-E)$, with $\Omega=\int d\mu(\x)$ denoting the volume
of the energy shell.  As a consequence of ergodicity, the Lyapunov
exponents $\lambda_m({\bf x})$ of almost all points ${\bf x}$ coincide with
the so-called Lyapunov exponents of the system $\lambda_m$, i.e., the
energy-shell averages of the local stretching rates $\chi_m(\x)$.

{\it Mixing} is a stronger requirement than ergodicity. A system is
mixing if, for sufficiently large times $t$, we may neglect classical
correlations between a phase-space point $\x(0)=\x_0$ and its
time-evolved $\x(t)$.  Loosely speaking, a particle at $\x(t)$ does
not ``feel'' where the trajectory has been at time 0.  More
rigorously, an average over $\x_0$ may be calculated by replacing
$\x(t)$ with a phase-space point $\x'$, and averaging separately over
$\x_0$ and $\x'$, as in
\begin{eqnarray}
\label{mixing}
&&\overline{g(\x_0)h(\x(t))}=\int\frac{d\mu(\x_0)}{\Omega}g(\x_0)h(\x(t))\nonumber\\
&&\ \ \ \ \ \xrightarrow[t\to\infty]{}  \ \ \
\int\frac{d\mu(\x_0)}{\Omega}g(\x_0)
\int\frac{d\mu(\x')}{\Omega}g(\x')=
\overline{g}\,\overline{h}\,.
\end{eqnarray}

Periodic orbits are exceptional in the sense that they cannot visit
the whole energy shell.  Consequently, periodic orbits of
inhomogeneously hyperbolic systems typically come with their own Lyapunov
exponents $\lambda_{\gamma m}$ different from the Lyapunov exponent of the
system $\lambda_m$.

However, ergodicity and mixing have important consequences on {\it
  ensembles} of {\it long} periodic orbits $\gamma$, weighted with the
square of the factor
\begin{equation}
 \label{amplitude_class}
 {\cal A}_\gamma\equiv\frac{T_\gamma}{\sqrt{|\det(M_\gamma-1)|}}=\frac{T_\gamma} {\prod_m
   2\sinh\frac{\lambda_{\gamma m} T_\gamma}{2}}\,,
\end{equation}
later to be identified with the absolute value of the so-called
stability amplitude $A_\gamma$. The factor ${\cal A}_\gamma$ depends both on
the period $T_\gamma$ and on the stability matrix $M_\gamma\equiv M(\x_0,T_\gamma)$;
it is independent of the initial point $\x_0$ chosen on $\gamma$.  The
second equality in (\ref{amplitude_class}) follows if we evaluate
$\det(M_\gamma-1)$ in a basis given by the stable and unstable directions
at $\x_0$.

Most importantly, the {\it sum rule of Hannay and Ozorio de Almeida}
\cite{HOdA} guarantees that
\begin{equation}
\label{hoda}
\left\langle\sum_\gamma{\cal A}_\gamma^2\,\delta(T-T_\gamma)\right\rangle=T
\end{equation}
where the angular brackets denote averaging over a small time window
around $T$.  To prove (\ref{hoda}), one shows that
$\frac{1}{T}\sum_\gamma{\cal A}_\gamma^2\, \delta(T-T_\gamma)$ can be identified with the
trace of the Frobenius-Perron operator, guiding the time evolution of
classical phase-space densities \cite{Haake,Gaspard}.  The latter
trace can be written as a sum over eigenvalues of the form ${\rm
  e}^{-\nu_i T}$.  The leading eigenvalue, with $\nu_1=0$ and thus ${\rm
  e}^{-\nu_1 t}=1$, corresponds to a stationary uniform distribution on
the energy shell, according to the Liouville measure
$\frac{d\mu(\x)}{\Omega}$.  The remaining eigenvalues are related to
phase-space distributions decaying with a rate given by the
corresponding Ruelle-Pollicott resonances $\nu_i>0$.  In the limit
$T\to\infty$, the sum rule (\ref{hoda}) is recovered if these resonances
are bounded away from zero, i.e., if the associated classical time
scales remain finite.

A generalization of the sum rule (\ref{hoda}), the so-called {\it
  equidistribution theorem} \cite{Equidistribution}, guarantees that
the above ensembles of periodic orbits behave ergodically in the
following sense: If an observable $F(\x)$ is averaged (i) along a
periodic orbit $\gamma$, according to
\begin{equation}
[F]_\gamma\equiv\frac{1}{T_\gamma}\int_0^{T_\gamma}\!
dt\,F(\x(t))\,,
\end{equation}
and (ii) over the ensemble of all such $\gamma$ with periods inside a
small window $\Delta T$, as in $\frac{1}{T}\langle\sum_\gamma{\cal A}_\gamma^2\,\delta
(T-T_\gamma)\ldots\rangle_{\Delta T}$, we obtain an energy-shell average
\begin{equation}
\label{equidistribution}
\frac{1}{T}\!\left\langle\!\sum_\gamma {\cal A}_\gamma^2\,\delta(T-T_\gamma)\big[F\big]_\gamma\!\right\rangle_{\!\!\!\Delta T}
=\int\frac{d\mu({\bf x})}{\Omega}F({\bf x})\equiv\overline{F({\bf x})}\,.
\end{equation}

In our semiclassical reasoning, we will invoke {\it ergodicity} twice:
First, periodic orbits will be counted using Hannay's and Ozorio de
Almeida's sum rule (\ref{hoda}), which also requires that all {\it
  resonances are bounded away form zero} and thus all classical time
scales are finite.  Second, we need the probability for a trajectory
starting at ${\bf x}(0)$ to pierce through a Poincar{\'e} section ${\cal
  P}$ (as defined in Subsection \ref{sec:hyperbolicity}) in a time
interval $(t, t+dt)$ with sufficiently large
$t$ and stable and unstable components of, e.g., ${\bf x}(t)-{\bf x}(0)$ %(or${\cal T}{\bf x}_t-{\bf x}(0)$)
lying in intervals $(s, s+ds)$, $(u, u+du)$.  That probability is
given by $\frac{d{s}d{u}dt}{\Omega}$, which is nothing but the uniform
Liouville measure, expressed in terms of stable and unstable
coordinates; for $f>2$, the corresponding probability reads
$\frac{1}{\Omega}\prod_{m=1}^{f-1}ds_m\prod_{m=1}^{f-1} du_m\, dt$.  To treat
$\x(0)$ and $\x(t)$ as uncorrelated, we need {\it mixing}; to apply
our reasoning to (ensembles of) periodic orbits, we have to invoke the
equidistribution theorem.  A hyperbolic system satisfying all the
above conditions will be referred to as ``fully chaotic''.

\subsection{Billiards}

Two-dimensional billiards are among the simplest systems that can
exhibit chaotic behavior. They consist of an area with zero potential,
surrounded by a -- sometimes complicated -- boundary. The area outside
that boundary is classically forbidden; it may be seen as having
infinite potential.  Inside the billiard, particles move on straight
lines, until they are reflected from the boundary according to the
reflection law known from geometrical optics.

The properties of a billiard are determined purely by the shape of the
boundary.  In this thesis, two kinds of billiards will occasionally
serve as examples: {\it semidispersing billiards} and the so-called
{\it cardioid billiard}. For both of them, hyperbolic and ergodic
behavior was rigorously established. For further literature on the
ergodic theory of billiards, we refer to \cite{Chernov} and references
therein.

\begin{figure}
\begin{center}
  \includegraphics[scale=0.4]{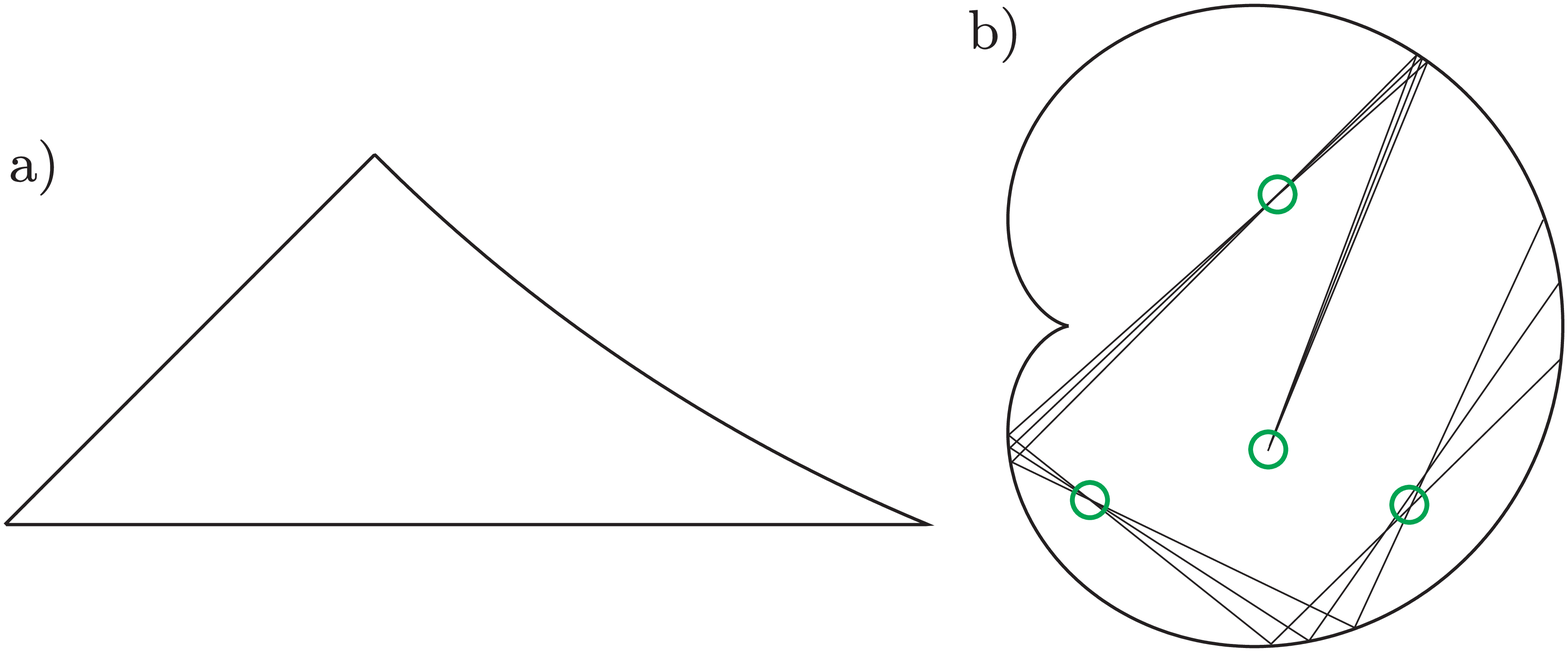}
\end{center}
\caption{Two billiard systems: a) the desymmetrized diamond billiard;
  b) the cardioid billiard, with a fan of trajectories focusing in a
  ``braid'' of mutually conjugate points (marked by circles).}
\label{fig:billiards_shape}
\end{figure}

In semidispersing billiards, the boundary consists of locally concave
and straight lines.  For instance, the desymmetrized diamond billiard,
see Fig. \ref{fig:billiards_shape}a, is surrounded by a part of a
circle and by two straight lines. It can be seen as originating from a
diamond-shaped billiard (the Sinai billiard with overlapping disks),
cut into eight equal pieces.\footnote{ In contrast to the Sinai
  billiard with separated disks, the case of overlapping disks has
  been studied rarely (~\cite{Artuso} being a notable exception). We
  will therefore briefly list the system parameters of the
  desymmetrized diamond.  The distance between the disks is chosen as
  one, and we choose their radius $r=0.541$ such that the interior
  angles become $\frac{\pi}{2}$, $\frac{\pi}{4}$ and $\frac{\pi}{8}$.
  With the circumference $C=0.671$ and the area $A=0.0157$, Santal{\'o}'s
  formula \cite{Santalo} gives the free path as $\bar{l}=\frac{\pi
    A}{C}=0.0735$. By averaging over the Lyapunov exponents of random
  non-periodic trajectories, we numerically obtain the Lyapunov
  exponent of the system as $\lambda = 4.31$.}

The cardioid billiard has a heart-shaped boundary, defined by the
equation $(q_1^2+q_2^2-q_1)^2-(q_1^2+q_2^2)=0$. The boundary is
locally convex; hence the cardioid belongs to the family of focusing
billiards.  A particular characteristic of this family is the
existence of conjugate points. A fan of trajectories (with an
infinitesimal opening angle) starting from one point in configuration
space may focus again in a second one after the next reflection. The
latter point is called conjugate to the initial one. There may even be
whole ``braids'' of mutually conjugate points, as in Fig.
\ref{fig:billiards_shape}b.

\subsection{Symbolic dynamics}

\begin{figure}
\begin{center}
  \includegraphics[scale=0.35]{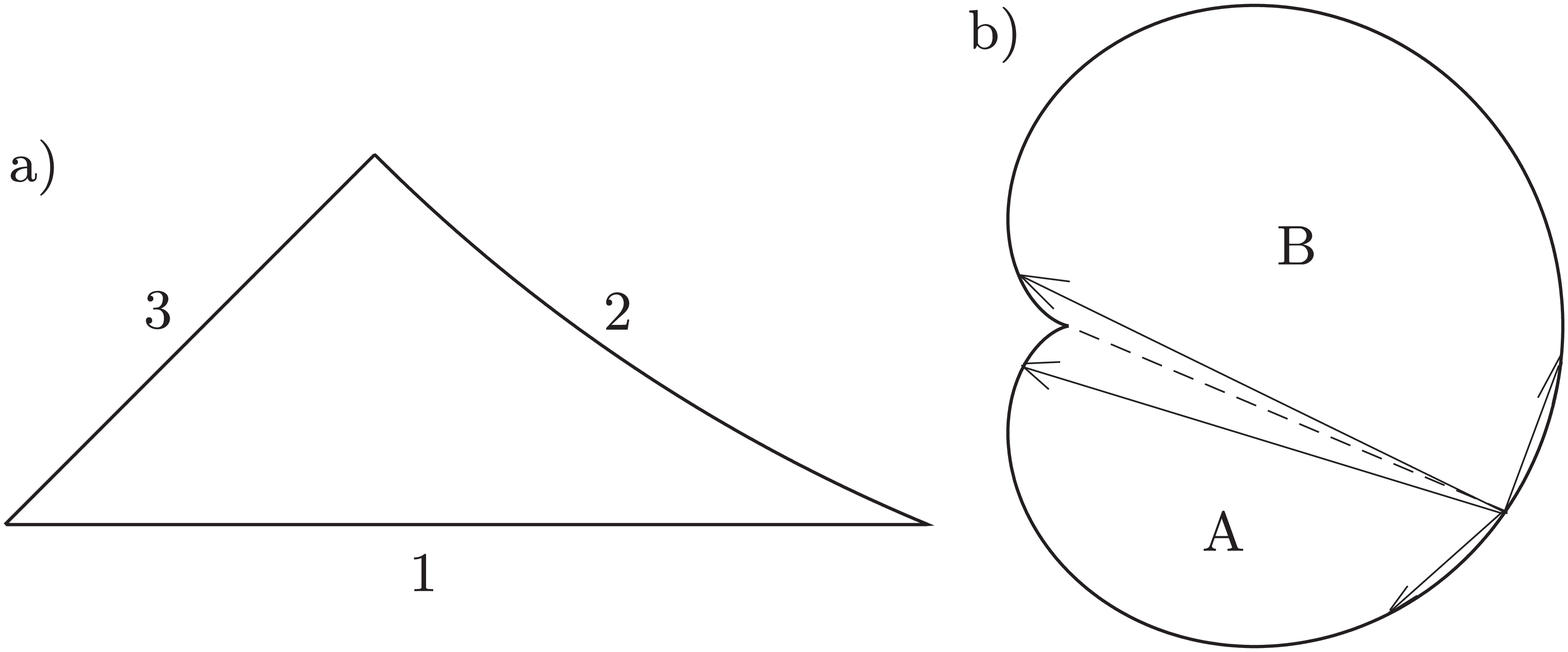}
\end{center}
\caption{Symbolic dynamics: a) In the desymmetrized diamond, symbols label boundary
  segments 1,2, and 3; b) in the cardioid, $A$ denotes ``clockwise''
  orbit segments and $B$ ``counter-clockwise'' ones, as explained in
  the text.}
\label{fig:billiards_sym}
\end{figure}

One of the nice features of the billiards introduced is symbolic
dynamics.  In systems with symbolic dynamics, periodic orbits are
fixed by certain sequences of numbers.  For instance, in {\it
  semidispersing billiards} these numbers denote the segments of the
boundary where the orbit is reflected \cite{Dispersing,Baecker}.  For
the desymmetrized diamond billiard, there will be three symbols for
the three parts of the boundary, see Fig.  \ref{fig:billiards_sym}a,
and the symbol sequence corresponding to an orbit will just enumerate
all reflections inside that orbit.  For each sequence of symbols,
there is at most one orbit; sequences without an associated orbit are
called ``pruned''.  Obviously, symbol sequences of {\it periodic}
orbits are defined modulo cyclic permutations of their members.

Periodic orbits of the cardioid can be described in an alphabet of two
symbols, representing the orbit segments between two reflections
\cite{Cardioid,Baecker}.  ``Clockwise'' segments are denoted by $A$.
More precisely, as sketched in Fig. \ref{fig:billiards_sym}b, the
symbol $A$ refers to clockwise motion in the vicinity of the boundary,
segments leading to a point just below the cusp, and everything in
between. Similarly, ``counter-clockwise'' segments are labelled by
$B$.

More generally, symbolic dynamics can be defined by introducing a
Poincar{\'e} section, e.g., consisting of all phase-space points with
configuration-space coordinates on the boundary of a billiard. This
section is then divided into several regions, and we assign one symbol
to each of them. The symbol sequence of an orbit now depends on the
points of piercing through that section; for each piercing, the symbol
of the corresponding region is added to the sequence. However, the
symbolic dynamics thus defined will be useful only if there is (at
most) one orbit per symbol sequence.

In time-reversal invariant systems, one can typically also define a
time-reversal operation acting on symbol sequences. In the example of
the desymmetrized diamond, time reversal simply means that the
ordering of symbols is reversed.  In the cardioid billiard, time
reversal both inverts the order of symbols and interchanges $A\leftrightarrow B$.

\section{Level density {\`a} la Gutzwiller and Weyl}

\label{sec:gutzwiller}

In the semiclassical limit, {\it quantum} spectra can be approximately
determined from information about the pertaining {\it classical}
dynamics.  The level density $\rho(E)=\sum_j\delta(E-E_j)$ of a bounded
quantum system ($E_j$ denoting the energy levels) may be split into a
local average $\overline{\rho}(E)$ and an oscillatory part $\rho_{\rm
  osc}(E)$ describing fluctuations around that average.  As shown by
Weyl, the {\it smooth part} $\overline{\rho}(E)$ is given by the number of
Planck cells $(2\pi\hbar)^f$ inside the energy shell; we thus obtain
\begin{equation}
\overline{\rho}(E)\sim\frac{\Omega(E)}{(2\pi\hbar)^f},
\end{equation}
with $\Omega(E)$ again denoting the volume of the energy shell.

On the other hand, the {\it oscillatory} contribution to the level
density depends on the classical {\it periodic orbits} of the system
in question. For the case of isolated periodic orbits, this relation
was discovered by Gutzwiller; his results mainly cover hyperbolic
systems (with all orbits isolated and unstable), but also exceptional
dynamics with isolated stable orbits. Gutzwiller showed that $\rho_{\rm
  osc}(E)$ is given by a sum over periodic orbits $\gamma$
\begin{equation}
\label{gutzwiller}
\rho_{\rm osc}(E)\sim\frac{1}{\pi\hbar}\operatorname{Re}\sum_\gamma A_\gamma{\rm
  e}^{{\rm i}S_\gamma/\hbar},
\end{equation}
each orbit contributing with a phase determined by its classical
(reduced) action $S_\gamma=\oint_\gamma{\bf p}\cdot d{\bf q}$ and with a stability
``amplitude''
\begin{equation}
\label{amplitude}A_\gamma=\frac{T_\gamma^{\rm
    prim}}{\sqrt{|\det(M_\gamma-1)|}}{\rm e}^{-{\rm i}\mu_\gamma\frac{\pi}{2}}.
\end{equation}
Here, $T_\gamma^{\rm prim}$ is the primitive period of $\gamma$; hence if $\gamma$
consists only of multiple repetitions of a shorter orbit, we have to
use the period of the latter.  Apart from the case of repetitions,
$|A_\gamma|^2$ coincides with the factor ${\cal A}_\gamma^2$, see Eq.
(\ref{amplitude_class}), appearing in the
sum rule (\ref{hoda}) of Hannay and Ozorio de Almeida and in the
equidistribution theorem (\ref{equidistribution}).  Since among all
orbits, such repetitions form a set of measure zero, we may safely
replace ${\cal A}_\gamma^2\to|A_\gamma|^2$ in the sum rules (\ref{hoda}) and
(\ref{equidistribution}).
%The denominator depends on the stability matrix $M_\gamma$ defined in ...,
%with $\det(M_\gamma-1)=\Lambda_\gamma+\Lambda_\gamma^{-1}$.

\begin{figure}
\begin{center}
  \includegraphics[scale=0.4]{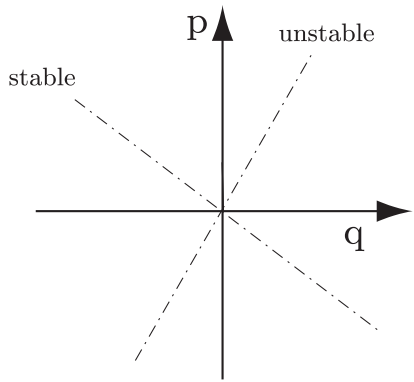}
\end{center}
\caption{Stable and unstable directions, visualized by straight dash-dotted lines,
  in a Poincar{\'e} section ${\cal P}$ parametrized by configuration-space
  and momentum coordinates $q$ and $p$.}
\label{fig:maslov}
\end{figure}

The so-called {\it Maslov index} $\mu_\gamma$ has an interesting geometrical
interpretation \cite{Littlejohn}, which will be exploited later. As in
Section \ref{sec:classical}, let us consider a Poincar{\'e} section ${\cal
  P}$ orthogonal to the orbit in an arbitrary point on $\gamma$.  For
simplicity, we assume two degrees of freedoms.  If we parametrize
${\cal P}$ by configuration-space and momentum coordinates $q$ and
$p$, the stable and unstable directions can be visualized by straight
lines through the origin, see Fig. \ref{fig:maslov}.  If we move
${\cal P}$ along the orbit, these lines will rotate around the origin,
returning to their initial position after each half-rotation.  Given
that the orbit is periodic, they both have to come back to their
initial position after one period. Hence, the number of (clockwise)
half-rotations along the orbit is integer and will be referred to as
the Maslov index $\mu_\gamma$ of that orbit.

We want to sketch the main ideas needed to derive (\ref{gutzwiller}).
The level density may be obtained from the trace of the time evolution
operator $\operatorname{Tr}\,\hat{U}(T)=\sum_j{\rm e}^{-{\rm i}E_j
  T/\hbar}$ as the imaginary part of the one-sided Fourier transform
$\frac{\rm i}{\pi\hbar}\int_0^{\infty}dT\,{\rm e}^{{\rm
    i}Et/\hbar}\operatorname{Tr}\,\hat{U}(T) $ (hence the name ``trace
formula'' for (\ref{gutzwiller})). We now express the diagonal
elements $\langle{\bf q}_0|\hat{U}(T)|{\bf q}_0\rangle$ as path integrals over
trajectories closed in configuration space with ${\bf q}(0)={\bf
  q}(T)={\bf q}_0$, the integrand ${\rm e}^{{\rm i}F[{\bf q}(t)]/\hbar}$
depending on the (full) action $F[{\bf q}(t)]= \int_0^T L({\bf
  q}(t),\dot{\bf q}(t))dt$ of the trajectory in question. This path
integral is evaluated in a stationary-phase approximation, leading to
a sum over classical orbits which start and end at ${\bf q}_0$ and
have stationary action, i.e., solve the canonical equations of motion.
If we subsequently integrate over ${\bf q}_0$ and perform the Fourier
transformation, further stationary-phase approximations lead to the
sum (\ref{gutzwiller}) over orbits periodic in phase space,
contributing with their reduced action $S[{\bf q}(t)]=F[{\bf q}(t)]-ET
=\oint{\bf p}\cdot d{\bf q}$.

\section{Universal spectral statistics}

In the semiclassical limit, fully chaotic quantum systems display {\it
  universal} properties, only {\it depending on their symmetries}.
One example stands out and will be the object of our investigation:
According to the Bohigas-Giannoni-Schmit (BGS) conjecture put forward
about two decades ago \cite{PreBGS,BGS}, highly excited energy levels
of generic fully chaotic systems have universal spectral statistics.
As a preparation, we need to investigate the symmetries of classical
and quantum dynamics, and show how to characterize the statistics of
energy levels.

\subsection{Symmetry classes}

To bar unnecessary difficulties, we will consider only dynamics
without conserved quantities; hence there may be no Hermitian
observables $\hat{A}$ commuting with the Hamiltonian $\hat{H}$. In
particular, this excludes any geometric symmetries, such as reflection
symmetry.

The systems may, however, be invariant with respect to {\it time reversal}.
Typically, if we traverse a trajectory with opposite sense, the
momentum will change its sign. This defines the conventional
time-reversal operator acting on phase-space coordinates ${\bf
  x}=({\bf q},{\bf p})$ as $\T({\bf q},{\bf p})=({\bf q},-{\bf p})$. A
given classical dynamics is called conventionally time-reversal
invariant if the Hamiltonian is even in the momentum, i.e., if $H({\bf
  q},-{\bf p})=H({\bf q},{\bf p})$. In this case each trajectory ${\bf
  x}(t)$ solving the canonical equations of motion yields a second
solution via time reversal, given by $\T{\bf x}(-t)$.

For the quantum dynamics, conventional time reversal amounts to
complex conjugation of the wavefunction $\T(\psi)=\psi^*$.  A Hamiltonian
is time-reversal invariant if it commutes with $\T$, i.e., if it is
real (and thus symmetric). In this case, it is easy to show that each
solution $\psi({\bf q},t)$ of the Schr{\"o}dinger equation $\hat{H}\psi({\bf
  q},t)=i\hbar\partial_t\psi({\bf q},t)$ gives rise to a second one, namely
$\T({\bf q},-t)=\psi^*({\bf q},-t)$.

A more general notion of time reversal includes operators $\T$ which
are both antilinear and antiunitary. Hence, for all
$a_1,a_2\in\mathbb{C}$ and all wavefunctions $\psi_1,\psi_2$, we must have
\begin{eqnarray}
\T(a_1\psi_1+a_2\psi_2)&=&a_1^*\T\psi_1+a_2^*\T\psi_2, \nonumber\\
\langle \T\psi_1|\T\psi_2\rangle&=&\langle\psi_1|\psi_2\rangle^*.
\end{eqnarray}
For example, $\T$ may refer to conventional time reversal, accompanied
by a reflection in configuration space. Systems with non-conventional
time-reversal invariance essentially show the same properties as
conventionally time-reversal invariant ones, and can even be brought
to conventional form by a suitable canonical transformation.

A time-reversal operator may square either to 1 or to $-1$, where the
latter case is possible only for systems with spin. Hamiltonian
dynamics can thus be divided into the three symmetry classes
introduced by Wigner and Dyson:
\begin{itemize}
\item the {\it unitary class} containing systems without time-reversal
  invariance,
\item the {\it orthogonal class} of dynamics invariant under a
  time-reversal operator $\T$ with $\T^2=1$,
\item and the {\it symplectic class} of $\T$-invariant (spin) systems
  with $\T^2=-1$.
\end{itemize}
Since we expect universal behavior only for fully chaotic dynamics
without conserved quantities, we restrict membership of the above
three classes to only such systems.

Recently, Wigner's and Dyson's ``threefold way'' was extended by seven
new classes  \cite{Tenfold} connected to further symmetries, such as symmetry with
respect to charge conjugation.  These new classes are
of experimental relevance, e.g., for normal-metal/superconductor
heterostructures and in quantum chromodynamics.

\subsection{Spectral statistics}

Level statistics can be characterized by the {\it two-point correlation
function} of the level density, proportional to $\rho_{\rm
  osc}(E+\frac{\epsilon}{2}) \rho_{\rm osc}(E-\frac{\epsilon}{2})$.  To obtain a
plottable function, two averages (to be denoted by $\langle\ldots\rangle$) are
necessary, like over windows of the center energy $E$ and the energy
difference $\epsilon$. Moreover, it is convenient to make the energy
difference dimensionless by referral to the mean level spacing
$\frac{1}{\overline{\rho}}$, setting $s\equiv\pi\overline{\rho}\epsilon$. We thus
define the two-point correlator as
\begin{equation}
\label{correlator}
R(s)\equiv\left\langle\frac{\rho_{\rm
      osc}(E+\frac{s}{2\pi\overline{\rho}}) \rho_{\rm
      osc}(E-\frac{s}{2\pi\overline{\rho}})}{\overline{\rho}(E)^2}\right\rangle\
      =\left\langle\frac{\rho(E+\frac{s}{2\pi\overline{\rho}}) \rho(E-\frac{s}{2\pi\overline{\rho}})}
      {\overline{\rho}(E)^2}\right\rangle-1\,, \end{equation}
where the last equality follows from $\langle\rho_{\rm osc}\rangle=0$.

The prime object of our investigation will be the Fourier transform
of $R(s)$, the so-called {\it spectral form factor}
\begin{equation}
\label{fourier}
K(\tau)=\frac{1}{\pi}\int_{-\infty}^{\infty}ds\, e^{2{{\rm i}} s\tau}R(s)\,.
\end{equation}
The variable $\tau>0$, conjugate to $s$, denotes the time measured in
units of the Heisenberg time
\begin{equation}
\label{heisenberg}
T_H=2\pi\hbar\overline{\rho}(E)=\frac{\Omega(E)}{(2\pi\hbar)^{f-1}}.
\end{equation}
By combining Eqs. (\ref{correlator}-\ref{heisenberg}), we may
represent the form factor as
\begin{equation}
\label{formfactor_def}
K(\tau)=\left\langle\int d\epsilon\,{\rm e}^{{\rm i}\epsilon\tau T_H/\hbar}\frac{\rho_{\rm
      osc}(E+\frac{\epsilon}{2}) \rho_{\rm
      osc}(E-\frac{\epsilon}{2})}{\overline{\rho}(E)}\right\rangle\,, \end{equation}
where we have replaced the average over the energy difference by an
average over a small window of $\tau$, and kept the average over $E$.
%where the average over the energy difference may be freely replaced with an average over a small
%window of $\tau$, as long as we keep two separate averages.
Since the study of high-lying states justifies the semiclassical
limit, we may take $\hbar\to 0,T_H\to\infty$, for fixed $\tau$.

Given full chaos, $K(\tau)$ is found to have a universal form, as
obtained by averaging over certain {\it ensembles} of random matrices
\cite{Stoeckmann,Haake,Wigner,Mehta}.  Choosing an arbitrary
orthonormal basis, the Hamiltonian may be written as an infinite
matrix.  For systems without symmetries (unitary class), we only know
that this matrix must be Hermitian, whereas for time-reversal
invariant dynamics with $\T^2=1$ (orthogonal class) it must be real
and symmetric.  Rather than considering an individual Hamiltonian, we
now average over the ensembles of all Hermitian or real symmetric
matrices, integrating over all independent matrix elements and taking
the limit of infinite matrix dimension $N\to\infty$.  The weight must be
chosen invariant under transformations that leave these sets of
matrices invariant; not surprisingly, these are orthogonal
transformations for the orthogonal class and unitary transformations
for the unitary class.  To furthermore guarantee matrix elements to be
uncorrelated, we need a Gaussian weight $\ARMT\,{\rm
  e}^{-\BRMT\operatorname{Tr}\,\hat{H}^2}$,
$\ARMT,\BRMT=\mbox{const}$.  Hence we speak of the {\it Gaussian Unitary}
and {\it Orthogonal Ensembles} (GUE and GOE); a {\it Gaussian Symplectic Ensemble}
(GSE) may be defined similarly.  For the three Gaussian ensembles,
random-matrix averages yield the following predictions for the
spectral form factor
\begin{eqnarray}
\label{formfactor}
K(\tau)=\begin{cases}
  \tau&
  \mbox{GUE,\ $\tau\leq 1$}\\
  1&
  \mbox{GUE,\ $\tau>1$}\\
  2\tau-\tau\ln(1+2\tau)=
  2\tau-2\tau^2+2\tau^3-\ldots&
  \mbox{GOE,\ $\tau\leq 1$}\\
  2-\tau\ln\frac{2\tau+1}{2\tau-1}&
  \mbox{GOE,\ $\tau>1$}\\
  \frac{\tau}{2}-\frac{\tau}{4}\ln(1-\tau)
  =\frac{\tau}{2}+\frac{\tau^2}{4}+\frac{\tau^3}{8}+\ldots&
  \mbox{GSE,\ $\tau\leq2$} \\
  1&
  \mbox{GSE,\ $\tau>2$;}
\end{cases}
\end{eqnarray}
the $\tau$ expansions for the orthogonal and the symplectic case
respectively converge for $0<\tau<\frac{1}{2}$ and $0<\tau<1$.

A further indicator of level statistics is the so-called {\it level spacing
distribution} $P(s)$, i.e., the distribution of differences $s$ between
neighboring energy levels, measured in units of the mean level
spacing.  Random-matrix theory (RMT) yields
\begin{equation}
\label{Ps}
P(s)\propto s^\beta{\rm e}^{-\alpha s^2}
\end{equation}
with $\beta=2,1,4$ for the GUE, GOE, and GSE, respectively; the factor
$\alpha$ depends on the symmetry class as well.  Most importantly, $P(s)\to
0$ for $s\to 0$ implies the energy levels of chaotic systems have a
tendency to repel each other.

According to the BGS conjecture, the level
statistics of {\it individual} chaotic dynamics is faithful to the
predictions, like (\ref{formfactor}) and (\ref{Ps}), of the pertaining
random-matrix ensembles.  A proof of this conjecture, and even the
precise assumptions required for a proof, have thus far remained a
challenge.  In the present thesis, we take up the challenge, and
derive the small-$\tau$ expansion of $K(\tau)$ for {\it individual}
systems; as our main assumptions, we employ ergodicity and
hyperbolicity of the classical dynamics.  Large $\tau$ will not be
considered.

Our approach will be based on periodic-orbit theory, following
previous work in \cite{Berry,Argaman,SR,Sieber}.  Earlier approaches
followed different strategies.  For example, in an ansatz known as
parametric level dynamics \cite{Haake}, the quantum spectrum is
modeled as a fictitious gas, with levels as particles; equilibration
then gives rise to universal statistics.  Field-theoretical methods
were employed in a non-perturbative setting in \cite{BallisticSigma},
and perturbatively in \cite{Aleiner,Disorder1,Disorder2}.

\section{Diagonal approximation}

Using Gutzwiller's trace formula, the form factor is expressed as a
double sum over orbits $\gamma$, $\gamma'$,
\begin{equation}
\label{doublesum}
K(\tau)\sim\frac{1}{T_H}\left\langle\sum_{\gamma,\gamma'}A_\gamma
A_{\gamma'}^*{\rm e}^{{\rm i}(S_\gamma-S_{\gamma'})/\hbar}
\delta\left(\tau T_H-\frac{T_\gamma+T_{\gamma'}}{2}\right)\right\rangle\,.
\end{equation}
To obtain this expression, we have to combine (\ref{gutzwiller}) and
(\ref{formfactor_def}), expand the action to linear order
$S_\gamma(E\pm\frac{\epsilon}{2})\approx S_\gamma(E)\pm T_\gamma(E)\frac{\epsilon}{2}$ and leave out
oscillatory terms $\propto\exp\left(\pm\frac{{\rm
      i}}{\hbar}(S_\gamma+S_{\gamma'})\right)$.

Most importantly, (\ref{doublesum}) implies that for $\hbar\to 0$, only
families of orbit pairs with small action difference
$|S_\gamma-S_{\gamma'}|\simlt\hbar$ can give a systematic contribution to the
form factor. For all others, the phase in (\ref{doublesum}) oscillates
rapidly, and the contribution is killed by the averages indicated.
Correlations of {\it quantum spectra} are thus related to correlations
among the actions of {\it classical orbits} \cite{Argaman}.

The simplest approximation is to keep only ``diagonal'' pairs of
coinciding ($\gamma'=\gamma$) and, for time $\cal T$-invariant dynamics,
mutually time-reversed ($\gamma'={\cal T}\gamma$) orbits, which obviously are
identical in action.  Indeed, Berry \cite{Berry} showed that these
pairs give rise to the leading term in the power series of $K(\tau)$.
Restricting ourselves to diagonal pairs, we obtain the single sum
\begin{equation}
\label{diagsum}
K_{\rm diag}(\tau)=\frac{\kappa}{T_H}\left\langle\sum_\gamma|A_\gamma|^2\delta(\tau
  T_H-T_\gamma)\right\rangle,
\end{equation}
with $\kappa=1$ in the unitary case. For $\T$-invariant dynamics belonging
to the orthogonal class, we also have to account for mutually
time-reversed orbits; hence we have to multiply with $\kappa=2$.  Using
the sum rule of Hannay and Ozorio de Almeida \cite{HOdA}, see Eq. (\ref{hoda}),
the sum in Eq.  (\ref{diagsum}) is easily evaluated to give
\begin{equation}
K_{\rm diag}(\tau)=\kappa\tau,
\end{equation}
as predicted by RMT; compare (\ref{formfactor}).

One may expect that higher-order contributions to the form factor
arise from further families of orbit pairs.  Indeed, using the
predictions for $K(\tau)$, Argaman et al.  \cite{Argaman} derived a
random-matrix expression for {\it classical} correlations between
actions of periodic orbits. For
several systems, the existence of additional correlations could be
confirmed numerically, motivating further studies in
\cite{Cohen,Tanner,Sano,Primack,Smilansky}.  The orbit pairs giving
rise to the $\tau^2$ contribution to the spectral form factor were
recently identified by Sieber and Richter. They will be the subject of
the following Chapter.

\section{Summary}

In this chapter, we introduced basic notions of classical and quantum
chaos.  {\it Classically}, fully chaotic systems are characterized by
hyperbolicity and ergodicity.  Due to hyperbolicity, small phase-space
separations may be decomposed into an unstable component (growing
exponentially in time) and a stable component (shrinking exponentially
in time).  Ergodicity implies that long trajectories uniformly fill
the energy shell.  Prominent examples are chaotic billiards, whose
orbits may be conveniently described by sequences of symbols.

On the {\it quantum-mechanical} side, energy levels of fully chaotic
systems have universal statistics, characterized for example by the
spectral form factor $K(\tau)$, the Fourier transform of the two-point
correlation function of the level density.  According to the so-called
BGS conjecture, the form factor depends only on whether the system in
question has no time-reversal ($\T$) invariance (unitary class), or is
$\T$ invariant with either $\T^2=1$ (orthogonal class) or $\T^2=-1$
(symplectic class); $K(\tau)$ is found to agree with averages over the
pertaining Gaussian ensembles of random matrices.

To explain the universal statistics of quantum spectra, we will employ
results from classical chaos.  Hyperbolicity allows to represent the
fluctuating part of the level density as a sum over periodic orbits,
with an amplitude depending on their classical stability and a phase
involving their classical action (divided by $\hbar$).  The form factor
may thus be written as a double sum over periodic orbits, the phase of
each pair given by the difference between actions (again divided by
$\hbar$).  As shown by Berry, the leading contribution to the $\tau$
expansion of $K(\tau)$ originates from pairs of orbits identical up to
time-reversal.

%% file: kapitel3.tex
\chapter{$\tau^2$ contribution to the spectral form factor}

\label{sec:tau2}

\section{Preliminaries}

%\enlargethispage{4mm}

%Motivated by an analogy to field theory \cite{Disorder},
The family of orbit pairs responsible for the next-to-leading order of
$K(\tau)$ was identified in Sieber's and Richter's seminal papers
\cite{SR,Sieber} for a homogeneously hyperbolic system, the so-called
Hadamard-Gutzwiller model (geodesic motion on a tesselated surface of
negative curvature with genus 2).  In each Sieber/Richter pair, one
orbit narrowly avoids one of the many small-angle self-crossings of
its partner; see Fig. \ref{fig:SR} or Fig. \ref{fig:SRbig}.

We have already sketched the phase-space description of these orbit
pairs.  Both orbits $\gamma$ and $\gamma'$ contain an ``encounter'' of two
almost time-reversed orbit stretches.  These encounter stretches
divide the remainder of $\gamma$ and $\gamma'$ into two ``loops''.  The two
orbits are distinguished only by different connections inside the
encounter. The partner $\gamma'$ almost coincides with $\gamma$ in one loop,
whereas it is nearly time-reversed with respect to $\gamma$ in the other
loop.  Sieber/Richter pairs may exist only in time-reversal invariant
systems, since it must be possible to traverse an orbit loop with
opposite sense of motion.

For systems with symbolic dynamics, Sieber/Richter pairs can be
characterized by their symbol sequences.  To do so, we assign a symbol
sequence to each encounter stretch and each loop.  Note that two {\it
  close} stretches or loops have the {\it same} symbol string since
they will, e.g., bounce from the same parts of the boundary with the
same ordering.  Likewise, two {\it approximately} time-reversed
stretches or loops are described by {\it exactly} time-reversed
sequences.
Thus, the two almost time-reversed stretches will have %mutually
%time-reversed
symbol sequences $\ES$ and $\bar{\ES}$, the overbar denoting time
reversal.  If we label the intervening loops by sequences $\LS$ and
$\RS$, the whole orbit $\gamma$ will be described by the sequence
$\LS\ES\RS\bar{\ES}$.  The partner $\gamma'$ must have the symbol string
of one loop, say $\RS$, reverted in time; the overall sequence will
thus read $\gamma'= \LS\ES\bar{\RS}\bar{\ES}$
\cite{ThePaper,Mueller,Diplom}.\footnote{ Time-reversal of the loop
  $\LS$ gives rise to a different partner
  $\overline{\gamma'}=\bar{\LS}\ES\RS\bar{\ES}$, time-reversed with
  respect to $\gamma'$.  The orbit pair $(\gamma,\overline{\gamma'})$ gives the
  same contribution to the form factor as $(\gamma,\gamma')$.}

To show that the contribution of these orbit pairs agrees with the
second term in the power series of the GOE form factor, $-2\tau^2$, two
key ingredients are needed.  First, Sieber and Richter showed that the
action difference between the two partner orbits is determined by the
crossing angle $\epsilon$ as $\Delta S\propto\frac{\epsilon^2}{2}+O(\epsilon^3)$.  Second, they
investigated the statistics of self-crossings.  The density of
crossing angles contains a term logarithmic in $\epsilon$, which (although
of subleading order in the orbit period) is crucial for spectral
universality. Upon integrating over $\epsilon$, the latter term indeed gives
rise to the anticipated contribution $-2\tau^2$.

This result was extended to general fully chaotic two-freedom systems
in \cite{Mueller}, connecting the treatment of self-crossings in
configuration space with an analysis of the invariant manifolds in
phase space.  Sieber's and Richter's approach was subsequently
reformulated purely in terms of phase-space coordinates by Spehner
\cite{Spehner} and by Turek and Richter \cite{Turek}.  This formulation
could, in an improved version, also be extended to systems with more
than two degrees of freedom, as shown in a joint publication with
these authors \cite{Higherdim}.

In contrast to the historical development, we here want to start with
a treatment in terms of phase-space separations since that language
will also be used when deriving higher-order contributions to the
spectral form factor. The results for configuration-space crossings of
\cite{SR,Sieber} and \cite{Mueller} will be reviewed in Appendix
\ref{sec:crossings}.  That Appendix will also contain numerical
investigations on billiards and more rigorous results on the
Hadamard-Gutzwiller model, which are easier explained in the language
of crossings.

Most of the reasoning in this and the following Chapters applies to
general fully chaotic dynamics. For complete generality, just two
points are missing: First, when showing that certain terms do not
contribute to the form factor, we assume ``homogeneously'' hyperbolic
dynamics, i.e., Lyapunov exponents of all orbits (and local stretching
rates of all phase-space points) coinciding. The necessary
modifications for general hyperbolic systems will be presented in
Appendix \ref{sec:maths}, for all orders of $K(\tau)$.  Second, for
simplicity we restrict ourselves to systems with $f=2$ degrees of
freedom. The results carry over to $f>2$ if we read the stable and
unstable coordinates $s$, $u$ as $(f-1)$-dimensional vectors; the
resulting changes will be listed in Appendix \ref{sec:multi}.

\section{Encounters}

\label{sec:sr_encounter}

\begin{figure}
\begin{center}
  \includegraphics[scale=0.4]{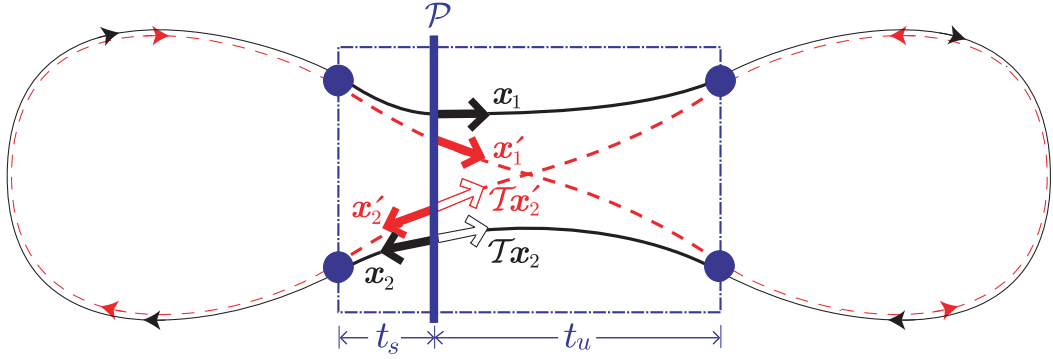}
\end{center}
\caption{
  Configuration-space sketch of a Sieber/Richter pair of orbits $\gamma$
  (full line) and $\gamma'$ (dashed line).  The orbits differ by their
  connections inside the encounter (thick lines, marked by a box).
  Both orbits pierce twice through a Poincar{\'e} section ${\cal P}$
  inside the encounter; the piercing points $\x_1,\x_2,\x_1',\x_2'$
  and the time-reversed piercing points $\T\x_2,\T\x_2'$ are marked by
  arrows pointing in the direction of motion.  The times $t_u$ and
  $t_s$ are the durations of the ``head'' and ``tail'' of the
  encounter.  }
\label{fig:SRbig}
\end{figure}

We will describe encounters by {\it phase-space coordinates}.  To do
so, we introduce a Poincar{\'e} surface of section ${\cal P}$ orthogonal
to the orbit at an arbitrary phase-space point ${\bf x}_{1}$ (passed
at time $t_1$) inside one of the encounter stretches, as in Fig.
\ref{fig:SRbig}.  The exact location of ${\cal P}$ inside the
encounter is not important.  The second stretch pierces through ${\cal
  P}$ at a time $t_2$ in an almost time-reversed point $\x_2$.  The
small difference $\T\x_2-\x_1$ can be decomposed in terms of the
stable and unstable directions at ${\bf x}_{1}$,
\begin{equation}
\T{\bf x}_{2}-{\bf x}_{1}=s\vs({\bf x}_{1})
+u\vu({\bf x}_{1})\;.\end{equation}
The stable and unstable components $s$, $u$ depend on the location
${\bf x}_{1}$ of the Poincar{\'e} section ${\cal P}$ chosen within the
encounter. If we shift ${\cal P}$ through the encounter, the stable
components will asymptotically decrease and the unstable components
will asymptotically increase with growing $t_1$, according to Eqs.
(\ref{linearize}) and (\ref{asymptotics}).

We can now refine our definition of an encounter.  To guarantee that
both stretches are almost mutually time-reversed, we demand both $|s|$
and $|u|$ to be smaller than a constant $c$.  The bound $c$ must be
chosen small enough for the motion around the two orbit stretches to
allow for the mutually linearized treatment (\ref{linearize});
however, the exact value of $c$ is not important for the following
considerations.

By definition, the encounter begins when the stable component $|s|$
falls below $c$, and ends when the unstable component $|u|$ reaches
$c$.  The {\it duration} of an encounter is thus obtained by summing
the durations of the ``head'' of the encounter (i.e., the time $t_u$
until the unstable component $|u|$ reaches $c$) and its ``tail''
(i.e., the time $t_s$ passed since the stable components $|s|$ has
fallen below $c$) as depicted in Fig. \ref{fig:SRbig}.  Given that the
unstable components grow exponentially in time, we have $|u|\e^{\lambda
  t_u}\sim c$ and thus
\begin{equation}
\label{sr_tu}
t_u\sim\frac{1}{\lambda}\ln\frac{c}{|u|}.
\end{equation}
Similarly, the ``tail'' has the duration
\begin{equation}
t_s\sim\frac{1}{\lambda}\ln\frac{c}{|s|}.
\end{equation}
The overall duration of the encounter is thus given by
\begin{equation}
\label{sr_tenc}
t_{\rm enc}=t_s+t_u\sim\frac{1}{\lambda}\ln\frac{c^2}{|su|}\,.
\end{equation}
Reassuringly, (\ref{linearize}) guarantees that the product $su$ and
hence the duration $t_{\rm enc}$ remain invariant if the Poincar{\'e}
section ${\cal P}$ is shifted through the encounter.

Note that we here described the {\it local} divergence inside the
encounter through the {\it global} Lyapunov of the system $\lambda$.  For
inhomogeneously hyperbolic systems, this is only an approximation (to
be avoided in Appendix \ref{sec:maths}), but rather accurate for
typical long encounters, which can be expected to explore the energy
shell approximately uniformly.

\section{Partner orbits}

Each encounter of two antiparallel stretches inside a periodic orbit
$\gamma$ gives rise to a partner orbit $\gamma'$. The partner $\gamma'$ is
distinguished from $\gamma$ by differently connecting the ``ports'' (i.e.,
the initial and final points) of the encounter stretches, marked by
dots in Fig. \ref{fig:SRbig}.  Moreover, $\gamma'$ has one, say, the
``right'' loop reversed in time.

\subsection{Partner piercings}

\label{sec:sr_manifolds}

The partner $\gamma'$ also pierces through the Poincar{\'e} section ${\cal P}$
in two almost mutually time-reversed phase-space points $\x_1'$ and
$\x_2'$, with $\x_1'\approx\T\x_2'\approx\x_1$.  These piercings are determined
by those of $\gamma$.

\begin{figure}
\begin{center}
  \includegraphics[scale=0.4]{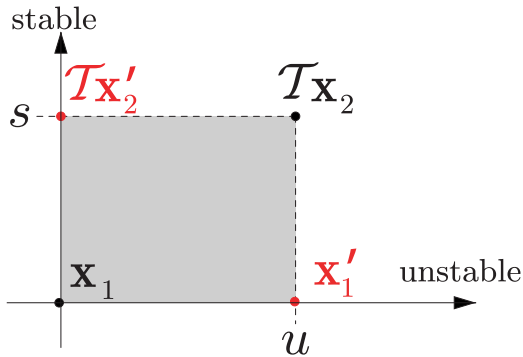}
  %\parbox{0.25cm}{}
  \parbox{7.9cm}{ \vspace{-4.4cm}

\caption{Piercing points of the original orbit $\gamma$ and its partner
  $\gamma'$ through a Poincar{\'e} section, parametrized by stable and
  unstable coordinates. The area of the rectangle coincides with the
  action difference $\Delta S$.  (Note that the Sieber/Richter pair in
  Fig. \ref{fig:SRbig} has an action difference $\Delta S=S_\gamma-S_{\gamma'}<0$
  and thus either $s$ or $u$ negative.)}
\label{fig:actiondiff_sr}}
\end{center}
\end{figure}

Let us first consider $\x_1'$.  The stretches passing through $\x_1'$
and $\T\x_2$ lead to (practically) the same port. Two trajectories
starting at $\x_1'$ and $\T\x_2$ thus approach each other for a long
time, at least until the end of the encounter and half-way through the
subsequent loop; in fact we shall see that the durations of the
relevant encounters diverge in the limit $\hbar\to 0$.\footnote{ The two
  trajectories will approach only for a short time in the exceptional
  case that (i) the Poincar{\'e} section ${\cal P}$ is placed close to the
  end of the encounter, and that (ii) the subsequent loop is short.
  (Vanishing loops are excluded and will be dealt with in Subsection
  \ref{sec:min_dist}.)  Since all possible locations of Poincar{\'e}
  sections will be treated on equal footing, the impact of such
  exceptional locations of ${\cal P}$ is negligible.}  Hence, the
difference between $\x_1'$ and $\T\x_2$ must be close to the stable
manifold, and their unstable coordinates practically coincide.
Similarly, the stretches passing through $\x_1'$ and $\x_1$ {\it
  start} from the same port and thus approach for large negative
times.  Hence, their difference is close to the unstable direction,
and the corresponding stable coordinates coincide.  We can draw the
location of the corresponding piercing points in our Poincar{\'e} section,
spanned by stable and unstable directions (Fig.
\ref{fig:actiondiff_sr}). If we choose $\x_1$ as the origin of that
section, $\T\x_2$ will have the stable coordinate $s$ and the unstable
coordinate $u$. The piercing point $\x_1'$ must have the same stable
coordinate as $\x_1$ and the same unstable coordinate as $\T\x_2$,
i.e.,
\begin{subequations}
\label{sol}
\begin{equation}
s_1'=0,\;u_1'=u.
\end{equation}
Similarly, one can show that $\T\x_2'$ has the coordinates
\begin{equation}
s_2'=s,\;u_2'=0.
\end{equation}
\end{subequations}
Thus, the piercings of $\gamma$ and $\gamma'$ together span a parallelogram in
${\cal P}$ \cite{Braun} (depicted as a rectangle in Fig.
\ref{fig:actiondiff_sr}).  The uniqueness of (\ref{sol}) implies that
there is exactly one partner orbit per encounter.

\subsection{Action difference}

We can now determine the difference between the actions of the two
partner orbits. Generalizing the results for configuration-space
crossings in \cite{SR,Sieber,Mueller}, we will show that the action
difference is just the symplectic area of the rectangle in Fig.
\ref{fig:actiondiff_sr} \cite{Spehner,Turek}.  Consider two segments
of the encounter stretches leading from the port on the upper left
side in Fig.  \ref{fig:SRbig} to the piercing point ${\bf x}_1$ of
$\gamma$, and to the piercing point ${\bf x}_1'$ of $\gamma'$, respectively.
Since the action variation brought about by a shift $d{\bf q}$ of the
final coordinate is ${\bf p}\cdot d{\bf q}$, the action difference
between the two segments will be given by $\Delta S^{(1)}=\int_{{\bf
    x}_1'}^{{\bf x}_1}{\bf p}\cdot d{\bf q}$.  The integration line may
be chosen to lie in the Poincar{\'e} section; then it coincides with the
unstable axis.  Repeating the same reasoning for the remaining
segments, we obtain the overall action difference $\Delta S\equiv
S_\gamma-S_{\gamma'}$ as the line integral $\Delta S=\oint{\bf p}\cdot d{\bf q}$ along
the contour of the parallelogram ${\bf x}'_1\to{\bf x}_1\to\T{\bf
  x}'_2\to\T{\bf x}_2\to{\bf x}'_1$, spanned by ${\bf x}_1'-{\bf
  x}_1={u}{\bf e}^u({\bf x}_1)$ and $\T{\bf x}_2'-{\bf x}_1={s}{\bf
  e}^s({\bf x}_1)$.  This integral indeed gives the symplectic area
\begin{equation}
\label{DS2}
\Delta S= {u}{\bf e}^u({\bf x}_1)\land{s}{\bf e}^s({\bf x}_1)
={s}{u}\,, \end{equation}
where we used the normalization in Eq. (\ref{norm}).  If the above
parallelogram is depicted as a rectangle, like in Fig.
\ref{fig:actiondiff_sr}, the action difference can simply be
interpreted as a geometric area.

\begin{figure}
\begin{center}
  \includegraphics[scale=0.585]{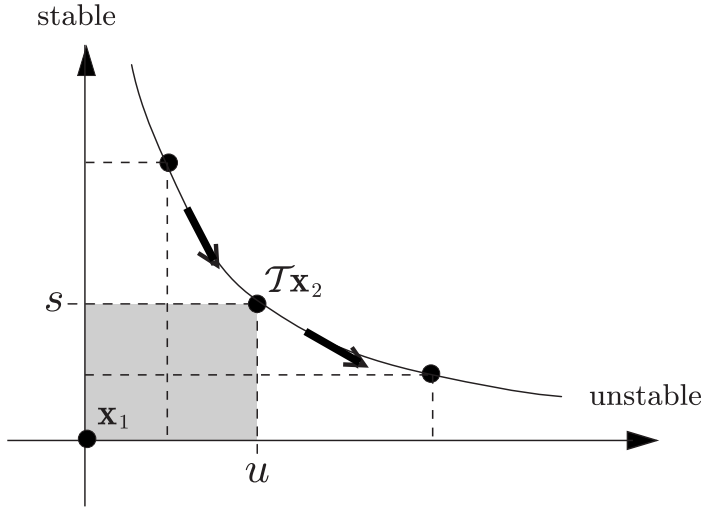}
  \parbox{5.5cm}{ \vspace{-6cm}

\caption{As ${\cal P}$ is shifted through the encounter,
  the piercing point $\T\x_2$ travels on a hyperbola with $\Delta S=su$
  constant.}
\label{fig:actiondiff_inv}
}
\end{center}
\end{figure}

We have already revealed $su$ as independent of the location of ${\cal
  P}$.  If we shift ${\cal P}$ through the encounter, $\T\x_2$ will
travel on a hyperbola with $\Delta S=su$ fixed, with unstable coordinates
growing and stable coordinates shrinking asymptotically; see Fig.
\ref{fig:actiondiff_inv}.  As $\T\x_2$ moves, the rectangle in ${\cal
  P}$ is deformed, but its symplectic area is conserved (as it should
be, given a Hamiltonian flow).

At this point, we can finally appreciate that the encounters relevant
for spectral universality have a duration of the order of the
Ehrenfest time. The form factor is determined by orbit pairs with an
action difference $\Delta S=su$ of order $\hbar$; hence indeed $t_{\rm
  enc}\sim\frac{1}{\lambda}\ln\frac{c^2}{|su|} =\frac{1}{\lambda}\ln\frac{c^2}{|\Delta
  S|}\sim \frac{1}{\lambda}\ln\frac{c^2}{\hbar}$.

\subsection{Stability amplitudes}

\label{sec:stability_amplitude}

It remains to be shown that the relative difference between the
stability amplitudes (see (\ref{amplitude_class}) and
(\ref{amplitude})) of $\gamma$,
\begin{equation}
\label{stability_amplitude}
A_\gamma=\frac{T_\gamma^{\rm prim}}{2\sinh\frac{\lambda_\gamma T_\gamma}{2}}
{\rm e}^{-{\rm i}\mu_\gamma\frac{\pi}{2}},
\end{equation}
and $\gamma'$ vanishes as the stable and unstable separations inside the
encounter go to zero.

First, we want to demonstrate that both orbits have the same {\it
  Maslov index}.  We use the geometrical interpretation of the Maslov
index introduced in Section \ref{sec:gutzwiller}: As the Poincar{\'e}
section ${\cal P}$ is shifted along the orbit, the stable and unstable
manifolds rotate around the origin, relative to the
configuration-space and momentum axes. The Maslov index of a periodic
orbit counts the number of clockwise half-rotations.  For our
argument, we also define a Maslov index for non-periodic pieces of
trajectory, as the sum of rotation angles of the stable and the
unstable manifold, divided by $2\pi$.  The Maslov index thus defined is
invariant under time reversal.  Obviously, time reversal leaves the
absolute value of a rotation angle invariant. The same is true for the
sense of rotation, since time reversal inverts the motion on the
Poincar{\'e} section in direction (turning a clockwise rotation into a
counter-clockwise one and vice versa), but also changes the sign of
the momentum (turning the sense of rotation back to the original
one).\footnote{Depending on conventions, time reversal may also invert
  the directions of the configuration-space and momentum axes inside
  ${\cal P}$, but this has no impact on the sense of rotation.} In
addition, time reversal exchanges the stable and unstable manifolds,
which also cannot affect the {\it sum} of their rotation angles.
Thus, the Maslov index is time-reversal invariant.  (An alternative
proof for the Maslov index of a periodic orbit is given in
\cite{Robbins}.)

The Maslov index of $\gamma$ is now obtained by summing the Maslov indices
of the encounter stretches and loops.  To obtain a partner orbit, we
invert the direction of motion on one loop, leaving $\mu_\gamma$ invariant,
and reconnect the orbit inside the encounter.  The latter
reconnections could at most lead to a small change of $\mu_\gamma$,
vanishing for phase-space separations going to zero.  Given that
$\mu_\gamma$ and $\mu_{\gamma'}$ have to be integer, reconnections cannot affect
the Maslov index at all; hence indeed $\mu_{\gamma'}=\mu_{\gamma}$.

Trivially, the primitive {\it periods} $T_{\gamma}^{\rm prim}$ and
$T_{\gamma'}^{\rm prim}$ approximately coincide, since the duration of
orbit loops and encounter stretches is invariant under time reversal,
and only slightly changed by reconnections.  (Incidentally, for
billiards the periods of $\gamma$ and $\gamma'$ are proportional to the almost
coinciding actions.)

The {\it Lyapunov exponents} $\lambda_\gamma$ and $\lambda_{\gamma'}$ exactly coincide
for homogeneously hyperbolic dynamics. For inhomogeneously hyperbolic
systems, recall that the Lyapunov exponent $\lambda_\gamma$ is obtained by
averaging the local stretching rate $\chi(\x(t))$ over the orbit $\gamma$,
i.e., $\lambda_\gamma T_\gamma=\int_0^{T_\gamma}dt\,\chi(\x(t))$.  In similar vein as
above, time reversal of an orbit loop leaves $\lambda_\gamma T_\gamma$ unchanged,
since $\chi(\x(t))$ is time-reversal invariant.  Reconnections inside
the encounter may only lead to a small change, vanishing for $s,u\to
0$.  Therefore, we indeed have $A_{\gamma'}\approx A_\gamma$.

In contrast to the action difference, the difference between stability
amplitudes or periods will never be referred to a small quantum scale.
Since the relative differences between the amplitudes and periods of
both partner orbits are small, we replace $A_{\gamma'}\to A_\gamma$, $T_{\gamma'}\to
T_\gamma$.  A rigorous justification of these approximations can be given
for the Hadamard-Gutzwiller model, where the small difference between
$A_{\gamma'}$ and $A_\gamma$ can be evaluated analytically; see Appendix
\ref{sec:notau3}.

\section{Necessity of non-vanishing loops}

\label{sec:min_dist}

It is important to consider only encounters with {\it stretches
  separated by non-vanishing loops}.  Only for such encounters, the
two stretches will begin and end in four different ports, and
reconnections between these ports give rise to a partner orbit.  To be
certain, we shall check in the following Subsection that our
prescription for determining partner orbits cannot be extended to work
in case of missing loops.  Afterwards, we will formulate the condition
of non-vanishing loops as a restriction on the piercing times $t_1$
and $t_2$.

\subsection{Almost self-retracing encounters}

\label{sec:stability_matrices}

\begin{figure}
\begin{center}
  \includegraphics[scale=0.33]{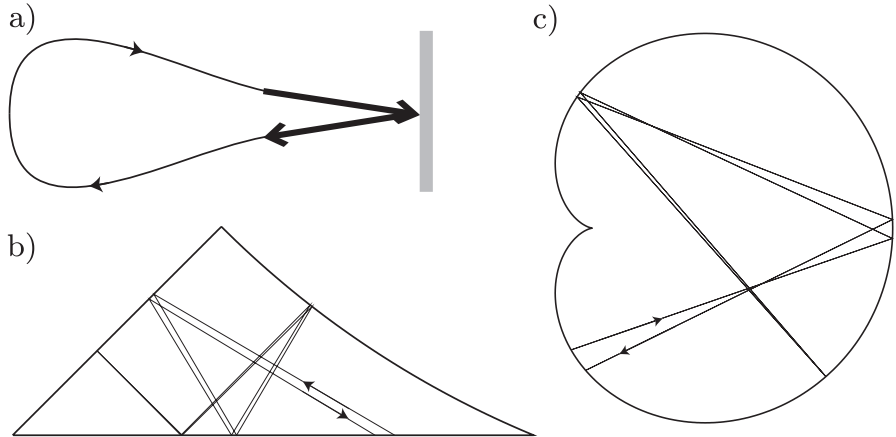}
\end{center}
\caption{Almost self-retracing encounters:
  a) schematic sketch, b) example for the desymmetrized diamond
  billiard, c) example for the cardioid billiard.}
\label{fig:retracer_examples}
\end{figure}

How do we have to imagine an encounter without an external loop?  If,
say, the right loop in Fig. \ref{fig:SRbig} is shrunk away, we will
obtain an encounter as depicted in Fig. \ref{fig:retracer_examples}a.
Somewhere within the encounter the orbit undergoes a nearly
self-retracing reflection from a hard wall.  After the reflection, the
particle will for some time travel close to the pre-collision
trajectory, such that technically an encounter of two antiparallel
orbit pieces is formed.  In some systems, such as the
Hadamard-Gutzwiller model, reflections like this never take place,
because there are no reflecting walls.  In contrast, almost
self-retracing encounters appear frequently in billiards, see Fig.
\ref{fig:retracer_examples}b-c for two examples.  Since there are only
two ports, the encounter effectively has to be considered as {\it one
  single stretch} folded back upon reflection, and no partner can be
obtained by reconnecting the two ports.

To verify this intuitive picture, let us first consider systems with
symbolic dynamics.
%Symbolic dynamics yields a surprisingly simple proof of this fact.
The symbol sequences of the two partner orbits
$\gamma=\LS\ES\RS\bar{\ES}$, $\gamma'=\LS\ES\bar{\RS}\bar{\ES}$ turn out equal
if the right loop with symbol sequence $\RS$ is absent.  Since pairs
of identical (or mutually time-reversed) orbits are already included
in the diagonal approximation, encounters without intervening loops
give no off-diagonal contribution to the form factor.

To generalize this result to systems without symbolic dynamics, we
first need to find an equation for the piercing points of the partner
orbit $\gamma'$ which remains valid even in absence of intervening loops.
Let us again consider a Poincar{\'e} section ${\cal P}$ placed somewhere
inside the encounter. This section divides the orbit into two parts
(respectively containing the left and right loops, if present, and the
``tail'' and ``head'' of the encounter) whose stability matrices will
be denoted by $L$ and $R$.  Note that we do not require these orbit
parts to be long.  We assume that $\gamma'$ approximately follows the
``left'' part of $\gamma$ in the same direction, and the ``right'' part
with opposite direction.  Clearly, this generalizes our previous
prescription for finding partners: Rather than reverting one loop in
time and changing connections between ports, we revert one orbit part;
to obtain a classical periodic orbit, we then must change connections
between the ends of the two orbit parts, and slightly deform the
resulting trajectory.  As in Fig. \ref{fig:SRbig}, the piercings of
$\gamma$ and $\gamma'$ will be denoted by $\x_1$, $\x_2$, $\x_1'$, and
$\x_2'$.  Since $\gamma$ and $\gamma'$ are close in the left part leading from
$\x_2$ to $\x_1$, we may linearize
\begin{subequations}
\label{partnereq0}
\begin{equation}
\x_1'-\x_1 = L (\x_2'-\x_2).
\end{equation}
The two orbits are almost mutually time-reversed in the right part,
leading from $\x_1$ to $\x_2$ in $\gamma$. The time-reversed of that part
leads from $\T\x_2$ to $\T\x_1$ in $\gamma$, and from $\x_1'$ to $\x_2'$
in $\gamma'$; it has the stability matrix $R^\T=\T R^{-1} \T$. We thus
obtain
\begin{equation}
\x_2'-\T \x_1 = R^\T (\x_1'-\T \x_2).
\end{equation}
\end{subequations}
With $\delta\x_1=\x_1'-\x_1$, $\delta\x_2=\x_2'-\x_2$, $\delta \x= \T\x_2-\x_1$,
(\ref{partnereq0}) simplifies to
\begin{eqnarray}
\delta\x_1&=&L\delta\x_2\nonumber\\
\delta\x_2+\T\delta\x&=&R^\T(\delta\x_1-\delta\x). \label{partnereq}
\end{eqnarray}
This system of equations uniquely determines the partner orbit $\gamma'$.
If the encounter stretches are separated by intervening loops, Eq.
(\ref{partnereq}) will entail the solution (\ref{sol}) derived
previously.\footnote{ To obtain (\ref{sol}), we represent $L$ and $R$
  in a basis given by the unstable and stable directions, i.e.,
  $L=\matr{\Lambda_L}{0}{0}{\Lambda_L^{-1}}$, $R=\matr{\Lambda_R}{0}{0}{\Lambda_R^{-1}}$,
  with $\Lambda_L$ and $\Lambda_R$ the stretching factors of the left and right
  orbit parts. If the stretches are separated by two intervening
  loops, both orbit parts will be long enough to invoke the limit
  $\Lambda_L,\Lambda_R\to\infty$, leading to the solution (\ref{sol}).  Starting from
  (\ref{partnereq}) one can also give an alternative proof for the
  partner $\gamma'$ being independent of the location of ${\cal P}$
  \cite{Mueller}.}

In contrast, if e.g. the ``right'' loop is absent, the right orbit
part will remain inside the encounter for its whole duration.  The
phase-space separations inside that part may thus be followed in a
linear approximation.  We may, for instance, linearize the equations
of motion around the trajectory leading from $\T\x_2$ to $\T\x_1$, with
the stability matrix $R^\T$.  During the same time, the piercing point
$\x_1$ is carried into $\x_2$. Linearizing with the help of $R^\T$, we
obtain (up to quadratic order in $\T\x_2-\x_1$)
\begin{eqnarray}
\x_2-\T\x_1=R^\T(\x_1-\T\x_2)
\end{eqnarray}
or, equivalently,
\begin{eqnarray}
\T\delta\x=-R^\T\delta\x.
\end{eqnarray}
Inserting the latter result into the second equation in
(\ref{partnereq}), we see that (\ref{partnereq}) has the trivial
solution $\delta\x_1=\delta\x_2=0$. Thus, the ``partner'' with time-reversed
right loop coincides with the initial orbit. Conversely, if the left
loop were absent, the corresponding ``partner'' would coincide with
the time-reversed of the initial orbit.  Again, we see that almost
self-retracing encounters do not yield off-diagonal orbit pairs and
therefore do not contribute to the spectral form factor.

\subsection{Minimal distances}

To give rise to a partner orbit, two encounter stretches need to be
separated by intervening loops. Such loops exist if the times $t_1$
and $t_2$ (when the stretches pass through the section ${\cal P}$)
observe certain minimal distances.  To show this, we first have to fix
notation.  The times $t_1$ and $t_2$ are well defined only modulo the
period of the orbit $T_\gamma$. In the sequel, it will be convenient to
take $0<t_1<T_\gamma$ and $t_1<t_2<t_1+T_\gamma$.

To allow for a non-vanishing loop, the orbit part to the right of
${\cal P}$ (leading from $t_1$ to $t_2$) must exceed twice the
duration of the head of the encounter.  Only then it is long enough to
contain (i) the head of the first encounter stretch with duration
$t_u$, (ii) a loop of positive duration, and (iii) a piece of the
second stretch again with duration $t_u$. Hence we have to demand that
\begin{equation}
\label{min_right}
t_1+2t_u<t_2
\end{equation}

Similarly, the orbit part to the left of ${\cal P}$ (from $t_2$ to
$t_1+T_\gamma$) must be longer than $2t_s$, to leave time for two
traversals of the ``tail'' of the encounter, separated by a
non-vanishing loop. We thus need to have
\begin{equation}
\label{min_left}
t_2<t_1+T_\gamma-2t_s
\end{equation}

Altogether (\ref{min_right}) and (\ref{min_left}) imply that the time
interval accessible to the second piercing is given by
\begin{equation}
\label{reduced_T}
(t_1+T_\gamma-2t_s)-(t_1+2t_u)=T_\gamma-2(t_s+t_u)=T_\gamma-2t_{\rm enc};
\end{equation}
i.e., it is reduced by the overall duration of both encounter
stretches, $2t_{\rm enc}$.

\section{Statistics of encounters}

\label{sec:sr_statistics}

To determine the contribution of Sieber/Richter pairs to the spectral
form factor, we need to count orbit pairs and thus encounters.  More
precisely, we have to determine the average number $P_T(\Delta S) d\Delta S$
of encounters inside orbits $\gamma$ with period $T$ leading to
Sieber/Richter partners $\gamma'$ with action difference $S_\gamma-S_{\gamma'}$
inside $(\Delta S,\Delta S+d\Delta S)$.  Given that the contribution of each orbit
pair is proportional to $|A_\gamma|^2$, $P_T(\Delta S)$ has to be understood
as averaged over the ensemble of all $\gamma$ with periods inside a time
window around $T$, weighted with $|A_\gamma|^2$.

To evaluate $P_T(\Delta S)$, we introduce an auxiliary density $w_T(s,u)$
of stable and unstable separations inside encounters.  Recall that
these separations depend on the location of our Poincar{\'e} section. The
density $w_T(s,u)$ will be normalized to guarantee that after
integration over $s$ and $u$ each encounter is counted exactly once.
Hence, we have to demand that
\begin{equation}
\label{normalization}
P_T(\Delta S)=\int_{-c}^c ds du\,w_T(s,u)\delta(\Delta S-su).
\end{equation}

To establish $w_T(s,u)$, we need only two ingredients: {\it
  ergodicity} and the necessity of having encounter stretches
separated by {\it non-vanishing loops}.  Note that even though
individual periodic orbits need not be ergodic, the equidistribution
theorem (Subsection \ref{sec:ergodicity}) allows to invoke ergodicity
when averaging over the ensembles of orbits introduced above.

Let us first fix one location of the section ${\cal P}$, at a point
$\x_1$ (passed at time $t_1$) somewhere along the orbit. We now have
to count all further piercings through this section in phase-space
points $\x_2$ (passed at times $t_2$) almost time-reversed with
respect to $\x_1$. Each of these piercing points will correspond to a
different encounter, with $\x_1$ forming part of one stretch, and
$\x_2$ belonging to the other one. Due to {\it ergodicity}, the
expected number of piercings at times inside $(t_2,t_2+dt_2)$ with
stable and unstable components of $\T\x_2-\x_1$ in intervals
$(s,s+ds)$ and $(u,u+du)$ is given by the Liouville measure, expressed
in terms of times and stable and unstable coordinates (see Subsection
\ref{sec:ergodicity})
\begin{equation}
\label{pierce}
\frac{1}{\Omega}dsdudt_2,
\end{equation}
$\Omega$ denoting the volume of the energy shell.

The total number of piercing points inside intervals $ds$ and $du$ on
our Poincar{\'e} section is obtained by integration over $t_2$.  To
guarantee that the encounter stretches are {\it separated by
  intervening loops} we have to restrict ourselves to
$t_1+2t_u<t_2<t_1+T-2t_s$, i.e., an interval of width $T-2t_{\rm
  enc}$.  Since the integrand is independent of $t_2$, the resulting
{\it density of piercing through one section} is simply given by
\begin{equation}
\label{return}
\frac{1}{\Omega}(T-2t_{\rm enc}).
\end{equation}

We have to keep into account {\it all encounters} along the orbit in
question, and hence all possible Poincar{\'e} sections. Given that ${\cal
  P}$ is placed at a point $\x_1$, passed at time $t_1$, we thus have
to integrate over all possible piercing times $0<t_1<T$, leading to a
factor $T$.  Note that ${\cal P}$ may be moved freely throughout each
encounter without changing the partner orbit. Hence, when integrating
over $t_1$ each encounter is counted for a time $t_{\rm enc}$.  To
avoid weighting each encounter with its duration, we have to divide
out $t_{\rm enc}$.

Still, each encounter is counted twice, since any of the two encounter
stretches may be considered as ``the first''. Both choices give
separate contributions to the above integrals.  Dividing out 2, the
desired density $w_T(s,u)$ is finally obtained as\footnote{ When
  extending to higher orders, it will be convenient to divide out an
  analogous overcounting factor in the generalization of
  (\ref{normalization}) rather than in the generalization of
  (\ref{wT}); compare Eqs. (\ref{density}) and (\ref{numorb}).}
\begin{equation}
\label{wT}
w_T(s,u)=\frac{T(T-2t_{\rm enc}(s,u))}{2\Omega t_{\rm enc}(s,u)}.
\end{equation}

We need to discuss two small corrections to (\ref{wT}).  First, for
loops shorter than a classical relaxation time $t_{\rm cl}$ describing
the decay of correlations, the two piercings $\x_1$ and $\x_2$ will be
correlated, leading to corrections to the piercing probability
(\ref{pierce}) and thus to (\ref{return}) and (\ref{wT}).  However,
these corrections are negligible for $\hbar\to 0$, when $t_{\rm cl}$
vanishes compared to $T\sim T_H$ and $t_{\rm enc}\sim T_E$.  Second, for
inhomogeneously hyperbolic systems, the formula (\ref{sr_tenc}) used
for the encounter duration $t_{\rm enc}(s,u)$ is only an
approximation.  We shall see in Appendix \ref{sec:maths} that the
contribution to the form factor remains unaffected if we avoid that
approximation.

\section{Contribution to the spectral form factor}

\label{sec:contribution_sr}

We can now determine the contribution to the spectral form factor
arising from the Sieber/Richter family of orbit pairs.  We start from
the double sum over orbits in (\ref{doublesum}) and, as explained in
Subsection \ref{sec:stability_amplitude}, replace $T_{\gamma'}\to T_\gamma$ and
$A_{\gamma'}\to A_\gamma$.  Organizing the sum over partners $\gamma'$ of an orbit
$\gamma$ of period $T_\gamma=\tau T_H$ as an integral over the density of action
differences $P_{\tau T_H}(\Delta S)$, we obtain
\begin{equation}
K_{\rm
  SR}(\tau)=\mbox{Re}\frac{2}{T_H}\left\langle\sum_\gamma|A_\gamma|^2\delta(\tau T_H-T_\gamma)
  \int\Delta S\,P_{\tau T_H}(\Delta S)\e^{{\rm i}\Delta S/\hbar}\right\rangle\,,
\end{equation}
where we multiplied with 2 since each encounter of $\gamma$ gives rise to
two mutually time-reversed partner orbits.

Evaluating the sum over $\gamma$ using the sum rule of Hannay and Ozorio
de Almeida (\ref{hoda}), and expressing $P_{\tau T_H}(\Delta S)$ via $w_{\tau
  T_H}(s,u)$ as in (\ref{normalization}), we are led to the following
integral over stable and unstable coordinates
\begin{equation}
\label{sr_int}
K_{\rm SR}(\tau)=2\tau\left\langle\int_{-c}^c ds du\,w_{\tau
    T_H}(s,u)\e^{{\rm i}su/\hbar}\right\rangle.
\end{equation}

The periods of the relevant orbits are of the order Heisenberg time
$T_H\sim\hbar^{-1}$, whereas the durations of the encounters are of the
order Ehrenfest time $T_E\sim\ln\frac{{\rm const.}}{\hbar}$.  The density
$w_{\tau T_H}(s,u)$ can thus be split into a leading term $\frac{\tau^2
  T_H^2}{2\Omega t_{\rm enc}(s,u)}$ of order $\frac{T_H^2}{T_E}$ and a
correction $-\frac{\tau T_H}{\Omega}$ of order $T_H$, both giving separate
contributions to the integral (\ref{sr_int}).

The contribution of the {\it leading term} is proportional to
\begin{equation}
\label{osc_integral_sr}
\left\langle\int_{-c}^cdsdu\frac{1}{t_{\rm enc}(s,u)}{\rm e}^{{\rm i} s u/\hbar}\right\rangle\,.
\end{equation}
We will see that (\ref{osc_integral_sr}) effectively vanishes, since
the integral over $s$ and $u$ oscillates rapidly in the semiclassical
limit and is therefore annihilated by averaging.  To show this, we
shall restrict ourselves to homogeneously hyperbolic systems, with
stretching factors $\Lambda(\x,t)={\rm e}^{\lambda t}$ for all $\x$ and $t$ (see
Subsection \ref{sec:hyperbolicity}); general hyperbolic dynamics will
be discussed in Appendix \ref{sec:maths}. We now split the integral in
two parts $I^+$ and $I^-$ corresponding to positive and negative
values of $u$.  Both $I^+$ and $I^-$ are evaluated by transforming to
new integration variables: (i) the duration of the encounter head
$t_u=\frac{1}{\lambda}\ln\frac{c}{|u|}$, and (ii) the stable coordinate in
the end of the encounter $s^{\rm e}= s{\rm e}^{-\lambda t_u}$. Using that
the unstable coordinate in the end of the encounter is given by $\pm c$,
we can write $\Delta S=\pm s^{\rm e}c$ and $t_{\rm
  enc}=\frac{1}{\lambda}\ln\frac{c^2}{|\Delta S|}
=\frac{1}{\lambda}\ln\frac{c}{|s^{\rm e}|}$.  The Jacobian of the above
transformation is given by $\lambda c$, and the resulting variables are
restricted to the ranges $-c<s^{\rm e}<c$ and $0<t_u<t_{\rm
  enc}(s^{\rm e})$.  We thus obtain
\begin{equation}
I^\pm=\lambda c\int_{-c}^cds^{\rm e}{\rm e}^{\pm{\rm i} s^{\rm e}c/\hbar}
\frac{1}{t_{\rm enc}(s^{\rm e})}\int_0^{t_{\rm enc}(s^{\rm e})}dt_u
=2\lambda \hbar\sin\frac{c^2}{\hbar}\,;
\end{equation}
note that the two occurrences of $t_{\rm enc}(s^{\rm e})$ mutually
cancel.  In the semiclassical limit, $\sin\frac{c^2}{\hbar}$ oscillates
rapidly as $\hbar$ or $c$ are varied. Both integrals therefore vanish
after averaging over either of these quantities.\footnote{Note that an
  average over $c$ is equivalent to the average over $E$ implied by
  $\langle\ldots\rangle$, because $c$ has to be regarded as a function of $E$: An
  increase of the energy leads to an increase of the momentum and thus
  of all symplectic products $\vect{\delta q_1}{\delta p_1}\land\vect{\delta q_2}{\delta
    p_2} = \delta q_1\delta p_2-\delta q_2 \delta p_1$.  Since we maintain the
  normalization ${\bf e}^u({\bf x})\land{\bf e}^s({\bf x})=1$ of the
  basis vectors of our Poincar{\'e} sections ${\cal P}$, all stable and
  unstable coordinates are increased.  For billiards, the increase in
  energy does not affect the shape of the trajectory, or the
  applicability of the linear approximation (\ref{linearize}).  Hence,
  the bound $c$ needed for mutual linearization is increased as well.
  For general systems, the relation between $c$ and $E$ is more
  involved, but $c$ remains a function of $E$.}

Consequently, the {\it correction term} -- stemming from our condition
of having encounter stretches separated by loops -- becomes important.
Since the latter correction is independent of $s$ and $u$, the
resulting contribution to the form factor is easily evaluated as
\begin{equation}
 K_{\rm
  SR}(\tau)=-2\tau^2\frac{T_H}{\Omega}\int_{-c}^c ds du\,\e^{{\rm i}
  su/\hbar}\xrightarrow[\hbar\to 0]{}-2\tau^2,
\end{equation}
where we met with the simple integral
\begin{equation}
\label{simple_integral}
\int_{-c}^c ds du\,\e^{{\rm i} su/\hbar}\xrightarrow[\hbar\to 0]{} 2\pi\hbar,
\end{equation}
and used that $T_H=\frac{\Omega}{2\pi\hbar}$. Indeed, $K_{\rm SR}(\tau)$
reproduces the next-to-leading contribution to the form factor for the
Gaussian Orthogonal Ensemble.  We have thus verified that {\it
  individual} fully chaotic dynamics are faithful to random-matrix
theory, at least up to quadratic order in $\tau$.

In the following Chapters, this result will be extended to arbitrary
order in $\tau$. Interestingly, the universal contributions to the form
factor will always originate from statistical corrections reflecting
the necessity of non-vanishing loops.

\section{Summary}

Following Sieber and Richter, we showed that the partner orbits $\gamma$,
$\gamma'$ responsible for the next-to-leading order of the spectral form
factor differ noticeably only by their connections inside an
``encounter'' of two almost time-reversed orbit stretches.  These
encounter stretches need to be separated by non-vanishing loops.  If
we place a Poincar{\'e} section inside the encounter, each stretch will
pierce through that section once. The separation between the two
piercings can be decomposed into stable and unstable components,
determining both the duration of the encounter (of order Ehrenfest
time) and the action difference between $\gamma$ and $\gamma'$.  Ergodicity
allows to determine the number of encounters in long orbits and to
evaluate their contribution to the form factor.

%% file: kapitel4.tex
\chapter{Orbit pairs responsible for $\tau^3$ and beyond}

\label{sec:tau3}

\stand

We set out to identify the families of orbit pairs responsible for all
orders of the $\tau$ expansion of $K(\tau)$. The key point is that long
orbits have a huge number of close self-encounters which may involve
arbitrarily many orbit stretches.  We thus have to consider pairs of
orbits differing in any number of encounters, with any number of
stretches.  Generalizing the language of the previous chapter, we
speak of an $l$-encounter whenever $l$ stretches of an orbit get close
in phase space (up to time-reversal).  The partner orbits are
distinguished only by their reconnections inside such encounters.  In
contrast, the orbit loops in between encounters are almost identical
or mutually time-reversed.  Like in case of 2-encounters, the relevant
encounters will turn out to have durations of the order of the
Ehrenfest time, while the orbit periods are of the order of the
Heisenberg time.

We will start with the example of orbit pairs differing either in two
2-encounters or in one encounter involving 3 orbit stretches; these
pairs, analogous to field theoretical diagrams discussed in
\cite{Disorder1,Disorder2} and orbit pairs in quantum graphs studied
in \cite{Berkolaiko}, give rise to the cubic term in $K(\tau)$.  We
shall then generalize to all other pairs in Subsection
\ref{sec:overview}.

\section{Pairs of 2-encounters}

\label{sec:22}

Let us first consider orbit pairs differing in a pair of 2-encounters.
The two stretches of each encounter may be either close in phase space
(depicted by nearly parallel arrows $\psr$ or $\doublearrow[3]$), or
almost mutually time-reversed (like in $\asr$ or $\doublearrow[2]$).
We already met with antiparallel 2-encounters when deriving the $\tau^2$
contribution to the form factor. We have seen that such encounters can
only exist in time-reversal invariant systems, since we have to
require that the time-reversed of a classical encounter stretch, or an
orbit loop, again solves the Hamiltonian equations of motion.

\begin{figure}
\begin{center}
  \includegraphics[scale=0.35]{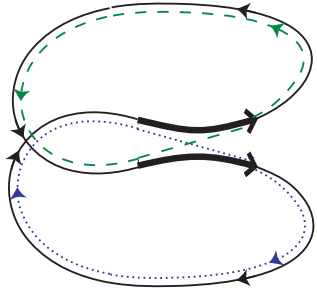} \parbox{1cm}{\ \ 
    \ \ \ } \parbox{7cm}{ \vspace{-4cm}
\caption{Reconnections inside a parallel 2-encounter yield
  a pseudo-orbit decomposing into two separate periodic orbits.}
\label{fig:2decompose}
}
\end{center}
\end{figure}

In contrast, parallel 2-encounters $\psr$ do not require time-reversal
invariance.  However, reconnections inside one single such encounter
will never lead to a partner orbit. As shown in Fig.
\ref{fig:2decompose}, we rather obtain two disjoint periodic orbits
(depicted by dashed and dotted lines, respectively).  Thus, the
partner can be seen as a ``pseudo-orbit'' decomposing into two
periodic orbits. Given that such pseudo-orbits are not admitted in the
Gutzwiller trace formula, they do not contribute to the spectral form
factor. However, we will see that reconnections inside parallel
2-encounters may well yield a periodic orbit if combined with
reconnections inside further encounters.

\begin{figure}
\begin{center}
  \includegraphics[scale=0.23]{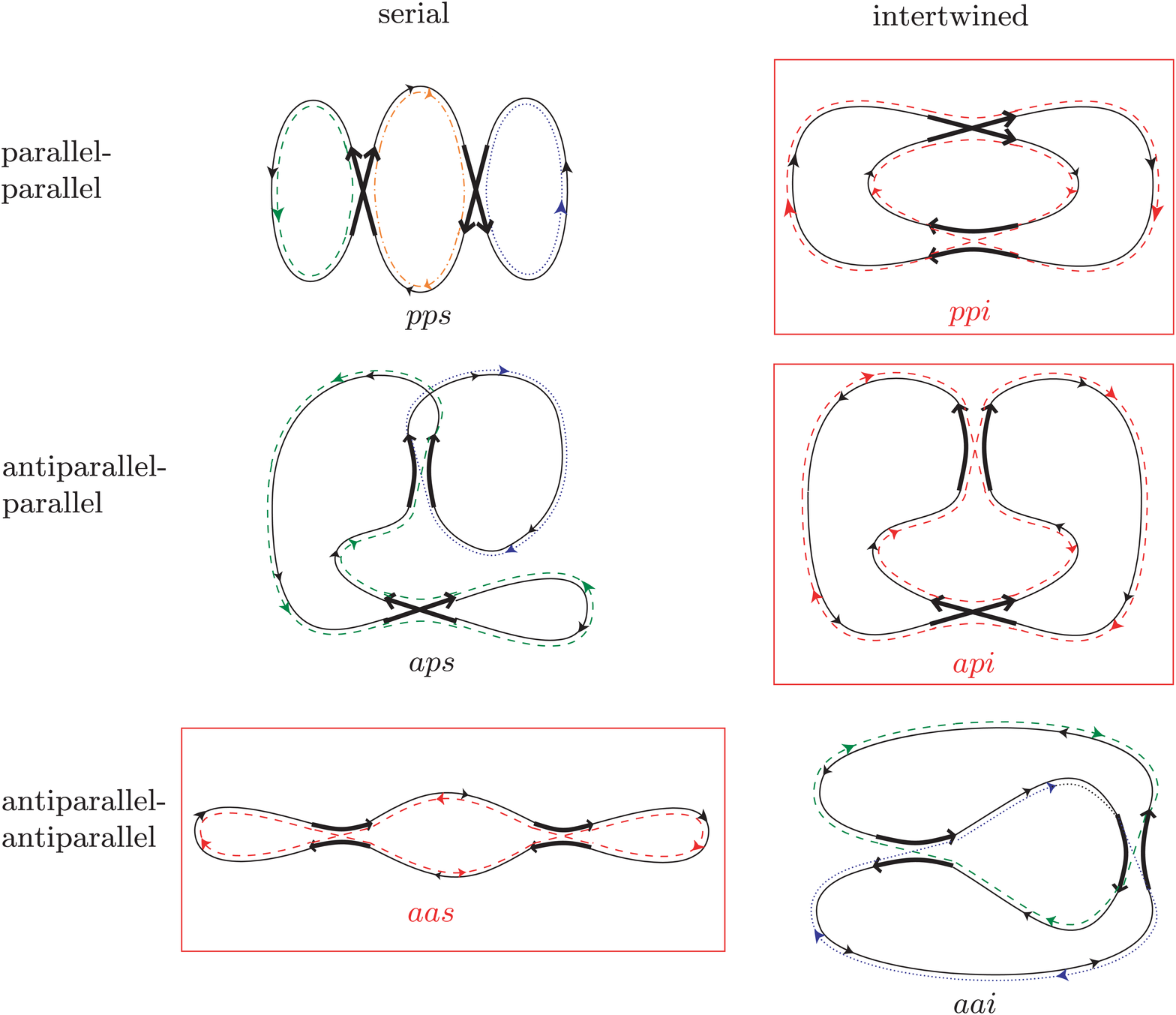}
\end{center}
\caption{Six classes of pairs of 2-encounters, distinguished by parallel vs. antiparallel
  orientation of stretches, and serial vs. intertwined ordering.
  Reconnections inside the encounters give rise to a connected partner
  orbit (dashed line) only for the three cases highlighted by boxes.
  Otherwise the partner decomposes into several orbits, respectively
  depicted by dashed, dotted, and dash-dotted lines.}\label{fig:2-2}
\end{figure}

Considering orbit pairs that differ in two 2-encounters, we may thus
allow for (i) both encounters being parallel, (ii) one being parallel
and one antiparallel, and (iii) two antiparallel 2-encounters.
Moreover, the encounters may be ordered in two different ways. In a
``serial'' ordering, the two stretches of each encounter follow each
other immediately after an intervening loop.  In an ``intertwined''
ordering, stretches of both encounters are traversed in alternation.
Altogether, this leaves six different ways of drawing pairs of
2-encounters, as shown in Fig. \ref{fig:2-2}.

Reconnections inside these encounters lead either to a connected
periodic orbit, or to a pseudo-orbit decomposing into several periodic
orbits (depicted by dashed, dotted, and dash-dotted lines in Fig.
\ref{fig:2-2}).  For example, two parallel encounters give rise to a
connected partner only if their ordering is intertwined. If they are
ordered in series, we obtain a pseudo-orbit decomposing into three
disjoint orbits; this pseudo-orbit does not contribute to the form
factor.  One easily sees that among the six cases depicted in Fig.
\ref{fig:2-2}, only three (marked by boxes) lead to a connected
partner orbit, namely
\begin{itemize}
\item two intertwined parallel encounters (abbreviated by {\it ppi}
  for {\it p}arallel-{\it p}arallel {\it i}ntertwined),
\item pairs of one {\it a}ntiparallel and one {\it p}arallel encounter
  with {\it i}ntertwined ordering ({\it api}), and
\item two {\it a}ntiparallel encounters in {\it s}eries ({\it aas}).
\end{itemize}
Time-reversal invariant systems allow for all three kinds of encounter
pairs. Apart from the partner orbits drawn in Fig.  \ref{fig:2-2}, we
then also need to account their time-reversed versions.  In contrast,
for systems without time-reversal invariance only parallel encounters,
and thus case {\it ppi}, need to be considered, giving rise only to
the one partner orbit depicted in \ref{fig:2-2}.

Incidentally, the partner orbit will always have a pair of encounters
of exactly the same class as the original orbit, i.e., {\it ppi}, {\it
  api}, and {\it aas}, respectively. However, for {\it api}
reconnections turn the antiparallel encounter into a parallel one and
vice versa.

To systematically classify orbit pairs, we have to number all (at
present four) encounter stretches in order of traversal by the
original orbit $\gamma$, starting with one arbitrary stretch.  The numbers
$1,2,3,4$ are then divided into groups, one corresponding to each
encounter, and we have to fix the mutual orientation of stretches
inside each group.  We are interested only in those divisions which
give rise to a connected partner orbit; they will be referred to as
``structures'' of orbit pairs.

For example, in the case {\it ppi}, we will have stretches 1 and 3
forming a parallel encounter, and stretches 2 and 4 forming one
further such encounter. This statement is true regardless of which
stretch is chosen as the first. For instance, if we rename stretch 1
as stretch 2, we have to cyclically permute the four numbers $1\to
2\to3\to4\to1$. Afterwards, we will still find one encounter of parallel
stretches 1 and 3 (the former stretches 2 and 4), and one parallel
encounter of stretches 2 and 4 (the former stretches 1 and 3).  In
this sense, all four stretches are indistinguishable.  We have thus
shown that {\it ppi} corresponds to one structure as defined above.

The situation becomes more complex for {\it api}. If one of the two
parallel stretches is chosen as the first, we obtain a parallel
encounter of stretches 1 and 3, and an antiparallel encounter of 2 and
4.  However, taking one stretch of the antiparallel encounter as a
reference, we obtain a parallel encounter of 2 and 4, and an
antiparallel encounter of 1 and 3.  Therefore, {\it api} encounters
have two different structures.  Both structures refer to the {\it same
  orbit pairs}, but differ in which of the stretches is taken as the
first. The two structures may also be understood as follows: If the
orbits $\gamma$ and $\gamma'$ are cut open inside the loop preceding the
``first'' stretch, each structure gives rise to topologically
different trajectory pairs. For {\it api}, the trajectories associated
to one structure will first traverse a stretch of the parallel
encounter, whereas trajectories of the second structure first traverse
a stretch of the antiparallel encounter. Intuitively, the appearance
of two different structures implies that {\it api} encounters are
``less symmetric'' than {\it ppi} encounters, since one encounter is
parallel, whereas the other one involves two antiparallel stretches.

We shall see that if a family of orbit pairs is associated with
several equivalent structures, all yield separate (and equal)
contributions to the form factor.\footnote{ If we count families of
  orbit pairs (like {\it ppi}, {\it api}, {\it aas}, $\ldots$) rather than
  structures, the contribution of each family depends on its
  symmetries.  In \cite{Tau3}, we thus had to introduce multiplicity
  factors ${\cal N}_{ppi}, {\cal N}_{api},{\cal N}_{aas}, \ldots$, which
  are not needed when summing over structures.}

Finally, for {\it aas} two among the four stretches are followed by
stretches of the same encounter. If one of these two is chosen as a
reference, we obtain two antiparallel encounters of stretches 1 and 2,
and stretches 3 and 4, respectively.  In contrast, if one of the other
two stretches is considered the first, there are antiparallel
encounters of stretches 1 and 4, and 2 and 3.  Again, we end up with
two equivalent structures.

Altogether, given time-reversal invariance, pairs of 2-encounters can
have 5 different structures, one single structure related to {\it
  ppi}, and two structures each for {\it api} and for {\it aas}.  In
systems without time-reversal invariance, only parallel encounters are
possible. Hence, only one among the above structures remains, the one
related to (intertwined) pairs of parallel encounters.

\section{Triple encounters}

\label{sec:3}

We now turn to pairs of orbits differing in one 3-encounter.  Here, in
both partner orbits, three stretches come close up to time reversal.
These stretches are separated by three intervening loops, which can be
arranged to form a ``cloverleaf'' (see Fig. \ref{fig:3}).

We have to distinguish between two special cases. First, all three
stretches may be close in phase space $\pc$. In this case, we speak of
a ``parallel cloverleaf'' ({\it pc}). Obviously, {\it pc} encounters
may exist even in absence of time-reversal invariance.

Partner orbits are again obtained by reshuffling intra-encounter
connections.  In principle, the ports can be reconnected in five
different ways, depicted by $\triplearrow[4]$\,, $\triplearrow[3]$\,,
$\triplearrow[5]$\,, $\triplearrow[2]$\,, and $\triplearrow[1]$\,.
However, in the first three cases only two stretches take part in the
reconnection whereas the third one remains unaffected. A potential
partner orbit could thus also be obtained by reconnections inside a
2-encounter.  This implies that the possibilities $\triplearrow[4]$\,,
$\triplearrow[3]$\,, and $\triplearrow[5]$ have to be disregarded; the
corresponding orbit pairs were already taken into account when summing
over Sieber/Richter pairs in Chapter \ref{sec:tau2}.  In similar vein,
we will always require reconnections to {\it involve all stretches of
  an encounter}.  Among the remaining two options, $\triplearrow[1]$
leads to a ``partner'' decomposing into three disjoint periodic
orbits; see Fig. \ref{fig:3}.  Only reconnections according to by
$\triplearrow[2]$\, give rise to a connected partner orbit, shown
in the same picture.\footnote{ In Fig. \ref{fig:3}, the orbit $\gamma$
  first goes through the upper stretch of $\pc$, then through the
  middle and lower stretches, and finally returns to the upper
  stretch.  If the stretches are traversed in different order, the
  reconnections leading to a connected partner would look like
  $\triplearrow[1]$ rather than $\triplearrow[2]$.}

\begin{figure}
\begin{center}
  \includegraphics[scale=0.22]{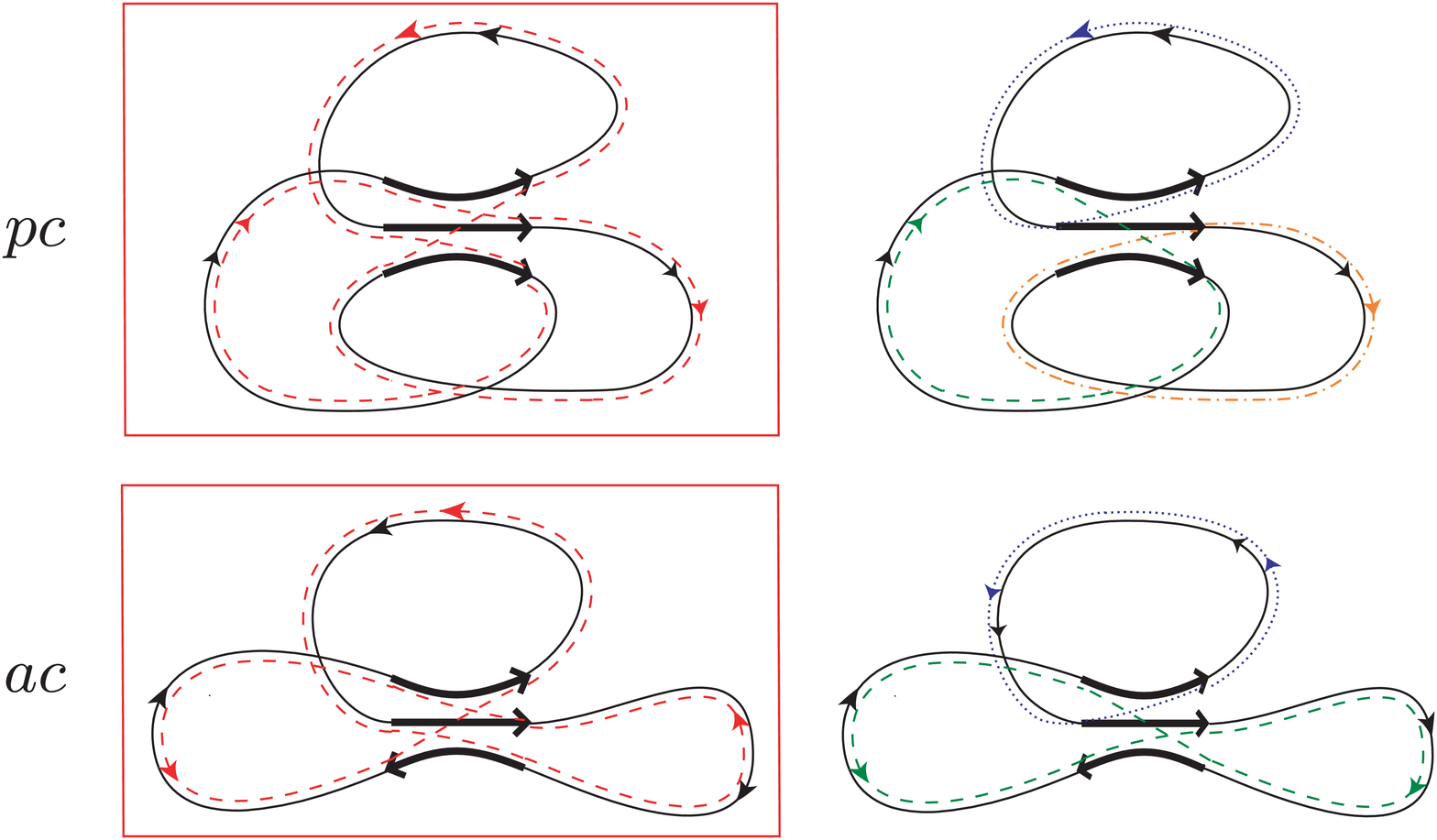}
\end{center}
\caption{
  Examples for 3-encounters of type {\it pc} and {\it ac}.  The
  reconnections on the left-hand side.  give rise to a connected
  partner orbit (dashed lines), whereas those on the right-hand side.
  yield a partner decomposing into two or three periodic orbits,
  respectively.  }\label{fig:3}
\end{figure}

One easily sees that {\it pc} encounters may have only one structure.
Again numbering all encounter stretches in order of traversal, we may
say that stretches 1,2, and 3 are close in phase space, irrespective
of which stretch is chosen as the first.

For time-reversal invariant dynamics, there is one further way of
forming 3-encounters. Instead of having three stretches close in phase
space, these encounters involve only two almost parallel stretches,
and a third one approximately time-reversed with respect to the
initial two. In this case, we speak of an ``{\it a}ntiparallel {\it
  c}loverleaf'' ({\it ac}), depicted by $\ac$.  Since {\it ac}
encounters exist only in time-reversal invariant systems, we may also
read {\it a} as ``{\it a}nti-unitary symmetry required''.

Again, we
obtain a connected partner orbit by reshuffling intra-encounter
connections according to $\tripline[2]$\,; compare Fig. \ref{fig:3}.
Here, we momentarily dropped the arrows; the encounter stretches of
the partner orbit may either be close or time-reversed with respect to
those of the original orbit.  By reasoning analogous to {\it pc}
encounters, we see that the reconnections $\tripline[4]$\,,
$\tripline[3]$\,, and $\tripline[5]$ could equivalently be described
in terms of a 2-encounter, and that $\tripline[1]$ would lead to a
pseudo-orbit, this time containing two separate periodic
orbits.\footnote{If the three stretches $\ac$ are traversed in an
  order different from Fig. \ref{fig:3}, not $\tripline[2]$, but
  $\tripline[1]$ leads to a connected partner.}  Again, the
time-reversed versions of the partners depicted in Fig. \ref{fig:3}
also form valid partner orbits.

Orbit pairs of type {\it ac} can be described by three equivalent
structures.  To show this, we again number the stretches in order of
traversal by $\gamma$, starting from an arbitrary one. If that reference
stretch is the one approximately time-reversed with respect to the
others, we have stretch 1 antiparallel to 2 and 3, which are mutually
close. Choosing a different reference, we are led to two further
structures, with 2 antiparallel to 1 and 3, and 3 antiparallel to 1
and 2.  Again, if the orbits $\gamma$ and $\gamma'$ are cut open in the loop
preceding the first stretch, each structure gives rise to
topologically different trajectory pairs.
%Note that if we had to consider
%more than one reconnection, we would have to count structures separately for each
%one of them, and thus obtain even more structures.

To summarize, systems without time-reversal invariance allow for one
``cloverleaf'' structure, related to {\it pc}, whereas in presence of
time-reversal invariance we observe 4 such structures, including 3
stemming from {\it ac} encounters.

\section{Symbolic dynamics}

\label{sec:tau3symbols}

Given symbolic dynamics, the orbit pairs introduced in the preceding
Sections can be characterized through their symbols sequences.
Within each encounter, stretches are either very close in phase space
and thus labelled by the same symbol sequence $\ES_{\al}$, or almost
mutually time-reversed and thus denoted by mutually time-reversed
sequences $\ES_{\al}$ and $\bar{\ES}_{\al}$.  Each of the intervening
loops comes with its own symbol sequence, to be denoted by
$\AS,\BS,\ldots$.  The symbol strings of the orbits $\gamma$ and $\gamma'$ are
built as alternating series of encounter and loop sequences.  If a
given loop $\AS$ of $\gamma$ connects, say, stretches with symbol
sequences $\ES_1$ and $\ES_2$, the corresponding loop of $\gamma'$ must
either connect stretches with the same sequences, as in
$\ldots\ES_{1}\AS\ES_2\ldots$, or appear in time-reversed form, like
$\ldots\bar{\ES}_2\bar{\AS}\bar{\ES}_1\ldots$.

For instance, two intertwined parallel 2-encounters $\ES_1\ldots\ES_1$ and
$\ES_2\ldots\ES_2$ entail the symbol sequence
$\gamma=\ES_1\AS\ES_2\BS\ES_1\CS\ES_2\DS$.  Reconnections yield a partner
with the symbol sequence $\gamma'=\ES_1\AS\ES_2\DS\ES_1\CS\ES_2\BS$.
Likewise, intertwined parallel and antiparallel 2-encounters
$\ES_1\ldots\ES_1$ and $\ES_2\ldots\bar{\ES}_2$ lead to
$\gamma=\ES_1\AS\ES_2\BS\ES_1\CS\bar{\ES}_2\DS$ and a partner orbit
$\gamma'=\ES_1\AS\ES_2\bar{\CS} \bar{\ES}_1\bar{\DS}\ES_2\BS$ with loops
$\CS$ and $\DS$ reversed in time.
Two serial antiparallel encounters $\ES_1\ldots\bar{\ES}_1$ and
$\ES_2\ldots\bar{\ES}_2$ can be combined into the sequences
$\gamma=\ES_1\AS\bar{\ES}_1\BS\ES_2\CS\bar{\ES}_2\DS$ and
$\gamma'=\ES_1\AS\bar{\ES}_1\bar{\DS}\ES_2\CS\bar{\ES}_2 \bar{\BS}$.

Parallel 3-encounters $\ES\ldots\ES\ldots\ES$ lead to orbit pairs
$\gamma=\ES\AS\ES\BS\ES\CS$, $\gamma'=\ES\AS\ES\CS\ES\BS$, whereas
$\ES\ldots\ES\ldots\bar{\ES}$ denotes an encounter with the third stretch
antiparallel to the first two. In the latter case, orbit pairs have
symbol sequences of the form $\gamma=\ES\AS\ES\BS\bar{\ES}\CS$,
$\gamma'=\ES\AS\ES\bar{\BS}\bar{\ES}\bar{\CS}$.

\section{Orbit pairs responsible for all orders in $\tau$}

\label{sec:overview}

\begin{figure}
\begin{center}
  \includegraphics[scale=0.26]{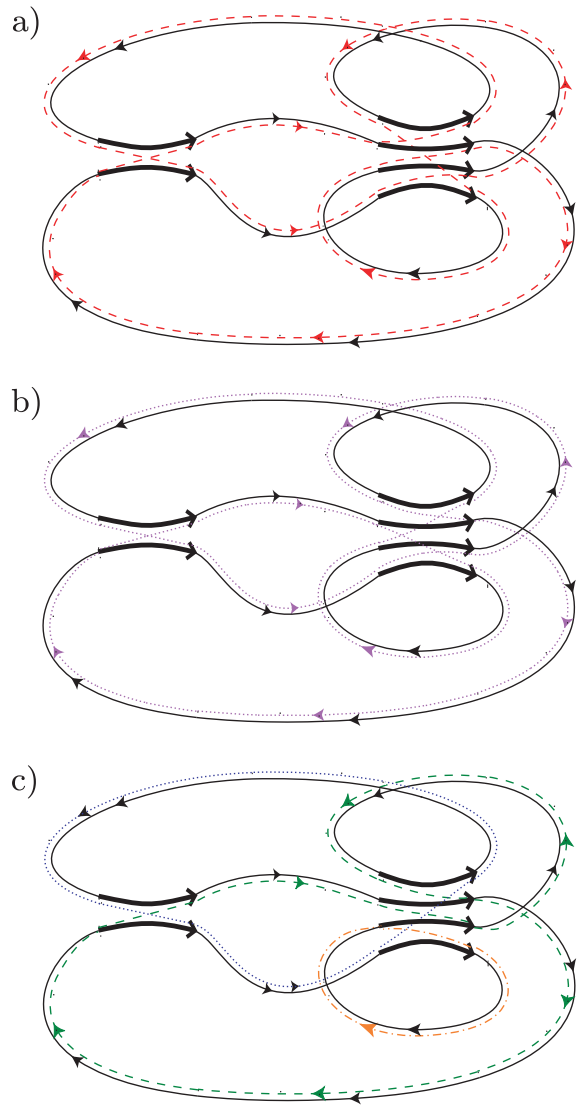}
\end{center}
\caption{a) Solid line: Periodic orbit $\gamma$ with one 4-encounter
  and one 2-encounter highlighted by bold arrows. Dashed line: Partner
  $\gamma'$ differing from $\gamma$ by connections in these encounters.  b)
  Dotted line: One further orbit, differing from $\gamma$ in the same set
  of encounters.  c) Other reconnections yield a pseudo-orbit
  decomposing into three periodic orbits (dashed, dotted, and
  dash-dotted).}
\label{fig:24big}
\end{figure}

To obtain all orders of the power series of $K(\tau)$, we have to
account for orbit pairs differing in any number of $l$-encounters,
with arbitrary $l\geq 2$.

Let us first sketch two more examples.  Fig. \ref{fig:24big}a
highlights two encounters inside a periodic orbit, one 2-encounter and
one 4-encounter.  The stretches involved can be reconnected in several
different ways. Some of these reconnections, like in Fig.
\ref{fig:24big}a or b, yield connected partner orbits.  Other
reconnections, as in Fig. \ref{fig:24big}c, give rise to a
pseudo-orbit decomposing into a number of disjoint periodic orbits,
and thus do not contribute to the form factor.  Since both encounters
involve only parallel stretches, orbit pairs as in Fig.
\ref{fig:24big}a and b exist both in systems with and without
time-reversal invariance.

\begin{figure}
\begin{center}
  \includegraphics[scale=0.2]{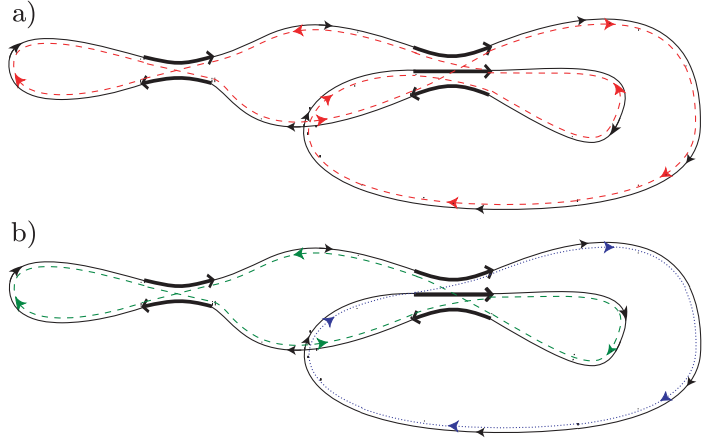}
\end{center}
\caption{
  a) Periodic orbit $\gamma$ with one 2- and one 3-encounter highlighted,
  and a partner $\gamma'$ obtained by reconnection; the encounters
  depicted exist only in time-reversal invariant systems. b) Again,
  different reconnections lead to a ''pseudo-orbit'', this time
  decomposing into two separate periodic orbits (dashed and dotted).}
\label{fig:23big}
\end{figure}

For time-reversal invariant dynamics, we also must allow for
encounters whose stretches get close only {\it up to time reversal}.
Correspondingly, loops inside mutual partner orbits may be related by
time reversal.  See Fig. \ref{fig:23big}a for an example of two orbits
differing in encounters $\asr$ and $\ac$. The pseudo-orbit obtained in
Fig. \ref{fig:23big}b does not contribute to the form factor.

We thus obtain a natural extension of Berry's diagonal approximation.
Instead of considering only pairs of orbits which exactly coincide (or
are mutually time-reversed), we employ {\it all} pairs whose members
are composed of {\it similar} (up to time reversal) loops.

We have to classify these orbit pairs.  The two orbits of each pair
differ in a {\bf number $v_l$ of $l$-encounters}; we shall assemble
these numbers to a ``vector'' $\vec{v}=(v_2,v_3,\ldots)$.  The total
number of encounters is given by $V(\vec{v})=\sum_{l\geq 2}v_l$.  The
number of orbit stretches involved in encounters, coinciding with the
number of intervening loops, reads $L(\vec{v})=\sum_{l\geq 2}lv_l$.

As we have seen for orbit pairs differing in two 2-encounters or in
one 3-encounter, orbit pairs related to the same $\vec{v}$ may have
different {\bf structures}.  In general, structures are defined as
follows: Starting from an arbitrarily chosen reference stretch, the
$L(\vec{v})$ encounter stretches are numbered in order of traversal by
$\gamma$.  Each structure now corresponds to one way of
\begin{enumerate}
\item[(i)] assembling the $L(\vec{v})$ labels into groups, each
  corresponding to one encounter,
\item[(ii)] fixing the mutual orientation of stretches in each group,
  like $\psr$ vs. $\asr$, or $\pc$ vs. $\ac$ (if the system is
  time-reversal invariant), and
\item[(iii)] reconnecting the stretches to form a non-decomposing
  partner orbit.
\end{enumerate}
The third point is particularly important if several reconnections in
the {\it same} set of encounters lead to different connected partner
orbits, as in Figs. \ref{fig:24big}a and b.  In this case, each such
reconnection gives rise to an orbit pair of a different structure.  We
did not meet this situation in the examples of Subsections
\ref{sec:22}-\ref{sec:tau3symbols}.  (Note that for time-reversal
invariant systems, orbit pairs $(\gamma,\gamma')$ and $(\gamma,\T\gamma')$
distinguished only by time-reversal of the partner orbit are
considered as belonging to the same structure.)

Again, some orbit pairs can be described in terms of several
equivalent structures, depending on which stretch is taken as the
first.  If we, however, imagine the orbits $\gamma,\gamma'$ cut open in the
loop preceding the first stretch, each structure gives rise to a
topologically different class of trajectory pairs.  Another
alternative definition of structures uses (numbered) {\it loops}
rather than encounter stretches: Each structure corresponds to a
different ordering -- and, given time-reversal invariance, different
sense of traversal -- of the loops of $\gamma$ inside the partner orbit
$\gamma'$.

To evaluate the spectral form factor, we need to determine the number
$N(\vec{v})$ of structures related to the same $\vec{v}$.

Apart from vectors $\vec{v}$ and structures, orbit pairs can be
characterized by the {\bf phase-space separations} between encounter
stretches.  In Chapter \ref{sec:tau2}, we introduced stable and
unstable coordinates $s,u$ measuring the separation between stretches
of a 2-encounter.  We have to define similar coordinates for arbitrary
$l$-encounters, and derive a density $w_T(s,u)$ of phase-space
separations analogous to (\ref{wT}). The double sum (\ref{doublesum})
over orbits defining the spectral form factor will then be written as
a sum of contributions from families of orbit pairs, with a weight
proportional to $N(\vec{v})w_T(s,u)$.

In the following Chapter, we will study the phase-space geometry of
encounters and determine the density $w_T(s,u)$.  The purely
combinatorial task of determining the number of structures
$N(\vec{v})$ for arbitrary $\vec{v}$ is attacked in Chapter
\ref{sec:combinatorics} with the help of permutation theory.  We will
then obtain expansions of $K(\tau)$ for individual chaotic systems, in
line with the respective predictions of the GUE and the GOE.

\section{Summary}

We define $l$-encounters as regions inside an orbit $\gamma$ where $l$
orbit stretches are mutually close up to time reversal. We consider
orbit pairs differing by reconnections inside an arbitrary number
$v_l$ of $l$-encounters; both partners must be connected periodic
orbits.  The orbit pairs are further classified by ``structures''
(i.e., ordering and mutual orientation of stretches, and
reconnections) and phase-space separations. The $\tau^3$-contribution
originates from orbit pairs differing in two 2-encounters (1 structure
for systems without ${\cal T}$ invariance, altogether 5 structures in
presence ${\cal T}$ invariance) or in one 3-encounter (1 structure
without ${\cal T}$ invariance, 4 structures with ${\cal T}$
invariance).

%% file: kapitel5.tex
\chapter{Phase-space geometry of encounters}

\stand

\label{sec:geometry}

In the present Chapter, we want to determine the contribution to the
form factor originating from all orbit pairs of a given structure. To
that end, we have to extend the results of Chapter \ref{sec:tau2}, and
describe the phase-space geometry of encounters with arbitrary number
of stretches $l$.

\section{Encounters}

\label{sec:encounter}

We first need to introduce suitable {\it stable and unstable
  coordinates}.  To do so, we again consider a Poincar{\'e} surface of
section ${\cal P}$ orthogonal to the orbit at an arbitrary phase-space
point ${\bf x}_{1}$ (passed at time $t_1$) inside one of the encounter
stretches.  The remaining stretches pierce through ${\cal P}$ at times
$t_j$ ($j=2,\ldots,l$) in points ${\bf x}_{j}$. If the $j$-th encounter
stretch is close to the first one in phase space, we must have ${\bf
  x}_{j}\approx{\bf x}_{1}$; if it is almost time-reversed with respect to
the first one, we have ${\cal T}{\bf x}_{j}\approx{\bf x}_{1}$.  In the
following, we shall use the shorthand ${\bf y}_{j}\approx{\bf x}_{1}$ with
${\bf y}_{j}$ either ${\bf x}_{j}$ or ${\cal T}{\bf x}_{j}$.

The small displacement ${\bf y}_{j}-{\bf x}_{1}$ can be decomposed
into components along the stable and unstable directions at $\x_1$,
\begin{equation}
\label{decompose}
{\bf y}_{j}-{\bf x}_{1}=\hat{s}_{j}{\bf e}^s({\bf x}_{1})
+\hat{u}_{j}{\bf e}^u({\bf x}_{1})\,.
\end{equation}
Thus, each encounter is described by $l-1$ pairs of stable coordinates
$\hat{s}_j$ and unstable coordinates $\hat{u}_j$. These coordinates
depend on the location of ${\cal P}$; as $t_1$ grows, stable
coordinates asymptotically decrease and unstable coordinates
asymptotically increase.  We will later subject $\hat{s}_j$,
$\hat{u}_j$ to a coordinate transformation yielding new coordinates to
be denoted by $s_j$, $u_j$; hence the ``hats'' in (\ref{decompose}).

By definition, an $l$-encounter lasts as long as {\it all} $l$
stretches are mutually close.  We thus have to demand the stable and
unstable differences $|\hat{s}_{j}|$, $|\hat{u}_{j}|$ of all stretches
from the first one to be smaller than our constant $c$.  As a
consequence, the mutual difference between two stretches different
from the first one is limited by $2c$.  ``Fringes'' where only a few
stretches remain close are not regarded as part of the encounter.

We can now generalize the formula (\ref{sr_tenc}) for the {\it
  duration} $t_{\rm enc}$ of an encounter.  Each of the unstable
coordinates $|\hat{u}_j|$ needs the time
$\frac{1}{\lambda}\ln\frac{c}{|\hat{u}_j|}$ to reach $c$. The encounter
ends as soon as the first of these components passes the threshold,
i.e., after a time
\begin{equation}
\label{tu}
t_u\sim\min_{j}\left\{\frac{1}{\lambda}\ln\frac{c}{|\hat{u}_{j}|}\right\}\,.
\end{equation}
Similarly, the time since the beginning of the encounter (i.e., since
the last of the stable coordinates has fallen below $c$) can be
estimated as
\begin{equation}
\label{ts}
t_s\sim\min_{j}\left\{\frac{1}{\lambda}\ln\frac{c}{|\hat{s}_{j}|}\right\}\,.
\end{equation}
As for 2-encounters, we will refer to $t_u$ and $t_s$ as the durations
of the ``head'' and the ``tail'' of the encounter. Both quantities sum
up to the overall duration of the encounter
\begin{equation}
\label{tenc}
t_{\rm enc}=t_s+t_u\sim\frac{1}{\lambda}\ln\frac{c^2}{\max_i\{|\hat{s}_i|\}
\max_j\{|\hat{u}_j|\}}\,;
\end{equation}
in view of Eq. (\ref{linearize}), $t_{\rm enc}$ remains invariant if
the Poincar{\'e} section ${\cal P}$ is shifted through the encounter.

An $l$-encounter involves $l$ different orbit stretches whose initial
and final phase-space points will be referred to as {\it ``entrance''}
and {\it ``exit ports''}.  If all encounter stretches are (almost)
parallel, as in $\pc$, all entrance ports are located on the same side
of the encounter, and the exit ports are located on the opposite side.
If the encounter involves mutually time-reversed orbit stretches like
$\ac$, this is no longer the case. Thus, it is useful to introduce the
following convention: All ports on the side where the (arbitrarily
chosen) first stretch begins are called ``left ports'', while those on
the opposite side are ``right ports''. For parallel encounters,
``entrance'' and ``left'' are synonymous, as well as ``exit'' and
``right''.

\section{Partner orbits}

\label{sec:partner_orbits}

\begin{figure}
\begin{center}
  \includegraphics[scale=0.37]{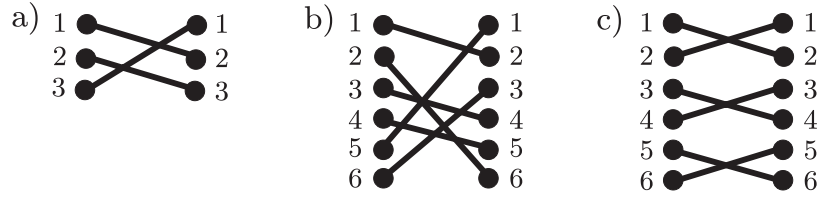}
\end{center}
\caption{Possible connections between left and right ports in partner orbits $\gamma'$. In c), the encounter splits into three pieces
  respectively containing the two upper, middle, and lower stretches.}
\label{fig:permutations}
\end{figure}

The partner orbits $(\gamma,\gamma')$ differ from one another only inside the
encounters, by their connections between left and right ports.  We
shall number these ports in order of traversal by $\gamma$, such that the
$j$-th encounter stretch of $\gamma$ connects left port $j$ to right port
$j$. Inside $\gamma'$, the left port $j$ is connected to a different right
port $k$; see Figs. \ref{fig:permutations}a and b for possible
reconnections in a 3- and a 6-encounter.

When reshuffling connections, we must mix {\it all} stretches of a
given encounter. In contrast, Fig. \ref{fig:permutations}c shows
reconnections only between stretches 1 and 2, 3 and 4, and stretches 5
and 6 of a 6-encounter which therefore decomposes into three separate
2-encounters.

\pagebreak

\subsection{Piercing points}

\begin{figure}
\begin{center}
  \includegraphics[scale=0.39]{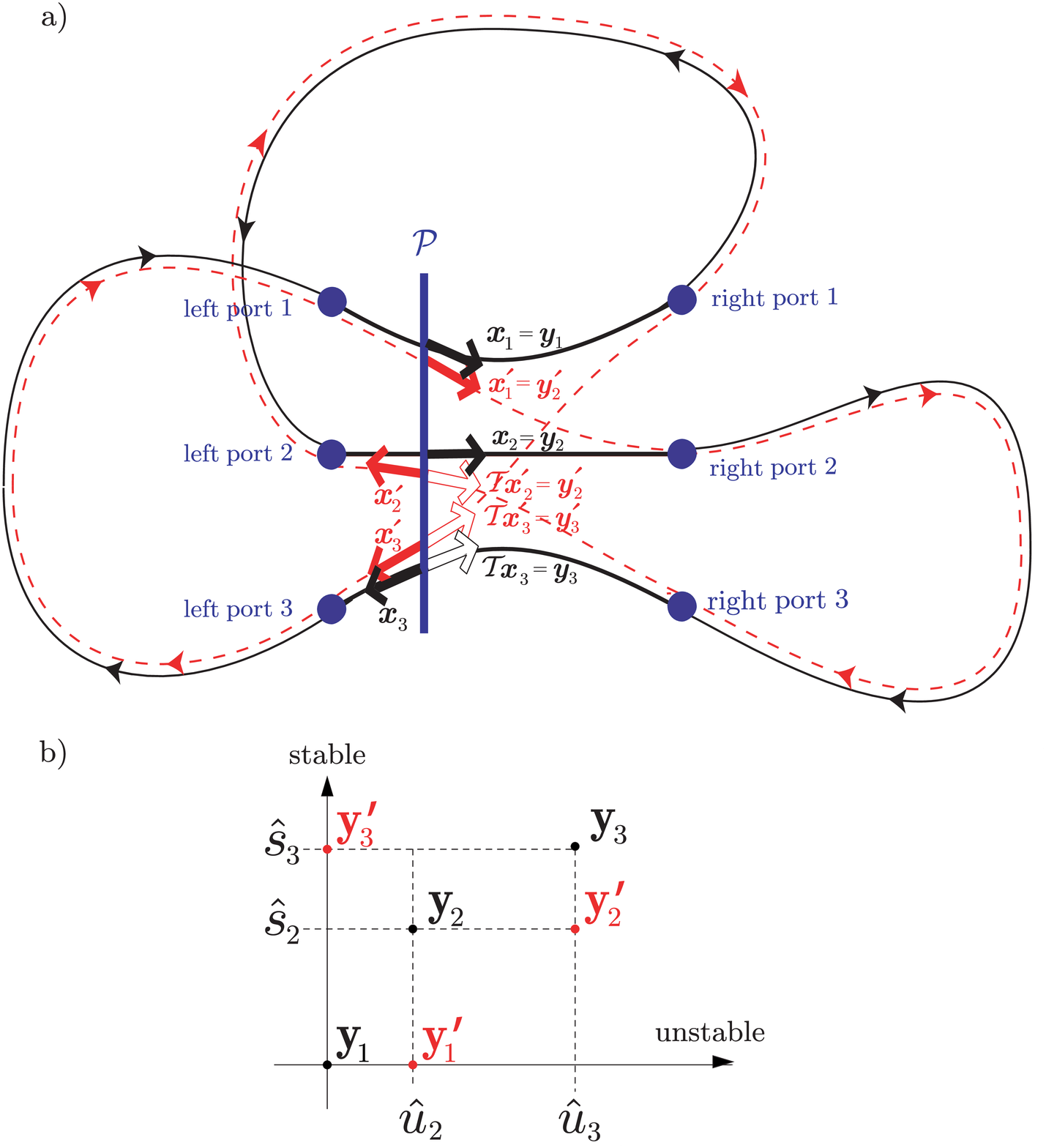}
\end{center}
\caption{Piercing points of orbits $\gamma$,$\gamma'$ differing in a 3-encounter;
  inside $\gamma'$, the left ports 1, 2, 3 are connected to the right
  ports 2, 3, 1, respectively. a) Sketch in configuration space. b)
  Stable and unstable coordinates in a Poincar{\'e} section ${\cal P}$.}
\label{fig:piercings3}
\end{figure}

\label{sec:manifolds}

%\enlargethispage{4mm}

The partner $\gamma'$ also pierces through our Poincar{\'e} section ${\cal
  P}$.  The piercing point of the stretch connecting left port $j$ to
right port $k$ will be denoted by $\x_j'$.  Similarly as for the
piercings of $\gamma$, the shorthand $\y_j'$ will represent either $\x_j'$
or $\T\x_j'$, depending on whether the stretch in question is parallel
or antiparallel with respect to the first one.  See Fig.
\ref{fig:piercings3}a for the example of a 3-encounter $\ac$.

Like in case of 2-encounters the piercing points of $\gamma'$ are
determined by those of $\gamma$.  In particular, the unstable coordinates
of a point $\y_j'$ always depend on the following right port. If two
stretches of $\gamma$ and $\gamma'$ lead to the same right port, they have to
approach each other at least for the duration of the encounter head;
for the relevant encounters this duration is of the order $T_E\to\infty$.
Since the phase-space difference shrinks in time, it may have only a
very small unstable component, meaning that the unstable coordinates
of $\y_j$ and $\y_j'$ practically coincide.  Likewise, the stable
coordinate of $\y_j'$ is determined by the previous left port, since
stretches with the same left port approach for large negative times.
If a stretch of $\gamma'$ connects left port $j$ to right port $k$, it
thus pierces through our Poincar{\'e} section with stable and unstable
coordinates approximately given by
\begin{equation}
\label{su_partner} \hat{s}_j'=\hat{s}_j,\ \ \ \
  \hat{u}_j'=\hat{u}_k\,;  \end{equation}
this generalizes the relation (\ref{sol}) for 2-encounters.

For the 3-encounter of Fig. \ref{fig:piercings3}a, the partner orbit
$\gamma'$ has left ports 1, 2, and 3 of $\gamma$ respectively connected to the
right ports 2, 3, and 1. The first stretch of $\gamma'$ connects left port
1 to right port 2 and pierces through ${\cal P}$ in ${\bf y}_1'$ with
$\hat{s}_1'=\hat{s}_1=0$, $\hat{u}_1'=\hat{u}_2$.  The second stretch
pierces through ${\cal P}$ in $\y_2'$ with $\hat{s}_2'=\hat{s}_2$,
$\hat{u}_2'=\hat{u}_3$, and the third one in $\y_3'$ with
$\hat{s}_3'=\hat{s}_3$, $\hat{u}_3'=\hat{u}_1=0$; see Fig. \ref{fig:piercings3}b.

\subsection{Action difference}

\label{sec:action_difference}

\begin{figure}
\begin{center}
  \includegraphics[scale=0.36]{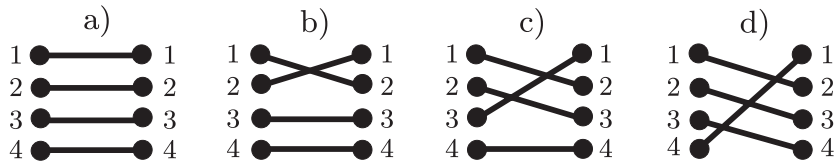}
\end{center}
\caption{Steps from connections in $\gamma$ (depicted in a) to those in $\gamma'$
  (shown in d), in each step interchanging right ports of two
  encounter stretches.}
\label{fig:steps}
\end{figure}

We have already seen that for $\gamma$, $\gamma'$ differing in one
2-encounter, the action difference is given by the product of the
stable and unstable separations between the two encounter stretches.
In order to generalize to $\gamma$, $\gamma'$ differing in an $l$-encounter,
we can imagine the partner orbit $\gamma'$ constructed out of $\gamma$ by
$l-1$ successive steps.  For a 3-encounter $\tripline[0]$, we may
first reconnect the two upper stretches as in $\tripline[4]$, and then
the stretches starting from the second and third left port as in
$\tripline[2]$.  One further example, involving a 4-encounter, is
given in Fig.~\ref{fig:steps}.  Each step interchanges the right ports
of two encounter stretches and contributes to the action difference an
amount given by the product of their stable and unstable separations.
At the same time, the two piercing points change their position as
discussed in Subsection \ref{sec:manifolds}.

This step-by-step process suggests a useful transformation of
coordinates.  Let $s_{j}$, $u_{j}$ denote the stable and unstable
differences between the two stretches affected by the $j$-th step.
Note that in contrast to $\hat{s}_{j}$, $\hat{u}_{j}$ the index $j$ no
longer represents encounter stretches $2,\ldots,l$ but steps $1,\ldots,(l-1)$.
Now, the change of action in each step is simply given by $
s_{j}u_{j}$.  Summing over all steps, we obtain a total action
difference
\begin{equation}
\label{DS}
\Delta S=\sum_{j=1}^{l-1}s_{j}u_{j}\,.
\end{equation}

The transformation leading from $\hat{s}_{j}$, $\hat{u}_{j}$ to
$s_{j}$, $u_{j}$ is linear and volume-preserving.  To explicitly
express $s_j$, $u_j$ as a function of $\hat{s}_j$ and $\hat{u}_j$, we
first consider reconnections as given above for $l=3$, or depicted in
Fig. \ref{fig:steps}d for $l=4$.  We proceed from Fig.
\ref{fig:steps}a to \ref{fig:steps}d in $l-1=3$ steps.  In the $j$-th
step, we change connections between left ports $j$ and $j+1$, and
right ports 1 and $j+1$.  Recall that stable and unstable coordinates
of piercing points are determined by the left and right ports,
respectively.  Thus, the separation between the stretches affected has
a stable component $s_j=\hat{s}_{j+1}-\hat{s}_j$ and an unstable
component $u_j=\hat{u}_{j+1}-\hat{u}_{1}=\hat{u}_{j+1}$.  The Jacobian
of the transformation $\hat{s},\hat{u}\to s,u$ is equal to 1.  All
other permissible reconnections can be brought to a form similar to
Fig.  \ref{fig:steps}d, albeit with different $l$, by appropriately
changing the numbering of stretches; hence they allow for the same
step-by-step procedure. (An interpretation of this procedure in the
context of permutations will be given in Subsection
\ref{sec:intro_unitary}.)

%\enlargethispage{4mm}

Due to the elegant form of Eq.  (\ref{DS}), it will be convenient to
use $s_{j}$, $u_{j}$ rather than $\hat{s}_{j}$, $\hat{u}_{j}$ in
defining the encounter regions, demanding all $|s_{j}|$, $|u_{j}|$ to
be smaller than our bound $c$;\footnote{ The mutual differences
  between two stretches $j'<j$ will now be limited by
  $|\hat{u}_{j}-\hat{u}_{j'}|=|u_{j-1}-u_{j'-1}|
  <|u_{j-1}|+|u_{j'-1}|<2c$ and $|\hat{s}_{j}-\hat{s}_{j'}|
  =\big|\sum_{j''=j'}^{j-1}(\hat{s}_{j''+1}-\hat{s}_{j''})\big|
  =\big|\sum_{j''=j'}^{j-1}s_{j''}\big|<lc$, allowing for mutually
  linearized treatment if $c$ is chosen sufficiently small.}  the
encounter duration (\ref{tenc}) is changed accordingly.

Employing Eq. (\ref{linearize}), one easily shows that $\Delta S$ remains
invariant if the Poincar{\'e} section ${\cal P}$ is shifted through the
encounter.  Moreover, if the orbits $\gamma$ and $\gamma'$ differ in several
encounters (labeled by $\alpha=1,\ldots,V$), the total action difference is
additive in their contributions, and each is given by Eq. (\ref{DS});
i.e., we have
\begin{equation}
\Delta S=\sum_{\alpha,j}s_{\alpha j}u_{\alpha j}\,.
\end{equation}
Like in case of Sieber/Richter pairs, the relevant orbit pairs with
$\Delta S\sim\hbar$ have stable and unstable coordinates of the order
$\sqrt{\hbar}$, and hence encounter durations (\ref{tenc}) of the order
of the Ehrenfest time.

Finally, the relative difference between the stability amplitudes of
the relevant partner orbits $\gamma$, $\gamma'$ can be neglected in the
semiclassical limit.  The arguments used in Subsection
\ref{sec:stability_amplitude} immediately carry over, given that the
two partner orbits $\gamma$, $\gamma'$ are close everywhere up to time
reversal.

\section{Necessity of non-vanishing loops}

\label{sec:loops}

%\enlargethispage{6cm}

Contributions to the form factor arise only from sets of encounters
whose stretches are {\it separated by non-vanishing loops, i.e., do
  not overlap}.  After all, we obtained partner orbits by reshuffling
connections between ports, where the orbit either enters an encounter
stretch coming from an intervening loop, or leaves the encounter to
follow an intervening loop.  This procedure requires separated
encounter stretches.

Like for Sieber/Richter pairs, we need to check whether our method of
``generating'' partner orbits can be extended to encounters with
overlapping stretches, i.e., whether with the above prescription we
miss any orbit pairs related to encounters.  This question is
investigated in detail in Appendix \ref{sec:overlap_appendix}, showing
that no such orbit pairs are missed, and that overlapping stretches
indeed do not contribute to the form factor.  Here we just want to
list the main ideas.

\begin{figure}
\begin{center}
  \includegraphics[scale=0.35]{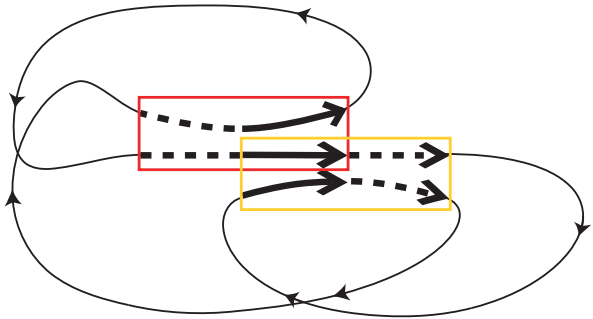}
\end{center}
\caption{Two 2-encounters (marked by boxes) overlap in one stretch and
  thus merge to a 3-encounter (solid bold arrows)}
\label{fig:overlap}
\end{figure}

First, assume that two stretches of {\it different encounters}
overlap.  In this case, the two encounters effectively merge, leaving
one larger encounter with more internal stretches, see Fig.
\ref{fig:overlap}.  The partners are thus seen as differing in one
larger encounter (e.g., a 3-encounter), rather than in two smaller
ones (e.g., two overlapping 2-encounters).\footnote{ Recall that we
  speak of ``overlap'' if two stretches are not separated by an
  intervening loop. In contrast, encounters just occupying the same
  region in phase space (as in Fig. \ref{fig:permutations}c) have to
  be considered as independent, as long as the are separated by
  intervening loops and reconnections are performed independently in
  each encounter.}

As in Subsection \ref{sec:min_dist}, overlapping {\it antiparallel}
stretches have to be regarded as one single stretch reflected at a
hard wall.  There is no way to reconnect such stretches to form a
partner orbit.

Finally, we shall see that {\it parallel} stretches can overlap (or
come very close in time) only if they follow multiple revolutions of a
shorter orbit $\tilde{\gamma}$.  In this case, several {\it different
  encounters} may lead to the {\it same partner orbit}. We will show
in Appendix \ref{sec:parallel_overlap} how to select one of these
encounters.  The resulting condition will be slightly more complicated
than just leaving out encounters with overlapping stretches but in the
semiclassical limit leads to the same contributions to the form
factor.

\section{Statistics of encounter sets}

\label{sec:statistics}

The statistics of close self-encounters inside periodic orbits can now
be established using the same two ingredients as in the case of
2-encounters: the {\it ergodicity} of the classical motion, and the
requirement of {\it non-vanishing loops}.

We will consider sets of encounters within orbit pairs $(\gamma,\gamma')$ with
fixed $\vec{v}$ and fixed structure.  Each of the $V$ encounters of
$\gamma$ is parametrized with the help of a Poincar{\'e} section ${\cal P}_\alpha$
($\alpha=1,\ldots,V$) crossing the orbit at an arbitrary phase-space point
inside the encounter, traversed at time $t_{\alpha1}$.  The orbit again
pierces through these sections at {times $t_{\alpha j}$} with
$j=2,\ldots,l_\alpha$ numbering the remaining stretches of the $\alpha$-th
encounter.  The first piercing may occur anywhere inside the orbit at
a time $0<t_{11}<T$, $T$ denoting the period.  The remaining $t_{\alpha
  j}$ follow in an order fixed by the structure at times $t_{11}<t_{\alpha
  j}<T+t_{11}$.  Each of the $v_l$ $l$-encounters is characterized by
$l-1$ pairs of {stable and unstable coordinates} $s_{\alpha j}$, $u_{\alpha
  j}$ ($j=1,\ldots,l-1$), which in total make for $2\sum_{l\geq
  2}(l-1)v_l=2(L-V)$ components.

We proceed to derive a {\it density $w_T(s,u)$ of phase-space
  separations $s$, $u$}.  To understand the normalization of
$w_T(s,u)$, recall that the separations $s$ and $u$ depend on the
location of the Poincar{\'e} sections ${\cal P}_\alpha$.  If ${\cal P}_\alpha$ is
shifted through the corresponding encounter, the stable and unstable
coordinates change while the contributions to the action difference
$\Delta S_{\alpha j}=s_{\alpha j}u_{\alpha j}$ remain invariant.  To make sure that
each encounter is counted exactly one, we demand that the integral $\int
d^{L-V}s\,d^{L-V}u\,w_T(s,u)\prod_{\alpha j}\delta(\Delta S_{\alpha j}-s_{\alpha j}u_{\alpha j})$
yields the density of encounter sets in an orbit $\gamma$ leading to orbit
pairs $(\gamma,\gamma')$ with given structure and action difference components
$\Delta S_{\alpha j}$.  We have to understand $w_T(s,u)$ as averaged over the
ensemble of all periodic orbits $\gamma$ with period $T$ in a given time
window, weighted with $|A_\gamma|^2$.  Again, such averaging allows us to
invoke ergodicity even for periodic orbits.  (A more careful
implementation will be given in Appendix \ref{sec:maths}.)

To determine $w_T(s,u)$, we need to count piercings through the
Poincar{\'e} sections ${\cal P}_\alpha$. We first assume that all times $t_{\alpha
  1}$ and thus all sections ${\cal P}_\alpha$ are fixed.  {\it Ergodicity}
then implies that for each section, the probability of a further
piercing occurring in a time interval $(t_{\alpha j},t_{\alpha j}+dt_{\alpha j})$
with stable and unstable coordinates inside $(\hat{s}_{\alpha
  j},\hat{s}_{\alpha j}+d\hat{s}_{\alpha j}) \times (\hat{u}_{\alpha j},\hat{u}_{\alpha
  j}+d\hat{u}_{\alpha j})$ is given by $\frac{1}{\Omega}d\hat{s}_{\alpha
  j}d\hat{u}_{\alpha j}dt_{\alpha j}$.  This probability holds for all
$l_\alpha-1$ additional piercings ($j=2,\ldots,l_\alpha$) through each of the $V$
sections.  Since the transformation leading from $\hat{s}$, $\hat{u}$
to $s$, $u$ is volume-preserving, the same uniform probability applies
to the components $s_{\alpha j}$, $u_{\alpha j}$.  We can therefore determine
the number $\rho_T(s,u)\,d^{L-V}s\,d^{L-V}u\,d^{L-V}t$ of {\it sets} of
$L-V$ piercings through our sections ${\cal P}_\alpha$ occurring in time
intervals $(t_{\alpha j},t_{\alpha j}+dt_{\alpha j})$, $j=2,\ldots,l_\alpha$, with stable
and unstable coordinates inside $(s_{\alpha j},s_{\alpha j}+ds_{\alpha j})$,
$(u_{\alpha j},u_{\alpha j}+du_{\alpha j})$, $j=1,\ldots,l_\alpha-1$; we may expect
$\rho_T(s,u)$ equal to $1/\Omega^{L-V}$.

However, recall that we are only interested in encounters separated by
{\it non-vanishing loops}.  To implement that restriction, we employ a
suitable characteristic function $\Theta_T(s,u,t)$ which vanishes if the
piercings described by $s$, $u$ and $t$ correspond to overlapping
stretches, and otherwise equals $1$. We thus obtain
\begin{equation}
\label{rho}
\rho_T(s,u,t)=\Theta_T(s,u,t)
\frac{1}{\Omega^{L-V}}\,.  \end{equation}

Proceeding towards $w_T(s,u)$ we integrate over the $L-V$ piercing
times $t_{\alpha j}$, $j\geq 2$, still for {\it fixed Poincar{\'e} sections}
${\cal P}_\alpha$. The integral yields a density for sets of $L-V$
piercings, characterized only by their stable and unstable coordinates
$s$, $u$.

To finally get to $w_T$, we must keep track of {\it all encounters}
along the orbits in question. To that end we have to consider all
possible positions of Poincar{\'e} sections and hence integrate over the
$V$ times $t_{\alpha1}$ (of the reference piercings) as well.  Doing so,
we count each encounter for a time $t_{\rm enc}^\alpha$, since we may move
each Poincar{\'e} section to any position inside the duration of the
encounter.  Each set of encounters is thus weighted with the product
of all durations $\prod_\alpha t_{\rm enc}^\alpha$.  In order to count each
encounter set exactly once, we divide out the factor $\prod_\alpha t_{\rm
  enc}^\alpha$, and arrive at the desired density
\begin{equation}
\label{wintegral}
w_T(s,u)=\frac{\int d^Lt\,\Theta_T(s,u,t)}
{\Omega^{L-V}\prod_\alpha t_{\rm enc}^\alpha}\,.
\end{equation}

Again, loops shorter than the classical relaxation time $t_{\rm cl}$
will entail small corrections to the uniform piercing probability and
therefore to (\ref{wintegral}); in the semiclassical limit these
corrections are negligible due to $t_{\rm cl}\ll T_E, T_H$.

It remains to evaluate the $L$-fold time integral in Eq.
(\ref{wintegral}).  The integration over $t_{11}$ runs from 0 to $T$;
it will be done as the last integral and then give a factor $T$. The
$L-1$ other $t_{\alpha j}$ must lie inside the interval
$(t_{11},t_{11}+T)$ and respect the ordering dictated by the structure
in question.

Moreover, we only consider encounter stretches separated by
intervening loops.  As a consequence, the piercing times $t_{\alpha j}$ of
subsequent stretches need to respect certain {\it minimal (time)
  distances}.  Let us first assume that all encounters are parallel
such that the orbit passes through each stretch from tail to head.  In
this case, the time between the piercings of two subsequent stretches
must be so large as to contain both the head of the first stretch and,
after a non-vanishing loop, the tail of the second stretch.  For
time-reversal invariant systems, encounter stretches may also be
time-reversed compared to the first stretch of the respective
encounter; in this case head and tail of the corresponding stretch
switch their roles.  If we sum up all minimal distances between
subsequent piercings, each stretch will appear in that sum once with
head and tail, and thus give a contribution $t_{\rm
  enc}^\alpha=t_s^\alpha+t_u^\alpha$.  Altogether, the minimal distances sum up to
the total duration of all encounter stretches $\sum_\alpha l_\alpha t_{\rm
  enc}^\alpha$, regardless of the structure considered.

The minimal distances effectively reduce the integration range, as we
may proceed to a new set of times $\tilde{t}_{\alpha j}$ obtained by
subtracting from $t_{\alpha j}$ both $t_{11}$ and the sum of minimal
distances between $t_{11}$ and $t_{\alpha j}$. The $\tilde{t}_{\alpha j}$ just
have to obey the same ordering as the times $t_{\alpha j}$, and lie in an
interval $(0,T-\sum_\alpha l_\alpha t_{\rm enc}^\alpha)$, where the subtrahend is
the total sum of minimal distances.  If we choose to label the $L-1$
times $\tilde{t}_{\alpha j}$ not by $\alpha$ and $j$, but in order of
traversal (indexed by one subscript $m=1,2,\ldots,L-1$), the integration
range may simply be written as
$0<\tilde{t}_1<\tilde{t}_2<\ldots<\tilde{t}_{L-1}<T-\sum_\alpha l_\alpha t_{\rm
  enc}^\alpha$.  We are therefore left with a trivial $(L-1)$-fold
integral over unity, yielding the density
\begin{eqnarray}
\label{density}
w_T(s,u)&=&\frac{T}{\Omega^{L-V}\prod_\alpha t_{\rm enc}^\alpha}
\int_0^{T-\sum_\alpha l_\alpha t_{\rm enc}^\alpha}d\tilde{t}_{L-1}
\int_0^{\tilde{t}_{L-1}}d\tilde{t}_{L-2}\ldots\int_0^{\tilde{t}_2}d\tilde{t}_1\nonumber\\
&=&\frac{T(T-\sum_\alpha l_\alpha t_{\rm enc}^\alpha)^{L-1}}
{(L-1)!\;\Omega^{L-V}\prod_\alpha t_{\rm enc}^\alpha}\,.
\end{eqnarray}
Crucially, $w_T(s,u)$ depends only on $\vec{v}$ but not on the
structure considered, and that fact strongly simplifies our treatment.

We can now determine the (average) number $P_T^{\vec{v}}(\Delta S)d\Delta S$
of partners differing from $\gamma$ in $v_l$ $l$-encounters, {\it with
  action difference in the interval $(\Delta S,\Delta S+d\Delta S)$}.\footnote{
  For simplicity, the two mutually time-reversed partners $\gamma'$,
  $\T\gamma'$ present in time-reversal invariant systems will still be
  counted as one.}  It is convenient to include each partner $\gamma'$ $L$
times, each time considering a different encounter stretch of $\gamma$ as
the first. Different such choices may lead to either different or
coinciding structures for the orbit pair $(\gamma,\gamma')$. In any case, the
separations $s_{\alpha j}$, $u_{\alpha j}$ (and the components $\Delta S_{\alpha j}=
s_{\alpha j}u_{\alpha j}$ used for normalizing $w_T(s,u)$) will differ, since
they depend on the numbering of encounters and stretches.  To take
into account all possibilities, we {\it sum over all structures}
related to $\vec{v}$ and {\it integrate over all phase-space
  separations} $s$, $u$ leading to the same overall action difference
$\Delta S=\sum_{\alpha j}s_{\alpha j}u_{\alpha j}$, and thus count each partner orbit
exactly $L$ times.  Given that $w_T(s,u)$ is the same for all
structures related to the same $\vec{v}$, summation over all
structures is equivalent to multiplication with the number
$N(\vec{v})$ of structures related to $\vec{v}$. Dividing out the
factor $L$, we end up with the number of partner orbits
\begin{eqnarray} \label{numorb}
P_T^{\vec{v}}(\Delta S)d\Delta S
= d\Delta S\frac{N(\vec v)}{L}\int d^{L-V}s\,d^{L-V}u\, \delta\Big(\Delta
S-\sum_{\alpha j} s_{\alpha j} u_{\alpha j}\Big) w_T(s,u) \,.
\end{eqnarray}

\section{Contribution of each structure}

\label{sec:contribution}

To determine the spectral form factor, we have to evaluate the double
sum over periodic orbits $\gamma$, $\gamma'$ in Eq. (\ref{doublesum}).  In
doing so, we will account for all families of orbit pairs whose
members are composed of loops similar up to time reversal, i.e., both
``diagonal'' pairs and orbit pairs differing in encounters.  We assume
that these are the only orbit pairs to give rise to a systematic
contribution (an assumption that will be further discussed in the
conclusions).  Neglecting the differences between the stability
amplitudes and periods of $\gamma$, $\gamma'$ as explained above, we many
simplify the double sum (\ref{doublesum}) as
\begin{equation}
\label{doublesumsimplified}
K(\tau)=\frac{1}{T_H}\left\langle\sum_{\gamma,\gamma'}|A_\gamma|^2
{\rm e}^{{\rm i}(S_\gamma-S_{\gamma'})/\hbar}
\delta\left(\tau T_H-T_\gamma\right)\right\rangle\,.
\end{equation}
The sum over $\gamma$ is evaluated using the rule of Hannay and Ozorio de
Almeida (\ref{hoda}).  The diagonal pairs contribute $\kappa\tau$, with
$\kappa=1$ in the unitary and $\kappa=2$ in the orthogonal case.  The sum over
partners $\gamma'$ differing from $\gamma$ in encounters can be performed with
the help of the density $P_T^{\vec{v}}(\Delta S)$, where $T=\tau T_H$.  We
thus find
\begin{equation}
\label{ksum_DS}
K(\tau)=\kappa \tau+\kappa \tau\left\langle\sum_{\vec{v}}\int d\Delta S\;
P_T^{\vec{v}}(\Delta S) {\rm e}^{{\rm i}\Delta S/\hbar}\right\rangle\,.
\end{equation}
The factor $\kappa$ in the second member is inserted since for
time-reversal invariant systems, each reconnection gives rise to two
mutually time-reversed partner orbits; a further factor $\tau T_H$ is
provided by the sum rule.  Substituting Eq.  (\ref{numorb}) for
$P_T^{\vec{v}}(\Delta S)$, we get
\begin{eqnarray}
\label{ksum}
K(\tau)=\kappa \tau+\kappa \tau\left\langle\sum_{\vec{v}} N(\vec{v})
\int d^{L-V}s\;d^{L-V}u \;
\frac{w_T(s,u)}{L}\;{\rm e}^{\frac{\rm i}{\hbar}\sum_{\alpha j}s_{\alpha j}u_{\alpha j}}\right\rangle\,.
\end{eqnarray}
Here, the orbit pairs with fixed $\vec{v}$, structures, and
separations $s,u$ appear with the weight
$N(\vec{v})\frac{w_T(s,u)}{L}$.

The integral over $s$ and $u$, multiplied with $\kappa \tau$, yields the
contribution to the form factor from each structure associated to
$\vec{v}$. The integral is surprisingly simple to do. Consider the
multinomial expansion of $(T-\sum_\alpha l_\alpha t_{\rm enc}^\alpha)^{L-1}$ in our
expression (\ref{density}) for the density $w_T(s,u)$.  We shall show
that only a single term of that expansion contributes, the one which
involves a product of all $t_{\rm enc}^\alpha$,
\begin{equation}
\frac{(L-1)!}{(L-V-1)!}T^{L-V-1}\prod_\alpha(-l_\alpha t_{\rm enc}^\alpha)\,,
\end{equation}
canceling with the corresponding product in the denominator of
(\ref{density}). The contributing term of $w_T(s,u)$ is thus given by
\begin{eqnarray}
\label{wcontr}
\frac{w_T^{\rm contr}}{L}&=&\frac{T\frac{(L-1)!}{(L-V-1)!}
T^{L-V-1}\prod_\alpha(-l_\alpha)}{L!\;\Omega^{L-V}}= h(\vec{v})
\left(\frac{T}{\Omega}\right)^{L-V}\,,\nonumber\\
h(\vec{v}) &\equiv&\frac{(-1)^V\prod_l l^{v_l}}{L(L-V-1)!}\,.
\end{eqnarray}
Due to the cancellation of $t_{\rm enc}^\alpha$, $w_T^{\rm contr}$ does
not depend on the stable and unstable coordinates and therefore the
remaining integral over $s$ and $u$ is easily calculated,
\begin{eqnarray}
\label{contribution}
&&\kappa \tau\int d^{L-V}s\;d^{L-V}u\;\frac{w_T^{\rm contr}}{L}\;
{\rm e}^{\frac{\rm i}{\hbar}\sum_{\alpha j}s_{\alpha j}u_{\alpha j}}\nonumber\\
&=&\kappa \tau h(\vec{v}) \left(\frac{T}{\Omega}\right)^{L-V}
\int_{-c}^c d^{L-V}s\;
d^{L-V}u\;{{\rm e}}^{\frac{\rm i}{\hbar}\sum_{\alpha j}s_{\alpha j}u_{\alpha j}}\nonumber\\
&\to& \kappa  h(\vec{v}) \;\tau^{L-V+1}\,;
\end{eqnarray}
here, we have met with the $(L-V)$-fold power of the integral
$\int_{-c}^c ds\;du\; {\rm e}^{{\rm i} s u/\hbar}\to 2\pi\hbar$, see Eq.
(\ref{simple_integral}), and used that
$2\pi\hbar\frac{T}{\Omega}=\frac{T}{T_H}=\tau$.  In the semiclassical limit, the
contributions of all other terms in the multinomial expansion vanish
for one of two possible reasons:

First, consider terms in which at least one encounter duration $t_{\rm
  enc}^\alpha$ in the {\it denominator} is not compensated by a power of
$t_{\rm enc}^\alpha$ in the numerator. The corresponding contribution to
the form factor is proportional to
\begin{equation}
\label{oscillating}
\left\langle\int_{-c}^c\prod_{j}ds_{\alpha j}du_{\alpha j}\frac{1}{t_{\rm enc}^\alpha}
{\rm e}^{\frac{\rm i}{\hbar}\sum_{j}s_{\alpha j}u_{\alpha j}}\right\rangle\,,
\end{equation}
similar to Eq. (\ref{osc_integral_sr}).  As shown in Appendix
\ref{sec:integral}, such integrals oscillate rapidly and effectively
vanish in the semiclassical limit, as $\hbar\to0$.

Second, there are terms with all encounter durations in the
denominator canceled but, say, $p$ factors $t_{\rm enc}^\alpha$ in the
{\it numerator} left uncanceled.  To show that such terms do not
contribute we employ a scaling argument.  Obviously, the considered
terms may involve only a smaller order of $T$ than $w_T^{\rm contr}$;
they are of order $T^{L-V-p}$.  However, $\Omega$ still appears in the
same order $\frac{1}{\Omega^{L-V}}$.  To study the scaling with $\hbar$, we
transform to variables $\tilde{s}_{\alpha j}=\frac{s_{\alpha j}}{\sqrt{\hbar}}$,
$\tilde{u}_{\alpha j}=\frac{u_{\alpha j}}{\sqrt{\hbar}}$, eliminating the
$\hbar$-dependence in the phase factor of Eq. (\ref{ksum}).  The
resulting expression is proportional to $\hbar^{L-V}$ due to the Jacobian
of the foregoing transformation, and proportional to $(\ln\hbar)^p$ due
to the $p$ remaining encounter durations $\sim\ln\hbar$.  Together with the
factor $\tau$ originating from the sum rule, the corresponding
contribution to the form factor scales like
\begin{eqnarray} \tau T^{L-V-p}\left(\frac{\hbar}{\Omega}\right)^{L-V}\!(\ln
  \hbar)^p \propto \tau\frac{T^{L-V-p}}{T_H^{L-V}}(\ln\hbar)^p
\propto
  \left(\frac{\ln\hbar}{T_H}\right)^p\tau^{L-V+1-p}, \end{eqnarray}
and thus disappears as $\hbar\to 0$, $T_H\propto\hbar^{-1}\to\infty$.

Therefore, the contribution to the form factor arising from each
structure with the same $\vec{v}$ is indeed determined by Eq.
(\ref{contribution}).  Remarkably, this result is due to a subleading
term in the multinomial expansion of $w_T(s,u)$, originating only from
the small corrections due to the ban of encounter overlap.  Summing
over all structures as in (\ref{ksum}), the spectral form factor is
determined as
\begin{equation}
\label{k}
K(\tau)=\kappa \tau+\kappa \sum_{\vec{v}}N(\vec{v})h(\vec{v}) \tau^{L-V+1}\,.
\end{equation}
The calculation of $K(\tau)$ is now reduced to the purely combinatorial
task of determining the numbers $N(\vec{v})$ of structures.  The
$n$-th term in the series $K(\tau)=\kappa \tau+\sum_{n\geq 2}K_n\tau^n$ is
exclusively determined by structures with $\nu(\vec{v})\equiv
L(\vec{v})-V(\vec{v})+1=n$,
\begin{equation}
K_n=\kappa\sum_{\vec{v}}^{\nu(\vec{v})=n}N(\vec{v})h(\vec{v})\,;
\end{equation}
it will be convenient to represent $K_n$ as\footnote{ We slightly
  depart from the notation in \cite{Letter}, where
  $\tilde{N}(\vec{v})$ was defined to include the denominator
  $(n-2)!$.}
\begin{eqnarray}
\label{kn}
  K_n&=&\frac{\kappa }{(n-2)!}\sum_{\vec{v}}^{\nu(\vec{v})=n}\tilde{N}(\vec{v})\,,
\\
\tilde{N}(\vec{v})&\equiv&N(\vec{v})\frac{(-1)^V\prod_l l^{v_l}}{L(\vec{v})}\,.
\label{tilde}
\end{eqnarray}

The present reasoning can be generalized to general hyperbolic systems
(avoiding the approximation (\ref{tenc}) for $t_{\rm enc}$) with
arbitrary number of degrees of freedoms, again yielding Eqs.
(\ref{kn}) and (\ref{tilde}); the changes arising are listed in
Appendix \ref{sec:general}.

\section{$\tau^3$-contribution to the spectral form factor}

We are now equipped to determine the cubic contribution to the
spectral form factor.  There are two kinds of orbit pairs contributing
to $\tau^3$, i.e., $n=L-V+1=3$: those with $v_2=2$ (and all other $v_l$
zero) and thus $L=4$ and $V=2$, and those with $v_3=1$ and thus $L=3$
and $V=1$.  These are precisely the orbit pairs described in Sections
\ref{sec:22} and \ref{sec:3}.  The corresponding proportionality
factors $h(\vec{v})$ read
\begin{equation}
h(v_2=2)=1\ \ \ \ \mbox{and}\ \ \ \ h(v_3=1)=-1\,;
\end{equation}
hence each structure respectively contributes $\kappa\tau^3$ or $-\kappa\tau^3$.
Without time-reversal invariance, there is one structure associated to
$v_2=2$ ({\it ppi}) and one structure related to $v_3=1$ ({\it pc}).
Their contributions mutually cancel.  Hence no correction to the
diagonal contribution $\tau$ arises, in agreement with the RMT
prediction for the GUE.  In presence of time-reversal invariance, we
found 5 structures with $v_2=2$ and 4 structures with $v_3=1$, leading
to an overall contribution $2\tau^3$, in line with the corresponding
terms of the GOE form factor.

For the fourth and higher orders of the spectral form factor, the task
of enumerating structures is too cumbersome to be done by hand. In the
following Chapter, we will rather analytically derive a recursion for
$N(\vec{v})$, which will provide us with all further Taylor
coefficients.

\section{Summary}

Reconnections inside an $l$-encounter can be performed in $l-1$ steps,
each involving two stretches. We thus characterize $l$-encounters by
$l-1$ stable and $l-1$ unstable coordinates $s_j$, $u_j$ measuring the
phase-space separations between the stretches involved in the $j$-th
step. The formulas for the action difference and the encounter
duration of Chapter \ref{sec:tau2} immediately generalize. For each
structure of orbit pairs, we determined the density $w_T(s,u)$ of
stable and unstable separations in orbits of period $T$.  To arrive at
this density, we (i) used ergodicity, and (ii) took into account only
encounter stretches separated by non-vanishing loops.  With the help
of $w_T(s,u)$, we determined the contribution to the form factor
arising from each structure, leaving only the combinatorial problem of
counting structures.  Interestingly, the latter contribution is due to
a correction term inside $w_T(s,u)$ which is subleading in $T$ and
originates from the necessity of non-vanishing loops.

%% file: kapitel6.tex
\chapter{Combinatorics}

\label{sec:combinatorics}

\section{Unitary case}

\label{combinatorics_unitary}

\subsection{Structures and permutations}

\label{sec:intro_unitary}

\stand To determine the combinatorial numbers $N(\vec{v})$, first for
systems without time-reversal invariance, we must relate structures of
orbit pairs to permutations.  Mathematically speaking, permutations
are bijective mappings between a finite number of discrete objects,
like the natural numbers $1,2,\ldots,L$.  We may denote a permutation of
$L$ objects by $\{a\to P(a), a=1\ldots L\}$ or $ P=\left({1\atop
    P(1)}{2\atop P(2)}{\ldots\atop } {L\atop P(L)}\right)$.

Each permutation contains one or more {\bf cycles}
\cite{Permutations}.  We may start with some object $a_1$ and note the
sequence of successors, $a_1\to a_2= P(a_1)\to a_3= P(a_2)\to\ldots$; if that
sequence first returns to the starting object $a_1$ after precisely
$L$ steps one says that the permutation in question is a single cycle,
denotable simply as $(a_1,a_2,\ldots,a_L)$.  If the above sequence returns
to the initial object $a_1$ before exhausting all elements $1,2,\ldots,L$,
i.e., after $l_1<L$ steps, the permutation falls into several cycles.
One of these cycles is given by $(a_1,a_2,\ldots a_{l_1})$.  A second
cycle can be obtained if we choose an arbitrarily element $b_1$ not
included in the first cycle, and follow the sequence of its successors
until it returns to $b_1$.  This procedure can be continued until all
elements $1,2,\ldots,L$ are grouped into cycles.  The permutation $P$ can
then be characterized by writing down all of its cycles, as in
$P=(a_1,a_2,\ldots,a_{l_1})(b_1,b_2,\ldots,b_{l_2})\ldots$.  A cycle is defined up
to cyclic permutations of its member objects. The number of objects in
a cycle is called the length of that cycle.  Of course, the lengths of
all cycles of $P$ sum up to the total number $L$ of permuted objects.

We now turn to applying the notion of cycles to {\bf self-encounters}
of long periodic orbits. An orbit $\gamma$ and its partner orbit(s) differ
in a set of encounters. The $L$ encounter stretches of $\gamma$ will be
labelled by $1,2,\ldots,L$ starting from an arbitrary reference stretch.
Inside $\gamma$ the $a$-th encounter stretch connects the $a$-th entrance
port and the $a$-th exit port; the permutation defining which entrance
port (upper line) is connected to which exit port (lower line) thus
trivially reads $P_{\rm enc}^{\gamma}=\left({1\atop 1}{2\atop 2}{\ldots\atop
    \ldots}{L\atop L}\right)$.

A partner orbit $\gamma'$ differing from $\gamma$ in the said encounters has
the ports differently connected: The $a$-th entrance is now connected
to an exit $P_{\rm enc}(a)\neq a$.  This reconnection can be expressed
in terms of a different permutation $ P_{\rm enc}= \left({1\atop
    P_{\rm enc}(1)}{2\atop P_{\rm enc}(2)} {\ldots\atop \ldots}{L\atop P_{\rm
      enc}(L)}\right)$; e.g., reconnections as in
Fig.~\ref{fig:permutations}b are described by the permutation $ P_{\rm
  enc}=\left({1\atop 2}{2\atop 6}{3\atop 4}{4\atop 5}{5\atop 1}{6\atop
    3}\right)$. Note that we refrain from indexing the latter
permutation by a superscript $\gamma'$.

A permutation $ P_{\rm enc}$ accounting for a single $l$-encounter is
a single cycle of length $l$, e.g. $(1,2,6,3,4,5)$ in the above
example.  This is easily understood as follows: To allow for
reconnections, the entrance port $a$ and the exit port $\PE(a)$ have
to form part of the same encounter of $\gamma$, and so do the
corresponding stretches $a$ and $\PE(a)$. Thus, acting on a stretch
$a$, $\PE$ always returns a stretch of the same encounter. By
repeatedly applying $\PE$, we obtain a list $a,\PE(a),\PE^2(a),
\PE^3(a),\ldots$ of stretches all belonging to the same encounter.
Continuing to iterate, we will finally return to the first element
$a$.  Then, we have enumerated all stretches of the encounter in
question. Missing stretches would belong to a different encounter
since they may not be involved in reconnections with any stretch on
the list. Moreover, by returning to the first element, our list has
turned into a cycle of the permutation $\PE$. Given an $l$-encounter,
that cycle must have $l$ elements.\footnote{ Note that our derivation
  of the action difference $\Delta S=\sum_{j=1}^{l-1} s_j u_j$ related to an
  $l$-encounter works only if the corresponding reconnections
  correspond to an $l$-cycle of $P_{\rm enc}$.  We first determined
  $\Delta S$ for encounters as depicted in Fig. \ref{fig:steps}, with port
  connections of the type $P_{\rm enc}=\left({1\atop 2}{2\atop 3}
    {\ldots\atop \ldots}{l\atop 1}\right)=(1,2,\ldots,l)$.  The result carries
  over to all $l$-encounters characterized by single-cycle
  permutations $(a_1,a_2,\ldots,a_l)$, if we simply change the naming of
  stretches.  However, if the encounter would correspond to, say, two
  cycles of $P_{\rm enc}$, no renaming would be able to bring it to
  the desired form.
%Our stepwise procedure has to be applied separately to both cycles,
%leading to altogether $l-2$ steps.
  Formulated in terms of permutation theory, splitting reconnections
  into $l-1$ steps each involving two stretches amounts to factoring a
  single-cycle permutation into $l-1$ permutations each transposing
  two elements; for the relevant mathematical literature, see
  \cite{Transpositions}.}  If there are no further encounters, $l$
will coincide with the total number of permuted elements $L$.

If $\PE$ has multiple cycles, there are several disjoint encounters,
and the connections between entrance and exit ports are reshuffled
separately within these encounters.  For example, Fig.
\ref{fig:permutations}c visualizes a permutation with three cycles
$(1,2)$, $(3,4)$, and $(5,6)$.  As already mentioned, reconnections
take place only between stretches 1 and 2, stretches 3 and 4, and
stretches 5 and 6, which thus have to be considered as three
independent encounters.

If $\gamma$ and $\gamma'$ differ in several encounters, the corresponding
permutation $ P_{\rm enc}$ has precisely one $l$-cycle corresponding
to each of the $v_l$ $l$-encounters, for all $l\geq 2$, the total number
of permuted objects being $L=\sum_{l\geq 2}lv_l$.

We also have to account for the {\bf orbit loops}.  The $a$-th loop
connects the exit of the $(a-1)$-st encounter stretch with the
entrance of the $a$-th one. These connections can be associated with
the permutation $ P_{\rm loop}=\left({1\atop 2}{2\atop 3}{\ldots\atop \ldots}
  {L\atop 1}\right)$, where the upper line refers to exit ports and
the lower line to entrance ports.  Obviously, $P_{\rm loop}$ consists
of a single cycle $(1,2,\ldots,L)$.  Since the loops of $\gamma$ and $\gamma'$
coincide, they are characterized by one and the same permutation
$P_{\rm loop}$.

As a whole, the {\bf orbit} $\gamma$ is now described by the product $
P^\gamma= P_{\rm loop} P_{\rm enc}^\gamma= P_{\rm loop}$.  That product gives
the order in which entrance ports (and thus loops) are traversed in
$\gamma$: Acting on the label of an entrance port $a$, $P_{\rm enc}^\gamma$
will return the following exit port; subsequent application of $P_{\rm
  loop}$ then gives the next entrance port.  The permutation $P^\gamma$ is
single-cycle -- as it should be, because $\gamma$ is a periodic orbit and
hence returns to the first entrance port only after traversing all
others.

Similarly, the sequence of entrance ports (or, equivalently, loops)
traversed by $\gamma'$ is represented by the product of ``loop'' and
``encounter permutations''
\begin{equation}
\label{porb}
 P= P_{\rm loop} P_{\rm enc}
\end{equation}
with the same $P_{\rm loop}$ as above.  We must demand $ P$ to be a
single cycle for $\gamma'$ to be a connected periodic orbit. If $P$
decomposes into several cycles, $\gamma'$ will decompose into several
disjoint periodic orbits, as in Fig.  \ref{fig:24big}c.

The permutations $P_{\rm enc}$ are in one-to-one correspondence to the
structures of orbit pairs introduced in Chapter \ref{sec:tau3}.  When
defining structures, we numbered the encounter stretches $1,2,\ldots,L$ in
order of traversal by $\gamma$, starting from an arbitrary reference
stretch -- just like in the present Chapter.  For systems without
time-reversal invariance, each structure corresponds to one way of (i)
dividing the labels $1,2,\ldots,L$ into groups (each corresponding to one
encounter), and (ii) producing a non-decomposing partner orbit by
reshuffling connections between these encounters.  The cycles of
$P_{\rm enc}$ are just the above groups. The ordering of elements
inside each cycle determines the connections inside $\gamma'$, which may
not decompose due to our restriction on $P$.

We shall denote by ${\cal M}(\vec{v})$ the set of permutations $
P_{\rm enc}$ which have $v_l$ $l$-cycles, for each $l\geq 2$, and upon
multiplication with $ P_{\rm loop}$ yield single-cycle permutations as
in Eq. (\ref{porb}).  The number of elements of ${\cal M}(\vec{v})$
coincides with the number $N(\vec{v})$ of structures related to
$\vec{v}$.

\subsection{Examples}

\label{sec:unitary_examples}

The numbers $N(\vec{v})$ can be determined numerically, by generating
all possible permutations $ P_{\rm enc}$ with $v_l$ $l$-cycles and
counting only those for which $ P= P_{\rm loop} P_{\rm enc}$ is
single-cycle. The $ P_{\rm enc}$'s contributing to the orders $n=3$,
$5$, and $7$ of the spectral form factor are shown in Table
\ref{tab:unitary}.

\begin{table}
\begin{center}
\begin{tabular}{|c|c|c|c|r|r|r|} \hline
order & $\vec{v}$ & $L$ & $V$ & $N(\vec{v})$ & $\tilde{N}(\vec{v})$ & contribution \\
\hline\hline
$\tau^3$
& $(2)^2$ &4 & 2 & 1\;\; & 1\; & $1\tau^3$\;\;\;\;\; \\ \hhline{~------}
& $(3)^1$ &3 & 1 & 1\;\; & $-1$\; & $-1\tau^3$\;\;\;\;\; \\ \hhline{~------}
\hhline{~------}
&&&&&0\;&0$\tau^3$\;\;\;\;\; \\
\hline\hline
$\tau^5$
& $(2)^4$ &8 & 4 & 21\;\; & 42\; & $7\tau^5$\;\;\;\;\; \\ \hhline{~------}
& $(2)^2(3)^1$ &7 & 3 & 49\;\; & $-84$\; & $-14\tau^5$\;\;\;\;\; \\ \hhline{~------}
& $(2)^1(4)^1$ &6 & 2 & 24\;\; & 32\; & $\frac{16}{3}\tau^5$\;\;\;\;\; \\ \hhline{~------}
& $(3)^2$ &6 & 2 & 12\;\; & 18\; & $3\tau^5$\;\;\;\;\; \\ \hhline{~------}
& $(5)^1$ &5 & 1 & 8\;\; & $-8$\; & $-\frac{4}{3}\tau^5$\;\;\;\;\; \\ \hhline{~------}
\hhline{~------}
&&&&&0\;&0$\tau^5$\;\;\;\;\; \\
\hline\hline
$\tau^7$
& $(2)^6$ &12 & 6 & 1485\;\; & 7920\; & $66\tau^7$\;\;\;\;\; \\ \hhline{~------}
& $(2)^4(3)^1$ &11 & 5 & 5445\;\; & $-23760$\; & $-198\tau^7$\;\;\;\;\; \\ \hhline{~------}
& $(2)^3(4)^1$ &10 & 4 & 3240\;\; & 10368\; & $\frac{432}{5}\tau^7$\;\;\;\;\; \\ \hhline{~------}
& $(2)^2(3)^2$ &10 & 4 & 4440\;\; & 15984\; & $\frac{666}{5}\tau^7$\;\;\;\;\; \\ \hhline{~------}
& $(2)^2(5)^1$ &9 & 3 & 1728\;\; & $-3840$\; & -32$\tau^7$\;\;\;\;\; \\ \hhline{~------}
& $(2)^1(3)^1(4)^1$ &9 & 3 & 2952\;\; & $-7872$\; & $-\frac{328}{5}\tau^7$\;\;\;\;\; \\ \hhline{~------}
& $(3)^3$ &9 & 3 & 464\;\; & $-1392$\; & $-\frac{58}{5}\tau^7$\;\;\;\;\; \\ \hhline{~------}
& $(2)^1(6)^1$ &8 & 2 & 720\;\; & 1080\; & $9\tau^7$\;\;\;\;\; \\ \hhline{~------}
& $(3)^1(5)^1$ &8 & 2 & 608\;\; & 1140\; & $\frac{19}{2}\tau^7$\;\;\;\;\; \\ \hhline{~------}
& $(4)^2$ &8 & 2 & 276\;\; & 552\; & $\frac{23}{5}\tau^7$\;\;\;\;\; \\ \hhline{~------}
& $(7)^1$ &7 & 1 & 180\;\; & $-180$\; & $-\frac{3}{2}\tau^7$\;\;\;\;\; \\ \hhline{~------}
\hhline{~------}
&&&&&0\;&0$\tau^7$\;\;\;\;\; \\
\hline
\end{tabular}
\end{center}
\caption{Permutations, and thus structures of orbit pairs, giving rise to orders
$\tau^3$, $\tau^5$, and $\tau^7$ of the form factor, for systems
without time-reversal invariance.
We represent $\vec{v}$ by $(2)^{v_2}(3)^{v_3}\ldots$. The order of each
contribution is given by $n=L-V+1$.
We see that contributions add up to zero for odd $n$,
whereas there are no permutations for even $n$.}
\label{tab:unitary}
\end{table}

\begin{figure}[t]
\begin{center}
  \includegraphics[scale=0.32]{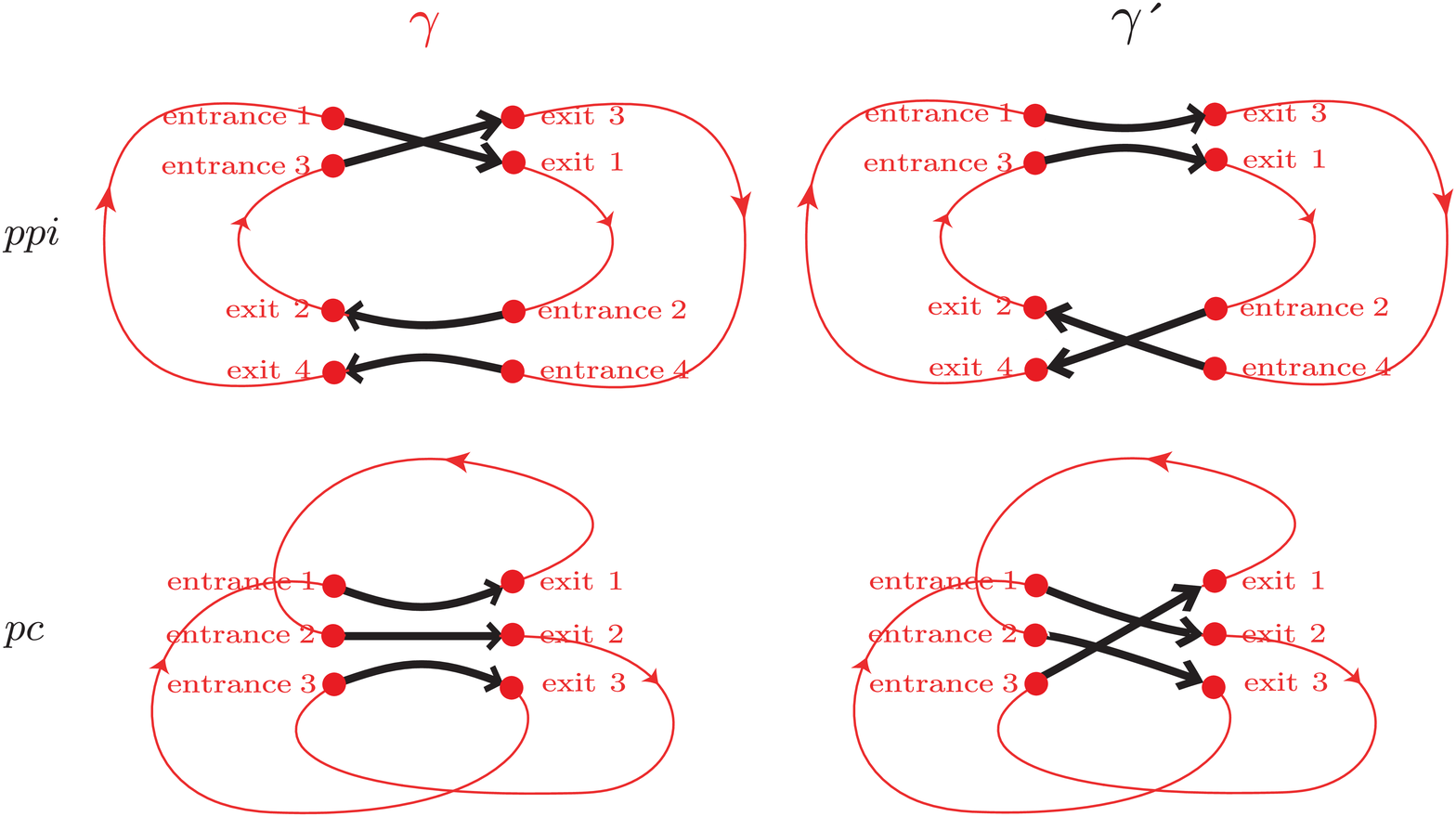}
\end{center}
\caption{
  Connections between entrance and exit ports for orbit pairs $(
  \gamma,\gamma')$ of structures {\it ppi} and {\it pc}, existing both for
  systems with and without time-reversal
  invariance.}\label{fig:parallel_example}
\end{figure}

Interestingly, no qualifying $ P_{\rm enc}$'s exist for even
$L-V+1=n$. For example, the only candidate for $n=2$ would be $ P_{\rm
  enc}=\left({1\atop 2}{2\atop 1}\right)$, describing reconnections
inside an encounter of two parallel orbit stretches $\psr$.  As shown
in Fig. \ref{fig:2decompose}, the corresponding partner decomposes
into two separate periodic orbits (corresponding to the cycles $(1)$
and $(2)$ of $ P= P_{\rm loop} P_{\rm enc}=\left({1\atop 1}{2\atop
    2}\right)$).

The same happens for all other permutations with $n$ even. This can be
proven based on the parities of the permutations involved. A
permutation is said to have parity 1 if it can be written as a product
of an even number of transpositions (i.e., permutations interchanging
two elements and leaving all others invariant), and to be of parity
$-1$ if it is a product of an odd number of transpositions. Parity is
given by $(-1)^{L-V}$, where $L$ is the number of permuted elements
and $V$ the number of cycles, and the parity of a product of
permutations equals the product of parities of the factors.  Since $
P$ and $ P_{\rm loop}$ both consist of one single cycle, they are of
the same parity.  Therefore, $ P= P_{\rm loop} P_{\rm enc}$ implies
that all allowed $ P_{\rm enc}$ need to have parity 1, i.e., $n=L-V+1$
must be odd.

For $n$ odd, the individual numbers $N(\vec{v})$ and
$\tilde{N}(\vec{v})$, see Eq. (\ref{tilde}), do not vanish, indicating
that there are orbit pairs contributing to the odd powers of $\tau$.
However, their contributions mutually cancel.  We see in Table
\ref{tab:unitary} that the $\tilde{N}(\vec{v})$ related to the same
$n$, and thus the corresponding contributions to $K(\tau)$, sum up to
zero. That cancellation is the reason why all off-diagonal
contributions to the spectral form factor vanish in the unitary case.

For example, the two permutations listed for $n=3$ correspond to the
structures {\it ppi} and {\it pc} introduced in Chapter
\ref{sec:tau3}.  For {\it ppi}, $\gamma'$ reconnects entrance and exit
ports according to the permutation $P_{\rm enc}=\left({1\atop
    3}{2\atop 4}{3\atop 1}{4\atop 2}\right)$; compare Fig.
\ref{fig:parallel_example}.  This permutation falls into two 2-cycles
$(1,3)$ and $(2,4)$ representing one parallel encounter of stretches 1
and 3, and one parallel encounter of stretches 2 and 4.  For ${\it
  pc}$, the encounter connections can be read off from Fig.
\ref{fig:parallel_example} as $ P_{\rm enc}=\left({1\atop 2}{2\atop
    3}{3\atop 1}\right)$, which contains one single 3-cycle $(1,3,2)$
representing a parallel encounter of stretches 1, 3, and 2.  We have
already seen that the resulting contributions mutually cancel.

In the following Subsections, we will show that the same cancellation
occurs for arbitrary $n$.  To that end we first need to derive a
recursion for $N(\vec{v})$.

\subsection{Recursion relation for $N(\vec{v})$}

\label{sec:unitary_recursion}

To find a recursion formula for $N(\vec{v})$, we will imagine one loop
(e.g. the one with index $L$) removed from both $\gamma$ and $\gamma'$ and
study the consequences on the encounters. We shall mostly reason with
permutations but the translation rule {\it cycle} $\to$ {\it encounter}
yields an interpretation for orbits. Readers wanting to skip the
reasoning may jump to the result in Eq. (\ref{rec2_unit}).

As a preparation, let us introduce a subset ${\cal M}(\vec{v},l)$ of
${\cal M}(\vec{v})$ containing all those permutations for which the
largest of the permuted numbers, i.e., $L(\vec{v})=\sum_k kv_k$ belongs
to a cycle of length $l$ (it is assumed that $v_l>0$).  The number of
elements in ${\cal M}(\vec{v},l)$ will be denoted by $N(\vec{v},l)$.
One can show that the sizes of ${\cal M}(\vec{v})$ and ${\cal
  M}(\vec{v},l)$ are related as
\begin{equation} \label{mengetomenge} N(\vec{v},l) = \frac{l
    v_l}{L(\vec{v})} N(\vec{v})\,.
\end{equation}
To derive (\ref{mengetomenge}), we use that all $P_{\rm enc}\in {\cal
  M}(\vec{v})$ can be brought to the form required for ${\cal
  M}(\vec{v},l)$ by applying a cyclic permutation. We have to consider
the cycle representation of $P_{\rm enc}$, and rename each element $a$
as $(a+\Delta a)\mod L$ with $\Delta a$ fixed.\footnote{ More formally, this
  renaming may be described as follows: Since $P_{\rm loop}$ maps each
  $a$ to $(a+1)\mod L$, a cyclic shift by $\Delta a$ steps corresponds to
  a similarity transformation using the $\Delta a$-fold power $P_{\rm
    loop}^{\Delta a}$, changing $P_{\rm enc}$ into $P_{\rm loop}^{\Delta
    a}P_{\rm enc}P_{\rm loop}^{-\Delta a}$.  The resulting permutation
  belongs to ${\cal M}(\vec{v})$ since (i) it has the same numbers
  $v_l$ of $l$-cycles as $P_{\rm enc}$, and (ii) multiplication with
  $P_{\rm loop}$ yields $P_{\rm loop}(P_{\rm loop}^{\Delta a}P_{\rm
    enc}P_{\rm loop}^{-\Delta a}) =P_{\rm loop}^{\Delta a}(P_{\rm loop}P_{\rm
    enc})P_{\rm loop}^{-\Delta a}$ which is similar to the single-cycle
  permutation $P=P_{\rm loop}P_{\rm enc}$ and thus single-cycle as
  well.}  For each $\Delta a$, a different element of $P_{\rm enc}$ will
be renamed as $L$.  We are interested only in those cyclic
permutations which place $L$ inside an $l$-cycle and therefore lead to
a member of ${\cal M}(\vec{v},l)$. For each $P_{\rm enc}\in{\cal
  M}(\vec{v})$, there are $lv_l$ such possibilities. Since each member
of ${\cal M}(\vec{v},l)$ can be accessed using $L(\vec{v})$ different
cyclic permutations, we have $l v_l N(\vec{v})=L(\vec{v})N(\vec{v},l)$
and thus (\ref{mengetomenge}).

\subsubsection{Mapping between permutations}

\begin{figure}[t]
\begin{center}
  \includegraphics[scale=0.44]{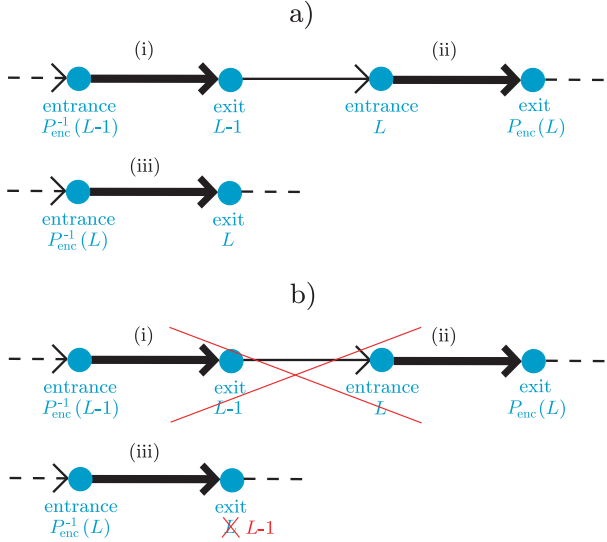}
  \parbox{1.5cm}{} \parbox{6.5cm}{ \vspace{-8cm}

\caption{
  a) Thick lines: Three encounter stretches of an orbit $\gamma'$,
  connecting entrance ports to exit ports. The stretches depicted may
  belong to either the same or different encounters. Thin line: Loop
  connecting exit $L-1$ to exit $L$.  In b), the above loop is
  removed, and exit $L$ is renamed as exit $L-1$.  }
\label{fig:recursion_unit}}
\end{center}
\end{figure}

We need a mapping that leads from a given permutation $ P_{\rm
  enc}\in{\cal M}(\vec{v},l)$ to a permutation $ Q_{\rm enc}$ of
smaller size, with a different cycle structure.  To motivate such a
mapping, we formally remove one orbit loop of $\gamma$ and $\gamma'$.  Let us
consider the sequence of entrance and exit ports traversed by the
partner orbit $\gamma'$, as depicted in Fig. \ref{fig:recursion_unit}a.
Entrance ports are connected to exit ports through encounter stretches
visualized by thick arrows, e.g. leading (i) from entrance $P_{\rm
  enc}^{-1}(L-1)$ to exit $L-1$, (ii) from entrance $L$ to exit
$P_{\rm enc}(L)$, and (iii) from entrance $P_{\rm enc}^{-1}(L)$ to
exit $L$.  It is easy to show that the stretches (i)-(iii) may not
mutually coincide.\footnote{ If (i)=(ii), we would have $L= P_{\rm
    enc}^{-1}(L-1)=P^{-1}P_{\rm loop}(L-1)=P^{-1}(L)$, i.e., $P$ would
  include a 1-cycle.  This would imply that $P$ either decomposes into
  several cycles, or there is only one permuted element and thus
  $P_{\rm enc}$ contains a 1-cycle.  (i)$\neq$(iii) follows trivially
  from $L-1\neq L$.  (ii)=(iii) would lead to $P_{\rm enc}(L)=L$, i.e.,
  $P_{\rm enc}$ containing a 1-cycle.}  In contrast, loops -- depicted
by thin arrows -- lead from exit ports to entrance ports, e.g. from
exit $L-1$ to entrance $L$.

We now remove the orbit loop with index $L$ leading from exit $L-1$ to
entrance $L$ (including the latter ports!). The remaining exit $L$ has
to be renamed as exit $L-1$; see Fig. \ref{fig:recursion_unit}b.  As a
consequence, entrance $P_{\rm enc}^{-1}(L-1)$ is now connected to exit
$P_{\rm enc}(L)$, and entrance $P_{\rm enc}^{-1}(L)$ is connected to
exit $L-1$. All other intra-encounter connections remain unaffected.
The resulting encounter permutation $Q_{\rm enc}$ thus acts on the
elements $a=1,2,\ldots,L-1$ as
\begin{equation}
\label{QPE_unitary}
 Q_{\rm enc}(a)=\begin{cases}
   P_{\rm enc}(L)&\text{if $a= P_{\rm enc}^{-1}(L-1)$}\\
  L-1&\text{if $a= P_{\rm enc}^{-1}(L)$}\\
   P_{\rm enc}(a)&\text{otherwise}\;.
\end{cases}
\end{equation}

Since we only removed an orbit loop, both $\gamma$ and $\gamma'$ remain
connected periodic orbits. The ordering of entrance ports in $\gamma$ and
$\gamma'$ is given by single-cycle permutations $Q^\gamma=Q_{\rm loop}$ and
$Q$, obtained from $P^\gamma=P_{\rm loop}$ and $P$ by simply removing the
entrance port $L$; in particular $Q_{\rm loop}=(1,2,\ldots,L-1)$. Thus,
$Q_{\rm enc}$ automatically satisfies the restriction of producing a
single-cycle permutation $Q=Q_{\rm loop}Q_{\rm enc}$.\footnote{
  Readers interested in a more formal proof can simply define $Q_{\rm
    enc}=Q_{\rm loop}^{-1}Q$ with $Q_{\rm loop}=(1,2,\ldots,L-1)$ and $ Q$
  differing from $ P$ only by mapping the predecessor of $L$, i.e., $
  P^{-1}(L)$, to the successor of $L$, i.e., $ P(L)$.
  To verify (\ref{QPE_unitary}), we then use that $ Q_{\rm loop}$
  differs from $ P_{\rm loop}$ only in the mapping of one number, same
  as for $ Q$ and $ P$.  Thus $ Q_{\rm enc}$ acts like $ P_{\rm enc}$
  on all but two numbers $a$.  These exceptional cases, given in the
  first two lines of Eq.  (\ref{QPE_unitary}), are checked by
  carefully applying the above definitions of $ Q_{\rm loop}$ and $ Q$
  as follows
\begin{eqnarray}
\label{check_unitary}
   Q_{\rm enc} P_{\rm enc}^{-1}(L-1)&=& Q_{\rm loop}^{-1} Q P^{-1} P_{\rm loop}(L-1)=
 Q_{\rm loop}^{-1} Q P^{-1}(L)\nonumber\\
  &=& Q_{\rm loop}^{-1} P(L)
  {\stackrel{(*)}{=}} P_{\rm loop}^{-1} P(L)= P_{\rm enc}(L)\nonumber\\
   Q_{\rm enc} P_{\rm enc}^{-1}(L)&=& Q_{\rm loop}^{-1} Q P^{-1} P_{\rm loop}(L)
  = Q_{\rm loop}^{-1} Q P^{-1}(1)\nonumber\\
&{\stackrel{(**)}{=}}& Q_{\rm loop}^{-1} P P^{-1}(1)
= Q_{\rm loop}^{-1}(1)=L-1 \,;
\end{eqnarray}
here, we used $ P(L)\neq 1$ for $(*)$, since otherwise $ P_{\rm enc}$
would have a 1-cycle (i.e., $ P_{\rm enc}(L)= P_{\rm loop}^{-1} P(L)=
P_{\rm loop}^{-1}(1)=L$), and $ P(L)\neq L$, since otherwise $ P$ would
have a 1-cycle. To check $(**)$, we need $ P^{-1}(1)\neq L$ (since $
P(L)\neq 1$) and $ P^{-1}(1)\neq P^{-1}(L)$.  These arguments are
analogous to those invoked when showing that the three stretches in
Fig. \ref{fig:recursion_unit}a may not mutually coincide.  }

We need to connect the cycle structures of $ Q_{\rm enc} $ and $
P_{\rm enc}$, in particular the corresponding vectors $\vec{v}$ and
the placement of the largest permuted number. To do so, we shall
assume $P_{\rm enc}\in{\cal M}(\vec{v},l)$ with arbitrary $l$ and
distinguish between two cases. (For the unitary form factor, we only
need $l=2$ and the first case, but the remaining considerations are an
essential preparation for the treatment of time-reversal invariant
systems.)

\subsubsection{First case: $L-1$ and $L$ belong to different cycles}

Let us first consider the case that {\it the element $L-1$ of the
  permutation $ P_{\rm enc}$ belongs to a different cycle than $L$},
say a $k$-cycle. ($k$ may be equal to $l$.) Hence, $ P_{\rm enc}$ has
the form
\begin{equation}
  \label{pcyc}  P_{\rm enc}=[\ldots] (L-1,a_2,a_3,\ldots a_k)
  (L,b_2,b_3,\ldots
  b_l) \end{equation}
where the two aforementioned cycles are written in round brackets, and
$[\ldots]$ stands for all other cycles.  Then, the $Q_{\rm enc}$ given in
(\ref{QPE_unitary}) differs from $P_{\rm enc}$ by mapping $ P_{\rm
  enc}^{-1}(L-1)=a_k$ to $ P_{\rm enc}(L)=b_2$, and $ P_{\rm
  enc}^{-1}(L)=b_l$ to $L-1$. It follows that the above $k$- and
$l$-cycles of $ P_{\rm enc}$ merge to a $(k+l-1)$-cycle of $Q_{\rm
  enc}$. We may write
\begin{equation} \label{qcyc}  Q_{\rm enc} =[\ldots]
  (L-1,a_2,a_3,\ldots a_k,b_2,b_3,\ldots b_l)
\end{equation}
where $[\ldots]$ is the same as in Eq. (\ref{pcyc}).  Compared to $ P_{\rm
  enc}$, $ Q_{\rm enc}$ has one $k$-cycle and one $l$-cycle less, but
one additional $(k+l-1)$-cycle.  The changed ``vector'' with $v_k\to
v_k-1,v_l\to v_l-1,v_{k+l-1}\to v_{k+l-1}+1$ will be denoted as
$\vec{v}^{\;[k,l\to k+l-1]}$.  In general, $\vec{v}^{[\alpha_1,\ldots,\alpha_m \to
  \beta_1,\ldots,\beta_n]}$, $m\geq0$, $n\geq0$ denotes the vector obtained from
$\vec{v}$ if we decrease all $v_{\alpha_i}$ by one, increase all
$v_{\beta_i}$ by one, and leave all other components unchanged.  If one
number appears several times on either the left or the right hand
side, the corresponding component of $\vec{v}$ is respectively
decreased or increased by the number of occurrences; if no $\beta_i$
appear on the right-hand side, no components of $\vec{v}$ are
increased.

The permutation $ Q_{\rm enc}$ belongs to the subset ${\cal
  M}(\vec{v}^{[k,l\to k+l-1]},k+l-1)$ since the largest permuted number
$L-1$ is included in a cycle with the length $k+l-1$.  We have seen
that the additional restriction of $Q=Q_{\rm loop}Q_{\rm enc}$ being
single-cycle is satisfied by construction.

Each $ P_{\rm enc}$ of the form (\ref{pcyc}) (with $k$ and $l$ fixed)
generates one member of the set ${\cal M}(\vec{v}^{[k,l\to
  k+l-1]},k+l-1)$.  Vice versa, for fixed $k$ the $ Q_{\rm enc}$ given
in Eq. (\ref{qcyc}) uniquely determines one $ P_{\rm enc}$ as given in
Eq. (\ref{pcyc}): For each $Q_{\rm enc}$, we may simply read off the
elements $a_2,\ldots,a_k,b_2,\ldots,b_l$ by comparison with (\ref{qcyc}) and
then rearrange them to form a permutation $P_{\rm enc}$ as in
(\ref{pcyc}).  Hence, there are
\begin{equation}
\label{no_1}
N(\vec{v}^{[k,l\to k+l-1]},k+l-1)\,
\end{equation}
members of ${\cal M}(\vec{v},l)$ structured like Eq. (\ref{pcyc}).

Physically, the present scenario is analogous to the merger of a $k$-
and an $l$-encounter into a $(k+l-1)$-encounter, by shrinking away an
intervening loop.

\subsubsection{Second case: $L-1$ and $L$ belong to the same cycle}

We now turn to the second scenario where {\it $L$ and $L-1$ belong to
  the same $l$-cycle of $ P_{\rm enc}$}.  In this case, the element
$L$ has to follow $L-1$ after a certain number $m$ of iterations of
$P_{\rm enc}$, i.e., $L=P_{\rm enc}^m(L-1)$.  The number $m$ must
satisfy $1\leq m\leq l-2$; $m=l-1$, and thus $L=P_{\rm
  enc}^{l-1}(L-1)=P_{\rm enc}^{-1}(L-1)$, is excluded since otherwise
the stretches (i) and (ii) of Fig. \ref{fig:recursion_unit}a would
coincide.  The permutation $ P_{\rm enc}$ will be of the form
\begin{equation}
  \label{pmcyc}  P_{\rm enc}=[\ldots] (L-1,a_2,a_3,\ldots a_{m},L,a_{m+2},
\ldots, a_l)\,.
\end{equation}
According to Eq. (\ref{QPE_unitary}), $ Q_{\rm enc}$ differs from $
P_{\rm enc}$ by mapping $ P_{\rm enc}^{-1}(L-1)=a_l$ to $ P_{\rm
  enc}(L)=a_{m+2}$ and mapping $ P_{\rm enc}^{-1}(L)=a_{m}$ to $L-1$.
The permutation $Q_{\rm enc}$ thus reads
\begin{equation}
  \label{breakup}  Q_{\rm enc} =[\ldots] (L-1,a_2,a_3,\ldots
  a_{m})(a_{m+2},\ldots, a_l)\,. \end{equation}
Here, the $l$-cycle of $ P_{\rm enc}$ is broken up into two cycles,
with the lengths $m$ and $l-m-1$. Since the largest number $L-1$ is
included in an $m$-cycle, $ Q_{\rm enc}$ belongs to ${\cal
  M}(\vec{v}^{[l\to m,l-m-1]},m)$.

In contrast to the first scenario, there are typically several $P_{\rm
  enc}$ producing the same $ Q_{\rm enc}$.  Indeed Eq. (\ref{breakup})
would not only result from Eq. (\ref{pmcyc}), but also from all
$l-m-1$ permutations $P_{\rm enc}$ obtained by cyclic permutation of
the last elements $a_{m+2},\ldots,a_l$ in Eq. (\ref{pmcyc}).  Besides,
$[\ldots]$ in $P_{\rm enc}$ contains $v_{l-m-1}$ cycles of length $l-m-1$.
If we transpose the content of one of these cycles with the
subsequence $a_{m+2},\ldots,a_l$ in Eq. (\ref{pmcyc}), the resulting
$P_{\rm enc}$ will lead to the same $Q_{\rm enc}$; in any of these
cases $Q_{\rm enc}$ will contain the $v_{l-m-1}$ cycles of length
$l-m-1$ included in the $[\ldots]$ of Eq. (\ref{pmcyc}) and one additional
such cycle given by $(a_{m+2},\ldots,a_l)$.  Thus, there are
$(l-m-1)(v_{l-m-1}+1)$ permutations $P_{\rm enc}$ producing the same
$Q_{\rm enc}$, due to the $v_{l-m-1}+1$ possibilities for transposing
(or keeping) $a_{m+2},\ldots,a_l$, and $l-m-1$ possibilities for cyclicly
permuting them.  Consequently, for each $m$ the subset of elements $
P_{\rm enc}\in{\cal M}(\vec{v},l)$ structured like Eq. (\ref{pmcyc}) is
$(l-m-1)(v_{l-m-1}+1)$ times larger than ${\cal M}(\vec{v}^{[l\to
  m,l-m-1]},m)$, i.e., it has the size
\begin{equation}
  (l-m-1)(v_{l-m-1}+1)N(\vec{v}^{[l\to m,l-m-1]},m) \,.
\end{equation}

The second scenario, too, has an interesting physical interpretation.
If $L-1$ and $L$ both belong to the same $l$-cycle, they must
correspond to two subsequent parallel stretches of the same encounter.
If we remove the intervening orbit loop, these stretches will overlap;
this overlap will be studied in more detail in Appendix
\ref{sec:parallel_overlap}.

\subsubsection{Resulting recursion}

We have decomposed ${\cal M}(\vec{v},l)$ into several subsets of size
$N(\vec{v}^{[k,l\to k+l-1]},k+l-1)$, $k\geq 2$, and further subsets of
size $(l-m-1)(v_{l-m-1}+1)N(\vec{v}^{[l\to m,l-m-1]},m)$, with
$m=1,\ldots,l-2$.  The size of ${\cal M}(\vec{v},l)$ thus reads
\begin{eqnarray}
\label{NL} N(\vec{v},l)&=&\sum_{k\geq
    2}N(\vec{v}^{[k,l\to k+l-1]},k+l-1)\nonumber\\
    &+&\sum_{m=1}^{l-2}(l-m-1)(v_{l-m-1}+1)N(\vec{v}^{[l\to m,l-m-1]},m)\,.
\end{eqnarray}
Using $N(\vec{v},l) = \frac{l v_l}{L} N(\vec{v})$, we find the desired
recursion for the number of structures $N(\vec{v})$,
\begin{eqnarray}
\label{Nrecur1}
\frac{lv_l}{L}N(\vec{v})&=&\sum_{k\geq 2}\frac{(k+l-1)(v_{k+l-1}+1)}{L-1}
N(\vec{v}^{[k,l\to k+l-1]}) \\
&+&\sum_{m=1}^{l-2}\frac{(l-m-1)(v_{l-m-1}+1)m v_m^{[l\to m,l-m-1]}}{L-1}
N(\vec{v}^{[l\to m,l-m-1]})\nonumber\,.
\end{eqnarray}
Alternatively, we may formulate our recursion in terms of the numbers
$\tilde{N}(\vec{v})=N(\vec{v})\frac{(-1)^V\prod_l l^{v_l}}{L(\vec{v})}$
determining the contributions to the spectral form factor,
\begin{eqnarray}
\label{recur1}
&&v_l \tilde{N} (\vec{v})+ \sum_{k\geq 2}(v_{k+l-1}+1)k \tilde{N}(
\vec{v}^{[k,l\to k+l-1]})\nonumber\\
&&+  \sum_{m=1}^{l-2}(v_{l-m-1} +1)v_{m}^{[l\to
  m,l-m-1]} \tilde{N}(\vec{v}^{[l\to m,l-m-1]})=0 \,.
\end{eqnarray}
Note that $v_{k+l-1}+1=v_{k+l-1}^{[k,l\to k+l-1]}$.  Of course, the
$k$th summand vanishes if there are no $k$-cycles present, i.e., if
$v_k=0$ and thus formally $v_k^{[k,l\to k+l-1]}=-1$.

To determine the form factor for systems without time-reversal
invariance, we need only the special case $l=2$. In this case, our
recursion strongly simplifies,
\begin{equation}\label{rec2_unit}
  v_2\tilde{N}(\vec{v})+\sum_{k\geq 2}v_{k+1}^{[k,2\to k+1]}k \tilde{N}(\vec{v}^{[k,2\to k+1]})=0\,,
\end{equation}
since only the first of the two above scenarios is possible.  That is,
a 2-cycle may only merge with a $k$-cycle to form a $(k+1)$-cycle, but
not split into two separate cycles.  Recall that $\vec{v}^{[k,2\to
  k+1]}$ is obtained from $\vec{v}$ by decreasing both $v_k$ and $v_2$
by one, and increasing $v_{k+1}$ by one.

\subsection{Spectral form factor}

\label{sec:formfactor_unitary}

We had expressed the Taylor coefficients of the form factor as a sum
over the combinatorial numbers $\tilde{N}(\vec{v})$,
\begin{equation} K_n=\frac{1}{(n-2)!}\sum_{\vec{v}}^{\nu(\vec{v})=n}\tilde{N}(\vec{v})\,,\quad n\geq2\,,
\end{equation}
see Eq. (\ref{kn}), where the sum runs over all $\vec{v}$ with $v_1=0$
which satisfy $\nu(\vec{v})\equiv L(\vec{v})-V(\vec{v})+1=n$; the condition
$v_1=0$ is not written out explicitly.  Our recursion relation for
$\tilde{N}(\vec{v})$ now translates into a recursion for $K_n$, albeit
a trivial one in the unitary case, implying that all $K_n$ except
$K_1$ vanish.
(Alternatively, one may use a rather involved explicit formula for
$N(\vec{v})$ \cite{JMueller}.)

To show this, consider the recursion (\ref{rec2_unit}) for
$\tilde{N}(\vec{v})$ and sum over $\vec{v}$ as above
\begin{equation}
\label{gue_sum}
\sum_{\vec{v}}^{\nu(\vec{v})=n}\left(v_2\tilde{N}(\vec{v})+\sum_{k\geq 2}v_{k+1}^{[k,2\to
    k+1]}k\tilde{N}(\vec{v}^{[k,2\to k+1]})\right)=0\,.
\end{equation}
Each of the sums
\begin{equation}
\label{partialsum}
\sum_{\vec{v}}^{\nu(\vec{v})=n}v_{k+1}^{[k,2\to
  k+1]}\tilde{N}(\vec{v}^{[k,2\to k+1]})
\end{equation}
may be transformed into a sum over the argument of $\tilde{N}$, i.e.,
$\vec{v}'=\vec{v}^{[k,2\to k+1]}$.  The vectors $\vec{v}'$ must satisfy
three restrictions.  First, by definition we must have $v_{k+1}'\geq 1$.
However, this restriction may be dropped because due to the prefactor
$v'_{k+1}$ terms with $v'_{k+1}=0$ do not contribute to
(\ref{partialsum}).  Second, $\nu(\vec{v})=n$ implies that
$\nu(\vec{v}')=n$, since going from $\vec{v}$ to $\vec{v}'$ decreases
both $L$ and $V$ by one and thus leaves $\nu$ invariant.  Third, just
like $\vec{v}$ the vectors $\vec{v}'$ must fulfill the implicit
condition $v_1'=0$.  We thus have to sum over all $\vec{v}'$ with
$\nu(\vec{v}')=n$ (and $v_1'=0$), and simplify (\ref{partialsum}) as in
\begin{equation}
\label{rule_special}
\sum_{\vec{v}}^{\nu(\vec{v})=n}v_{k+1}^{[k,2\to
  k+1]}\tilde{N}(\vec{v}^{[k,2\to k+1]})=\sum_{\vec{v}'}^{\nu(\vec{v}')=n}v'_{k+1}\tilde{N}(\vec{v}')\,.
\end{equation}
Applying this rule to all terms in Eq. (\ref{gue_sum}) and dropping
the primes, we obtain
\begin{equation}
\sum_{\vec{v}}^{\nu(\vec{v})=n}\Big(v_2+\sum_{k\geq2}v_{k+1}k\Big)\tilde{N}(\vec{v})=0\,.
\end{equation}
Since the term in parentheses is just $\sum_{l\geq
  2}v_l(l-1)=L-V=\nu-1=n-1$ we have
\begin{equation} (n-1)\sum_{\vec{v}}^{\nu(\vec{v})=n}\tilde{N}(\vec{v})= (n-1)!\,K_n=0\,,
\end{equation}
for all $n\geq 2$.

We see that all Taylor coefficients except $K_1$ vanish: orbit pairs
differing in encounters yield no net contribution to the form factor,
but mutually compensate. Only the ``diagonal'' orbit pairs remain, and
indeed lead to the same small-time form factor as predicted by
random-matrix theory for the GUE.

\section{Orthogonal case}

\label{combinatorics_orthogonal}

\subsection{Structures and permutations}

\label{orthogonal_intro}

For time-reversal invariant systems, too, structures of orbit pairs
can be related to permutations. Given time-reversal invariance, the
partners of an orbit $\gamma$ may involve loops of both $\gamma$ and its
time-reversed $\overline\gamma$.  All permutations used will thus
simultaneously describe $\gamma$ and $\overline\gamma$, or both one partner
$\gamma'$ and its time-reversed $\overline{\gamma'}$.  Compared to the unitary
case, the permuted elements will be doubled in number.

\begin{figure}[t]
\begin{center}
  \includegraphics[scale=0.5]{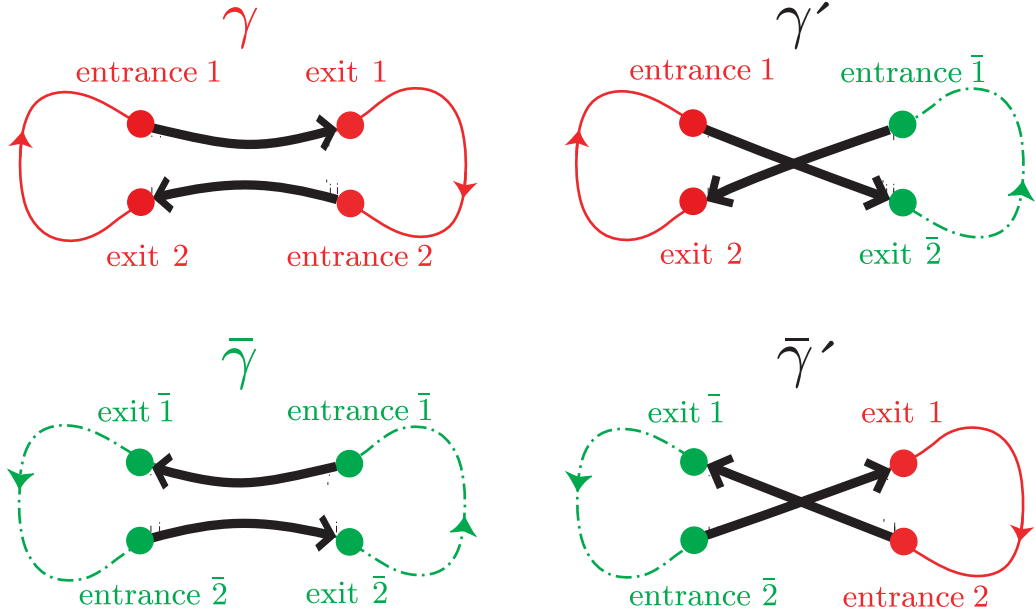}     %was 0.47
\end{center}
\caption{Entrance-to-exit connections for a
  Sieber/Richter pair, in an orbit $\gamma$, its time-reversed
  $\overline{\gamma}$, and the partners $\gamma',\overline{\gamma'}$.  Loops of
  $\gamma$ are depicted by full lines, loops of $\overline{\gamma}$ by
  dash-dotted lines.  }
\label{fig:2notation}
\end{figure}

The above orbits differ in a certain number of encounters.  Again, the
encounter stretches of $\gamma$ will be numbered in order of traversal as
$1,2,\ldots,L$.  Each of these stretches leads from an entrance to an exit
port, both to be labelled by the same index as the corresponding
encounter stretch.  The encounter stretches of $\overline\gamma$ are
time-reversed with respect to those of $\gamma$, and will be labelled by
$\overline{1},\overline{2},\ldots, \overline{L}$, with $\overline a$
denoting the time-reversed of the $a$th stretch of $\gamma$.  Stretch
$\overline{a}$ of $\overline{\gamma}$ leads from entrance port
$\overline{a}$ to exit port $\overline{a}$.  These ports coincide with
those of $\gamma$ in configuration space, but are time-reversed and have
entrance and exit swapped. The exit port $\overline{a}$ of
$\overline{\gamma}$ is the time-reversed of the entrance port $a$ of $\gamma$,
and entrance port $\overline{a}$ of $\overline{\gamma}$ is the
time-reversed of the exit port $a$ of $\gamma$.  See Fig.
\ref{fig:2notation} for the example of a 2-encounter.

The {\bf intra-encounter connections} of $\gamma$ and $\overline{\gamma}$ are
represented by the trivial permutation $P_{\rm enc}^\gamma=\left({1\atop
    1}{2\atop 2}{\ldots\atop \ldots}{L\atop L}{\overline{1}\atop
    \overline{1}}{\overline{2}\atop \overline{2}}{\ldots\atop
    \ldots}{\overline{L} \atop \overline{L}}\right)$ indicating that each
entrance port (upper line) is connected to an exit (lower line) with
the same index.

The corresponding connections in $\gamma'$, $\overline{\gamma'}$ are given by
a different permutation $\PE$.  In the example of a Sieber/Richter
pair, Fig.  \ref{fig:2notation}, the partner $\gamma'$ connects the
entrance 1 of $\gamma$ to the exit $\overline{2}$ of $\overline{\gamma}$, and
the entrance $\overline{1}$ of $\overline{\gamma}$ to the exit 2 of $\gamma$.
The time-reversed partner $\overline{\gamma'}$ connects the entrance
$\overline 2$ to exit 1, and entrance 2 to exit $\overline 1$.  We
thus write $P_{\rm enc}=\left({1\atop \overline{2}}{2\atop
    \overline{1}}{\overline{1}\atop 2} {\overline{2}\atop 1}\right)$.
(Note that the sequence of columns in $P_{\rm enc}$ may be ordered
arbitrarily!)
%We shall mostly order such
%that the first lines in $P_{\rm enc}$ and $P_{\rm enc}^\gamma$ coincide;
%columns describing $\gamma'$ and $\overline{\gamma'}$ may thus become mutually
%interspersed.

In general, time-reversal invariance imposes the following restriction
on $P_{\rm enc}$: If a stretch connects entrance $a$ to exit $b$, the
time-reversed stretch must lead from entrance $\overline b$ (the
time-reversed of exit $b$) to exit $\overline a$ (the time-reversed of
entrance $a$). This restriction also applies if $a$ or $b$ belong to
$\overline\gamma$ and thus contain an overbar, e.g. $a=\overline p$; we
then define $\overline{\overline{p}}=p$. In any case, {\it if $P_{\rm
    enc}$ maps $a$ to $b$, it has to map $\overline{b}$ to
  $\overline{a}$}, with $a,b$ standing for elements out of $1,\ldots,
L,\overline 1,\ldots,\overline L$.  This restriction on $P_{\rm enc}$ will
be referred to as \textit{``time-reversal covariance''}. Writing
%be referred to as {\it ``time-reversal covariance''}.  Writing
$\T(a)\equiv\overline a$, a given $P_{\rm enc}$ is time-reversal covariant
if $\T P_{\rm enc}\T=P_{\rm enc}^{-1}$ such that $\T P_{\rm
  enc}\T(b)=\T P_{\rm enc}(\overline{b})=\T(\overline{a})=a= P_{\rm
  enc}^{-1}(b)$ holds for arbitrary $b$. Obviously, for Sieber/Richter
pairs the $P_{\rm enc}$ given above is time-reversal covariant.

Again, $l$-encounters are related to cycles of $\PE$. If we take into
account both $\gamma$ and $\overline{\gamma}$, each encounter contains $2l$
stretches -- $l$ stretches belonging to $\gamma$ and $l$ stretches
belonging to $\overline{\gamma}$.  Alternatively, the $2l$ stretches may
be grouped into $l$ stretches pointing ``from left to right'' (i.e.,
parallel to the first stretch inside $\gamma$ belonging to the
corresponding encounter), and $l$ stretches leading from right to
left.  The stretches of each group are mutually close in phase space,
and approximately time-reversed with respect to those of the other
group.  The two groups may mix stretches of $\gamma$ and $\overline{\gamma}$;
only for parallel encounters all stretches of $\gamma$ point from left to
right and all stretches of $\overline{\gamma}$ point from right to left.
The partner orbits $\gamma'$, $\overline{\gamma'}$ can be obtained by
reconnecting stretches {\it inside these two groups}, without having
to revert stretches or ports in time.

Each of the two groups corresponds to one $l$-cycle of $\PE$.  To show
this, we reason similarly to Subsection \ref{sec:intro_unitary}.  Two
stretches with labels $a$ and $P_{\rm enc}(a)$ must belong to the same
group, such that the entrance of $a$ can be connected to the exit of
$P_{\rm enc}(a)$.  Starting with an arbitrary label $a$, iteration
yields a complete list $a,\PE(a),\PE^2(a),\PE^3(a),\ldots$ of stretches
involved in mutual reconnections and thus belonging to the same group.
After $l$ iterations, we are led back to the initial element $a$ and
thus obtain an $l$-cycle.  Consequently, each $l$-encounter
corresponds to a pair of ``twin'' $l$-cycles.  (An encounter
associated with more cycles would decompose into several encounters
with independent reconnections, similarly to the encounters depicted
in Fig.  \ref{fig:permutations}c for the unitary case.)

The stretches of both groups are mutually time-reversed.  If one group
contains a stretch leading from entrance $a_i$ to exit $a_{i+1}$, the
stretch leading from entrance $\overline{a_{i+1}}$ to exit
$\overline{a_i}$ is part of the other group.  This behavior is
mirrored by the two associated cycles, which must be of ``mutually
time-reversed'' form $(a_1,a_2,\ldots,a_{l-1},a_l)$,
$(\overline{a_l},\overline{a_{l-1}},\ldots,
\overline{a_2},\overline{a_1})$.  For example, the permutation $P_{\rm
  enc}=\left({1\atop \overline{2}}{2\atop
    \overline{1}}{\overline{1}\atop 2} {\overline{2}\atop 1}\right)$
describing reconnections in Sieber/Richter pairs has two mutually
time-reversed cycles $(1,\overline{2})$ and $(2,\overline{1})$, with
$(1,\overline{2})$ representing the stretches leading from left to
right, and $(2,\overline{1})$ representing the stretches directed from
right to left.

The {\bf orbit loops} are associated with $P_{\rm loop}=\left({1\atop
    2}{2\atop 3} {\ldots\atop \ldots}{L\atop 1}{\overline{2}\atop
    \overline{1}}{\overline{3} \atop \overline{2}}{\ldots\atop
    \ldots}{\overline{1}\atop \overline{L}}\right)$, since if one loop of
$\gamma$ leads from the exit of the $(a-1)$-st stretch to the entrance of
the $a$-th one, its time-reversed must go from exit $\overline{a}$ to
entrance $\overline{a-1}$.  The loops remain unchanged in the partner
orbits $\gamma'$, $\overline{\gamma'}$ and are thus always described by the
same $P_{\rm loop}$.

The product $P^\gamma=P_{\rm loop} P_{\rm enc}^\gamma=P_{\rm loop}$ specifies
the ordering of entrance ports along the two {\bf orbits} $\gamma$ and
$\overline{\gamma}$. That $P^\gamma$ has two cycles $(1,2,\ldots,L)$ and
$(\overline{L},\overline{L-1},\ldots,\overline{1})$, one each for $\gamma$ and
$\overline{\gamma}$.

The ordering of entrance ports along the partners $\gamma'$,
$\overline{\gamma'}$ is given by the product $P=P_{\rm loop}P_{\rm enc}$.
To obtain two connected partner orbits $\gamma'$ and $\overline{\gamma'}$, we
now have to demand $P$ to consist of only two $L$-cycles, listing the
entrance ports in $\gamma'$ and $\overline{\gamma'}$, respectively.  This
provides a restriction on $P_{\rm enc}$ analogous to the unitary case.

The two cycles of $P$ are not independent.  The second cycle
containing the entrance ports of $\overline{\gamma'}$ can also be
interpreted as the list of exit ports of $\gamma'$, time-reversed and
written in reverse order.  Since an entrance $a_i$ is connected by a
loop to the exit $b_i\equiv P_{\rm loop}^{-1}(a_i)$, the two cycles of $P$
must be of the form $(a_1,a_2,\ldots,a_L)$ and
$(\overline{b_L},\overline{b_{L-1}},\ldots, \overline{b_1})$.  It is easy
to show that this form follows immediately from time-reversal
covariance and the existence of two cycles, and thus does not provide
a further restriction on $P_{\rm enc}$.

Each of the permutations $P_{\rm enc}$ corresponds to one structure of
orbit pairs as defined in Section \ref{sec:overview}.  The division of
labels $1,2,\ldots,L,\overline{1}, \overline{2},\ldots,\overline{L}$ into
pairs of cycles determines how stretches are divided among encounters.
By further dividing labels (and thus stretches) of one encounter in
two ``mutually time-reversed'' cycles, we fix the mutual orientation
of stretches inside each encounter.  Finally, the ordering of labels
inside cycles determines the reconnections leading to $\gamma'$,
$\overline{\gamma'}$ which have to be connected periodic orbits if $P$
falls in two $L$-cycles.

To establish a recursion for the numbers of structures $N(\vec{v})$,
we may consider the set ${\cal M}(\vec{v})$ of permutations $P_{\rm
  enc}$ acting on $1,2,\ldots,L,\overline{1},\overline{2},\ldots,\overline{L}$
which (i) are time-reversal covariant, (ii) have $v_l$ pairs of
$l$-cycles for all $l\geq2$, and (iii) lead to a permutation $P=P_{\rm
  loop} P_{\rm enc}$ consisting of two $L$-cycles.  Each element
$P_{\rm enc}$ of the set ${\cal M}(\vec{v})$ stands for one of the
structures related to $\vec{v}$.  We need to calculate the number
$N(\vec{v})$ of elements of ${\cal M}(\vec{v})$.

\subsection{Examples}

\begin{table}
\begin{center}
\begin{tabular}{|c|c|c|c|r|r|r|} \hline
order & $\vec{v}$ & $L$ & $V$ & $N(\vec{v})$ & $\tilde{N}(\vec{v})$ & contribution \\
\hline\hline
$\tau^2$
& $(2)^1$ &2 & 1 & 1 & $-1$ & $-2\tau^2$\;\;\;\; \\ \hhline{~------}
\hhline{~------}
&&&&&$-1$&$-2\tau^2$\;\;\;\; \\
\hline\hline
$\tau^3$
& $(2)^2$ &4 & 2 & 5 & 5 & $10\tau^3$\;\;\;\; \\ \hhline{~------}
& $(3)^1$ &3 & 1 & 4 & $-4$ & $-8\tau^3$\;\;\;\; \\ \hhline{~------}
\hhline{~------}
&&&&&1&2$\tau^3$\;\;\;\; \\
\hline\hline
$\tau^4$
& $(2)^3$ &6 & 3 & 41 & $-\frac{164}{3}$ & $-\frac{164}{3}\tau^4$\;\;\;\; \\ \hhline{~------}
& $(2)^1(3)^1$ &5 & 2 & 60 & 72 & $72\tau^4$\;\;\;\; \\ \hhline{~------}
& $(4)^1$ &4 & 1 & 20 & $-20$ & $-20\tau^4$\;\;\;\; \\ \hhline{~------}
\hhline{~------}
&&&&&$-\frac{8}{3}$&$-\frac{8}{3}\tau^4$\;\;\;\; \\
\hline\hline
$\tau^5$
& $(2)^4$ &8 & 4 & 509 & 1018 & $\frac{1018}{3}\tau^5$\;\;\;\; \\ \hhline{~------}
& $(2)^2(3)^1$ &7 & 3 & 1092 & $-1872$ & $-624\tau^5$\;\;\;\; \\ \hhline{~------}
& $(2)^1(4)^1$ &6 & 2 & 504 & 672 & $224\tau^5$\;\;\;\; \\ \hhline{~------}
& $(3)^2$ &6 & 2 & 228 & 342 & $114\tau^5$\;\;\;\; \\ \hhline{~------}
& $(5)^1$ &5 & 1 & 148 & $-148$ & $-\frac{148}{3}\tau^5$\;\;\;\; \\ \hhline{~------}
\hhline{~------}
&&&&&12&4$\tau^5$\;\;\;\; \\
\hline\hline
$\tau^6$
& $(2)^5$ &10 & 5 & 8229 & $-\frac{131664}{5}$ & $-\frac{10972}{5}\tau^6$\;\;\;\; \\ \hhline{~------}
& $(2)^3(3)^1$ &9 & 4 & 23160 & 61760 & $\frac{15440}{3}\tau^6$\;\;\;\; \\ \hhline{~------}
& $(2)^2(4)^1$ &8 & 3 & 12256 & $-24512$ & $-\frac{6128}{3}\tau^6$\;\;\;\; \\ \hhline{~------}
& $(2)^1(3)^2$ &8 & 3 & 10960 & $-24660$ & $-2055\tau^6$\;\;\;\; \\ \hhline{~------}
& $(2)^1(5)^1$ &7 & 2 & 5236 & 7480 & $\frac{1870}{3}\tau^6$\;\;\;\; \\ \hhline{~------}
& $(3)^1(4)^1$ &7 & 2 & 4396 & 7536 & $628\tau^6$\;\;\;\; \\ \hhline{~------}
& $(6)^1$ &6 & 1 & 1348 & $-1348$ & $-\frac{337}{3}\tau^6$\;\;\;\; \\ \hhline{~------}
\hhline{~------}
&&&&&$-\frac{384}{5}$&$-\frac{32}{5}\tau^6$\;\;\;\; \\
\hline
\end{tabular}
\end{center}
\caption{Permutations, and thus structures of orbit pairs, giving rise to orders
$\tau^2$ to $\tau^6$ of the form factor, for systems
with time-reversal invariance; notation as in Table \ref{tab:unitary}.
The results coincide with the predictions of RMT for the GOE.}
\label{tab:orthogonal}
\end{table}

Again, the numbers $N(\vec{v})$ can be determined by numerically
counting permutations. From the results shown in Table
\ref{tab:orthogonal}, we see that indeed the form factor of the
Gaussian Orthogonal Ensemble is reproduced semiclassically.

The $\tau^2$ contribution comes from pairs of orbits differing in one
antiparallel 2-encounter, with the ``encounter permutation'' $P_{\rm
  enc}=\left({1\atop \overline{2}}{2\atop
    \overline{1}}{\overline{1}\atop 2} {\overline{2}\atop 1}\right)$.

We have already seen that the $\tau^3$ contribution originates from four
structures related to 3-encounters and five structures related to
pairs of 2-encounters. If all encounters are parallel (structures {\it
  ppi} and {\it pc}), the partner $\gamma'$ is obtained by reconnecting
the ports of $\gamma$, and determined by the permutations given in
Subsection \ref{sec:unitary_examples}.  If we also want to describe
the time-reversed orbits $\overline{\gamma}$ and $\overline{\gamma'}$, we have
to extend the permutations given as required by time-reversal
covariance, i.e., each mapping $a\to b$ must be complemented by a
connection $\overline{b}\to\overline{a}$. This yields $P_{\rm
  enc}=\left({1\atop 3}{2\atop 4}{3\atop 1}{4\atop
    2}{\overline{1}\atop \overline{3}}{\overline{2}\atop
    \overline{4}}{\overline{3}\atop\overline{1}}{\overline{4}
    \atop\overline{2}}\right)
=(1,3)(\overline{3},\overline{1})(2,4)(\overline{4},\overline{2})$ for
{\it ppi} and $P_{\rm enc}=\left({1\atop 2}{2\atop 3}{3\atop
    1}{\overline{1}\atop \overline{3}}{\overline{2}\atop
    \overline{1}}{\overline{3}\atop\overline{2}}\right)
=(1,2,3)(\overline{3},\overline{2},\overline{1})$ for {\it pc}.

\begin{figure}
\begin{center}
  \includegraphics[scale=0.28]{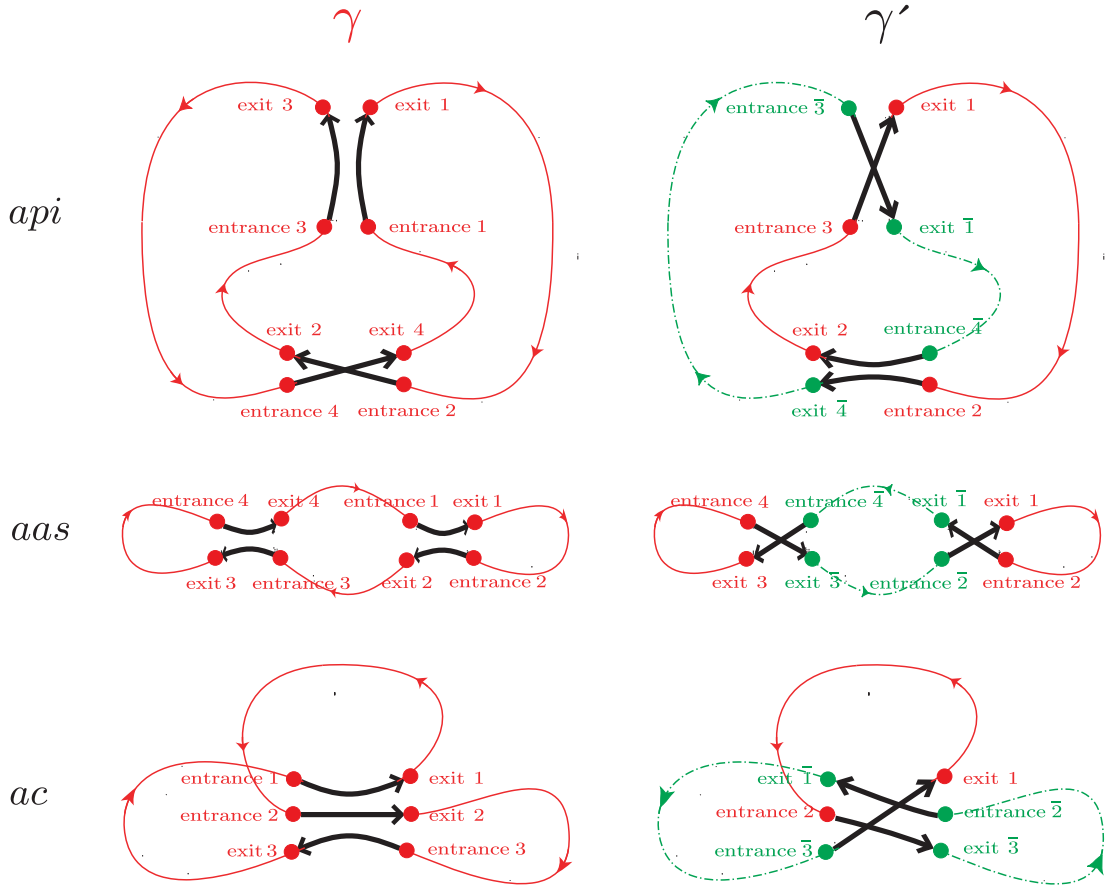}
\end{center}
\caption{
  Connections between entrance and exit ports of orbit pairs
  $(\gamma,\gamma')$ existing only for systems with time-reversal invariance.
  The partner $\gamma'$ mixes loops of $\gamma$ (full lines) and its
  time-reversed $\overline{\gamma}$ (dash-dotted lines).  (See
  \ref{fig:antiparallel_example} for orbit pairs which do not require
  time-reversal invariance.)  For each of the families {\it api}, {\it
    aas}, and {\it ac} we depicted one of several equivalent
  structures; the others are obtained by cyclic permutation of the
  port labels.}
\label{fig:antiparallel_example}
\end{figure}

For the remaining cases {\it api}, {\it aas}, and {\it ac}, the
original orbit $\gamma$ and one of its partners $\gamma'$ are depicted in Fig.
\ref{fig:antiparallel_example}; $\gamma'$ includes ports and loops of both
$\gamma$ and its time-reversed $\overline{\gamma}$ (with loops respectively
depicted by full and dash-dotted lines).  Similarly to {\it ppi}, each
parallel encounter of two stretches $a$ and $b$ leads to two cycles
$(a,b)(\overline{b},\overline{a})$ representing the stretches $a$, $b$
pointing from left to right, and the stretches $\overline{b}$,
$\overline{a}$ pointing from right to left.  In contrast, an
antiparallel 2-encounter of a stretch $a$ and a later stretch $b$
leads to $(a,\overline{b})(b,\overline{a})$ since $a$ and
$\overline{b}$ are directed from left to right, whereas $b$ and
$\overline{a}$ are directed from right to left (like in the above
example of a Sieber/Richter pair).  The two structures related to {\it
  api} (Section \ref{sec:22}) can now immediately be translated into
permutations.  A parallel encounter of stretches 1 and 3, complemented
with an antiparallel encounter of stretches 2 and 4, is represented by
the permutation $P_{\rm
  enc}=(1,3)(\overline{3},\overline{1})(2,\overline{4})(4,\overline{2})=
\left({1\atop 3}{2\atop \overline{4}}{3\atop 1}{4\atop
    \overline{2}}{\overline{1}\atop \overline{3}}{\overline{2}\atop
    4}{\overline{3}\atop\overline{1}} {\overline{4}\atop 2}\right)$.
Indeed, Fig. \ref{fig:antiparallel_example} shows that $\gamma'$ connects
the entrance and exit ports as $3\to1$, $2\to\overline{4}$,
$\overline{3}\to \overline{1}$, and $\overline{4}\to 2$; the remaining
connections belong to $\overline{\gamma'}$.  Different numbering of
stretches leads to a different structure. That structure corresponds
to the permutation $P_{\rm
  enc}=(1,\overline{3})(3,\overline{1})(2,4)(\overline{4},\overline{2})=\left({1\atop
    \overline{3}}{2\atop 4}{3\atop \overline{1}}{4\atop
    2}{\overline{1}\atop 3}{\overline{2}\atop
    \overline{4}}{\overline{3}\atop 1}{\overline{4}\atop
    \overline{2}}\right)$, and involves an antiparallel encounter of
stretches 1 and 3, and a parallel encounter of stretches 2 and 4.

For {\it aas}, antiparallel encounters of stretches 1 and 2, and
stretches 3 and 4 lead to the permutation $P_{\rm
  enc}=(1,\overline{2})(2,\overline{1})(3,\overline{4})(4,\overline{3})=
\left({1\atop \overline{2}}{2\atop \overline{1}}{3\atop
    \overline{4}}{4\atop \overline{3}}{\overline{1}\atop
    2}{\overline{2}\atop 1}{\overline{3}\atop 4}{\overline{4}\atop
    3}\right)$.  Again, the corresponding port connections are
displayed in Fig.  \ref{fig:antiparallel_example}.  If we order
stretches differently, we obtain $P_{\rm
  enc}=(1,\overline{4})(4,\overline{1})(2,\overline{3})(3,\overline{2})
=\left({1\atop \overline{4}}{2\atop \overline{3}}{3\atop
    \overline{2}}{4\atop \overline{1}}{\overline{1}\atop
    4}{\overline{2}\atop 3}{\overline{3}\atop 2}{\overline{4}\atop
    1}\right)$.

Orbit pairs of type {\it ac} are described by three equivalent
structures and thus permutations.  Encounters with stretch 3
antiparallel to stretches 1 and 2 correspond to $P_{\rm
  enc}=(1,2,\overline{3})(3,\overline{2}, \overline{1})=\left({1\atop
    2}{2\atop \overline{3}}{3\atop \overline{2}}{\overline{1}\atop
    3}{\overline{2}\atop \overline{1}}{\overline{3}\atop 1}\right)$,
with cycles respectively standing for the stretches traversed from
left to right, and from right to left.  Two further permutations are
obtained by cyclic permutation of 1,2,3 as well as
$\overline{1},\overline{2},\overline{3}$.

\subsection{Excursion: Left and right ports}

\label{sec:leftright}

%\enlargethispage{0.5cm}%!!!!

So far, we have described intra-encounter connections by permutations
$P_\Enc$ mapping between entrance and exit ports, labelled by
$1,2,\ldots,L,\overline{1},\overline{2}, \ldots,\overline{L}$.  Alternatively,
we might consider permutations $p_{\rm enc}$ describing which left
port (upper line) is connected to which right port (lower line).  When
defining $p_{\rm enc}$, we can identify mutually time-reversed ports
coinciding in configuration space and let $p_{\rm enc}$ act only on
the $L$ numbers $1,2,\ldots,L$.  Such a description would nicely fit with
the treatment of Chapter \ref{sec:geometry}. For instance, Eq.
(\ref{su_partner}), stating that stable and unstable coordinates of a
piercing point are respectively fixed by the associated left and right
ports, could be formulated as $\hat{s}_j'\approx\hat{s}_j$,
$\hat{u}_j'\approx\hat{u}_{p_{\rm enc}(j)}$.  Moreover, each $l$-encounter
corresponds to just one $l$-cycle of $p_{\rm enc}$ rather than a pair
of cycles.\footnote{ The identification of $l$-encounters with
  $l$-cycles of $p_{\rm enc}$ can be justified along the lines of
  Subsection \ref{sec:intro_unitary}.  Moreover, similarly as for the
  unitary case (compare the footnote in Subsection
  \ref{sec:intro_unitary}) our derivation of the action difference $\Delta
  S$ only works for for $l$-encounters which are characterized by a
  single cycle $(a_1,a_2,\ldots,a_l)$ of $p_{\rm enc}$ and therefore can
  be brought to the form $(1,2,\ldots,l)$ by simple renaming.}  However,
it is rather difficult to formulate our recursion for $N(\vec{v})$ in
terms of $p_{\rm enc}$, which is why we use $P_{\rm enc}$ in the
remainder of this thesis.

In the present excursion (which may be skipped by the impatient
reader), we will investigate the relation between $p_{\rm enc}$ and
$P_{\rm enc}$.  To translate between the two, we first consider the
special case when all stretches in every encounter are parallel, i.e.,
directed from left to right in $\gamma$ and from right to left in
$\overline\gamma$.  Then, in the partner $\gamma'$ the left port $a$ of $\gamma$
is connected to the right port $p_\Enc(a)$, and in $\overline{\gamma'}$
the right port $\overline{p_\Enc(a)}$ of $\overline{\gamma}$ has to be
connected to the left port $\overline{a}$.  The permutation
\begin{equation}\label{quasiuni}
  \tilde P_\Enc=\left({1\atop p_\Enc(1)}{\ldots\atop \ldots}\;\;{L\atop p_\Enc(L)}
  {\overline{p_\Enc( 1)}\atop \overline 1}{\ldots\atop \ldots}\;\;
  {\overline{p_\Enc( L)}\atop \overline L}\right)
\end{equation}
thus gives the mapping between entrances (i.e., left ports of $\gamma$ and
right ports of $\overline \gamma$) and exits (i.e., right ports of $\gamma$
and left ports of $\overline{\gamma}$).  Half of the cycles of
$\tilde{P}_\Enc$ are those of $p_\Enc$, another half are their
time-reversed twins.

Now consider the general case when the motion over some of the
encounter stretches is directed from right to left in the original
orbit $\gamma$ and from left to right in the time reversed $\overline \gamma$;
we shall call such stretches ``reverted''.  Eq.  (\ref{quasiuni})
remains true provided the elements in the upper line are interpreted
as left ports of $\gamma$ and right ports of $\overline{\gamma}$ (opposite for
the lower line).  However, we are interested in the encounter
permutation $P_\Enc$ which maps entrance to exit ports.  For a
reverted stretch, e.g., a left port of $\gamma$ is an exit rather than an
entrance.  To convert this port into an entrance, we have to consider
the time-reversed stretch, leading from left to right in
$\overline{\gamma}$.  To convert all elements in the upper line to
entrances and those in the lower line to exits, we have to exchange
$a\leftrightarrow\overline a$ (i.e., apply time reversal) for all ports of reverted
stretches $a$.  That exchange of elements amounts to the permutation
$\Sigma(a)=\overline a,\Sigma(\overline a)=a$ for all reverted stretches while
otherwise $\Sigma(a)= a,\Sigma(\overline a)=\overline a$.  The encounter
permutation $P_\Enc$ is related to $\tilde{P}_\Enc$ by
\begin{equation}
P_\Enc=\Sigma \tilde P_\Enc \Sigma\,.
\end{equation}
Since $\Sigma$ is idempotent ($\Sigma^2=1$) the transformation leading from
$\tilde{P}_{\rm enc}$ to $P_{\rm enc}$ is a similarity transformation
and thus does not change the cycle structure \cite{Permutations}. Each
$l$-cycle of $p_\Enc$ gives rise to two mutually time reversed
$l$-cycles of $\tilde{P}_\Enc$ and therefore to two cycles of $P_{\rm
  enc}$.

\subsection{Recursion relation for $N(\vec{v})$}

\label{sec:orthogonal_recursion}

We are now equipped to establish a recursion relation for $N(\vec{v})$
in much the same way as in the unitary case.  We will finally be led
to a recursion for $K_n$, given in Eq.  (\ref{welcome_back}).

First of all, we recover Eq. (\ref{mengetomenge}),
$N(\vec{v},l)=\frac{v_l l}{L(\vec{v})}N(\vec{v})$ using exactly the
same arguments as in the unitary case.  Each permutation $P_{\rm
  enc}\in{\cal M}(\vec{v})$ can be converted into a member of ${\cal
  M}(\vec{v},l)$ by $lv_l$ cyclic permutations and each member of
${\cal M}(\vec{v},l)$ is accessed through $L(\vec{v})$ permutations.
(The labels with overbar $\overline{1},\overline{2},\ldots,\overline{L}$
have to be permuted in the same way as $1,2,\ldots,L$.)

\subsubsection{Mapping between permutations}

\begin{figure}
\begin{center}
  \includegraphics[scale=0.48]{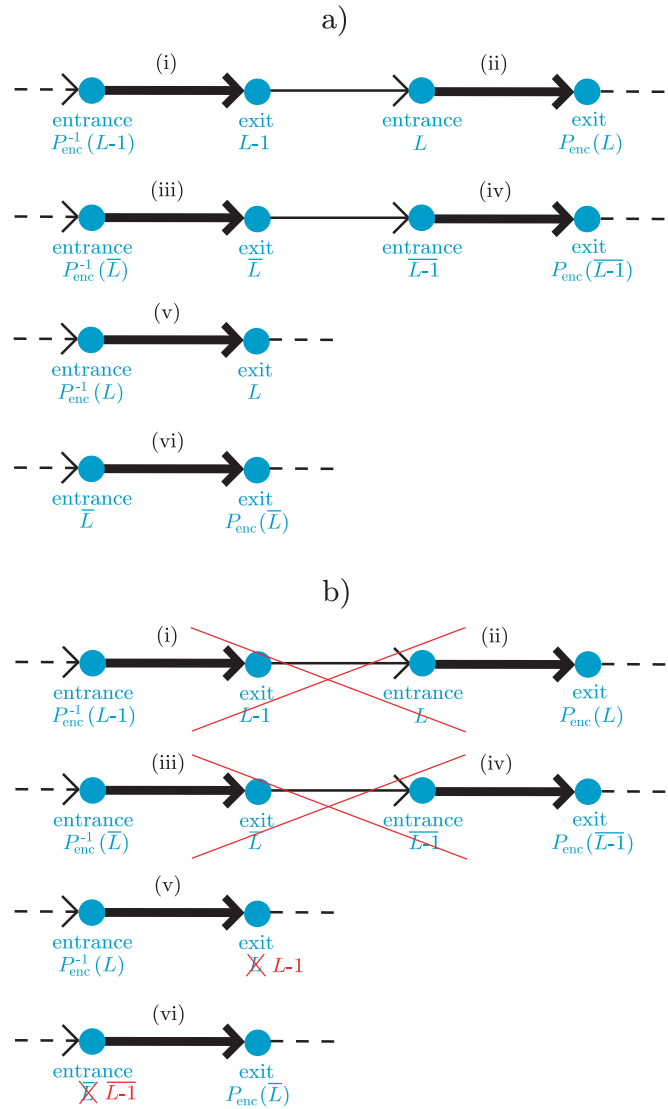}
\end{center}
\caption{
  a) Thick lines: Six encounter stretches of orbits $\gamma'$ and
  $\overline{\gamma'}$, connecting entrance ports to exit ports. The
  stretches depicted may belong to either the same or different
  encounters, and to either $\gamma'$ or $\overline{\gamma'}$.  Thin line:
  Loop connecting exit $L-1$ to entrance $L$, and its time-reversed
  leading from exit $\overline{L}$ to entrance $\overline{L-1}$.  In
  b), the two above loops are removed, exit $L$ is renamed as exit
  $L-1$, and entrance $\overline{L}$ is renamed as entrance
  $\overline{L-1}$.}\label{fig:recursion_orth}
\end{figure}

To define our recursion, we again imagine one orbit loop of $\gamma$ and
$\gamma'$ removed.  Fig. \ref{fig:recursion_orth} visualizes the ordering
of entrance and exit ports along the orbit $\gamma'$ {\it and its
  time-reversed} $\overline{\gamma'}$, with thick arrows denoting
encounter stretches and thin arrows denoting loops.  It is easy to
show that all six stretches (i)-(vi) depicted in Fig.
\ref{fig:recursion_orth} must be different (except for the special
case $P_{\rm enc}(\overline{L-1})= L$ and thus $P_{\rm
  enc}^{-1}(L-1)=\overline{L}$, which will be investigated
later).\footnote{ If (i)=(ii) or (iii)=(iv) the orbits $\gamma$ and
  $\overline{\gamma}$ would contain just one stretch and one loop, i.e.,
  we would have $L=1$. If (i)=(iv) the elements $L-1$ and
  $\overline{L-1}$ would belong to the same cycle of $P_{\rm enc}$
  rather than to mutually time-reversed cycles, same for (ii)=(iii) or
  (v)=(vi) and elements $L$ and $\overline{L}$.  (i)=(vi) and (iv)=(v)
  were excluded above. (ii)=(v) would imply that $P_{\rm enc}$ has a
  1-cycle $(L)$, same for (iii)=(vi) and $(\overline{L})$.  The
  remaining cases are trivial.}

Starting from a permutation $P_{\rm enc}$ we may define a permutation
$Q_{\rm enc}$ acting on the elements $1,2,\ldots,L-1,\overline{1},
\overline{2},\ldots,\overline{L-1}$ by eliminating both the loop leading
from exit $L-1$ to entrance $L$ {\it and its time-reversed} leading
from entrance $\overline{L}$ to exit $\overline{L-1}$. The remaining
exit $L$ has to be renamed as exit $L-1$, and entrance $\overline{L}$
is renamed as entrance $\overline{L-1}$.  The resulting changes are
depicted in Fig.  \ref{fig:recursion_orth}b. A look at the remaining
thick arrows reveals the new encounter permutation as
\begin{equation}
\label{permutation_orthogonal}
Q_{\rm enc}(a)=\begin{cases}
  P_{\rm enc}(L)&\text{if $a=P_{\rm enc}^{-1}(L-1)$}\\
  P_{\rm enc}(\overline{L-1})&\text{if $a=P_{\rm enc}^{-1}(\overline{L})$}\\
  L-1&\text{if $a=P_{\rm enc}^{-1}(L)$}\\
  P_{\rm enc}(\overline{L})&\text{if $a=\overline{L-1}$}\\
  P_{\rm enc}(a)&\text{otherwise}\,.
\end{cases}
\end{equation}

One can easily show that $Q_{\rm enc}$ is a permissible encounter
permutation. First, note that the second and fourth line extend Eq.
(\ref{QPE_unitary}) as required to make $Q_{\rm enc}$ {\it
  time-reversal covariant}.  The mapping $P_{\rm enc}^{-1}(L-1)\to
P_{\rm enc}(L)$ is thus complemented by $\overline{P_{\rm
    enc}(L)}=P_{\rm enc}^{-1}(\overline{L})\to \overline{P_{\rm
    enc}^{-1}(L-1)}= P_{\rm enc}(\overline{L-1})$, and $P_{\rm
  enc}^{-1}(L)\to L-1$ implies $\overline{L-1}\to\overline{P_{\rm
    enc}^{-1}(L)}=P_{\rm enc}(\overline{L})$.  All other mappings $a\to
b$ are complemented by $\overline b\to\overline a$ due to the
time-reversal covariance of $P_{\rm enc}$.

Second, by construction reconnections according to $Q_{\rm enc}$ must
lead to a {\it connected partner orbit}.  The respective ordering of
entrance ports in $\gamma$, $\gamma'$ and $\overline\gamma$, $\overline{\gamma'}$ is
given by permutations $Q^\gamma=Q_{\rm loop}$ and $Q$.  These permutations
are obtained from $P^\gamma=P_{\rm loop}$ and $P$ by removing the entrance
ports $L$ and $\overline{L-1}$ (forming part of the deleted loops) and
renaming entrance $\overline{L}$ as $\overline{L-1}$.  Thus, $Q_{\rm
  loop}$ is given by $Q_{\rm loop}=(1,2,\ldots,L-1),(\overline{L-1},\ldots,
\overline{2},\overline{1})$ and $Q=Q_{\rm loop}Q_{\rm enc}$ indeed
consists of two cycles respectively representing $\gamma'$ and
$\overline{\gamma'}$.\footnote{ Without appealing to the picture of
  encounters and loops, we can again {\it define} $Q_{\rm loop}$ and
  $Q$ as above, and set $Q_{\rm enc}=Q_{\rm loop}^{-1}Q$.  It is easy
  to show that the two cycles of $Q$ satisfy the same relation as
  those of $P$ and may thus indeed be interpreted as lists of entrance
  ports of two time reversed orbits.  The $\QL^{-1}$ thus defined
  differs from $\PL^{-1}$ by mapping 1 to $L-1$, and $\overline{L-1}$
  to $\overline{1}$; $Q$ differs from $P$ by mapping $P^{-1}(L)$ to
  $P(L)$, $P^{-1}(\overline{L-1})$ to $P(\overline{L-1})$,
  $P^{-1}(\overline{L})$ to $\overline{L-1}$, and $\overline{L-1}$ to
  $P(\overline{L})$.  The first and third line of Eq.
  (\ref{permutation_orthogonal}) can be checked as done for the
  unitary case in Eq. (\ref{check_unitary}).  Note that step $(*)$ of
  (\ref{check_unitary}) now also requires $P(L)\neq\overline{L-1}\Leftrightarrow
  P_{\rm enc}(L)\neq\overline{L}$ ($L$ and $\overline{L}$ belong to
  different cycles of $\PE$!), and $(**)$ requires
  $P^{-1}(1)\neq\overline{L-1}\Leftrightarrow\PE(\overline{L-1})\neq L$ (demanded
  above).  The second and fourth line of
  (\ref{permutation_orthogonal}) follow from
\begin{eqnarray}
\label{check_orthogonal}
\QE\PE^{-1}(\overline{L})&=&\QL^{-1}QP^{-1}\PL(\overline{L})
=\QL^{-1}QP^{-1}(\overline{L-1}) \nonumber\\
&=&\QL^{-1}P(\overline{L-1})
{\stackrel{(***)}{=}}\PL^{-1}P(\overline{L-1})
=\PE(\overline{L-1})\nonumber\\
\QE(\overline{L-1})&=&\QL^{-1}Q(\overline{L-1})
=\QL^{-1}P(\overline{L})  \nonumber\\
&{\stackrel{(****)}{=}}&\PL^{-1}P(\overline{L})
=\PE(\overline{L})
\end{eqnarray}
where $(***)$ requires $P(\overline{L-1})\neq\overline{L-1}$ ($P$ has no
1-cycles) and $P(\overline{L-1})\neq 1\Leftrightarrow\PE(\overline{L-1})\neq L$
(demanded above), and $(****)$ follows from
$P(\overline{L})\neq\overline{L-1}\Leftrightarrow\PE(\overline{L})\neq\overline{L}$
($\PE$ has no 1-cycles) and $P(\overline{L})\neq 1\Leftrightarrow\PE(\overline{L})\neq
L$ ($L$ and $\overline{L}$ belong to different cycles).  Moreover, it
is important to check that
\begin{eqnarray}
\QE\PE^{-1}(\overline{1})=\QL^{-1}QP^{-1}\PL(\overline{1})
=\QL^{-1}QP^{-1}(\overline{L})
=\QL^{-1}(\overline{L-1})
=\overline{1}\,;
\end{eqnarray}
$Q_{\rm enc}$ coincides with $P_{\rm enc}$ in the mapping of the
element $P_{\rm enc}^{-1}(\overline{1})$ because two of the six
changes pointed out above mutually compensate.}

\subsubsection{Cycle structure of $Q_{\rm enc}$}

When analyzing the cycle structure of $Q_{\rm enc}$, we now have to
distinguish between {\it three} cases, the first two paralleling the
treatment of Subsection \ref{sec:unitary_recursion}.  Note however a
factor~2 appearing in the second case. For each $Q_{\rm enc}\in{\cal
  M}(\vec{v}^{[l\to m,l-m-1]},m)$, there are now twice as many, namely
$2(l-m-1)(v_{l-m-1}+1)$ related $P_{\rm enc}\in {\cal M}(\vec{v},l)$
structured like Eq. (\ref{pmcyc}), since $Q_{\rm enc}$ also remains
unaffected by time reversal of $a_{m+2},\ldots,a_l$ in Eq. (\ref{pmcyc}).
The second and fourth line in Eq. (\ref{permutation_orthogonal}) make
sure that merging or splitting of cycles is mirrored by the respective
twins.

A {\bf third possibility} appears because \textbf{the cycles involving
  $L$ and $L-1$ may be twins}, and hence belong to the same encounter.
Since the twin cycles are mutually time-reversed there is one cycle
containing both $L$ and $\overline{L-1}$, and another one containing
$\overline{L}$ and $L-1$.  Assume that inside the first cycle, the
element $\overline{L-1}$ follows $L$ after $m$ iterations, i.e.,
\begin{equation}
\label{miterations}
\overline{L-1}=P_{\rm enc}^m(L)\,,
\end{equation}
with $1\leq m\leq l-1$.  Then $P_{\rm enc}$ can be written as
\begin{eqnarray}
  \label{PPE_orth}
  P_{\rm enc}=&&[\ldots](L,a_2,\ldots,a_{m},\overline{L-1},a_{m+2},\ldots,a_l)\nonumber\\
  &&(\overline{a_l},\ldots,\overline{a_{m+2}},L-1,
  \overline{a_{m}},\ldots,\overline{a_2},\overline{L})\,.
\end{eqnarray}
Due to Eq. (\ref{permutation_orthogonal}), $Q_{\rm enc}$ differs from
$P_{\rm enc}$ by mapping
\begin{eqnarray}
    Q_{\rm enc}(\overline{a_{m+2}})=a_2\,,\quad\quad\quad
    Q_{\rm enc}(\overline{a_2})=a_{m+2}\,,\nonumber\\
    Q_{\rm enc}(a_l)=L-1\,,\quad\quad\quad
    Q_{\rm enc}(\overline{L-1})=\overline{a_l}\,.
\end{eqnarray}
The initial pair of twin cycles of $P_{\rm enc}$ is transformed to a
pair of twin $(l-1)$-cycles forming part of
\begin{eqnarray}\label{QPE_orth}
    Q_{\rm enc}=&&[\ldots](a_2,\ldots,a_{m},\overline{L-1},\overline{a_l},\ldots,
    \overline{a_{m+2}})\nonumber\\
    &&(a_{m+2},\ldots,a_l,L-1,\overline{a_{m}},\ldots,\overline{a_2})\,;
\end{eqnarray}
again $[\ldots]$ represents the unaffected cycles.  Given that the largest
number permuted by $Q_{\rm enc}$, i.e., $L-1$, is included in one of
the above $(l-1)$-cycles, we have $Q_{\rm enc}\in{\cal M}(\vec{v}^{[l\to
  l-1]},l-1)$.  Conversely, for any $Q_{\rm enc}$ as in
(\ref{QPE_orth}) and each $1\leq m\leq l-1$, there is exactly one related
$P_{\rm enc}$.  Given one $Q_{\rm enc}$ we may read off $a_2,\ldots,a_l$
by comparing with Eq.  (\ref{QPE_orth}) and recombine them to form a
permutation $P_{\rm enc}$ as in Eq. (\ref{PPE_orth}).
We thus see that each of the $l-1$ subsets of ${\cal M}(\vec{v},l)$
with $\overline{L-1}=P_{\rm enc}^m(L)$ is in one-to-one correspondence
to ${\cal M}(\vec{v}^{[l\to l-1]},l-1)$ and thus has an equal number
\begin{equation}
N(\vec{v}^{[l\to l-1]},l-1)
\end{equation}
of elements.

Physically, the appearance of $L$ and $L-1$ in ``mutually
time-reversed'' cycles signals that the corresponding orbit stretches
belong to the same encounter, but are mutually time-reversed. Our
recursion step can be interpreted as removing a loop separating two
antiparallel encounter stretches, and thus merging the two stretches
into one (as described in Section \ref{sec:min_dist} and Appendix
\ref{sec:antiparallel_overlap}).

\subsubsection{Special cases}

\begin{figure}[t]
\begin{center}
  \includegraphics[scale=0.47]{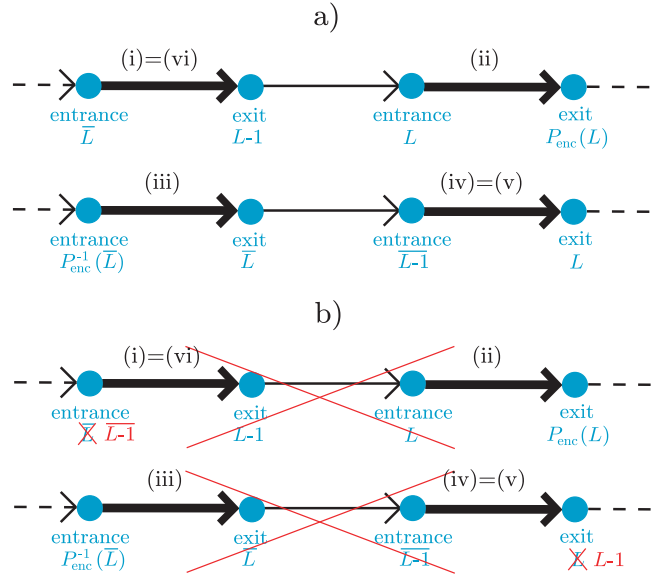}
\end{center}
\caption{
  a) Encounter stretches (thick lines) and loops (thin lines) of $\gamma'$
  and $\overline{\gamma'}$ in the special case $\PE(\overline{L-1})=L$ and
  $\PE^{-1}(L-1)=\overline{L}$.  The stretches (i) and (vi), and
  stretches (iv) and (v) of Fig. \ref{fig:recursion_orth}a coincide.
  In b), the loops $L-1\to L$ and $\overline{L}\to\overline{L-1}$ are
  removed and entrance $\overline L$ and exit $L$ are respectively
  renamed as entrance $\overline{L-1}$ and exit $L-1$.
}\label{fig:recursion_special}
\end{figure}

To be safe, we need to check that Eq. (\ref{QPE_orth}) remains valid
in the two special cases $m=l-1$ and $m=1$.  First assume that the
elements $a_{m+2},\ldots,a_l$ and $\overline{a_l},\ldots,\overline{a_{m+2}}$
in Eq. (\ref{PPE_orth}) are removed, which formally corresponds to
$m=l-1$.  In this case, Eq. (\ref{miterations}) implies
$\overline{L-1}=\PE^{l-1}(L)=\PE^{-1}(L)$ and thus
$\PE(\overline{L-1})=L$ and $\PE^{-1}(L-1)=\overline{L}$.  Our
caricature of port connections in Fig.  \ref{fig:recursion_orth}a has
to be modified, since the stretches (iv) and (v) coincide, as well as
(i) and (vi); see Fig. \ref{fig:recursion_special}a.  If we again
remove the loops leading from exit $L-1$ to entrance $L$ and from exit
$\overline{L}$ to entrance $\overline{L-1}$, rename exit $L$ as $L-1$,
and rename entrance $\overline{L}$ as $\overline{L-1}$ (Fig.
\ref{fig:recursion_special}b), we can immediately read off the new
encounter permutation
\begin{eqnarray}
Q_{\rm enc}(a)=\begin{cases}
  L-1&\text{if $a=P_{\rm enc}^{-1}(\overline{L})$}\\
  P_{\rm enc}(L)&\text{if $a=\overline{L-1}$}\\
  P_{\rm enc}(a)&\text{otherwise}\,.
\end{cases}
\end{eqnarray}
Our conclusions remain unaffected. The permutation $\QE$ now differs
from $\PE$ by mapping $\QE(\overline{a_2})=L-1$ and
$\QE(\overline{L-1})=a_2$. We thus recover Eq. (\ref{QPE_orth}), again
with $a_{m+2},\ldots,a_l$ and $\overline{a_l},\ldots,\overline{a_{m+2}}$
removed.

On the other hand, if $a_2,\ldots,a_m$ and
$\overline{a_m},\ldots,\overline{a_2}$ are removed from (\ref{PPE_orth})
(corresponding to $m=1$ and $\PE(L)=\overline{L-1}$), the recursion
(\ref{permutation_orthogonal}) remains in force and implies
\begin{eqnarray}
    Q_{\rm enc}(\overline{a_{m+2}})&=&\overline{L-1}\,,\quad\quad\quad
    Q_{\rm enc}(L-1)=a_{m+2}\,,\nonumber\\
    Q_{\rm enc}(a_l)&=&L-1\,,\quad\quad\quad
    Q_{\rm enc}(\overline{L-1})=\overline{a_l}\,,
\end{eqnarray}
leading to Eq. (\ref{QPE_orth}) with the above elements absent.  Thus,
both special cases were correctly taken into account.

\subsubsection{Resulting recursion}

We have seen that ${\cal M}(\vec{v},l)$ falls into subsets similar to
the unitary case, time-reversal invariance making for the factor 2
explained above, and for $l-1$ additional subsets of size
$N(\vec{v}^{[l\to l-1]},l-1)$. The various sizes combine to the
orthogonal analog of the recursion relation (\ref{NL}),
\begin{eqnarray}
\label{recursion_nvl}
N(\vec{v},l)&=&\sum_{k\geq 2}N(\vec{v}^{[k,l\to k+l-1]},k+l-1)\nonumber\\
&+&\sum_{m=1}^{l-2}2(l-m-1)(v_{l-m-1}+1)
N(\vec{v}^{[l\to m,l-m-1]},m)\nonumber\\
&+&(l-1)N(\vec{v}^{[l\to l-1]},l-1)\,,
\end{eqnarray}
which using Eq. (\ref{mengetomenge}) may be written as
\begin{eqnarray}
\label{Nrecur2}
\frac{lv_l}{L}N(\vec{v})&=&\sum_{k\geq 2}\frac{(k+l-1)(v_{k+l-1}+1)}{L-1}
N(\vec{v}^{[k,l\to k+l-1]}) \nonumber\\
&+&\sum_{m=1}^{l-2}\frac{2(l-m-1)(v_{l-m-1}+1)m v_m^{[l\to m,l-m-1]}}{L-1}
N(\vec{v}^{[l\to m,l-m-1]})\nonumber\\
&+&\frac{(l-1)^2 (v_{l-1}+1)}{L-1}N(\vec{v}^{[l\to l-1]})\,.
\end{eqnarray}
The shorthand $\tilde{N}(\vec{v})=N(\vec{v})\frac{(-1)^V\prod_l
  l^{v_l}}{L}$ leads to
\begin{eqnarray}
\label{recursion_orthogonal}
\label{recur2}
&&v_l\tilde{N}(\vec{v})+\sum_{k\geq 2}(v_{k+l-1}+1)k
\tilde{N}(\vec{v}^{[k,l\to k+l-1]})\nonumber\\
&&+\sum_{m=1}^{l-2}2(v_{l-m-1}+1)v_m^{[l\to m,l-m-1]}
\tilde{N}(\vec{v}^{[l\to m,l-m-1]})
\nonumber\\
&&-(l-1)(v_{l-1}+1)\tilde{N}(\vec{v}^{[l\to l-1]})=0\,;
\end{eqnarray}
recall that $v_{k+l-1}+1=v_{k+l-1}^{[k,l\to k+l-1]}$ and
$v_{l-1}+1=v_{l-1}^{[l\to l-1]}$.

\subsection{Spectral form factor}

Similarly as in Subsection \ref{sec:formfactor_unitary} we now turn
the recursion relation for $\tilde{N}(\vec{v})$ into one for the
Taylor coefficients $K_n$.  As a preparation we generalize our rule
(\ref{rule_special}).  For all similar sums over $\vec{v}$ with fixed
$\nu(\vec{v})=n$ and $v_1=0$ we find
\begin{equation}
\label{rule}
\sum_{\vec{v}}^{\nu(\vec{v})=n}\!\!f(\vec{v}^{[\alpha_1,\alpha_2,\ldots\to \beta]})\tilde{N}(\vec{v}^{[\alpha_1,\alpha_2,\ldots\to \beta]})=\!\!\!\!\sum_{\vec{v}'}^{\nu(\vec{v}')=n'}\!\!\!f(\vec{v}')\tilde{N}(\vec{v}')\,,
\end{equation}
with integers $\alpha_i\geq 2$, $\beta\geq 2$ and $f$ any function of $\vec{v}'$
vanishing for $v'_\beta=0$.  We need to sum over all $\vec{v}'$ with
$n'=\nu(\vec{v}')=\nu(\vec{v}^{[\alpha_1,\alpha_2,\ldots\to \beta]})=n-\sum_i
(\alpha_i-1)+(\beta-1)$, where we used that removing an $\alpha_i$-cycle
decreases $L$ by $\alpha_i$, $V$ by 1, and thus $\nu=L-V+1$ by $\alpha_i-1$;
similarly, adding a $\beta$-cycle increases $\nu$ by $\beta-1$.  Eq.
(\ref{rule}) follows in the same way as Eq.  (\ref{rule_special}),
i.e., by switching to $\vec{v}'=\vec{v}^{[\alpha_1,\alpha_2,\ldots\to \beta]}$ as the
new summation variable and dropping the restriction $v'_\beta\geq 1$
($\vec{v}'$ with $v'_\beta=0$ do not contribute due to the vanishing of
$f$).  One similarly shows that the foregoing rule holds even without
any conditions on $f$ if $\beta$ is removed, i.e., if no new cycles are
created. It is convenient to abbreviate the right-hand side of Eq.
(\ref{rule}) with the help of
\begin{equation}
\label{shorthand}
S_{n'}[f]\equiv\sum_{\vec{v}'}^{\nu(\vec{v}')=n'}f(\vec{v}')\tilde{N}(\vec{v}') \,
\end{equation}
for arbitrary $f$; we note that $K_n=\frac{2}{(n-2)!} S_{n}[1]$. Thus
equipped we turn to the three special cases $l=1,2,3$ of our recursion
relations which we shall need below.

\subsubsection{$l=1$}

The case $l=1$ involves permutations with 1-cycles, appearing only in
intermediate steps of our calculation.  If the element $L$ forms a
1-cycle, it may simply be removed from a permutation without affecting
the other cycles, which corresponds to a transition
$\vec{v}\to\vec{v}^{[1\to ]}$.  The number of permutations with given
$\vec{v}$ and $L$ forming part of a 1-cycle thus coincides with the
number of permutations related to $\vec{v}^{[1\to]}$, i.e.,
$N(\vec{v},1)=N(\vec{v}^{[1\to]})$ and equivalently
\begin{equation}
\label{case_one}
\tilde{N}(\vec{v})=-\frac{L(\vec{v}^{[1\to ]})}{v_1}\tilde{N}(\vec{v}^{[1\to ]})\,.
\end{equation}

\subsubsection{$l=2$}

For $l=2$ (and $v_1=0$) the recursion (\ref{recursion_orthogonal})
boils down to
\begin{eqnarray}
\label{case_two}
v_2\tilde{N}(\vec{v})+\sum_{k\geq 2}\left(v_{k+1}^{[k,2\to
    k+1])}k\tilde{N}(\vec{v}^{[k,2\to k+1]})\right)
    -\tilde{N}(\vec{v}^{[2\to 1]})=0\,,
\end{eqnarray}
where only the last term is new compared to the unitary case; to
determine its prefactor we have used that $(l-1)(v_{l-1}+1)=1$ for
$l=2$, $v_1=0$.  We can bring this term to a form free from 1-cycles
by invoking Eq. (\ref{case_one}) and thus $\tilde{N}(\vec{v}^{[2\to
  1]})=-L(\vec{v}^{[2\to]}) \tilde{N}(\vec{v}^{[2\to]})$, to get
\begin{eqnarray}
\label{case_two_L}
v_2\tilde{N}(\vec{v})+\sum_{k\geq 2}\left(v_{k+1}^{[k,2\to
    k+1])}k\tilde{N}(\vec{v}^{[k,2\to k+1]})\right)
    +L(\vec{v}^{[2\to ]})\tilde{N}(\vec{v}^{[2\to ]})=0\,.
\end{eqnarray}
Again, we sum over all $\vec{v}$ with $v_1=0$ and $\nu(\vec{v})=n$.
The sum over the first two terms of (\ref{case_two_L}) may be
evaluated as in the unitary case. Using the rule (\ref{rule}) and the
shorthand (\ref{shorthand}), the sum over the third term leads to
\begin{equation}
\sum_{\vec{v}}^{\nu(\vec{v})=n}L(\vec{v}^{[2\to]})\tilde{N}(\vec{v}^{[2\to]})
=S_{n-1}[L(\vec{v})]
\end{equation}
where we used that removal of a 2-cycle decreases $\nu$ by 1.
Altogether, we obtain
\begin{eqnarray}
\label{result_two_a}
&& S_{n}\Bigg[v_2+\sum_{k\geq
    2}v_{k+1}k\Bigg]+ S_{n-1}[L(\vec{v})]\nonumber\\
    &&=(n-1) S_{n}[1]+ S_{n-1}[L(\vec{v})]=0\,.
\end{eqnarray}

A further relation is obtained by multiplying Eq. (\ref{case_two_L})
with $L(\vec{v})-1=L(\vec{v}^{[k,2\to k+1]})=L(\vec{v}^{[2\to]})+1$,
\begin{eqnarray}
&& (L(\vec{v})-1)v_2\tilde{N}(\vec{v})+\sum_{k\geq 2}\left(L(\vec{v}^{[k,2\to k+1]}) v_{k+1}^{[k,2\to
    k+1])}k\tilde{N}(\vec{v}^{[k,2\to k+1]})\right)\nonumber\\
    &&+L(\vec{v}^{[2\to ]})(L(\vec{v}^{[2\to ]})+1)\tilde{N}(\vec{v}^{[2\to ]})=0\,.
\end{eqnarray}
Summing over $\vec{v}$ with the help of (\ref{rule}) and
(\ref{shorthand}), we obtain
\begin{equation}
S_n\Bigg[(L(\vec{v})-1)v_2+\sum_{k\geq 2}L(\vec{v})v_{k+1}k\Bigg]+S_{n-1}[L(\vec{v})(L(\vec{v})+1)]=0
\end{equation}
and thus
\begin{equation}
  (n-1) S_{n}[L(\vec{v})]- S_{n}[v_2]+ S_{n-1}[L(\vec{v})(L(\vec{v})+1)]=0\,.
\end{equation}
This equation can be simplified if we eliminate $S_n[L(\vec{v})]$ with
the help of Eq. (\ref{result_two_a}) and replace $n\to n-2$,
\begin{equation}
\label{result_two_b}
-(n-2)(n-3) S_{n-1}[1]= S_{n-2}[v_2]- S_{n-3}[L(\vec{v})(L(\vec{v})+1)]\,.
\end{equation}

\subsubsection{$l=3$}

Finally, we consider the special case $l=3$ (and $v_1=0$) of
(\ref{recursion_orthogonal}),
\begin{eqnarray}
\label{case_three}
&&
v_3\tilde{N}(\vec{v})+\sum_{k\geq 2}\left(v_{k+2}^{[k,3\to
    k+2]}k\tilde{N}(\vec{v}^{[k,3\to k+2]})\right)\nonumber\\
    &&
    +4\tilde{N}(\vec{v}^{[3\to 1,1]})-2v_2^{[3\to
  2]}\tilde{N}(\vec{v}^{[3\to 2]})=0\,.
\end{eqnarray}
The third term originates from the sum over $m$ in
(\ref{recursion_orthogonal}) which only draws a contribution from
$m=1$, with the prefactor $2(v_1+1)v_1^{[3\to 1,1]}=4$.  The 1-cycles
are eliminated using the identity $\tilde{N}(\vec{v}^{[3\to
  1,1]})=\frac{1}{2}L(\vec{v}^{[3\to
  ]})(L(\vec{v}^{[3\to]})+1)\tilde{N}(\vec{v}^{[3\to]})$, which follows
by twice applying Eq. (\ref{case_one}) to $\vec{v}^{[3\to 1,1]}$.  We
thus find
\begin{eqnarray}
\label{case_three_b}
&&
v_3\tilde{N}(\vec{v})+\sum_{k\geq 2}\left(v_{k+2}^{[k,3\to
    k+2]}k\tilde{N}(\vec{v}^{[k,3\to k+2]})\right)\nonumber\\
    &&
    +2L(\vec{v}^{[3\to
  ]})(L(\vec{v}^{[3\to]})+1)\tilde{N}(\vec{v}^{[3\to]})-2v_2^{[3\to
  2]}\tilde{N}(\vec{v}^{[3\to 2]})=0\,.
\end{eqnarray}
It is easy to see that Eq. (\ref{case_three_b}) connects combinatorial
numbers related to three different orders of $K(\tau)$.  If $\vec{v}$
and thus $\vec{v}^{[k,3\to k+2]}$ contribute to order $n$,
$\vec{v}^{[3\to]}$ contributes to $n-2$, and $\vec{v}^{[3\to 2]}$
contributes to $n-1$.  Summing over $\vec{v}$, we find
\begin{eqnarray}
 S_{n}\Bigg[v_3+\sum_{k\geq
    2}v_{k+2}k\Bigg]+2 S_{n-2}[L(\vec{v})(L(\vec{v})+1)]
    -2 S_{n-1}[v_2]=0\,.
\end{eqnarray}
This expression can be simplified using that $v_3+\sum_{k\geq 2}v_{k+2}k
=\sum_{l\geq
  2}v_l(l-2)=L(\vec{v})-2V(\vec{v})=2(\nu(\vec{v})-1)-L(\vec{v})$,
which leads to
\begin{eqnarray}
 2(n-1) S_{n}[1]- S_{n}[L(\vec{v})]+2 S_{n-2}[L(\vec{v})(L(\vec{v})+1)]
-2 S_{n-1}[v_2]=0\,.
\end{eqnarray}
Finally applying Eq. (\ref{result_two_a}) to eliminate
$S_n[L(\vec{v})]$, substituting $n\to n-1$, and dividing by 2, we
proceed to
\begin{eqnarray}
\label{result_three}
\frac{n-1}{2} S_{n}[1]+(n-2) S_{n-1}[1]
= S_{n-2}[v_2]- S_{n-3}[L(\vec{v})(L(\vec{v})+1)]\,.
\end{eqnarray}

\subsubsection{Final result}

Upon comparing the recursion relations (\ref{result_two_b}) and
(\ref{result_three}), obtained for the cases $l=2$ and $l=3$, we find
the coefficients $ S_{n}[1]=\frac{(n-2)!}{2}K_n$ and $
S_{n-1}[1]=\frac{(n-3)!}{2}K_{n-1}$ related as $\frac{n-1}{2}
S_{n}[1]=-(n-2)^2 S_{n-1}[1]$ or
\begin{equation}
\label{welcome_back}
(n-1)K_n=-2(n-2)K_{n-1}\,.
\end{equation}
Eq. (\ref{welcome_back}) applies to all $n\geq 3$, but not to $n=2$,
because the diagonal term $K_1$ is unrelated to our permutation
treatment.

An initial condition is provided by the Sieber/Richter result for
orbits differing in one 2-encounter, $K_2=-2$.  Thus started, our
recursion yields the Taylor coefficients
\begin{equation} K_n=\frac{(-2)^{n-1}}{n-1}
\end{equation}
coinciding with the random-matrix result from the GOE. Universal
behavior is thus ascertained for the small-time form factor of fully
chaotic dynamics from the orthogonal symmetry class, at least in the
limit $\tau\ll 1$.  The resulting series converges for $\tau<\frac{1}{2}$
and remains valid, by analytic continuation, up to the singularity at
$\tau=1$.

\section{Summary}

We have shown that structures of orbit pairs $\gamma$, $\gamma'$ are in
one-to-one correspondence to permutations.  Encounters are described
by permutations $P_{\rm enc}$ determining how the ports of $\gamma$ are
reconnected in $\gamma'$, i.e., which entrance port is connected to which
exit port. In absence of time-reversal invariance, $P_{\rm enc}$ has
one cycle per encounter.  Loops are characterized by another
permutation $P_{\rm loop}$ fixing the entrance port following each
exit port. The ordering of entrance ports along $\gamma'$ is thus given by
the product $P_{\rm loop}P_{\rm enc}$ which must be single-cycle due
to the periodicity of $\gamma'$.

For time-reversal invariant systems, we consider permutations mapping
between ports of both $\gamma$ and its time reversed $\overline{\gamma}$. In
this case encounters correspond to pairs of ``mutually time-reversed''
cycles, and $P_{\rm loop}P_{\rm enc}$ has one cycle related to $\gamma'$
and one cycle related to $\overline{\gamma'}$.

The numbers of structures $N(\vec{v})$ can now be determined from a
recursion, motivated by the physical picture of ``removing loops''
between encounters. This recursion translates into a recursion for the
Taylor coefficients of $K(\tau)$ which indeed yields small-time form
factors in agreement with the GUE and the GOE.

%% file: kapitel7.tex
\chapter{Relation  to the $\sigma$ model}

\stand

\label{sec:sigma}

\section{Introduction}

The so-called nonlinear $\sigma$ model \cite{Sigma,Efetov} is a convenient
framework for calculating spectral correlators or, more generally,
averaged products of Green functions.  In particular, the
zero-dimensional variant of the $\sigma$ model can be used to implement
averages over random matrices of the Wigner/Dyson ensembles, as
required to evaluate the corresponding spectral form factors $K(\tau)$.
Similarly, for disordered systems, the $\sigma$ model has been used to
implement averages over impurity potentials. Steps towards a ballistic
$\sigma$ model for {\it individual} systems have been presented in
\cite{BallisticSigma}.

The $\sigma$ model proved of great heuristic value for our semiclassical
approach: We were led to the correct combinatorics of families of
orbit pairs by a perturbative analysis of the $\sigma$ model.  Such a
perturbative treatment yields the spectral form factor as power series
in the time $\tau$, analogous the series extracted from Gutzwiller's
semiclassical periodic-orbit theory in the preceding Chapters.  In
order to gain intuition for our semiclassical analysis, the
perturbative treatment of the $\sigma$ model had to be carried to all
orders in $\tau$. To our knowledge, this task has never been attacked
explicitly before, since within random-matrix theory the whole
spectral form factor is accessible through a non-perturbative
approach.  The analogy of periodic-orbit expansions to perturbation
series might prove fruitful for future applications of periodic-orbit
theory, and that possibility motivates the following exposition.

\begin{figure}
\begin{center}
  \includegraphics[scale=0.7]{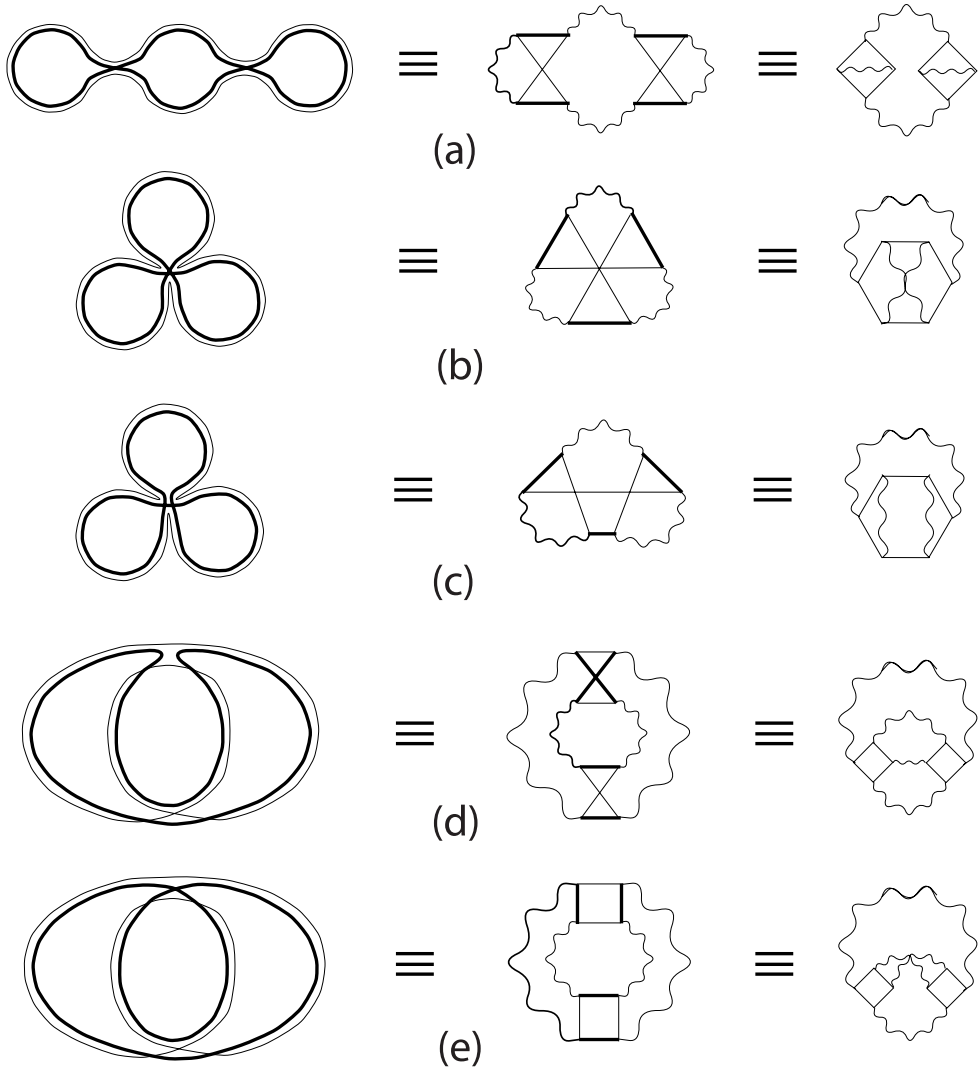}
\end{center}
\caption{Families of orbit pairs (left column) and the corresponding disorder diagrams
  (middle and right column), responsible for the cubic order of
  $K(\tau)$. The Figure is taken from \cite{Disorder1}. The cases
  (a)-(e) respectively correspond to the families {\it aas}, {\it pc},
  {\it ac}, {\it api}, and {\it ppi} discussed in Chapter
  \ref{sec:tau3}.  (There are slight inconsistencies in the cases (d)
  and (e).)  In the middle column, diagrams are arranged such as to
  stress the analogy to orbits; the right column follows usual
  field-theoretical conventions.}
\label{fig:disorder}
\end{figure}

We will mostly work with the $\sigma$ model in a random-matrix setting.
However, it will be worthwhile to at least qualitatively point out
similarities to the diagrammatic expansion yielding $K(\tau)$ for
disordered systems.  These similarities were first noticed in
\cite{Disorder1,Disorder2}, in a previous attempt to recover the form
factor for individual chaotic dynamics.  Fig. \ref{fig:disorder},
taken from \cite{Disorder1}, displays the diagrams responsible for the
cubic contribution to the form factor. We see that these diagrams are
in one-to-one correspondence to the families of orbit pairs discussed
in Chapter \ref{sec:tau3} and depicted, in a slightly different style,
in the left column. The shapes drawn by straight lines are known as
``Hikami boxes''; these boxes form a particular kind of vertex.
Hikami boxes are analogous to self-encounters of periodic orbits, with
each $l$-encounter leading to a Hikami box with $2l$ ``ports''
depicted as corners.  The straight lines correspond to encounter
stretches of the two partner orbits; the thick and thin straight lines
in the middle column of Fig. \ref{fig:disorder} can respectively be
associated with stretches of the partner orbits drawn in the left
column.  The Hikami boxes related to parallel and antiparallel
2-encounters are known as diffusons and Cooperons.  (In contrast to
the interpretation in terms of classical orbits, in field theory
diffusons are often depicted by antiparallel arrows, whereas Cooperons
are denoted by parallel arrows.)  Outside the Hikami boxes, the ports
are connected by (curly) propagator lines -- obviously the analog of
orbit loops.  Each diagram should be taken to represent all possible
orbit pairs obtainable by reconnections in encounters of the same
topology as the Hikami boxes depicted.

In field theory, vertices are essentially points (even though they are
depicted in a different manner in Fig. \ref{fig:disorder}) -- in
contrast to periodic-orbit theory where the related encounters have a
non-zero duration, of the order of the Ehrenfest time $T_E\propto
\ln\frac{\rm const}{\hbar}$.  Of course, the relevant encounter durations
are vanishingly small compared to the typical loop durations ($\sim
T_H\propto \hbar^{-f+1}$); nevertheless, we may say that self-encounters give
internal (phase-space) structure to vertices.

In the remainder of this Chapter, the relation between periodic-orbit
theory and the $\sigma$ model will be explored from a random-matrix
perspective.  We shall first introduce the bosonic replica variant of
the zero-dimensional nonlinear $\sigma$ model. In Subsection
\ref{sec:sigmamodel_formfactor}, the resulting spectral form factor
$K(\tau)$ will be expanded perturbatively into a sum over vectors
$\vec{v}$, analogous to the sum obtained from periodic-orbit theory.
Wick's theorem can be used to recursively evaluate each summand, with
the help of contraction rules introduced in Subsections
\ref{sec:con_rules} and \ref{sec:con_rules_special}. These contraction
rules have a striking similarity to semiclassics: In Subsection
\ref{sec:analogy} we will reveal both expansions for $K(\tau)$ as
identical, by relating structures of orbit pairs to so-called ``full
contractions'' in the $\sigma$ model. In Subsection
\ref{sec:rec_contractions} we shall see that the recursion for the
numbers of structures $N(\vec{v})$ in Chapter \ref{sec:combinatorics}
is mirrored by an analogous recursion based on Wick contractions in
the $\sigma$ model.

\section{Background: The nonlinear $\sigma$ model}

We proceed to give a brief introduction to the nonlinear $\sigma$ model,
in a random-matrix setting. In particular, we will show that the
bosonic replica variant of the $\sigma$ model yields the
$\frac{1}{s}$-expansion of the two-point correlator
\begin{equation}
\label{corr}
R(s)=\left\langle\frac{\rho(E+\frac{s}{2\pi\overline{\rho}}) \rho(E-\frac{s}{2\pi\overline{\rho}})}
      {\overline{\rho}(E)^2}\right\rangle-1\,,
\end{equation}
see also Eq. (\ref{correlator}), as an integral over matrices $B$,
\begin{equation}\label{rs}
         R(s) \sim -\frac{1}{2} {\rm Re}\ {\rm lim}_{r \to 0} \frac{1}{r^2}
        \partial^2_{s^+}\,
        {(s^+)}^{-\kappa  r^2}{\rm e}^{is^+r}%\nonumber\\&&
        \int {d[B]} \,{\rm e}^{(2{{\rm i}}/\kappa ) \sum_{l=1}^{\infty}
         {(s^+)}^{1-l}{\rm tr}(B {B^\dagger})^l}-\frac{1}{2}\,;
\end{equation}
an analogous formula for $K(\tau)$ is obtained by Fourier transforming
according to (\ref{fourier}).  On the right-hand side, $s^+$ must be
read as $s+{{\rm i}}\delta'$ with $\delta'\downarrow 0$.  The matrices $B, {B^\dagger}$ are $r \times
r$ for the GUE and $2r \times 2 r$ for the GOE, and the measure $d[B]$ will
be defined in Eq. (\ref{measure}); as before, the factor $\kappa $ takes
the respective values 1 and 2 for the unitary and the orthogonal
cases.  The above formula yields only the non-oscillatory part of
$R(s)$, needed to recover the power series of $K(\tau)$; the dropping of
oscillatory terms is signaled by $\sim$.  Eq. (\ref{rs}) will later
serve as the starting point for our perturbative treatment of the
spectral form factor, paralleling the previous semiclassical approach.

\subsection{Average over the GUE}

To derive the random-matrix expression for the two-point correlator
(\ref{rs}), we first consider the GUE. The average $\langle\ldots\rangle$ over small
intervals of $E$ and $s$ in (\ref{corr}) is thus replaced by an
average $\averagedots$ over all independent matrix elements $H_{\mu
  \nu}$ of Hermitian $N\times N$ matrices, with $N\to\infty$ and a Gaussian
weight defined by
\begin{equation}
\label{baraverage}
\averagedots\equiv\int d[H]\;\ARMT\,{\rm e}^{-\BRMT \Tr H^2}\ldots\equiv\int\prod_{\mu<\nu}(d\mbox{Re}H_{\mu\nu}\,d\mbox{Im}H_{\mu\nu})
\prod_\mu d H_{\mu\mu}\;\ARMT\,{\rm e}^{-\BRMT \Tr H^2}\ldots\,
\end{equation}
We choose $\ARMT$ and $\BRMT$ such that the Gaussian weight is
normalized and, moreover, the average of each squared matrix element
is given by $\average{|H_{\mu\nu}|^2}=\frac{\lambda^2}{N}$ with $\lambda$
constant.  In particular, this choice implies $\BRMT=\frac{N}{2\lambda^2}$.
All other products of matrix elements $H_{\mu\nu}H^{*}_{\mu'\nu'}$,
$(\mu,\nu)\neq(\mu',\nu')$ have to average to zero regardless of $\ARMT$ and
$\BRMT$.  In the following, we shall always consider the vicinity of
$E=0$, where the average level density is given by
$\overline{\rho}=\frac{N}{\pi\lambda}$.

\subsubsection{Green functions}

It is convenient to express $\rho(E)$ and $R(s)$ in terms of the
advanced and retarded Green functions
\begin{equation}
G_\pm(E)=(E\pm {{\rm i}}\delta-H)^{-1}\,,
\end{equation}
where we implicitly take $\delta\downarrow 0$.  Writing the diagonal elements of
$G_\pm(E)$ as
\begin{equation}
\frac{1}{(E-E_n)\pm{{\rm i}}\delta}=P\frac{1}{E-E_n}\mp{{\rm i}}\pi\delta(E-E_n)\,
\end{equation}
with $P$ denoting the principal value, one can show that the level
density is simply given by
\begin{equation}
\label{rhotrace}
\rho(E)=\frac{{\rm i}}{2\pi}(\Tr\, G_+(E)-\Tr\, G_-(E))\,.
\end{equation}
Eqs. (\ref{corr}), (\ref{baraverage}), and (\ref{rhotrace}) entail the
corresponding expression for the correlator\footnote{ When inserting
  (\ref{rhotrace}) into (\ref{corr}), averaged products of two
  advanced or two retarded Green functions can be evaluated as
  $\average{\Tr\, G_\pm(E)\,\Tr\, G_\pm(E')}
  \stackrel{(*)}{=}\average{\Tr\, G_\pm(E)}\;\average{\Tr\, G_\pm(E')}
  \stackrel{(**)}{=}-\pi^2\overline{\rho}^2$, where $(*)$ is demonstrated
  in \cite{Haake}.  The step $(**)$ follows from $\average{\sum_n
    P\frac{1}{E-E_n}}=0$ in the vicinity of $E=0$, given that the
  average level density is symmetric around 0.  The above averages
  finally give rise to a contribution $+\frac{1}{2}$ to $R(s)$,
  combining to $-\frac{1}{2}$ with the $-1$ of Eq. (\ref{corr}).  }
\begin{eqnarray}
\label{RsCs}
R(s)&=&\frac{1}{2(\pi\overline{\rho})^2}\mbox{Re}\,C(s)-\frac{1}{2}
\end{eqnarray}
with $C(s)$ defined by
\begin{eqnarray}
\label{Cs}
C(s)&\equiv&\average{\Tr\, G_+\left(E+\frac{s}{2\pi\overline{\rho}}\right)
\Tr\, G_-\left(E-\frac{s}{2\pi\overline{\rho}}\right)}\,.
\end{eqnarray}

\subsubsection{Generating function}

Rather than expressing $R(s)$ through traces of $G_\pm(E)$, it is
technically simpler to work with determinants.  To translate between
the two, we use the {\it ``replica trick''}: For $r$ real and $f(E)\equiv\det
G_\pm(E)^{-1}$, we have
\begin{eqnarray}
    \label{tr_det}
    {\rm tr}\,G_\pm(E)&=&\partial_E {\,\rm Tr}\ln(E\pm{{\rm i}}\delta-H)=\partial_E {\,\rm Tr}\ln(G_\pm(E)^{-1}) \nonumber\\
    &=&\partial_E \ln f(E)
    = \frac{f'(E)}{f(E)}
    =\lim_{r\to 0}f(E)^{-r-1}f'(E) \nonumber\\
    &=&-\lim_{r\to 0}\frac{1}{r}\partial_E f(E)^{-r}
    =- \lim_{r\to 0
    }{1\over r}\partial_E\det G_\pm(E)^r\,.
\end{eqnarray}
As usual, we assume (\ref{tr_det}) valid even if the so-called replica
index $r$ is restricted to the integer numbers.  This assumption
obviously demands mathematical justification, which is beyond the
scope of this thesis.

Invoking the replica trick for both traces in (\ref{Cs}), we represent
$C(s)$ as the two-fold derivative of a {\it generating function} ${\cal
  Z}(\epsilon_+, \epsilon_-)$ involving determinants of Green functions,
\begin{eqnarray}
\label{generating}
 C(s)&=&\lim_{r\to 0}\frac{1}{r^2}
\left[\frac{\partial^2  {\cal Z}(\epsilon_+,\epsilon_-)}{\partial \epsilon_+
\partial\epsilon_-}\right]_{\epsilon_{\pm}
 =\pm \frac{\epsilon}{2}=\pm\frac{s}{2\pi\overline{\rho}}}\nonumber\\
 {\cal Z}(\epsilon_+,\epsilon_-)&=&\average{\det G_-(E+\epsilon_+)^r
 \det G_+(E+\epsilon_-)^r}\,.
\end{eqnarray}

Crucially, such determinants allow for the {\it integral representation}
\begin{equation}
\label{determinant}
\det{G}_\pm=\int d[\phi]\exp\left(\pm{{\rm i}}\phi^\dagger G_\pm^{-1}\phi\right)
\end{equation}
where the integration runs over $N$-dimensional complex vectors $\phi$
with the measure defined by $d[\phi]=\prod_{\mu=1}^N\frac{d{\rm Re}\phi\,d{\rm
    Im}\phi}{\pi}$.  The sign in the exponent of (\ref{determinant}) is
chosen such as to make the integral convergent.  The individual
determinants in Eq. (\ref{generating}) can now be numbered by
$\alpha=1,\ldots, 2r$ (with $\alpha=1,\ldots, r$ corresponding to the advanced and
$\alpha=(r+1),\ldots, 2r$ to the retarded Green functions), and each
represented as an integral
\begin{equation}
    \det G_\pm(E+\epsilon_\pm)=\int d[\psi^\alpha]\,\exp\left(\pm{{\rm i}}
    \psi^{\alpha\dagger}(E+(\epsilon_\pm\pm{{\rm i}}\delta)-H)\psi^\alpha\right)\,;
    \label{det_gaussian}
\end{equation}
over $\psi^\alpha=\{\psi^\alpha_\mu\}$, $\mu=1,\ldots, N$.

All $2r$ integrals together may be written as
\begin{eqnarray}
{\cal Z}(\epsilon_+,\epsilon_-)&=&\average{\int d[\psi]\exp\Big({{\rm i}}\sum_{\alpha=1,\ldots,2r}
{\psi^\alpha}^\dagger\Lambda^\alpha(E+\hat{\epsilon}^\alpha-H)\psi^\alpha\Big)}
\end{eqnarray}
with
\begin{eqnarray}
\label{explain}
d[\psi]&=&\prod_{\alpha=1}^{2r} d[\psi^\alpha]\nonumber\\
\Lambda^\alpha&=&\begin{cases}
1 &\mbox{for $\alpha=1,\ldots, r$}\\
-1 &\mbox{for $\alpha=(r+1),\ldots, 2r$}
\end{cases}\nonumber\\
\hat{\epsilon}^\alpha&=&\begin{cases}
\epsilon_++{{\rm i}}\delta &\mbox{for $\alpha=1,\ldots ,r$}\\
\epsilon_--{{\rm i}}\delta &\mbox{for $\alpha=(r+1),\ldots, 2r$\;.}
\end{cases}
\end{eqnarray}
In the following, we will drop the index $\alpha$, and simply write
\begin{eqnarray}
\label{replica}
{\cal Z}(\epsilon_+,\epsilon_-)=\average{\int d[\psi]\exp\left({{\rm i}}
\psi^\dagger\Lambda(E+\hat{\epsilon}-H)\psi\right)}\,.
\end{eqnarray}
The resulting $H$ now has to be read as a block-diagonal matrix with
$2r$ identical $N\times N$ blocks containing the matrix elements of the
Hamiltonian.  $\Lambda$ and $\hat{\epsilon}$ turn into diagonal matrices with
diagonal elements given by (\ref{explain}), i.e., $\Lambda=\diag({\mathbf
  1}_{rN},-{\mathbf 1}_{rN})$, $\hat\epsilon=\diag((\epsilon_++{{\rm i}}\delta){\mathbf
  1}_{rN},(\epsilon_--{{\rm i}}\delta) {\mathbf 1}_{rN})$, where ${\mathbf 1}_{rN}$ is
the $rN$-dimensional unit matrix.  (Analogous $2r\times 2r$ matrices will
later be denoted by the same symbols $\Lambda$, $\hat{\epsilon}$.)

\subsubsection{Average over $H$ for fixed $\psi$}

The replica representation (\ref{replica}) has an important advantage:
If we exchange the $\psi$ integration and the average over $H$ implied
by $\averagedots$, we obtain a simple Gaussian integral
\begin{equation}
\label{haverage}
 {\cal Z}(\epsilon_+,\epsilon_-)
=\int d[\psi]\exp({{\rm i}}\psi^\dagger\Lambda(E+\hat{\epsilon})\psi)
\int d[H] \ARMT\exp\left(-\frac{N}{2\lambda^2}\Tr H^2-{{\rm i}}\psi^\dagger\Lambda H\psi\right)\,.
\end{equation}
To evaluate this integral, for fixed $\psi$, we first need eliminate to
the term linear in $H$.  This can be achieved by writing $-{{\rm i}}\psi^\dagger\Lambda
H\psi=-{{\rm i}}\Tr\, HX$ with $X_{\mu\nu}=\sum_{\alpha}{\psi_\nu^\alpha}^*\Lambda^{\alpha} \psi_\mu^\alpha$
(or, in short, $X_{\mu\nu}={\psi_\nu}^\dagger\Lambda \psi_\mu$) and transforming
$H+\frac{{{\rm i}}\lambda^2}{N}X\to H$.  Through this transformation, the exponent
obtains a new summand $-\frac{\lambda^2}{2N}\Tr\, X^2=
-\frac{\lambda^2}{2N}\tr\YRMT^2$, where $\tr$ denotes a trace over $\alpha$,
and $\YRMT^{\alpha\beta}=\sum_\mu\psi_\mu^\alpha{\psi_\mu^\beta}^*\Lambda^{\beta}$.
The remaining Gaussian integral $\int
d[H]\ARMT\exp\left(-\frac{N}{2\lambda^2}\tr H^2\right)$ gives unity, due to
the normalization of the Gaussian weight. We thus find
\begin{equation}
\label{quartic}
{\cal Z}(\epsilon_+,\epsilon_-)=\int d [\psi]
\exp\left({\rm i} \psi^{\dagger}\Lambda(\hat
  \epsilon+E)\psi-\frac{\lambda^2}{2N} {\rm tr\,}\YRMT^2\right)\,.
\end{equation}

\subsubsection{Hubbard--Stratonovich transformation}

To proceed, we need to eliminate the term $\propto {\rm tr}\,\YRMT^2$ {\it
  quartic} in the integration variables $\psi$ by a so-called
Hubbard--Stratonovich transformation.  To this end, we again introduce
an additional integration, and multiply $\mathcal{Z}$ by a Gaussian
integral $\int d[Q] \,\exp\left(\frac{N}{2}{\rm tr}\,Q^2\right) $, where
$Q=\{Q^{\alpha\beta}\}$ is a $2r$--dimensional anti-Hermitian matrix and $\int
d[Q]$ denotes integration over all its independent matrix elements.
That integral is convergent, since the exponent may be written as
$-\frac{N}{2}\tr({{\rm i}} Q)^2$ with ${{\rm i}} Q$ Hermitian, and yields an
$r$-fold power, going to 1 in the replica limit $r\to 0$.  Denoting
equivalence in the replica limit by $\sim$ and transforming $Q\to Q - \lambda
\YRMT/N$, we are led to
\begin{eqnarray}
\label{Qintegral}
        \mathcal{Z}(\epsilon_+,\epsilon_-)
      \! \! &\sim&\! \!\!\!\!\int\!\! d[Q]\!\int\!\! d[\psi]\exp
\!
\left({\textstyle{\frac{N}{2}}}{\rm tr}\,Q^2+{\rm i}
\psi^{\dagger}\Lambda(\hat \epsilon+E+ {\rm i}\lambda Q)\psi\right)\,,
\end{eqnarray}
with the quartic term removed.

The $\psi$ integral has now become easy: Each of the $N$ independent
Gaussian integrations over $\psi_\mu=\{\psi_\mu^\alpha\}$ yields a determinant
that can be incorporated into the exponent via ``$\det=\exp\tr\ln$'',
to get
\begin{eqnarray}
\label{ampli_act}
\mathcal{Z}(\epsilon_+,\epsilon_-)        &\sim& \!\!\!\!\int \!\!d[Q] \,\exp\left({\textstyle{\frac{N}{2}}}{\rm
tr}\,Q^2-N{\rm\, tr\,ln}(\hat \epsilon+E+ {\rm i}\lambda Q)\right)\;\;\;\;\;\;\,
\nonumber\\
        &=& \!\!\!\!\int \!\!d[Q] \,\exp\left(-N{\rm\, tr\,ln}\left(1+\frac{\hat \epsilon}{E+ {\rm i}\lambda Q}\right)\right)
\nonumber\\
&&\;\;\;\;\;\times\exp\left(N\left\{\frac{1}{2}{\rm
tr}\,Q^2-{\tr\,\ln}(E+ {\rm i}\lambda Q)\right\}\right)\,.
        \;\;\;\;\;\;
\end{eqnarray}
The generating function is thus expressed as an integral in replica
space. The only reminder left of the original matrices $H$ is their
dimension $N$.

\subsubsection{Saddle-point approximation}

The integral over $Q$ has to be evaluated in the limit of infinite
matrix dimension $N\to\infty$.  Since $\hat{\epsilon}$ is of the order of the
mean level spacing and thus proportional to $\frac{1}{N}$, the
exponential in the second line of Eq. (\ref{ampli_act}), the so-called
``amplitude'', may be approximated as
$\exp(-\tr\frac{N\hat{\epsilon}}{E+{\rm i}\lambda Q})$ by expanding the logarithm.

In the following exponential, the large parameter $N\to\infty$ is not
compensated by a factor $\hat{\epsilon}$. Thus, a saddle-point approximation
is called for. The remainder of the exponent, the ``action''
$S(Q)\equiv\frac{1}{2}{\rm tr}\,Q^2-{\tr\,\ln}(E+ {\rm i}\lambda Q)$, becomes
stationary for
\begin{equation}
\label{saddle}
Q=\frac{{\rm i}\lambda}{E+{\rm i}\lambda Q}\,,
\end{equation}
i.e., $Q=Q^{-1}$ if we restrict ourselves to $E=0$. This condition is
satisfied by all diagonal matrices with entries $\pm 1$.  Among these,
one can show that only the saddle point $\Lambda=\mbox{diag}({\mathbf
  1}_r,-{\mathbf 1}_r)$ is relevant for the short-time form factor
\cite{Weidenmueller}.

The restriction $Q=Q^{-1}$ is also fulfilled by all matrices $Q_{\rm
  s}=T\Lambda T^{-1}$ similar to $\Lambda$. To make the subsequent integrals
convergent we have to restrict ourselves to pseudounitary
transformation matrices $T\in U(r,r)$, $T\Lambda T^\dagger=\Lambda$ \cite{SymmSpaces}.
Several $T$ lead to the same $Q_{\rm s}$: Since transformations from
$U(r)\times U(r)$ commute with $\Lambda$, $T$ may be multiplied with any such
transformation without affecting the saddle point $Q_{\rm s}$.  The
saddle-point manifold of all possible $Q_{\rm s}$ can thus be
identified with the set of equivalence classes of possible $T$,
denoted by ${\rm U}(r,r)/({\rm U}(r)\times{\rm U}(r))$.  (This set can be
interpreted as a so-called ``symmetric space'', see
\cite{SymmSpaces}.)

To evaluate (\ref{Qintegral}), we have to expand the action to
quadratic order in the deviation $Q-Q_{\rm s}$ from the saddle-point
manifold, and split integration into one integral over the points
$Q_{\rm s}$ on the saddle-point manifold, and one integral over the
deviations $Q-Q_{\rm s}$.  Approximating the amplitude by its value at
the saddle point $Q_{\rm s}$, we then find (in somewhat sloppy
notation)
\begin{eqnarray}
\mathcal{Z}(\epsilon_+,\epsilon_-)&\sim&
\int d[Q_{\rm s}] \,\exp\left(-{\tr\,}\frac{N \hat \epsilon}{E+ {\rm i}\lambda Q_{\rm s}}\right)\nonumber\\
&&\times\int d[Q-Q_{\rm s}]\exp\left(N\Big\{S(Q_{\rm s})+\frac{1}{2}S''(Q_{\rm s})(Q-Q_{\rm s})^2\Big\}\right)\,.
\end{eqnarray}
The integral over $Q-Q_{\rm s}$ (also including the stationary value
of the action) leads, once more, to an $r$-fold power, converging to 1
in the limit $r\to 0$.  We are thus left with an integral over the
saddle-point manifold involving only the amplitude
$\exp(-\tr\frac{N\hat{\epsilon}}{E+{\rm i}\lambda Q_{\rm s}})$,
\begin{eqnarray}
\mathcal{Z}(\epsilon_+,\epsilon_-)        &\sim&
     \int d[Q_{\rm s}]\exp\left(-\tr\frac{N\hat{\epsilon}}{E+{\rm i}\lambda Q_{\rm s}}\right)\,.
\end{eqnarray}

The latter amplitude can be simplified if we use (\ref{saddle}) and,
furthermore, decompose the $2r\times 2r$ matrices $\hat{\epsilon}$ and $\Lambda$ into
blocks of size $r\times r$, as in
\begin{eqnarray}
-\tr\frac{N\hat{\epsilon}}{E+{\rm i} \lambda Q_{\rm s}}&\stackrel{(\ref{saddle})}{=}&\frac{{\rm i} N}{\lambda}\tr \hat{\epsilon}\,Q_{\rm s}\nonumber\\
&=&\frac{{\rm i} N}{\lambda}\tr\matr{\epsilon_++{\rm i}\delta}{0}{0}{\epsilon_--{\rm i}\delta}
\matr{Q_{{\rm s},11}}{{*}}{{*}}{Q_{{\rm s},22}}\nonumber\\
&=&\frac{{\rm i} N}{\lambda}\tr\matr{(\epsilon_++{\rm i}\delta)Q_{{\rm s},11}}{{*}}{{*}}{(\epsilon_--{\rm i}\delta)Q_{{\rm s},22}}
\nonumber\\
&=&\frac{{\rm i} N}{\lambda}\left((\epsilon_++{\rm i}\delta)\tr Q_{{\rm s},11}
+(\epsilon_--{\rm i}\delta)\tr Q_{{\rm s},22}\right) \nonumber\\
&=&\frac{{\rm i} N}{2\lambda}\Big((\epsilon_++\epsilon_-)(\tr Q_{{\rm s},11}+\tr Q_{{\rm s},22})\nonumber\\
&&\;\;\;\;\;\;\;+(\epsilon_+-\epsilon_-+2{\rm i}\delta)(\tr Q_{{\rm s},11}-\tr Q_{{\rm s},22})\Big)\nonumber\\
&=&\frac{{\rm i} s^+}{2}\tr \Lambda Q_{\rm s}\,.
\end{eqnarray}
In the last step, we used that $\tr Q_{{\rm s},11}+\tr Q_{{\rm
    s},22}=\tr Q_{\rm s}=\tr(T\Lambda T^{-1})=\tr \Lambda=0$ and substituted
$s^+\equiv(\epsilon_+-\epsilon_-+2{\rm i}\delta)\pi\overline{\rho}
=(\epsilon_+-\epsilon_-+2{\rm i}\delta)\frac{N}{\lambda}$.  Altogether, we thus obtain
\begin{eqnarray}
\label{saddle_integral}
       \mathcal{Z}(\epsilon_+,\epsilon_-)&\sim&\int d [Q_{\rm s}]\, \exp{\left(\frac{{\rm i}s^+}{2}
      {\rm tr}\Lambda  Q_{\rm s}\right)}\,.
      \label{finZ}
\end{eqnarray}

\subsubsection{Rational parametrization}

The integral (\ref{saddle_integral}) over the saddle-point manifold
can be performed if we parametrize $Q_{\rm s}$ by $r\times r$ matrices $B$
as
\begin{equation}\label{parametrization}
 Q_{\rm s}=(1-W)\Lambda (1-W)^{-1}\,;\;\;\;\;\;
W=\frac{1}{\sqrt{s^+}}\left({0\atop B^\dagger}{B\atop 0}\right)\,,
\end{equation}
where the scaling factor $\frac{1}{\sqrt{s^+}}$ makes for notational
convenience. The matrices $1-W$ are proportional to the pseudounitary
transformations $T$ mentioned above; the ($B$-dependent) prefactor
cancels in (\ref{parametrization}).  In this so-called rational
parametrization, the measure $d[Q_{\rm s}]$ may be written as
\begin{equation}
\label{measure}
d[Q_{\rm s}]\sim(s^+)^{-r^2}d[B] \equiv (s^+)^{-r^2}\prod_{i,j=1,\ldots,r}(d{\rm Re}B_{ij}\,d{\rm Im}B_{ij})
\end{equation}
The exponent appearing in the generating function (\ref{finZ}) can be
parametrized by inserting (\ref{parametrization}) and expanding as in
$(1-W)^{-1}= \sum_{n=0}^\infty W^n$.  We then find $\tr\Lambda Q_{\rm
  s}=2r+4\sum_{l=1}^\infty(s^+)^{-l}\tr(B B^\dagger)^l$, and the whole generating
function is expressed as
\begin{equation}
\label{finfinZ}
 \mathcal{Z}(\epsilon_+,\epsilon_-)\sim(s^+)^{-\kappa r^2}\, {\rm e}^{{\rm i}s^+r}
\int {d[B]} \,{\rm e}^{(2{{\rm i}}/\kappa ) \sum_{l=1}^{\infty}
         (s^+)^{1-l}{\rm tr}(B {B^\dagger})^l}\,,
\end{equation}
where the factor $\kappa=1$ has been sneaked in for later use.

Using Eqs. (\ref{RsCs}), (\ref{generating}), and (\ref{finfinZ}), it
is easy to check that the resulting two-point correlator $R(s)$ is
indeed of the form (\ref{rs}). We only have to realize that, when
applied to functions of $s^+=(\epsilon_+-\epsilon_-+2{\rm i}\delta)\pi\overline{\rho}$, the
double derivative $\frac{\partial^2}{\partial \epsilon_+\partial\epsilon_-}$ of (\ref{generating})
acts like $-\pi^2\overline{\rho}^2\frac{\partial^2}{\partial {s^+}^2}$.  Inserting
$\epsilon_\pm=\pm\frac{\epsilon}{2}=\pm\frac{s}{2\pi\overline{\rho}}$, $s^+$ now takes the
form $s^+=s+{\rm i}\delta'$ with $\delta'=2\pi\overline{\rho}\delta\downarrow 0$.

\subsection{Average over the GOE}

\label{sec:sigma_orthogonal}

For time-reversal invariant systems, we have to integrate over all
real symmetric $N\times N$ matrices, as in
\begin{equation}
\averagedots\equiv\int d[H]{\rm e}^{-\BRMT \Tr H^2}\ldots\equiv\int\prod_{\mu\leq\nu}dH_{\mu\nu}
\ARMT{\rm e}^{-\BRMT \Tr H^2}\ldots\;.
\end{equation}
The constants $\ARMT$ and $\BRMT$ appearing in the Gaussian weight are
chosen such that all products $H_{\mu\nu}^2 =H_{\mu\nu}H_{\nu\mu}$ with
$\mu\neq\nu$ average to $\frac{\lambda^2}{N}$, implying $\BRMT=\frac{N}{4\lambda^2}$
divided by two compared to the unitary case; the squared diagonal elements then
average to $\frac{2\lambda^2}{N}$.  The quadratic exponent is most
conveniently represented as $\frac{N}{4\lambda^2}\Tr
H^2=\frac{N}{8\lambda^2}\sum_{\mu\nu}H_{\mu\nu}(H_{\nu\mu}+ H_{\mu\nu})$.\footnote{
  If instead we work with $\frac{N}{4\lambda^2}\sum_{\mu\nu}H_{\mu\nu}H_{\nu\mu}$,
  the matrix $X$ arising when eliminating the linear term of
  (\ref{haverage}) needs to be symmetrized before transforming
  $H+\mbox{const}\,X\to H$, to comply with $H$ being real symmetric.
  This leads to the same changes as outlined in the text.}  After
averaging over $H$, the summand $H_{\mu\nu}H_{\mu\nu}$ leads to an
additional term in the exponent of ${\cal Z}(\epsilon_+,\epsilon_-)$, Eq.
(\ref{quartic}), again quartic in the integration variables $\psi$.
Rather than $\tr \YRMT^2=\Tr\,
X^2=\sum_{\mu\nu}\psi_\mu^\dagger\Lambda\psi_\nu\psi_\nu^\dagger\Lambda\psi_\mu$, we now obtain
\begin{equation}
\sum_{\mu\nu}\psi^\dagger_\mu \Lambda
\psi_\nu\left(\psi^{\dagger}_\nu \Lambda \psi_\mu+\psi^{\dagger}_\mu \Lambda \psi_\nu\right)=
\sum_{\mu\nu}2\Psi_\mu^\dagger
\Lambda \Psi_\nu\Psi^\dagger_\nu \Lambda \Psi_\mu\,,
\end{equation}
with the $4r$--component vectors $\Psi$ and $\Psi^\dagger$ defined by
\begin{equation}
    \Psi\equiv\frac{1}{\sqrt 2}\left(
    \begin{array}{c}
        \psi\cr \psi^\ast
    \end{array}\right)\ \ \ \Rightarrow\ \ \ \Psi^\dagger= \frac{1}{\sqrt 2}(\psi^\dagger,\psi^T)\,.
    \label{eq:Psi_def}
\end{equation}
Note that, apart from being mutually adjoint, the vectors $\Psi$ and
$\Psi^\dagger$ have to fulfill the symmetry relation
\begin{equation}
    \Psi^\dagger = \Psi^T
    \sigma_{1},
    \label{eq:Psi_sym}
\end{equation}
where the Pauli matrix $\sigma_{1}=\matr{0}{1}{1}{0}$ acts in the
two--component space of Eq.~(\ref{eq:Psi_def}).

The following reasoning essentially carries over from the unitary
case, if we replace $\psi\to\Psi$ and thus double the dimension of all
matrices involved.  Due to Eq. (\ref{eq:Psi_sym}) these matrices must
satisfy additional symmetry relations: The matrices $Q$ used to
decouple the quartic terms must be subject to the constraint $Q=\sigma_{1}
Q^T \sigma_{1}$.  The saddle points $Q_{\rm s}=T\Lambda T^{-1}$ will comply
with that restriction if $T$ is pseudo-orthogonal, i.e., $T^T=\sigma_1
T^{-1}\sigma_1$.  As a consequence, our $2r$-dimensional
matrices $B$ need to fulfill the condition
\begin{equation}
\label{sigma_time_reversal}
B^\dagger=-\sigma_1 B^T\sigma_1\,.
\end{equation}
Thus, we now have to integrate over $2r^2$ independent matrix elements, and the
measure $d[Q_{\rm s}]$ is proportional to
$(s^+)^{-2r^2}$ rather than $(s^+)^{-r^2}$.  We finally recover
(\ref{rs}) with $\kappa=2$, the two additional factors arising from the
changes in $d[Q_{\rm s}]$ and in $\BRMT$.

\section{Expansion of the two-point correlator and the form factor}
\label{sec:sigmamodel_formfactor}

We proceed to convert the $\sigma$ model expression for $R(s)$, Eq.
(\ref{rs}), into an expansion closely resembling our semiclassical
treatment.  In the limit $s\to\infty $ the principal contribution to the
exponent in (\ref{rs}) comes from the quadratic term $(2{{\rm
    i}}/\kappa)\, \mbox{tr}\,BB^\dagger$.  It is thus convenient to express the
$B$ integral in Eq. (\ref{rs}) as a Gaussian
average, like in%
\begin{equation}\label{sigma_quadratic_av}
  \big< f(B, {B^\dagger}) \big>         \equiv \int {d[B]}\,
        f(B, {B^\dagger}) \ {\rm e}^{(2{{\rm i}}/\kappa )  {\rm tr}B {B^\dagger}}\,,
\end{equation}
over the remaining exponential; setting $M\equiv BB^\dagger$ the latter
exponential can be expanded as
\begin{eqnarray}\label{expan1}
\exp \left(\frac{2{{\rm i}}}{\kappa }\sum_{l\geq 2} {s^+}^{1-l}{\rm tr }\,
M^l\right) &=&\sum_{V=0}^\infty
\frac{1}{V!}\left(\frac{2{{\rm i}}}{\kappa }\right)^V
\left(\sum_{l\geq2} {s^+}^{1-l}{\rm tr}\,M^l \right)^V \nonumber\\
&=&\sum_{\vec{v}} \frac{1}{\prod_{l\geq 2}v_l!}\left(\frac{2
{{\rm i}}}{\kappa }\right)^V{s^+}^{V-L}
 \prod_{l\geq2}
(\mbox {tr}\,M^l)^{v_l},
\end{eqnarray}
where in the last step we performed a multinomial expansion of
$(\sum_{l\geq2} {s^+}^{1-l}{\rm tr }\, M^l)^V$.  The summation extends over
integers $v_2,v_3,\ldots,v_l,\ldots$ each of which runs from zero to infinity,
and we write $\vec{v}=(v_2,v_3,\ldots)$ just like in our semiclassical
analysis.  The total number of traces in the summand $\vec v$ is
$V(\vec v) =\sum_{l\geq 2} v_l$, and again we define $L(\vec{v})=\sum_{l\geq
  2}lv_l$.  With Eqs. (\ref{sigma_quadratic_av}) and (\ref{expan1}),
the two-point correlator of Eq. (\ref{rs}) turns into
\begin{eqnarray}
\label{Rs_expanded}
R(s)&\sim&-\frac{1}{2}\mbox{Re}\left\{\lim_{r\to
0}\frac{1}{r^2}\partial_{s^+}^2{s^+}^{-\kappa  r^2}
{\rm e}^{{\rm i} s^+r}
\right.\nonumber\\&&\left.\times\sum_{\vec{v}}
\frac{1}{\prod_{l\geq 2}v_l!}\left(\frac{2
{{\rm i}}}{\kappa }\right)^V{s^+}^{V-L} \left\langle\prod_{l\geq2}(\mbox
{tr}M^l)^{v_l}\right\rangle\right\}-\frac{1}{2}.
\end{eqnarray}

The leading term $\vec{v}=0$ corresponds to
$\left<1\right>=\left(\mathrm{const}\right)^{\kappa r^2}\stackrel{{r\to
    0}}{\longrightarrow} 1$.  The resulting contribution to the two-point
correlator is given by
\begin{eqnarray}
\!&&\!\!-\frac{1}{2}\mbox{Re} \;{\rm lim}_{r \to 0}
\frac{1}{r^2} \partial^2_{s^+} {s^+}^{-\kappa r^2}{\rm e}^{{\rm i} s^+ r}\nonumber\\
\!&=&\!\! -\frac{1}{2}\mbox{Re} \;{\rm lim}_{r \to 0}
\frac{1}{r^2}
\left(\!(-\kappa r^2)(-\kappa r^2\!-\!1){s^+}^{-2}\!+\!2(-\kappa r^2){\rm i} r{s^+}^{-1}
\!\!+\!\!({\rm i} r)^2\right)\!{s^+}^{-\kappa r^2}
{\rm e}^{{\rm i} s^+ r}\nonumber\\
\!&=&\!\!\frac{1}{2}-\mbox{Re}\frac{\kappa
}{2{s^+}^2}\,,
\end{eqnarray}
where $\frac{1}{2}$ compensates the summand $-\frac{1}{2}$ in
(\ref{Rs_expanded}).  Upon Fourier transforming according to Eq.
(\ref{fourier}), the remaining term will bring about a contribution
$\kappa \tau$ to the spectral form factor, reproducing the diagonal part
both in the unitary and orthogonal cases.

For all other $\vec{v}$, the operations of taking the second
derivative by $s^+$ and going to the limit $r\to 0$ commute, meaning
that the factor ${s^+}^{-\kappa r^2}{\rm e}^{{\rm i} s^+ r}$ in
(\ref{Rs_expanded}) can be disregarded.  We thus obtain
\begin{equation}
R(s)\sim-\frac{1}{2}\mbox{Re}\left\{\frac{\kappa}{{s^+}^2}+
\sum_{\vec{v}\neq 0}
\frac{1}{\prod_{l\geq 2}v_l!}\left(\frac{2
{{\rm i}}}{\kappa }\right)^V\!\partial_{s^+}^2{s^+}^{V-L} \lim_{r\to
0}\frac{1}{r^2}\left\langle\prod_{l\geq2}(\mbox
{tr}M^l)^{v_l}\right\rangle\right\}.
\end{equation}
Fourier transformation yields a similar expression for the spectral
form factor. Using $ \frac{1}{\pi}\int_{-\infty}^{\infty}ds{\rm e} ^{2{\rm
    i}s\tau}{\rm Re}\left[{\rm
    i}^{-n+1}(s^+)^{-n-1}\right]=\frac{(-2)^n}{n!}\tau^n\,$, for $\tau>0$,
the Taylor coefficients of $K(\tau)=\kappa \tau+\sum_{n\geq 2}K_n\tau^n$ are
determined as
\begin{equation}
  \label{sigmak}
  K_n=\frac{\kappa }{(n-2)!}\sum_{\vec{v}}^{L(\vec{v})-V(\vec{v})+1=n}
  \frac{(-1)^V(-2{\rm i})^{L(\vec{v})}}{\kappa ^{V+1}\prod_{l\geq 2} v_l!}
\lim_{r\to 0}\frac{1}{r^2}\left\langle\traceprodM\right\rangle\,.
\end{equation}
The sum over $\vec{v}$ closely resembles the corresponding
periodic-orbit result (\ref{kn}).  The form factor is determined by
averaged trace products $\langle\traceprodM\rangle$, taking the place of the
numbers of structures $N(\vec{v})$ in periodic-orbit theory.  The
traces of $M^l$ will be called $l$-traces, to stress the analogy to
$l$-encounters of periodic orbits.

\section{Contractions}

\subsection{Contraction rules}

\label{sec:con_rules}

Similarly to the numbers of structures, the averaged trace products
$\langle\traceprodM\rangle$, $M=BB^\dagger$ can be evaluated recursively, with the
help of Wicks's theorem: As any Gaussian average, each
$\langle\traceprodB\rangle$ can be written as a sum over contractions.  For the
GUE, we have to single out one matrix $B$ in the integrand and then
take into account contractions with all matrices $B^\dagger$, as in
\begin{eqnarray}
\big<\tr(BB^\dagger BB^\dagger BB^\dagger)\big>
&=&
\big<\tr(
\contraction[2ex]{}{B}{}{B^\dagger}{}
BB^\dagger BB^\dagger BB^\dagger
)\big>
+
\big<\tr(
\contraction[2ex]{}{B}{B^\dagger B}{B^\dagger}{}
BB^\dagger BB^\dagger BB^\dagger
)\big> \nonumber\\
&+&
\big<\tr(
\contraction[2ex]{}{B}{B^\dagger B B^\dagger B}{B^\dagger}{}
BB^\dagger BB^\dagger BB^\dagger
)\big>
\end{eqnarray}
For the GOE, the chosen $B$ must be contracted with the remaining
matrices $B$ as well.

For each of the summands, the two matrices connected by the
contraction line can be eliminated by integrating over the elements of
contracted matrices as if the remaining elements were absent.  For the
GUE, two contracted matrices $B$ and $B^\dagger$ can be removed using two
simple rules,
\begin{subequations}
\label{sigma_con}
\begin{eqnarray}
             \big< {\rm tr} \ \contraction[2ex]{}{  {\mathbf  B}}{X\ {\rm tr}\ }
 {      {\mathbf B} } { B}
         X \ {\rm tr} \ { {  B^\dagger}} Y [\ldots] \big>
          &=&
        - \frac{1}{2 {{\rm i}}} \big< {\rm tr} ( X \ Y )  [\ldots]
                 \big>  \\
             \big< {\rm tr} \ \contraction[2ex]{}{  {\mathbf B}}{}{(X)
                  {\mathbf B} }
                 { B} X { { B^\dagger}} Y [\ldots] \big> &=&
                - \frac{1}{2 {{\rm i}}} \big< {\rm tr} X \ {\rm tr} Y
[\ldots] \big> \,,\ \
\end{eqnarray}
to be justified below.  Here, $X$ and $Y$ are products of $B$'s and
$B^\dagger$'s providing all traces on the left-hand side with alternating
sequences $BB^\dagger BB^\dagger\ldots BB^\dagger$; e.g., in Eq. (\ref{sigma_con}a) we
need $X$ beginning and ending with $B^\dagger$.  Then, the same alternating
structure will hold for all traces on the right-hand side.  We may
thus express all quantities in terms of $M=BB^\dagger$.  Each contraction
eliminates one $B$ and one $B^\dagger$ and thus reduces the number of $M$'s
by 1.

In case of the GOE, two more rules arise for contractions of $B$ with
$B$:
\begin{eqnarray}
             \big< {\rm tr} \ \contraction[2ex]{}{
{\mathbf B}}{X\ {\rm tr}\ }{      {\mathbf B} }
        { B} X \ {\rm tr}\ { B} Y[\ldots] \big>
         &=&
                - \frac{1}{2 {\rm i}}\big<{\rm tr}( X \ {Y^\dagger} )
[\ldots]\big>
\\
                \big< {\rm tr} \   \contraction[2ex]{}
{ {\mathbf B}}{}{(X)      {\mathbf B} }
        B X B Y [\ldots]\big> &=&
 - \frac{1}{2 {\rm i}} \big<  {\rm tr} ( X \ {Y^\dagger} )[\ldots] \big>\,.
\end{eqnarray}
\end{subequations}
Again, the only possible ordering of $B,{B^\dagger}$ on either side is
alternation $B {B^\dagger} B {B^\dagger} \ldots$.  Contractions between two matrices
$B^\dagger$ are described by analogous rules, with $B$ replaced by $B^\dagger$.

\subsubsection{Derivation}

We briefly outline the derivation of these contraction rules. In an
index notation, the Gaussian average (\ref{sigma_quadratic_av})
invoked in the unitary case reads
\begin{equation}
\label{averageindex}
\big\langle\ldots\big\rangle=\int d[B]{\rm e}^{2{\rm i}\sum_{\alpha\alpha'}
|B_{\alpha\alpha'}|^2}\ldots\,.
\end{equation}
For the orthogonal case, the factor 2 has to be removed if we
interpret $\sum_{\alpha\alpha'}$ as a sum over all $\alpha,\alpha'$; compare Eq.
(\ref{sigma_quadratic_av}).  However, in the latter case only one half
of the matrix elements $B_{\alpha\alpha'}$ are independent integration
variables --- a consequence of the time reversal relation
(\ref{sigma_time_reversal}).  We may thus equivalently restrict the
sum to independent variables only, and keep the prefactor 2. For both
the unitary and the orthogonal cases Gaussian integration now yields
\begin{equation}
\label{average}
\big\langle B_{\alpha\alpha'} {B^\dagger}_{\beta'\beta}\big\rangle =
  -{1\over 2{\rm i}}\delta_{\alpha\beta} \delta_{\alpha'\beta'}\,,
\end{equation}
and thus, by Wick's theorem, the prototypical contraction rule
\begin{equation}
\label{prototype}
\big<\ldots
\contraction[2ex]{}{B}{{}_{\alpha \alpha'}\ldots}{B}{}
B_{\alpha\alpha'}\ldots B^\dagger_{\beta'\beta}
\ldots\big>
=-\frac{1}{2{\rm i}}\delta_{\alpha\beta}\delta_{\alpha'\beta'}\big<\ldots\big>\,.
\end{equation}
From this relation we obtain (using the summation convention)
\begin{eqnarray}
  \big< {\rm tr}  \contraction[2ex]{}{  {\mathbf  B}}{(\;X)}
 {      {\mathbf B} } { B}
         X \ {\rm tr} { {  B^\dagger}} Y [\ldots]\big> &=&
\big< %%preamble
  \contraction[2ex]{}{  {\mathbf  B}}{(\; X_{\beta\gamma})} {  {\mathbf B} } {%Positioners
  B_{\alpha\beta}%%First contracted
  }
         X_{\beta\alpha}%%Insertion
          { {B^\dagger_{\gamma\delta}%%Second contracted
         }}
         Y_{\delta\gamma} [\ldots]\big> \nonumber\\%final part
& = &-{1\over 2{\rm i}}\big\langle X_{\beta
\alpha} Y_{\alpha\beta}[\ldots]\big\rangle\nonumber\\
& =& -{1\over 2{\rm i}} \big\langle{\rm
tr}(X \,Y)[\ldots]\big\rangle,
\end{eqnarray}
which is contraction rule (\ref{sigma_con}a). Rule (\ref{sigma_con}b)
is proven in the same manner.  Rule (\ref{sigma_con}d) for the GOE
results from
\begin{eqnarray}
  \big\langle{\rm tr}\contraction[2ex]{}{{B}}{}{(X){B}}{B}X{{ B}} Y[\ldots]\big\rangle
  &=&   \big\langle {\rm tr}(
  (\sigma_1 X)  %%preamble
  \contraction[2ex]{}{B}{(X \sigma_1)\,(\sigma_1\,\,}
  {  } %%Positioners
  {B}%%First contracted
        (Y \sigma_1)\,(\sigma_1 %%Insertion
          { {B%%Second contracted
         }}
          \sigma_1))[\ldots]\big\rangle \nonumber\\
  &=&%final part
  \big\langle
  (\sigma_1 X)_{\alpha\beta}%%preamble
  \contraction[2ex]{}{  {\mathbf  B}}{((Y \,\;\sigma_1)_{\gamma \delta} (\sigma_1 )\!\!} {  {\mathbf B} } {%Positioners
  B_{\beta\gamma}%%First contracted
  }
        (Y \sigma_1)_{\gamma \delta} (\sigma_1 %%Insertion
           { {  B%%Second contracted
         }}
         \sigma_1)_{\delta \alpha}[\ldots]\big\rangle \nonumber\\%final part
  &\stackrel{\protect~(\ref{sigma_time_reversal})}{=}&
   - \big\langle
  (\sigma_1 X)_{\alpha\beta} %%preamble
  \contraction[2ex]{}{  {\mathbf  B}}{((Y \sigma_1)_{\gamma \delta})} {\;  {\mathbf B} } {%Positioners
  B_{\beta\gamma}%%First contracted
  }
       (Y \sigma_1)_{\gamma \delta}%%Insertion
          { {B^\dagger_{\alpha\delta}%%Second contracted
         }}
        [\ldots]\big> \nonumber\\ %final part
%  &=&-\big\langle(\sigma_1 X)_{\alpha\beta}(Y\sigma_1)_{\alpha\beta}[\ldots]\big>\nonumber\\
  &=&+{1\over 2{\rm i}}\big\langle
(\sigma_1 X)_{\alpha\beta}  ( \sigma_1 Y^T)_{\beta\alpha}[\ldots]\big\rangle\nonumber\\
  &=&+{1\over 2{\rm i}}
\big\langle{\rm tr} ( X  ( \sigma_1 Y^T \sigma_1))[\ldots]\big\rangle\nonumber\\
&=&-{1\over 2{\rm i}}\big\langle{\rm
tr}(X\,Y^\dagger)[\ldots]\big\rangle\,,
\end{eqnarray}
where in the last step we used that $Y$ contains an odd number of
matrices $B$ and $B^\dagger$ such that the time reversal relation
(\ref{sigma_time_reversal}) implies $\sigma_1 Y^T \sigma_1= - Y^\dagger$. The proof
of rule (\ref{sigma_con}c) proceeds along the same lines.

\subsection{Contraction steps leading to traces of unit matrices}

\label{sec:con_rules_special}

By applying the rules (\ref{sigma_con}a-d) we can, step by step,
eliminate all matrices $B$ and $B^\dagger$ until we are finally left only
with traces of unit matrices.  These unit matrices come into play only
when contracting neighboring $B$'s and $B^\dagger$'s belonging to the same
trace. We then have to apply rule (\ref{sigma_con}b) with $X$ or $Y$
absent, or equivalently $X$ or $Y$ replaced by a unit matrix $\mathbf
1$; that matrix must be of size $r\times r$ in the unitary case and of size
$2r\times 2r$ in the orthogonal case.  We thus have to deal with the three
special cases $X=\mathbf 1\neq Y$, $Y=\mathbf 1\neq X$, and $X=Y=\mathbf
1$ of Eq. (\ref{sigma_con}b).  Using $\tr\mathbf 1=\kappa r$, we see that
these cases lead to the relations
\begin{subequations}
\label{finstep}
\begin{eqnarray}
  \big\langle\tr
    \contraction{}{B}{}{B^\dagger}{}
    BB^\dagger
  Y[\ldots]\big\rangle
  &=& -\frac{1}{2{\rm i}}\big\langle\tr\mathbf 1\tr Y[\ldots]\big\rangle
  = -\frac{\kappa r}{2{\rm i}}\big\langle\tr Y[\ldots]\big\rangle\\
  \big\langle\tr
    \contraction{}{B}{X}{B^\dagger}{}
    BXB^\dagger
  [\ldots]\big\rangle
  &=& -\frac{1}{2{\rm i}}\big\langle\tr X\tr\mathbf 1[\ldots]\big\rangle
  = -\frac{\kappa r}{2{\rm i}}\big\langle\tr X[\ldots]\big\rangle\\
  \big\langle\tr
    \contraction{}{B}{}{B^\dagger}{}
    BB^\dagger
  [\ldots]\big\rangle
  &=& -\frac{1}{2{\rm i}}\big\langle(\tr\mathbf 1)^2[\ldots]\big\rangle
  \,\,\,\,\,= -\frac{\kappa^2 r^2}{2{\rm i}}\big\langle [\ldots]\big\rangle\,.
\end{eqnarray}
\end{subequations}
In contrast, for the rules (\ref{sigma_con}a,c,d) the overall number
of matrices $B$ and $B^\dagger$ in $X$ must be odd and thus $X\neq\mathbf 1$;
the same applies to $Y$.  Hence these rules do not make for further
special cases.

The last two matrices $B$ and $B^\dagger$ will always be removed through a
step as in (\ref{finstep}c), yielding a factor $r^2$.  All {\it
  intermediate} steps involving contractions between neighboring
matrices $B$ and $B^\dagger$ further increase the power in $r$. Hence, such
contractions do not survive the replica limit $\lim_{r\to
  0}\frac{1}{r^2}$ and therefore do not contribute to the form factor.

\subsection{Analogy between full contractions and orbit pairs}

\label{sec:analogy}

To elucidate the relation between orbit pairs and the $\sigma$ model, the
contraction rules (\ref{sigma_con}a-d) and (\ref{finstep}a-c) for
trace products $\langle\traceprodB\rangle$ can be put to use in two ways: In the
present Subsection, we will reveal orbit pairs as topologically
equivalent to ``full contractions'' in the $\sigma$ model.  In Subsection
\ref{sec:rec_contractions}, we will translate the above contraction
rules into a recursion analogous to the one determining the numbers of
structures $N(\vec{v})$.

\subsubsection{Full contractions}

By iteratively applying Wick's theorem, each trace product can be
written as a sum over ``full contractions'', with each $B$ and $B^\dagger$
connected to another matrix through a contraction line. In the unitary
case each such line must connect one $B$ and one $B^\dagger$. We may thus
write, e.g.,
\begin{eqnarray}
\big<\tr(BB^\dagger BB^\dagger BB^\dagger)\big>
&=&
\big<\tr(
%1
\contraction[2ex]{}{B}{}{B^\dagger}{}
\contraction[2ex]{BB^\dagger}{B}{}{B^\dagger}{}
\contraction[2ex]{BB^\dagger BB^\dagger}{B}{}{B^\dagger}{}
BB^\dagger BB^\dagger BB^\dagger
)\big>
+
\big<\tr(
%2
\contraction[2ex]{}{B}{}{B^\dagger}{}
\contraction[3ex]{BB^\dagger}{B}{BB^\dagger}{B^\dagger}{}
\contraction[2ex]{BB^\dagger B}{B^\dagger}{}{B}{}
BB^\dagger BB^\dagger BB^\dagger
)\big>\nonumber\\
&+&
\big<\tr(
%3
\contraction[3ex]{}{B}{B^\dagger B}{B^\dagger}{}
\contraction[2ex]{B}{B^\dagger}{}{B}{}
\contraction[2ex]{BB^\dagger BB^\dagger}{B}{}{B^\dagger}{}
BB^\dagger BB^\dagger BB^\dagger
)\big>
+ \pcc\nonumber\\
&+&
\big<\tr(
%5
\contraction[3ex]{}{B}{B^\dagger B B^\dagger B}{B^\dagger}{}
\contraction[2ex]{B}{B^\dagger}{}{B}{}
\contraction[2ex]{BB^\dagger B}{B^\dagger}{}{B}{}
BB^\dagger BB^\dagger BB^\dagger
)\big>
+
\big<\tr(
%6
\contraction[4ex]{}{B}{B^\dagger B B^\dagger B}{B^\dagger}{}
\contraction[3ex]{B}{B^\dagger}{BB^\dagger}{B}{}
\contraction[2ex]{BB^\dagger}{B}{}{B^\dagger}{}
BB^\dagger BB^\dagger BB^\dagger
)\big>\,.
\label{fullcontr}
\end{eqnarray}
In the orthogonal case, contraction lines may also connect two $B$'s,
or two $B^\dagger$'s.

Each of these full contractions can be evaluated step by step, each
time eliminating the matrices connected by one contraction line using
the above rules.  If one of the intermediate steps involves a
contraction between neighboring $B$'s and $B^\dagger$'s of the same trace,
the corresponding full contraction will be at least of cubic order in
$r$ and thus vanish in the replica limit $\lim_{r\to 0}\frac{1}{r^2}$.
For all full contractions surviving the replica limit, the number of
$M=BB^\dagger$ can be brought to one through $L-1$ applications of rules
(\ref{sigma_con}), each yielding a factor $-\frac{1}{2{\rm i}}$. Finally
applying (\ref{finstep}c), we pick up a factor $-\frac{\kappa^2
  r^2}{2{\rm i}}$, multiplied with $\langle 1\rangle\xrightarrow[r\to 0]{}1$. Thus,
each such full contraction effectively yields $\frac{\kappa^2
  r^2}{(-2{\rm i})^{L(\vec{v})}}$.  In the example of Eq.
(\ref{fullcontr}), it is easy to show that only the full contraction
\begin{equation}
\label{survivor}
\pcc
\end{equation}
survives.  In general, for each trace product $\langle\traceprodM\rangle$,
$M=BB^\dagger$, the number of full contractions surviving in the replica
limit (or, in short, the ``number of contractions'') will be denoted
by $N_c(\vec{v})$.  In that limit, the respective trace product is
given by
\begin{equation}
\label{Ncdef}
\lim_{r\to 0}\frac{1}{r^2}\left\langle\traceprodM\right\rangle=
\frac{\kappa^2}{(-2{\rm i})^{L(\vec{v})}}N_c(\vec{v})\,.
\end{equation}
The Taylor coefficients of the spectral form factor (\ref{sigmak}) are
thus expressed as
\begin{equation}
  \label{sigmakNc}
  K_n=\frac{\kappa }{(n-2)!}\sum_{\vec{v}}^{L(\vec{v})-V(\vec{v})+1=n}
  \frac{(-1)^V}{\kappa ^{V-1}\prod_{l\geq 2} v_l!}N_c(\vec{v})\,.
\end{equation}

\subsubsection{Analogy to orbit pairs}

We will show that the surviving full contractions are directly related
to structures of orbit pairs $(\gamma,\gamma')$.  In particular, survival in
the replica limit will be revealed as analogous to $\gamma$ and $\gamma'$
being non-decomposing periodic orbits.

The $l$-traces $\tr(BB^\dagger)^l$ can be seen as equivalent to {\bf
  $l$-encounters}, with each $BB^\dagger$ standing for one stretch of the
original orbit $\gamma$.  The matrices $B$ and $B^\dagger$ are identified with
left and right ports. We assume that the ``stretches'' inside each
trace are ordered in such a way that $\gamma'$ connects each left port to
the right port of the {\it preceding} stretch; the ordering of
stretches is thus opposite to the permutations $P_{\rm enc}$ or the
$p_{\rm enc}$ defined in Subsection \ref{sec:leftright}.  If we
represent intra-encounter connections by lower lines, the connections
inside $\gamma$ are depicted as in
\begin{equation}
\label{gammaconnect}
\big\langle\tr(
\lcontraction[2ex]{}{B}{}{B^\dagger}{}
BB^\dagger
\lcontraction[2ex]{}{B}{}{B^\dagger}{}
BB^\dagger
\ldots
\lcontraction[2ex]{}{B}{}{B^\dagger}{}
BB^\dagger
)
\tr(
\lcontraction[2ex]{}{B}{}{B^\dagger}{}
BB^\dagger
\ldots)\big\rangle\,,
\end{equation}
whereas the connections inside $\gamma'$ are given by
\begin{equation}
\label{primeconnect}
\big\langle\tr(
{
\lcontraction[2ex]{B}{B^\dagger}{}{B}
\lcontraction[2ex]{B B^\dagger B\ldots}{B^\dagger}{}{B}
\lcontraction[3ex]{}{B}{B^\dagger B\ldots B^\dagger B}{B}
B B^\dagger B \ldots B^\dagger BB^\dagger
}
)
\tr(
{
\lcontraction[2ex]{B}{B^\dagger}{}{B}
\lcontraction[3ex]{}{B}{B^\dagger B\ldots}{B^\dagger}
B B^\dagger B \ldots B^\dagger
}
)\big\rangle\,.
\end{equation}
The (upper) contraction lines are analogous to {\bf loops}.

Each full contraction yields topological pictures of both {\bf orbits}
$\gamma$ and $\gamma'$. We only have to combine loops defined by the
contraction lines with intra-encounter connections as in Eqs.
(\ref{gammaconnect}) and (\ref{primeconnect}).  For instance, the full
contraction of Eq. (\ref{survivor}) leads to periodic orbits $\gamma$ and
$\gamma'$ topologically equivalent to
\begin{equation}
\label{connected_trace}
  \big<\tr(
  %4
  \contraction[4ex]{}{B}{B^\dagger B}{B^\dagger}
  \contraction[3ex]{B}{B^\dagger}{B B^\dagger}{B}
  \contraction[2ex]{BB^\dagger}{B}{B^\dagger B}{B^\dagger}
  \lcontraction[2ex]{}{B}{}{B^\dagger}
  \lcontraction[2ex]{BB^\dagger}{B}{}{B^\dagger}
  \lcontraction[2ex]{BB^\dagger BB^\dagger}{B}{}{B^\dagger}
  BB^\dagger BB^\dagger BB^\dagger
  )\big>
  \;\;\;\;\mbox{and}\;\;\;\;
  \big<\tr(
  %4
  \contraction[4ex]{}{B}{B^\dagger B}{B^\dagger}
  \contraction[3ex]{B}{B^\dagger}{B B^\dagger}{B}
  \contraction[2ex]{BB^\dagger}{B}{B^\dagger B}{B^\dagger}
  \lcontraction[2ex]{B}{B^\dagger}{}{B}
  \lcontraction[2ex]{BB^\dagger B}{B^\dagger}{}{B}
  \lcontraction[3ex]{}{B}{B^\dagger BB^\dagger B}{B^\dagger}
  BB^\dagger BB^\dagger BB^\dagger
  )\big>\,.
\end{equation}

To understand better the relation between full contractions and
orbits, imagine two partners accessible from the same $\gamma$ by {\it
  different reconnections} inside the {\it same encounters}, and
translate both orbit pairs into full contractions.  The different
reconnections entail different orderings of $BB^\dagger$'s inside each
trace, given that the ordering of ``stretches'' $BB^\dagger$ must be
opposite to the permutation $P_{\rm enc}$. When changing the order
$BB^\dagger$'s, the contraction lines must be carried along. Thus, both
orbit pairs lead to different full contractions with different upper
contraction lines.

Time-reversal invariant systems allow for more full contractions than
systems without time-reversal invariance.  Like in our semiclassical
treatment, the {\it unitary} case involves only loops connecting right
ports to left ports, and thus matrices $B^\dagger$ to matrices $B$.  The
sense of motion on $\gamma$ and $\gamma'$ may thus be fixed in a way that both
orbits run from left ports ($B$'s) to right ports ($B^\dagger$'s) inside
encounters, and from right to left ports inside loops.  Loops
connecting two left ports (two $B$'s) or two right ports (two
$B^\dagger$'s) are possible only in the {\it orthogonal} case.  In this
case ``left'' and ``right'' are arbitrary labels for the two sides of
each encounter.\footnote{ We here depart from the conventions of
  Subsection \ref{sec:encounter} where the left-hand side of an
  encounter was defined such as to include the entrance port of the
  corresponding first stretch.}

In general, $\gamma$ and $\gamma'$ may be either connected periodic orbits or
pseudo-orbits decomposing into a number of disjoint components.
However, we will show that full contractions {\it surviving the
  replica limit} correspond to {\it connected periodic orbits} $\gamma$
and $\gamma'$.  We have already seen that the surviving contraction
(\ref{survivor}) leads to connected $\gamma$ and $\gamma'$ as in
(\ref{connected_trace}); this contraction corresponds to the structure
{\it pc}, given that it involves a 3-trace, does not require
time-reversal invariance, and has both orbits connected.  In contrast,
the remaining five full contractions of Eq. (\ref{fullcontr}) are
killed in the replica limit and lead to decomposing $\gamma$ or $\gamma'$; for
instance,
\begin{equation}
\big<\tr(
\contraction[4ex]{}{B}{B^\dagger B B^\dagger B}{B^\dagger}{}
\contraction[3ex]{B}{B^\dagger}{BB^\dagger}{B}
\contraction[2ex]{BB^\dagger}{B}{}{B^\dagger} BB^\dagger BB^\dagger BB^\dagger)\big>
\end{equation}
yields a decomposing $\gamma'$ topologically equivalent to
\begin{equation}
\big<\tr(\contraction[4ex]{}{B}{B^\dagger B B^\dagger B}{B^\dagger}{}
\contraction[3ex]{B}{B^\dagger}{BB^\dagger}{B}
\contraction[2ex]{BB^\dagger}{B}{}{B^\dagger}
\lcontraction[2ex]{B}{B^\dagger}{}{B}
\lcontraction[2ex]{BB^\dagger B}{B^\dagger}{}{B}
\lcontraction[3ex]{}{B}{B^\dagger BB^\dagger B}{B^\dagger}
BB^\dagger BB^\dagger BB^\dagger
)\big>
\,.
\end{equation}

To generalize this example, let us consider an arbitrary full
contraction, and apply the contraction rules to remove two matrices
connected by a contraction line; we then want to check whether the
numbers of disjoint component orbits inside $\gamma$ and $\gamma'$ are
changed.  For contraction rules (\ref{sigma_con}a-d) with
$X,Y\neq\mathbf 1$, the numbers of disjoint orbits are preserved. For
example, adding the intra-encounter connections of $\gamma$, Eq.
(\ref{gammaconnect}), to rule (\ref{sigma_con}a) and cyclicly
permuting the elements inside the traces, we obtain
\begin{equation}
\label{contraction_conn}
\big\langle\tr
\contraction[2ex]{}{B}{X\,\tr Y\!\!}{B^\dagger}
\lcontraction[2ex]{}{B}{\!\!\!}{X}
\lcontraction[2ex]{BX\,\tr}{Y\,\,\,}{\!\!\!\!\!}{B^\dagger}
BX\,\tr Y B^\dagger
[\ldots]\big\rangle=
-\frac{1}{2{\rm i}}\big\langle
\lcontraction[2ex]{}{Y\,}{\!\!\!\!}{X}
YX
[\ldots]\big\rangle\,.
\end{equation}
The three lines on the left-hand side connect the last matrix $B$ in
$Y$, two matrices $B^\dagger$ and $B$, and the first matrix $B^\dagger$ in $X$.
These matrices and lines belong to the same component orbit of $\gamma$.
Applying contraction rule (\ref{sigma_con}a) as on the right-hand side
of Eq. (\ref{contraction_conn}), the two matrices connected by the
upper contraction line are removed, and the final $B$ of $Y$ is
connected directly to the first $B^\dagger$ in $X$.  We thus eliminated two
matrices related to our component orbit, but did not change the number
of component orbits inside $\gamma$.  The same applies to $\gamma'$, where
adding the intra-encounter connections of Eq. (\ref{primeconnect})
yields
\begin{equation}
\big\langle\tr
\contraction[2ex]{}{B}{X\,\tr Y\!\!}{B^\dagger}
\lcontraction[2ex]{}{B}{\,}{X}
\lcontraction[2ex]{BX\,\tr}{}{Y\!\!}{B^\dagger}
BX\,\tr Y B^\dagger
[\ldots]\big\rangle=
-\frac{1}{2{\rm i}}\big\langle\tr
\lcontraction[2ex]{}{}{Y\,}{X}
YX
[\ldots]\big\rangle\,,
\end{equation}
the only difference to (\ref{contraction_conn}) being that the
affected component of $\gamma'$ contains the last $B^\dagger$ in $X$ and the
first $B$ in $Y$.  Similar reasoning applies to rules
(\ref{sigma_con}b-d), if $X,Y\neq\mathbf 1$.

In contrast, the number of disjoint orbits is reduced when eliminating
contractions between neighboring $B$'s and $B^\dagger$'s according to the
special cases (\ref{finstep}a-c).  In particular, (\ref{finstep}a)
removes a component orbit $\contraction{}{B}{}{B^\dagger}
\lcontraction{}{B}{}{B^\dagger} BB^\dagger$ from $\gamma$,
\begin{equation}
  \big\langle\tr
    \contraction{}{B}{}{B^\dagger}
    \lcontraction{}{B}{}{B^\dagger}
    B B^\dagger
  Y[\ldots]\big\rangle
  = -\frac{\kappa r}{2{\rm i}}\big\langle\tr Y[\ldots]\big\rangle\,,
\end{equation}
whereas (\ref{finstep}b) removes a component of $\gamma'$,
\begin{equation}  \big\langle\tr\,
    \contraction{}{B}{X}{B^\dagger}
    \lcontraction{}{B}{X}{B^\dagger}
    B X B^\dagger
  [\ldots]\big\rangle
  = -\frac{\kappa r}{2{\rm i}}\big\langle\tr X[\ldots]\big\rangle\,,
\end{equation}
and (\ref{finstep}c) eliminates one component from both $\gamma$ and
$\gamma'$.

We can now prove that survival of the replica limit implies connected
orbits $\gamma$ and $\gamma'$.  Using the above contraction rules, all full
contractions are brought to a form $\propto\big\langle\tr
\contraction{}{B}{}{B^\dagger}{} BB^\dagger \big\rangle$, corresponding to a pair of
non-decomposing $\gamma$ and $\gamma'$ both represented by $\big\langle\tr
\contraction{}{B}{}{B^\dagger}{} \lcontraction{}{B}{}{B^\dagger}{} BB^\dagger
\big\rangle$.  If the initial full contractions involve either $\gamma$ or
$\gamma'$ decomposing into several disjoint orbits, some of the
intermediate steps have to reduce the number of component orbits;
these steps must involve cases (\ref{finstep}a-c) and thus kill the
contribution in the replica limit. In contrast, full contractions
corresponding to connected $\gamma$ and $\gamma'$ are reduced to $\big\langle\tr
\contraction{}{B}{}{B^\dagger}{} BB^\dagger \big\rangle$ without invoking
(\ref{finstep}a-c) and thus survive the replica limit.  Therefore
surviving full contractions indeed correspond to structures of pairs
of connected periodic orbits.

\subsubsection{Relation between $N(\vec{v})$ and $N_c(\vec{v})$}

The number $N_c(\vec{v})$ of contractions with fixed $\vec{v}$ thus
coincides with the number of structures $N(\vec{v})$, up to a
combinatorial factor.  To determine this factor, let us first consider
the {\it unitary} case.  In order to translate a surviving full
contraction into a structure of orbit pairs as defined in Chapter
\ref{sec:tau3}, one of the $L(\vec{v})$ ``stretches'' $BB^\dagger$ has to
be singled out as the first. Taking into account all $N_c(\vec{v})$
contractions, this leaves $L(\vec{v})N_c(\vec{v})$ possibilities.  The
structure obtained remains unchanged if we cyclicly permute the
elements $BB^\dagger$ in each $l$-trace (including their contraction
lines). Likewise the same structure arises from all $v_l!$ possible
orderings of $l$-traces inside our full contractions. Consequently,
each of the $N(\vec{v})$ structures is obtained \pagebreak

\noindent
$\prod_l l^{v_l} v_l!$ times.\footnote{ To see this explicitly, let us
  determine the permutation $P_{\rm enc}$ related to a given full
  contraction.  We number the ``stretches'' $BB^\dagger$ in order of
  traversal by $\gamma$, starting from an arbitrary reference stretch.
  Each trace gives rise to one cycle, obtained by replacing each
  $BB^\dagger$ with its number and reverting the order of numbers; as
  mentioned above the matrix products $BB^\dagger$ inside each trace are
  written in order opposite to $P_{\rm enc}$.  Now, cyclic
  permutations inside one trace correspond to cyclic permutations
  inside one cycle of $P_{\rm enc}$, and reordering traces to
  reordering cycles, obviously leaving $P_{\rm enc}$ invariant. Since
  we want to order traces by increasing size, traces of different size
  may not be interchanged.  } We thus have $\prod_l l^{v_l}v_l!
N(\vec{v})=L(\vec{v})N_c(\vec{v})$, or
\begin{equation}
\label{NNc_unitary}
N(\vec{v})=\frac{L(\vec{v})}{\prod_l l^{v_l}v_l!}N_c(\vec{v})\,.
\end{equation}

In the {\it orthogonal} case, we also need to account for the
directions of motion.  When translating full contractions to
structures of orbit pairs, we do not only have $L(\vec{v})$ choices
for a first stretch, but also two different ways to fix the sense of
motion on $\gamma$.  For all $N_c(\vec{v})$ contractions, this leaves
altogether $2L(\vec{v})N_c(\vec{v})$ possibilities.  (The different
choices for the direction of motion on $\gamma'$ do not lead to an
additional factor 2, since the pairs $(\gamma,\gamma')$ and
$(\gamma,\overline{\gamma'})$ are described by the same structure.)  On the
other hand, the structure remains unaffected by taking the adjoint of
the matrix product under a trace, and thus interchanging the ``left''
and ``right'' sides of the corresponding encounter; thus each
structure is obtained $2^V\prod_l l^{v_l} v_l!$ rather than $\prod_l l^{v_l}
v_l!$ times.\footnote{ Again, we can consider the corresponding
  permutations $P_{\rm enc}$.  The cycles representing stretches
  traversed from left to right are constructed as in the unitary case,
  but have the numbers of all $BB^\dagger$ traversed from right to left
  marked by an overbar, to denote time reversal; the opposite holds
  for their ``twin'' cycles.  Exchanging left and right interchanges
  the two twins and does not affect $P_{\rm enc}$.}  In general, the
numbers of contractions and the numbers of structures are thus related
by
\begin{equation}
\label{NNc}
N(\vec{v})=\frac{L(\vec{v})}{\kappa^{V-1}\prod_l l^{v_l}v_l!}N_c(\vec{v})\,
\end{equation}
with $\kappa=1$ and 2 respectively applying to the unitary and orthogonal
cases.  Crucially, (\ref{NNc}) implies that the series expansions of
$K(\tau)$ obtained form semiclassics, Eq. (\ref{kn}), and the $\sigma$
model, Eq. (\ref{sigmakNc}), coincide term by term.

\subsection{Recursion formula for the number of contractions}

\label{sec:rec_contractions}

Our recursion for the numbers of structures $N(\vec{v})$ has an
interesting field-theoretical interpretation.  The contraction rules
(\ref{sigma_con}a-d) can be turned into a recursion relation for the
trace products $\langle\traceprodB\rangle$ or, equivalently, $N_c(\vec{v})$
directly paralleling the results for $N(\vec{v})$, and indeed
inspiring some of the reasoning for $N(\vec{v})$.  To define that
recursion, let us select a trace $\mbox{tr}(BB^\dagger)^l$ inside the above
product (assuming $v_l>0$), and a matrix $B$ inside. We must contract
that $B$ with all other suitable matrices inside our trace product.
Three possibilities arise paralleling those met in Chapter
\ref{sec:combinatorics}, with the translation {\it trace $\leftrightarrow$
  encounter $\leftrightarrow$ cycle}.

\subsubsection{First case: contractions between different traces}

First, we take up the contractions between our selected $B$ in
$\mbox{tr}(BB^\dagger)^l$ and all suitable matrices in some {\it other}
trace $\mbox{tr}(BB^\dagger)^k$.  If we contract with the first $B^\dagger$ in
$\mbox{tr}(BB^\dagger)^k$, rule (\ref{sigma_con}a) implies that
\begin{equation}
\label{case1}
\bigg\langle\tr\Big(
\contraction[2ex]{}{B}{B^\dagger(BB^\dagger)^{l-1})\tr\Big(B}{B^\dagger}{}
BB^\dagger(BB^\dagger)^{l-1}\Big)\tr\Big(BB^\dagger
(BB^\dagger)^{k-1}\Big)[\ldots]\bigg\rangle
=-\frac{1}{2{\rm i}}\bigg\langle\tr(BB^\dagger)^{k+l-1}[\ldots]\bigg\rangle\,,
\end{equation}
i.e., one $k$-trace and one $l$-trace disappear while a
$(k+l-1)$-trace is born.  Contractions with all further $B^\dagger$ in
$\mbox{tr}(BB^\dagger)^k$ can be brought to the same form as in
(\ref{case1}), by cyclic permutation.  Consequently, for $k\neq l$ the
contractions with all $kv_k$ matrices $B^\dagger$ in $k$-traces
$\mbox{tr}(BB^\dagger)^k$ give the same result.  If $k=l$, we need to
exclude the $k$ possible contractions with $B^\dagger$'s of the same trace.
In general, we thus get $k(v_k-\delta_{kl})$ contractions like
(\ref{case1}).

In the orthogonal case we must also invoke rule (\ref{sigma_con}c) for
contractions with $k(v_k-\delta_{kl})$ matrices $B$ in traces
$\mbox{tr}(BB^\dagger)^k$,
\begin{equation}
\bigg\langle\tr\Big(
\contraction[2ex]{}{B}{B^\dagger(BB^\dagger)^{l-1}\Big)\tr\Big(}{B}{}
BB^\dagger(BB^\dagger)^{l-1}\Big)\tr\Big(BB^\dagger
(BB^\dagger)^{k-1}\Big)[\ldots]\bigg\rangle
=-\frac{1}{2{\rm i}}\bigg\langle\tr(BB^\dagger)^{k+l-1}[\ldots]\bigg\rangle \,,
\end{equation}
which again all contribute identically.

Each time, one $k$-trace and one $l$-trace disappear and one
$(k+l-1)$-trace is added to the trace product.  The vector $\vec v$
therefore changes according to $v_k\to v_k-1,\,v_l\to v_l-1,\,
v_{k+l-1}\to v_{k+l-1}+1$; using the same notation as in our
semiclassical analysis we write $\vec{v}'=\vec v^{[k,l\to k+l-1]}$. The
overall number of matrices $M=BB^\dagger$ is decreased by 1 such that $L\to
L-1$.  From each of the $\kappa k(v_k-\delta_{kl})$ contractions, the trace
product $\langle\traceprodB\rangle$ receives a contribution
$-\frac{1}{2{\rm i}}\langle\traceprodprime\rangle$.  If we formulate our recursion in
terms of numbers of contractions, the denominator $-2{\rm i}$ cancels due
to the factor $(-2{\rm i})^{L(\vec{v})}$ appearing in our formula for the
numbers of contractions (\ref{Ncdef}).  Thus, the above contractions
provide $N_c(\vec{v})$ with a contribution
\begin{equation}
\kappa k(v_k-\delta_{kl}) N_c(\vec{v}^{[k,l\to k+l-1]})\,.
\end{equation}
The present case is analogous to the merger of two encounters, or
cycles, into one.

\subsubsection{Second case: contractions between $B$, $B^\dagger$ inside the same trace}

Next, we turn to ``internal'' contractions between the selected $B$
and matrices $B^\dagger$ in the same trace $\mbox{tr}(BB^\dagger)^l$.  As
explained above, contractions with $B^\dagger$'s immediately preceding or
following the selected $B$ increase the order in $r$ and lead to a
result that vanishes in the replica limit.  For all other
contractions, we can apply rule (\ref{sigma_con}b), as in
\begin{equation}
\bigg\langle \tr\Big(
\contraction[2ex]{}{B}{(B^\dagger B)^m}{B^\dagger}{}
B(B^\dagger B)^m B^\dagger(BB^\dagger)^{l-m-1}\Big)[\ldots]\bigg\rangle
=-\frac{1}{2{\rm i}}\bigg\langle\tr(BB^\dagger)^m\tr(BB^\dagger)^{l-m-1}[\ldots]\bigg\rangle\,,
\end{equation}
with $m$ running 1,2,$\ldots, l-2$. Thus, one $l$-trace disappears and two
traces, of $(BB^\dagger)^m$ and $(BB^\dagger)^{l-m-1}$, are added; the vector
$\vec{v}$ then changes to $\vec v^{[l\to m,l-m-1]}$.  Again, the factor
$-\frac{1}{2{\rm i}}$ disappears if we formulate our recursion in terms of
numbers of contractions $N_c(\vec{v})$.  From each of the $l-2$
contractions, $N_c(\vec{v})$ thus receives a contribution
\begin{equation}
N_c(\vec v^{[l\to m,l-m-1]})\,.
\end{equation}

In terms of periodic orbits, contraction lines between one $B$ and one
$B^\dagger$ inside the same trace correspond to loops connecting the left
and right ports of two parallel stretches of the same encounter. The
present recursion steps corresponds to the removal of one such loop,
like in the second case of Subsection \ref{sec:unitary_recursion}.

\subsubsection{Third case: contractions between $B$, $B$ inside the same trace}

For the orthogonal case rule (\ref{sigma_con}d) yields further
contractions: Pairing the selected $B$ with all other $l-1$ matrices
$B$ appearing in the same trace, we obtain
\begin{equation}
\bigg\langle\tr\Big(
\contraction[2ex]{}{B}{B^\dagger(BB^\dagger)^m}{B}{}
BB^\dagger(BB^\dagger)^m BB^\dagger(BB^\dagger)^{l-m-2}\Big)[\ldots]\bigg\rangle
=-\frac{1}{2{\rm i}}\bigg\langle\tr(BB^\dagger)^{l-1}[\ldots]\bigg\rangle\,,
\end{equation}
where $m$ may take any value between 0 and $l-2$.  Thus,
$\langle\traceprodB\rangle$ picks up $l-1$ contributions of trace products with
one $\tr(BB^\dagger)^l$ replaced by $\tr(BB^\dagger)^{l-1}$, and thus $\vec{v}$
replaced by $\vec{v}^{[l\to l-1]}$. Correspondingly, $N_c(\vec{v})$
gains a contribution
\begin{equation}
(l-1)N_c(\vec v^{[l\to l-1]})\,.
\end{equation}
The present scenario is analogous to the merger of two antiparallel
stretches of the same encounter.

\subsubsection{Resulting recursion}

Summing up all contributions, we arrive at the recurrence for
$N_c(\vec{v})$, for any $l$ with $v_l>0$
\begin{eqnarray} \label{ncrec}
N_c(\vec{v})&=&\kappa \sum_{k\geq 2} k(v_k-\delta_{kl}) N_c(\vec{v}^{[k,l\to
k+l-1]}) %\nonumber\\&+&
+\sum_{m=1}^{l-2}N_c(\vec v^{[l\to m,l-m-1]})  \nonumber\\
&&+\;(\kappa -1)(l-1)N_c(\vec v^{[l\to l-1]})\,.
\end{eqnarray}

The recurrence relation (\ref{ncrec}) reflects a single contraction
step according to the rules (\ref{sigma_con}a-d) and gives a sum of
terms each containing one matrix $M=BB^\dagger$ less than the original
trace.  Repeated such contraction steps give a sum of an ever
increasing number of terms.  After $L(\vec v)-1$ steps each of these
summands reads $N_c(\vec{v}')$ with $v_1'=1$, $v'_l=0$ for $l\geq 2$,
which due to (\ref{Ncdef}) and (\ref{finstep}c) just equals unity.
The number of contractions $N_c(\vec{v})$ thus gives the number of
terms in the sum at the final stage.  This is reassuring, given that
each of these terms needs to correspond to one surviving full
contraction.

Most importantly, using that the numbers of contractions
$N_c(\vec{v})$ and the numbers of structures $N(\vec{v})$ are related
by Eq. (\ref{NNc}), we see that the recursion relations for both
quantities, (\ref{Nrecur1},\ref{Nrecur2}) and (\ref{ncrec}), coincide
-- just like the corresponding expansions of $K(\tau)$.

\section{Summary}

The nonlinear $\sigma$ model of quantum field theory provides a convenient
way of implementing random-matrix (or disorder) averages.
Perturbative implementations of the $\sigma$ model reveal a striking
analogy to our semiclassical procedure: Families of orbit pairs can be
interpreted as diagrams in field theory, encounters correspond to
vertices, and loops to propagator lines.  To explore these relations
quantitatively, we introduced the bosonic replica version of the
zero-dimensional $\sigma$ model, in its rational parametrization.  The
latter parametrization gives the spectral form factor as an integral
over matrices $B$, $B^\dagger$, and allows for a perturbative expansion in
powers of $BB^\dagger$.  The $l$-encounters correspond to factors
$\tr(BB^\dagger)^l$. Orbit loops are related to contraction lines between
$B$'s and $B^\dagger$'s.  Only terms {\it quadratic} in the so-called
replica index $r$ contribute, corresponding to pairs of {\it
  connected} periodic orbits $\gamma$, $\gamma'$.  For all $\vec{v}$, the $\sigma$
model leads to the same contributions to the form factor as our
semiclassical analysis.  The recursion obtained from our permutation
treatment (reducing the number of encounter stretches) is mirrored by
Wick contractions (reducing the number of $B$'s and $B^\dagger$'s).

%% file: kapitel8.tex
\chapter{Conclusions and outlook}

\stand

\label{sec:conclusions}

Within the semiclassical frame of periodic-orbit theory, we have
studied the spectral statistics of {\it individual} fully chaotic
dynamics. Central to our work are pairs of orbits which differ only
inside close self-encounters.  These orbit pairs yield series
expansions of the spectral form factor $K(\tau)$ for systems with and
without time-reversal invariance.  To all orders in $\tau$, our series
agree with the corresponding predictions of random-matrix theory for
the Gaussian Orthogonal and Unitary Ensembles.  Note that we do not
require any averaging over ensembles of systems.  Moreover, we find a
close analogy between semiclassical periodic-orbit expansions and the
perturbative treatment of the nonlinear $\sigma$ model.

Important questions about universal spectral fluctuations remain open.
The precise {\it conditions} for a system to be faithful to
random-matrix theory still need to be established.  When evaluating
the contributions to the form factor originating from orbit pairs
differing in encounters, we used ergodicity and hyperbolicity as our
main assumptions.  Furthermore, we required that all classical
resonances are bounded away from zero (and thus all classical time
scales are negligible compared to $T_E$ and $T_H$), and that the
dynamics is mixing (see Appendix \ref{sec:maths}).

To make sure that a given fully chaotic system has a universal
(small-time) form factor, we need to impose one further restriction:
The (small-$\tau$) contributions of all orbit pairs unrelated to close
self-encounters must mutually cancel.  On the one hand, this will
concern pairs of seemingly unrelated orbits with ``random'', but
possibly very small, action differences. In fact, the vast majority of
near-degeneracies in the action spectrum is of the latter kind, as can
be seen from the Poissonian statistics of the corresponding
nearest-neighbor distribution \cite{Smilansky}.  It is indeed
plausible to assume that such ``random'' contributions to the form
factor simply average out.

However, deviations from universality can arise from system-specific
families of orbit pairs.  For dynamics exhibiting arithmetic chaos,
strong degeneracies in the periodic-orbit spectrum give rise to
system-specific contributions to the form factor; hence the systems in
question deviate from random-matrix theory \cite{Bogomolny}.  Similar
exceptions are given in \cite{KeatingExc,DelandeExc}.  On the other
hand, for the Sinai billiard and the Hadamard-Gutzwiller model,
system-specific families of orbit pairs found in \cite{Primack} and
\cite{HeuslerPhD}, respectively, do not prevent universality.  In
order to formulate the precise conditions for the BGS conjecture, one
has to clarify when non-universal contributions may occur.

Some further points call for {\it mathematically more rigorous
  justification}.  Most importantly, a better justification is needed
for neglecting the difference between stability amplitudes and periods
of the partner orbits.  Physical reasons for the irrelevance of these
differences were given in Subsection \ref{sec:stability_amplitude}.
However, a full proof is available only for Sieber/Richter pairs in
the Hadamard-Gutzwiller model; see Appendix \ref{sec:notau3}.  In
Appendix \ref{sec:maths}, mathematicians might question the
application of the equidistribution theorem to a highly singular
observable depending on the orbit period.

Physically, the perhaps most urgent challenge is to go {\it beyond the
  range of small $\tau$}.  The series expansions obtained here are valid
in the limit of small $\tau$.  In the orthogonal case, the expansion
converges for $\tau<\frac{1}{2}$.  The corresponding linear and
logarithmic expressions for the form factor of the unitary and
orthogonal classes remain applicable, by analytic continuation, up to
the next singularity. As we know from random-matrix theory, that
singularity is located at $\tau=1$.  However, neither the locus of the
singularity nor the functional form of $K(\tau)$ for larger times are
directly accessible through the reasoning presented here.

It remains an interesting challenge to see whether semiclassical
methods can be extended to that regime.  Could large-time correlations
be related to {\it different pairs of orbits}, not affecting the
small-time form factor?  Indeed, starting from the full random-matrix
form factor, Argaman et al. \cite{Argaman} derived a weighted density
of classical action differences needed to recover universal short-time
and large-time statistics. Using orbit pairs differing in encounters, we
could reproduce
only those parts of Argaman et al.'s ``action correlation
function'' related to short-time
statistics \cite{LongPaper}.  The classical origin of the remaining
terms is still unknown.  In a related approach, an analogy to number
theory was proposed in \cite{Keating}: The Hardy-Littlewood conjecture
on prime numbers can be cast into the language of semiclassics, by
identifying prime numbers with periodic orbits.  Like the
action-correlation function, this relation seems to imply further
classical correlations between orbits.

Alternatively, one might try to improve the periodic-orbit approach by
explicitly taking into account the {\it unitarity} of the associated
quantum dynamics.  For instance, Bogomolny and Keating
\cite{Bootstrap} incorporated ``by hand'' the information that energy
eigenvalues are real. They approximated the level staircase $N(E)$
(i.e., the number of eigenvalues below $E$) using Gutzwiller's trace
formula, and assumed eigenvalues to be located wherever $N(E)$ takes
half-integer values. The resulting ``bootstrapped'' level density,
when inserted into the double sum over orbits, could yield a better
approximation for large-time correlations. However, the resulting
double sum has as yet been evaluated only within the diagonal
approximation.  Further hope to address the large-$\tau$ behavior of
$K(\tau)$ by incorporating unitarity can be drawn from the following
observation: In a discrete-time formulation, $K(\tau)$ can be expressed
in terms of traces of the time-evolution operator; as shown in
\cite{Haake} unitarity allows to obtain all traces from those
corresponding to $\tau<\frac{1}{2}$, accessible within the present
framework.

A further line of reasoning aims for a more direct connection to the
{\it $\sigma$ model}.  A ballistic $\sigma$ model for individual chaotic
systems was proposed in \cite{BallisticSigma}.  However, the
perturbative evaluation of that model seemed to be restricted to the
analog of the diagonal approximation.  In \cite{Jan}, the relation to
orbit pairs differing in encounters was used to extract off-diagonal
corrections.  For the spectral statistics of quantum graphs, a
$\sigma$-model description was justified by Gnutzmann and Altland
\cite{GnutzAlt}.  They revealed spectral averaging as equivalent to
averaging over an ensemble of graphs, with members distinguished only
by their boundary conditions; the latter ensemble average could be
implemented through a nonlinear $\sigma$ model.  By construction, any such
direct mapping to the $\sigma$ model also covers large-time statistics.
See \cite{Berkolaiko} for earlier works on quantum graphs yielding the
first three orders of $K(\tau)$.

For a full understanding of universal spectral statistics, further
statistical quantities such as {\it higher-order correlation
  functions} or the {\it level-spacing distribution} need to be
considered. The related small-time statistics should involve
correlations between more than two orbits, but not bring about
fundamental difficulties.  Large-time statistics -- or, equivalently,
the behavior of the level-spacing distribution for small energy
differences $s$ -- poses a challenge similar to the large-time form
factor. A solution of this problem would be highly desirable, since
only a treatment of small $s$ could finally lead to a semiclassical
understanding of level repulsion.

Such open questions not withstanding, we expect the ideas presented
here to carry over to all {\it remaining symmetry classes}, i.e., the
symplectic class and the new classes proposed in \cite{Tenfold}, and a
rich variety of {\it applications in mesoscopic physics}.

For instance, the same orbit pairs as described here have been
employed in \cite{Keppeler,Heusler,BolteHarrison,LongPaper} to show
that time-reversal invariant {\it spin} systems are faithful to
random-matrix theory. The GOE applies for integer quantum number $S$
and time-reversal operators $\T$ squaring to $+1$, whereas the GSE
requires half-integer spin and $\T^2=-1$.  While the relevant pairs of
orbits are the same as without spin, their contribution to the form
factor is changed: The state of the system is given by a spinor with
$2S+1$ components such that the van Vleck propagator of the spinless
system has to be multiplied by a $(2S+1)\times(2S+1)$ matrix representing
spin evolution.  For both $\gamma$ and $\gamma'$, the corresponding matrices
lead to additional factors in the double sum over periodic orbits
which, taken into account correctly, reproduce the pertinent RMT
results.  Thus, universal (small-time) behavior is ascertained for all
three Wigner/Dyson symmetry classes.

Likewise, periodic orbits allow to study the {\it crossover between
  symmetry classes}, as shown in \cite{Nagao,Turek,NagaoTau3} for the
GOE/GUE transition caused by a weak magnetic field. In
\cite{NagaoTau3} the cubic contribution to the form factor was
determined from exactly the same families of orbit pairs as introduced
in Chapter \ref{sec:tau3}.  This time, the contribution of each orbit
pair is modified by an additional action term depending on the
magnetic field. Ultimately, all encounter stretches or loops traversed
by $\gamma$ and $\gamma'$ in mutually time-reversed direction lead to
additional factors, shrinking exponentially as either the duration of
the stretch or loop, or the squared magnetic field are increased.

Fluctuations involving {\it matrix elements} of observables are
accounted for if we multiply the contribution of each periodic orbit
with an integral of the observable along the orbit in question
\cite{Higherdim}.

\begin{figure}
\begin{center}
  \includegraphics[scale=0.18]{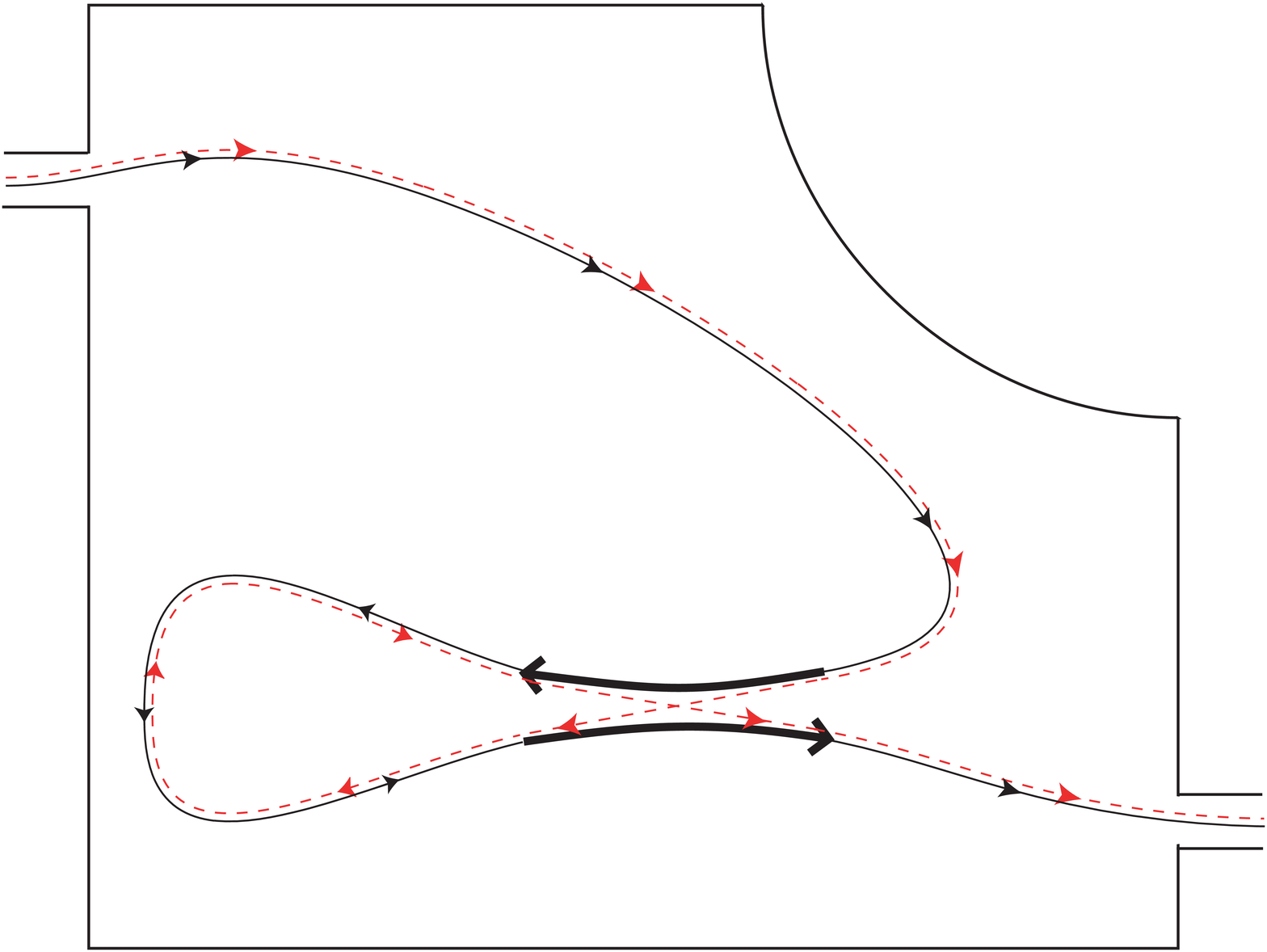}
%\parbox{1cm}{\ \ \ \ \ }
%\parbox{4cm}{
%\vspace{-4cm}
\end{center}
\caption{A Richter/Sieber pair of trajectories,
  contributing to conductance. The two trajectories connect leads
  attached to a chaotic cavity, and differ inside an antiparallel
  2-encounter.}
\label{fig:conductance_rs}
\end{figure}

A lot of further interesting applications become accessible if the
idea of encounter reconnections is applied to {\it different types of
  trajectories}.  Doing so, it should be possible to go beyond the
``threefold way'' and extend the present results to the seven {\it new
  symmetry classes} \cite{Tenfold}, of experimental relevance for
normal-metal/superconductor heterostructures and in quantum
chromodynamics. First steps are taken in \cite{GnutzSeif}.  Here, the
main quantity of interest is no longer the spectral form
factor, but the level density proper. Classical orbits in a
normal-conducting cavity attached to a superconductor have a peculiar
form: Upon reflection from the surface of the superconductor,
electrons are converted into holes, and vice versa.  The analog of,
say, a Sieber/Richter pair will be {\it one single periodic orbit}
consisting of two subsequent parts.  These parts differ (i) by
reconnections inside a 2-encounter, and (ii) because one part
describes the motion of an electron, whereas the other one describes
the motion of a hole.

{\it Transport} properties such as conductance, shot noise, or delay
times \cite{RichterSieber,Schanz,Puhlmann,Zaitsev,Whitney} provide
another rich field of experimentally relevant applications.  In the
latter cases, the trajectories to be considered are no longer
periodic, but connect two leads attached to a chaotic cavity.
Conductance is determined by pairs of such trajectories
\cite{RichterSieber}; in Fig.  \ref{fig:conductance_rs} we have
depicted two trajectories differing in a 2-encounter.  To study shot
noise, one also has to take into account {\it quadruples} of
trajectories \cite{Schanz}.

While previous results were restricted to the lowest orders in series
expansions of the quantities in question, our machinery of encounters
and permutations, together with intuition drawn from field theory
should allow to attack the full expansion.  Moreover, the present
formalism permits to elegantly treat general fully chaotic dynamics,
and thus go beyond the model systems (like quantum graphs or the
Hadamard-Gutzwiller model) employed in many of the above works.

Finally, our semiclassical treatment, applying to {\it individual}
systems, should be capable of describing {\it system-specific
  behavior} and deviations from random-matrix theory, as long as these
deviations are related to pairs of orbits differing in encounters.
Deviations have been observed both for very small times (the
corresponding range of $\tau$ shrinking to zero as $\hbar\to 0$), and around
$\tau=1$ \cite{Deviations}.  At least, we can explain the failure of
random-matrix theory for $\tau<\frac{T_E}{T_H}$: In that regime, the
present orbit pairs no longer exits, since the relevant encounter
durations of order $T_E$ would exceed the orbit periods $\tau T_H$.
Deviations from universality would also be of interest in the context
of localization, and for dynamics with mixed phase space.

%% file: anhanga.tex
\chapter{Self-crossings in configuration space}

\label{sec:crossings}

\stand

Sieber's and Richter's seminal works \cite{SR,Sieber} on the $\tau^2$
contribution to spectral form factor were formulated in terms of
self-crossings in configuration space, rather than close
self-encounters in phase space. In \cite{Mueller}, we extended this
approach to general fully chaotic systems with two degrees of freedom
and a Hamiltonian of the form $H({\bf q},{\bf p})=\frac{{\bf
    p}^2}{2m}+V({\bf q})$, by taking into account the geometry of the
stable and unstable manifolds. In the following, we will first briefly
review the analytical results of \cite{Mueller}, stressing the
connection to the treatment in Chapter \ref{sec:tau2}.  Thus prepared,
we will then move on to numerical experiments and a more careful
investigation of correction terms in the Hadamard-Gutzwiller model
(the latter not included in \cite{Mueller}); for both extensions the
language of self-crossings in configuration space seems to be better
suited than the phase-space language employed in Chapter
\ref{sec:tau2}.

\section{Overview}

\label{sec:crossings_overview}

\begin{figure}
\begin{center}
  \includegraphics[scale=0.3]{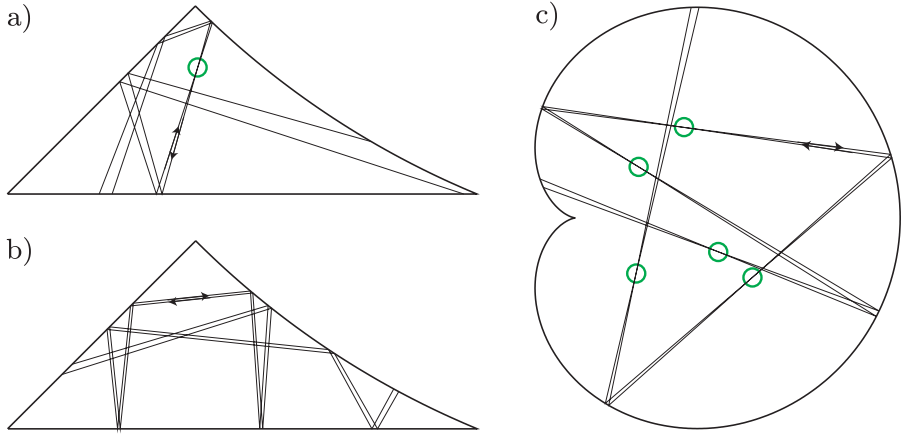}
\end{center}
\caption{Configuration-space projections of phase-space encounters
  in the desymmetrized diamond and the cardioid billiard: a)
  containing one self-crossing (in a circle), b) narrowly avoiding a
  self-crossing, c) containing a ``braid'' of self-crossings (marked
  by circles) close to mutually conjugate points.  }
\label{fig:crossings}
\end{figure}

In (two-dimensional) configuration space, an antiparallel 2-encounter
involves a self-crossing with a small angle $\epsilon$ as in Fig.
\ref{fig:crossings}a, or a narrowly avoided crossing as depicted in
Fig. \ref{fig:crossings}b.  In systems with conjugate points, there
can even be a ``braid'' of several small-angle crossings close to
mutually conjugate points: Each crossing can be seen as the starting
point of two close-by trajectories; these trajectories may meet again
in points (almost) conjugate to the initial crossing and thereby form
new crossings. An example for such a braid of self-crossings in the
cardioid billiard is shown in Fig. \ref{fig:crossings}c.  These braids
(as well as the almost self-retracing encounters discussed in Section
\ref{sec:min_dist}) did not show up in the Hadamard-Gutzwiller model;
they were first observed in \cite{Diplom} and treated analytically in
\cite{Mueller}.

Regardless of the system considered, in case of two dimensions at
least one of the two partners in each Sieber/Richter pair involves a
self-crossing. An investigation of self-crossings may therefore serve
as a basis for determining the contribution of Sieber/Richter pairs to
the spectral form factor.

To measure the separation between the two stretches of an encounter,
we may use the small crossing angle $\epsilon$ rather than the stable and
unstable coordinates $s$ and $u$. We can translate between both sets
of coordinates by considering a Poincar{\'e} section ${\cal P}$ orthogonal
to one of the encounter stretches at the location of a
configuration-space crossing.  This Poincar{\'e} section may be
parametrized by configuration-space and momentum coordinates.  The two
stretches pierce through ${\cal P}$ in two almost mutually
time-reversed phase-space points $\x_1$ and $\x_2$.  Since ${\cal P}$
is placed at a crossing, the separation $\T\x_2-\x_1$ must have a
vanishing configuration-space component. The momentum component may be
written as $\pm|{\bf p}|\sin\epsilon$, or $\pm|{\bf p}|\epsilon$ in the limit of small
crossing angles. Here $|{\bf p}|$ is the absolute value of the
momentum part of either $\x_1$ or $\x_2$, and ``$\pm$'' has to be
inserted because the angle $\epsilon$ will always be taken as positive.  We
thus obtain $\T\x_2-\x_1=\vect{0}{\pm|{\bf p}|\epsilon}$, the upper and lower
lines respectively referring to configuration space and momentum
space.

The separation $\vect{0}{\pm|{\bf p}|\epsilon}$ can be decomposed into one
stable and one unstable part.  Dropping the subscript of $\x\equiv \x_1$,
we will characterize the stable and unstable directions $\vs(\x)$,
$\vu(\x)$ at $\x$ by the ratios of their momentum and
configuration-space components $B_s(\x)=\frac{e^{s}_p(\x)}{e^{s}_q(\x)}$
and $B_u(\x)=\frac{e^{u}_p(\x)}{e^{u}_q(\x)}$; these ratios are
sometimes referred to as ``curvatures'' of the invariant manifolds
\cite{Chernov}.  The stable part of $\vect{0}{\pm|{\bf p}|\epsilon}$, denoted
by $s\vs(\x)$ in Chapter \ref{sec:tau2}, is now given by
\begin{equation}
\mp\frac{|{\bf p}|\epsilon}{B_u(\x)-B_s(\x)}\vect{1}{B_s(\x)}\,,
\end{equation}
whereas the unstable part reads
\begin{equation}
\pm\frac{|{\bf p}|\epsilon}{B_u(\x)-B_s(\x)}\vect{1}{B_u(\x)}\,;
\end{equation}
as required, these parts are proportional to
$\vs(\x)\propto\vect{1}{B_s(\x)}$ and $\vu(\x)\propto\vect{1}{B_u(\x)}$,
respectively, and sum up to $\vect{0}{\pm|{\bf p}|\epsilon}$.

As in Chapter \ref{sec:tau2}, the {\bf action difference} $\Delta
S=S_\gamma-S_{\gamma'}$ between the partner orbits $\gamma$ and $\gamma'$ ($\gamma$ now
containing the crossing) is given by the symplectic product of the
stable and unstable parts, i.e., by
\begin{equation}
\label{DS_cfg}
\Delta
S(\x,\epsilon)=\frac{|{\bf p}|^2\epsilon^2}{B_u(\x)-B_s(\x)}\,.
\end{equation}
This result was originally derived in \cite{Mueller} using slightly
different methods.  In the special case of the Hadamard-Gutzwiller
model, we have $B_u=-B_s= m\lambda$ and thus reproduce Sieber's and
Richter's result $\Delta S=\frac{|{\bf p}|^2\epsilon^2}{2m\lambda}$ \cite{SR,Sieber}.

Since crossing angles and stable and unstable coordinates are
proportional, the {\bf encounter duration} of Eq. (\ref{sr_tenc}) can
be written as
\begin{equation}
t_{\rm enc}(\x,\epsilon)\sim\frac{2}{\lambda}\ln\frac{C(\x)}{\epsilon} \,.
\end{equation}
The precise form of the proportionality factor $C(\x)$ is irrelevant
for the following considerations. (Using that $t_{\rm
  enc}=\frac{1}{\lambda}\ln\frac{c^2}{|\Delta S|}$, see Eqs.
(\ref{sr_tenc}) and (\ref{DS2}), it is easy to show that $C(\x)$ depends on
the bound $c$ and on the stable and unstable directions at $\x$ via
$C(\x)=\frac{c}{|{\bf p}|}\sqrt{|B_u(\x)-B_s(\x)|}$).  Note that we may
not neglect the $\x$ dependence of $C(\x)$, since $t_{\rm enc}(\x,\epsilon)$
must be the same for all crossings inside a given encounter, even
though these crossings involve different angles $\epsilon$.  The durations
$t_s$ and $t_u$ of the ``tail'' and ``head'' of the encounter can be
expressed as similar logarithmic functions of $\epsilon$, albeit with
different $\x$ dependent proportionality factors. For instance, $t_u$
may be written as\footnote{Here, the proportionality factor is given
  by $C_u(\x)=\frac{c}{|{\bf p}|}|(B_u(\x)-B_s(\x))e^{s}_q(\x)|$.}
\begin{equation}
\label{crossings_tu}
t_u(\x,\epsilon)\sim\frac{1}{\lambda}\ln\frac{C_u(\x)}{\epsilon}\,.
\end{equation}

To count orbit pairs, we need to determine the {\bf density of
  self-crossings} $P_T(\x,\epsilon)$, normalized such that integration over
an interval of angles and a region inside the energy shell yields the
corresponding number of self-crossings of one orbit of period close to
$T$.  Note that if we integrate over the whole energy shell, each
crossing will be counted twice -- once for each of the two almost
time-reversed phase-space points.  We include only crossings between
encounter stretches separated by non-vanishing loops.  Similarly as in
Chapter \ref{sec:tau2}, $P_T(\x,\epsilon)$ must be understood as averaged
over all orbits with periods inside a small window around $T$.

To determine $P_T(\x,\epsilon)$, we start from a density of crossings with
{\it angles} $\epsilon$, {\it points} $\x\equiv\x_1$ traversed at {\it time}
$t_1$, and the point almost time-reversed with respect to $\x$ being
traversed at {\it time} $t_2$.  This density is defined such that
integration over a region inside the energy shell and over intervals
of $\epsilon$, $t_1$, and $t_2$ gives the pertaining number of crossings,
averaged over the same ensemble of orbits as above.  Using the
ergodicity of the flow, one can show that this auxiliary density is
given by
\begin{equation}
\label{crossing_xett}
\frac{2|{\bf p}|^2\sin\epsilon}{m\Omega^2}\,.
\end{equation}
Here, $\frac{1}{\Omega}$ represents the Liouville density on the energy
shell (see Subsection \ref{sec:ergodicity}), whereas the probability
of finding a crossing in time intervals $(t_1,t_1+dt_1)$,
$(t_2,t_2+dt_2)$ with angle inside an interval $(\epsilon,\epsilon+d\epsilon)$ is given
by $\frac{2|{\bf p}|^2\sin\epsilon}{m\Omega}d\epsilon dt_1dt_2$.  In particular, the
factor $\sin\epsilon$ implies that orthogonal pieces of trajectory have a
much higher probability to intersect than almost parallel ones.  The
derivation follows the lines of the Appendix of \cite{SR}.

The desired density $P_T(\x,\epsilon)$ of phase-space points and crossing
angles only is now obtained by integrating over $t_1$ and $t_2$.  To
include only encounters whose stretches are separated by intervening
loops, we use the same restrictions on $t_1$, $t_2$ as in the
phase-space calculation.  We thus obtain
\begin{equation}
\label{crossing_xe}
P_T(\x,\epsilon)=\frac{2|{\bf p}|^2\sin\epsilon}{m\Omega^2}T(T-2t_{\rm enc}(\x,\epsilon)),
\end{equation}
with the correction term $\propto t_{\rm enc}$ due to the necessity of
intervening loops.

\begin{figure}
\begin{center}
  \includegraphics[scale=0.5]{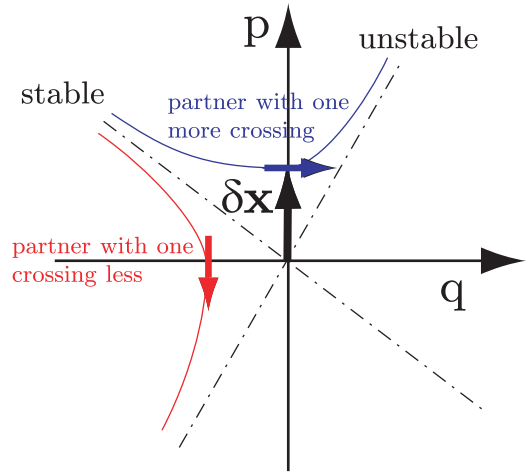}
\end{center}
\caption{Poincar{\'e} section orthogonal to one of the
  encounter stretches of either $\gamma$ or $\gamma'$, with lines indicating
  the stable and unstable directions and the vector $\delta{\bf x}$
  pointing to the time-reversed of the other stretch.  The picture
  shows the moment in which $\delta{\bf x}$ traverses the $p$-axis, i.e.,
  the stretches cross in configuration space.  The asymptotic motion
  of $\delta{\bf x}$ with respect to the invariant manifolds is indicated
  by arrows.}
\label{fig:poincare}
\end{figure}

When evaluating the contribution to the spectral form factor, we have
to make sure that each orbit pair is counted exactly once, even though
the pertaining encounter may involve arbitrarily many self-crossings.
To this end, we show that for general two-dimensional hyperbolic
Hamiltonians of the form $H({\bf q},{\bf p})=\frac{{\bf
    p}^2}{2m}+V({\bf q})$ the numbers of crossings in the partner
orbits $\gamma$ and $\gamma'$ differ by one, {\bf the orbit with larger action
  containing one more crossing}.  This trivially applies to systems
without conjugate points, where the partner with larger action
contains one crossing and the other none.  Our proof relies on an
argument of winding numbers. We follow one of the two encounter
stretches and study, in a Poincar{\'e} section orthogonal to the orbit,
three quantities, the directions of the stable and unstable manifolds
(which locally can be visualized as straight lines through the origin)
and the small phase-space vector $\delta{\bf x}=\T\x_2-\x_1$ pointing to
the time reversed of the other encounter stretch; see Fig.
\ref{fig:poincare}. As we move along the orbit, these lines and
vectors rotate around the origin, as in the treatment of the Maslov
index in Sections \ref{sec:gutzwiller} and
\ref{sec:stability_amplitude}.  The encounter stretches cross in
configuration space each time that $\delta{\bf x}$ rotates through the
$p$-axis.  Note that for Hamiltonians of the above form the $p$-axis
may be traversed only in clockwise direction, given that
$\dot{q}=\frac{p}{m}>0$ in the upper and $<0$ in the lower half plane.

For the orbit with larger action, our formula for the action
difference demands that whenever a crossing occurs we have $B_u({\bf
  x})-B_s({\bf x})>0$, i.e., the unstable manifold has a higher slope
in the Poincar{\'e} section than the stable one; $\delta{\bf x}$ is thus
located between the stable manifold (on the counter-clockwise side)
and the unstable manifold (on the clockwise side).  The opposite
applies to the partner with smaller action.  For both orbits, the
motion of $\delta{\bf x}$ is now given as a superposition of the rotation
of the invariant manifolds and a motion from the stable towards the
unstable manifold (since the stable components shrink and the unstable
components grow).  The motion towards the unstable manifold has
clockwise sense for the partner with larger action, and
counter-clockwise sense for the partner with smaller action.  For the
orbit with larger action, $\delta{\bf x}$ thus performs one more clockwise
half-rotation around the orbit and therefore crosses the $p$-axis one
more time.  Consequently, the latter orbit indeed contains one more
crossing.

To count each orbit pair exactly once, we have to weight crossings of
the partners with larger and smaller action with respective positive
and negative signs, by multiplying their contributions with
$\mbox{sign}(B_u({\bf x})-B_s({\bf x}))$.  As a result, only one
contribution per orbit pair remains effective.

We can now evaluate the {\bf contribution of Sieber/Richter pairs}
$(\gamma,\gamma')$ to the double sum for $K(\tau)$, see Eq. (\ref{doublesum}),
replacing the sum over partners $\gamma'$ by a sum over the self-crossings
of $\gamma$. The latter sum may be written as an integral over $\epsilon$ and
$\x$ with the density $P_T(\x,\epsilon)\mbox{sign}(B_u({\bf x})-B_s({\bf
  x}))$, as in
\begin{eqnarray}
\label{ksr_cr}
K_{\rm SR}(\tau)&=&\frac{2}{T_H}\left\langle\sum_\gamma |A_\gamma|^2\delta(\tau T_H-T_\gamma)\right.
\int d\mu(\x) \,\mbox{sign}(B_u({\bf x})-B_s({\bf x}))\nonumber\\
&&\left. \ \ \ \ \times\int_{\epsilon>0}d\epsilon\,P_{\tau T_H}({\bf x},\epsilon)\cos\frac{\Delta S({\bf x},\epsilon)}{\hbar}\right\rangle,
\end{eqnarray}
As before, we have replaced $A_{\gamma'}\to A_\gamma$, $T_{\gamma'}\to T_\gamma$, and a
factor 2 accounts for the time-reversed partner $\T\gamma'$.  (Two further
factors mutually cancel: A factor 2 is needed because there is
effectively one crossing per orbit pair -- instead of one encounter in
each $\gamma$ and $\gamma'$. A factor $\frac{1}{2}$ compensates the inclusion
of two almost time-reversed phase-space points $\x$ for each
crossing.)

Similar the density $w_{\tau T_H}(s,u)$ of Chapter
\ref{sec:tau2}, $P_{\tau T_H}(\x,\epsilon)$ (compare Eq.
(\ref{crossing_xe})) falls into a leading term $\propto T_H^2$ and a
subleading correction $\propto T_H t_{\rm enc}\sim T_E T_H$.  When evaluating
the $\epsilon$ integral in (\ref{ksr_cr}), the leading term yields a
contribution scaling like $\hbar^{-1}$ in the semiclassical limit, and
proportional to $\int d\epsilon\,\sin\epsilon\,\cos\frac{|{\bf
    p}|^2\epsilon^2}{(B_u(\x)-B_s(\x))\hbar}$.  If we approximate $\sin\epsilon\approx\epsilon$
for small angles, it is easy to show that the integral oscillates
rapidly as $\hbar\to 0$ and thus vanishes after averaging.

The $\epsilon$ integral of the correction term gives $-\tau\frac{|B_u({\bf
    x})-B_s({\bf x})|}{2m\lambda\Omega}$.  The absolute value is compensated by
multiplication with $\mbox{sign}(B_u({\bf x})-B_s({\bf x}))$.
Integrating over the energy shell, and applying the sum rule of Hannay
and Ozorio de Almeida (\ref{hoda}), we are led to
\begin{eqnarray}
\label{ksr}
K_{\rm SR}(\tau)
=-2\tau^2\frac{\,\overline{B_u-B_s}\,}{2m\lambda},\label{preff}
\end{eqnarray}
where $\averagedots$ denotes an average over the energy shell.

Now, ergodic theory needs to be invoked to relate the directions of
the invariant manifolds to the Lyapunov exponent. The local stretching
rate of a hyperbolic system depends on the normalization chosen for
the stable and unstable direction $\vs(\x)$ and $\vu(\x)$.  One of
these choices \cite{Gaspard} leads to
$\chi(\x)=\frac{B_u(\x)}{m}$.\footnote{ This identity was written in
  \cite{Gaspard} as $\chi({\bf x})=\mbox{tr}\left[\frac{\partial^2 H}{\partial{\bf
        q}\partial{\bf p}}+\frac{\partial^2 H}{\partial^2{\bf p}}{\sf C}({\bf
      x})\right]$, where the matrix ${\sf C}({\bf x})$ relates the
  momentum and configuration-space components of unstable deviations
  as $d{\bf p}={\sf C}({\bf x})d{\bf q}$.  If we consider
  two-dimensional systems with a Hamiltonian of the form $H({\bf
    q},{\bf p})=\frac{{\bf p}^2}{2m}+V({\bf q})$, and evaluate the
  trace in coordinates orthogonal and parallel to the orbit, we see
  that $\chi({\bf
    x})=\frac{1}{m}\hspace{0.05cm}\mbox{tr}\hspace{0.075cm}{\sf
    C}({\bf x})=\frac{B_u(\x)}{m}$.}  Since the energy-shell average
of $\chi(\x)$ coincides with the Lyapunov exponent of the system, we
infer that $\overline{B_u}=m\lambda$.  Moreover $B_s(\T{\bf x})=-B_u({\bf
  x})$ implies that $\overline{B_s} = -\overline{ B_u}=-m\lambda$.  (Proofs
for the special case of semi-dispersing billiards can also be found in
\cite{ErgBilliards}).  From this, we immediately obtain the universal
leading off-diagonal contribution to the spectral form factor
\begin{eqnarray}
K_{\rm SR}(\tau)=-2\tau^2.
\end{eqnarray}

\section{Numerical results}

We have numerically investigated the statistics of self-crossings for
two billiards, the desymmetrized diamond billiard and the cardioid
billiard.  For technical reasons, we considered {\it non-periodic}
orbits, starting at time 0 and ending at time $T$.  Each self-crossing
is traversed at two times $t_1$, $t_2$ with $0<t_1<t_2<T$.

For both billiards, we want to concentrate on just one plot,
displaying prominently a key idea of the present approach: The
relevant encounters must have their stretches separated by intervening
loops.  The time interval between $t_1$ and $t_2$ is large enough to
contain a non-vanishing loop only if time difference $\delta t=t_2-t_1$
exceeds the logarithmic threshold $2t_u(\x,\epsilon)\sim
\frac{2}{\lambda}\ln\frac{C_u(\x)}{\epsilon}$; see Eqs.  (\ref{min_right}) and
(\ref{crossings_tu}).  Since in case of non-periodic trajectories
there can be no second loop, the traversal times do not have to obey
any further restrictions (like Eq. (\ref{min_left}) for periodic
orbits).  To illustrate the threshold for $\delta t$, we consider a
combined density $P_T(\x,\epsilon,\delta t)$ of crossing points, angles, {\it
  and time differences} in non-periodic orbits of duration $T$.
Similarly to the preceding Subsection, an analytical prediction can be
made if we integrate over the auxiliary density in Eq.
(\ref{crossing_xett}).  We then obtain
\begin{eqnarray}
\label{crossing_xet}
P_T(\x,\epsilon,\delta t)
&=& \int_0^T dt_2\int_0^{t_2} dt_1\,\delta(\delta t-(t_2-t_1))\Theta(\delta t-2 t_u(\x,\epsilon))
\frac{2|{\bf p}|^2\sin\epsilon}{m\Omega^2}\nonumber\\
&=&\frac{2|{\bf p}|^2\sin\epsilon}{m\Omega^2}(T-\delta t)\Theta(\delta t-2 t_u(\x,\epsilon))\,,
\end{eqnarray}
where the $\Theta$ function originates from the minimal time difference
$2t_u(\x,\epsilon)$.

Similarly as in Section \ref{sec:sr_statistics} we expect slight
deviations from the ergodic crossing probability (\ref{crossing_xett})
and thus from (\ref{crossing_xet}) for encounter stretches separated
by loops shorter than the classical relaxation time $t_{\rm cl}$,
since in this case both stretches are statistically correlated; such
deviations will concern time differences $\delta t$ just slightly
exceeding the minimum $2t_u(\x,\epsilon)$.

Densities of other sets of parameters were investigated in
\cite{Diplom} in a preliminary version and in \cite{Mueller} in final
form.

\subsubsection{Desymmetrized diamond billiard}

\begin{figure}[t]
\begin{center}\includegraphics[scale=0.4]{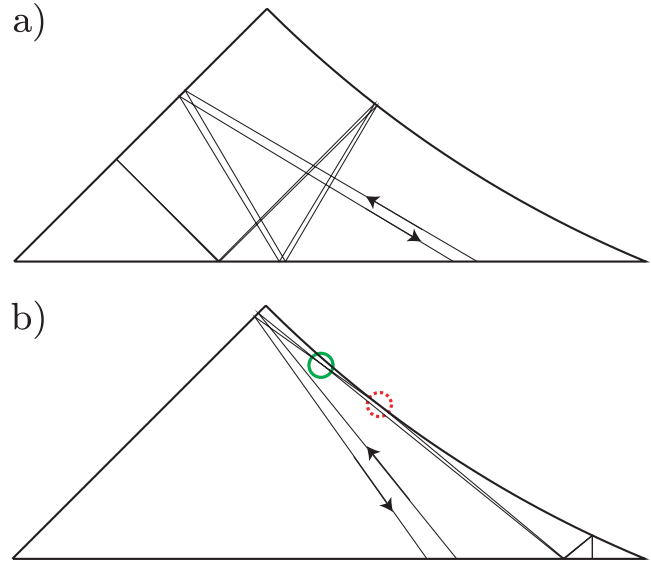}
\end{center}
\caption{Exceptional encounters in the desymmetrized diamond billiard:
  a) an almost self-retracing encounter without small-angle crossings,
  b) an example for a crossing (in the full circle) related to the
  tangential singularity; in the dotted circle, one stretch narrowly
  misses the boundary whereas the other one undergoes a glancing
  reflection.  }
\label{fig:crossings_diamond}
\end{figure}

\begin{figure}\begin{center}
    \includegraphics[scale=0.42]{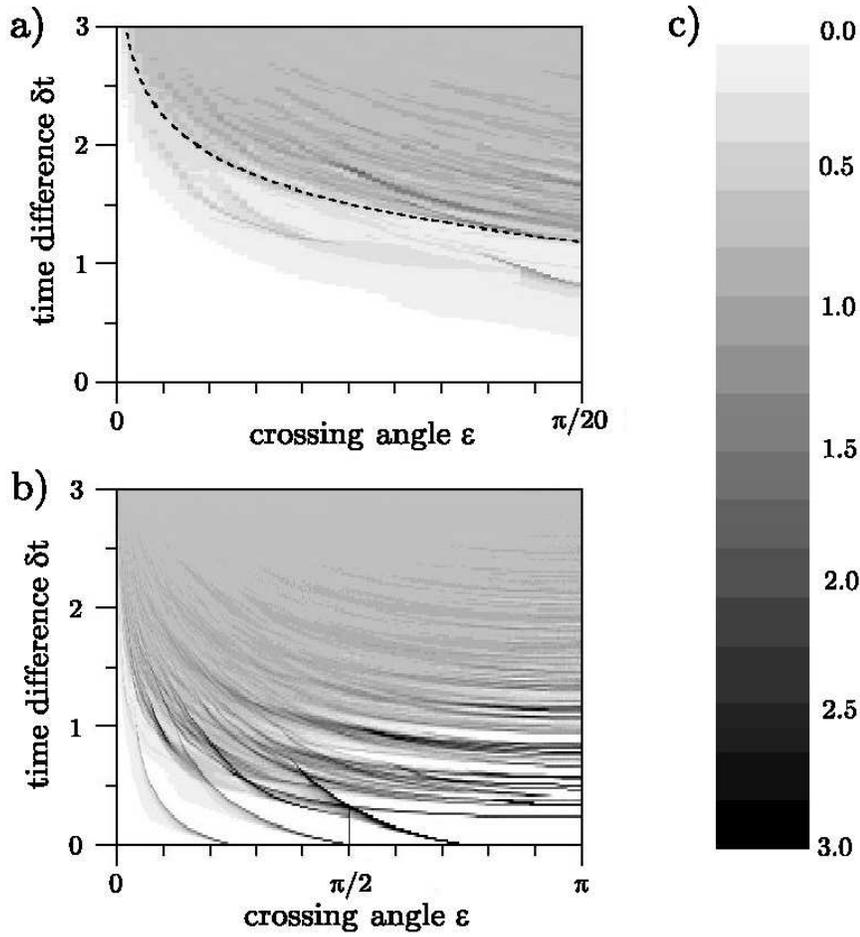}
\end{center}\caption{Plot of the combined density of
  time differences $\delta t$ and crossing angles $\epsilon$ in the
  desymmetrized diamond billiard: a) for $\epsilon<\frac{\pi}{20}$, b) for
  all angles, for non-periodic orbits of duration $T=10$.
  Normalization as defined in the text; the resulting scale is shown
  in c).  For small angles, we observe a threshold logarithmic in
  $\epsilon$, as indicated by a dashed line.}
  \label{fig:diamond_results}\end{figure}

We have first checked Eq. (\ref{crossing_xet}) for the desymmetrized
diamond billiard.  Here {\it all} self-crossings belong to encounters
separated by loops.  Almost self-retracing encounters as in Fig.
\ref{fig:crossings_diamond}a may not involve self-crossings, since
these would have to be conjugate to the reflection point. Hence,
encounters without a partner are automatically excluded from our
statistics.  The same applies to all systems without conjugate points.

Consider Fig. \ref{fig:diamond_results} for a density plot of crossing
angles and time differences, for crossing points $\x$ anywhere on the
energy shell.  The plot depicts the density $P_T(\x,\epsilon,\delta t)$, after
dividing out $\sin\epsilon$ and integrating over the energy shell.  The
results were obtained by averaging over $2\times 10^7$ non-periodic
trajectories with random initial conditions and duration $T=10$,
measured in dimensionless coordinates with mass and velocity equal to
one.

For small angles, the existence of a minimal time difference depending
logarithmitically on the crossing angle can be verified by a glance at
Fig. \ref{fig:diamond_results}a.  Here, the dashed line represents the
minimal time difference $2t_u(\x,\epsilon)\sim
\frac{2}{\lambda}\ln\frac{C_u(\x)}{\epsilon}$, with $C_u(\x)$ replaced by a
constant obtained through numerical fitting.  Sufficiently far above
the logarithmic curve, the density is almost uniform.  In agreement
with Eq. (\ref{crossing_xet}), it decays linearly towards larger $\delta
t$.  Below the dashed line, the density of crossings diminishes fast
before vanishing completely inside the white region.

In the vicinity of the logarithmic curve, i.e., for short loops,
deviations from the ergodic crossing probability of Eq.
(\ref{crossing_xett}) make for system-specific inhomogeneities.
Furthermore, the threshold is smeared out since, rather than exactly
following the dashed line, the minimal time difference weakly depends
on the location on the energy shell through the factor $C_u(\x)$ in
Eq.  (\ref{crossings_tu}).

Most of the crossings significantly below the latter line are due to a
system-specific effect, related to the tangential singularity of the
billiard flow: The linear approximation for the separation between the
two encounter stretches is already violated if one stretch is
reflected at the circular part of the boundary, and the other stretch
narrowly avoids the circle; see Fig.  \ref{fig:crossings_diamond}b for
an example.  The encounter should be considered to end at this
reflection point, even if the phase-space separation between the two
stretches is still very small.  Hence the duration of the ``head'' of
the encounter $t_u$ is much shorter than
$\frac{1}{\lambda}\ln\frac{C_u(\x)}{\epsilon}$.  The time difference $\delta t$
between two traversals of the crossing may thus remain far below the
logarithmic threshold, like in the example of Fig.
\ref{fig:crossings_diamond}b.  However, our numerical results indicate
that the resulting effect on the crossing distribution is minute,
since these exceptional crossings are much less numerous than generic
ones.

At larger angles, which in the semiclassical limit have no impact on
the form factor, Fig. \ref{fig:diamond_results}b shows further
system-specific structures.  Discrete lines at $\epsilon=\pi$ correspond to
periodic orbits, since by ``crossing'' itself at an angle $\pi$, an
orbit simply closes in phase space.  These lines are deformed and
broadened when going to smaller angles.  Thus, most ``orbit parts''
(i.e., pieces of trajectory between two traversals of a crossing)
close to $\delta t\approx 2t_u$ are obtained by deformation of the shortest
periodic orbits.  In contrast, the families of ``orbit parts''
starting at $\delta t=0$ are related to the corners of the billiard
\cite{Diplom,Mueller}.

\subsubsection{Cardioid billiard}

\begin{figure}[t]
\begin{center}
  \includegraphics[scale=0.36]{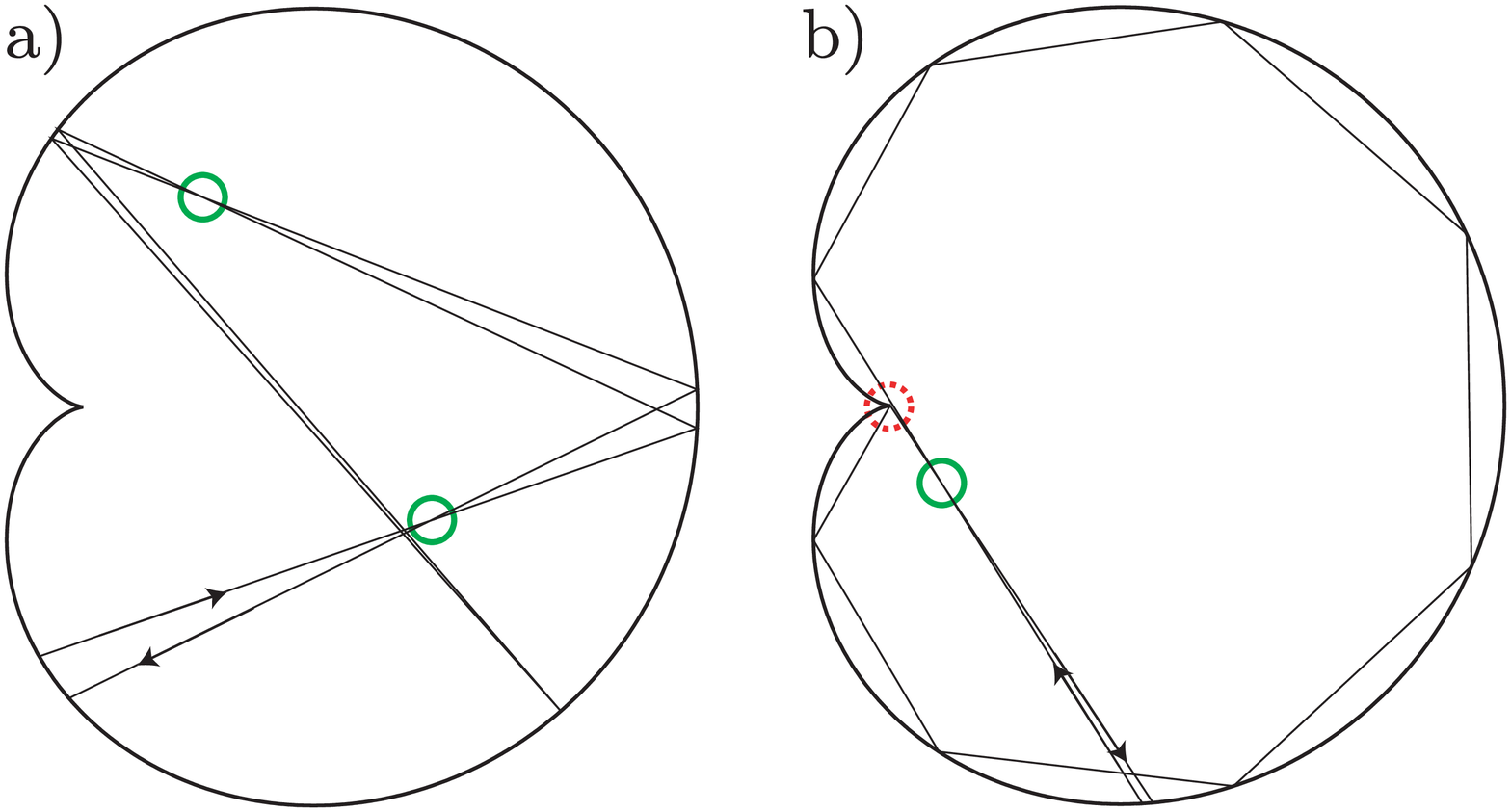}
\end{center}
\caption{Encounters in the cardioid billiard:
  a) an almost self-retracing encounter containing crossings (marked
  by circles) almost conjugate to the reflection point, b) example for
  a cusp-related crossing; in the dotted circle, only one of the two
  stretches is reflected.  }
\label{fig:crossings_cardioid}
\end{figure}

As a second example, we consider the cardioid billiard.  In the
cardioid, almost self-retracing encounters may involve crossings, as
seen in Fig.  \ref{fig:crossings_cardioid}a.  The locations of these
crossings are conjugate to each other, and to the point where the
orbit is reflected from the wall with an almost right angle: The two
traversals of each crossing limit a fan of trajectories with a small
opening angle which gathers again in the remaining crossings and in
the reflection point.  Since there is no associated partner orbit,
these crossings have to be excluded from our statistics, using a
criterion based on symbolic dynamics. We determine the symbol sequence
of the orbit and look for the pair of mutually time-reversed
subsequences $\ldots\ES\ldots\bar{\ES}\ldots$, representing the encounter in
question.  We consider only crossings for which $\ES$ and
$\overline{\ES}$ are separated by a subsequence $\RS$ representing an
intervening loop; the sequence $\RS$ may not be time-reversal
invariant, since otherwise it could be included in $\ES$.

An example for a further class of system-specific crossings is given
in Fig.  \ref{fig:crossings_cardioid}b.  The depicted encounter ends
abruptly when one of the two stretches is reflected close to the cusp
of the billiard, while the second stretch narrowly avoids the cusp.
The following loop remains close to the boundary of the billiard and
undergoes several almost glancing reflections.  Since the orbit in
question comes extremely close to the classically forbidden region,
the applicability of the Gutzwiller trace formula is highly
questionable.  We thus eliminate such crossings from our statistics as
well \cite{Mueller}.

\begin{figure}\begin{center}
    \includegraphics[scale=0.42]{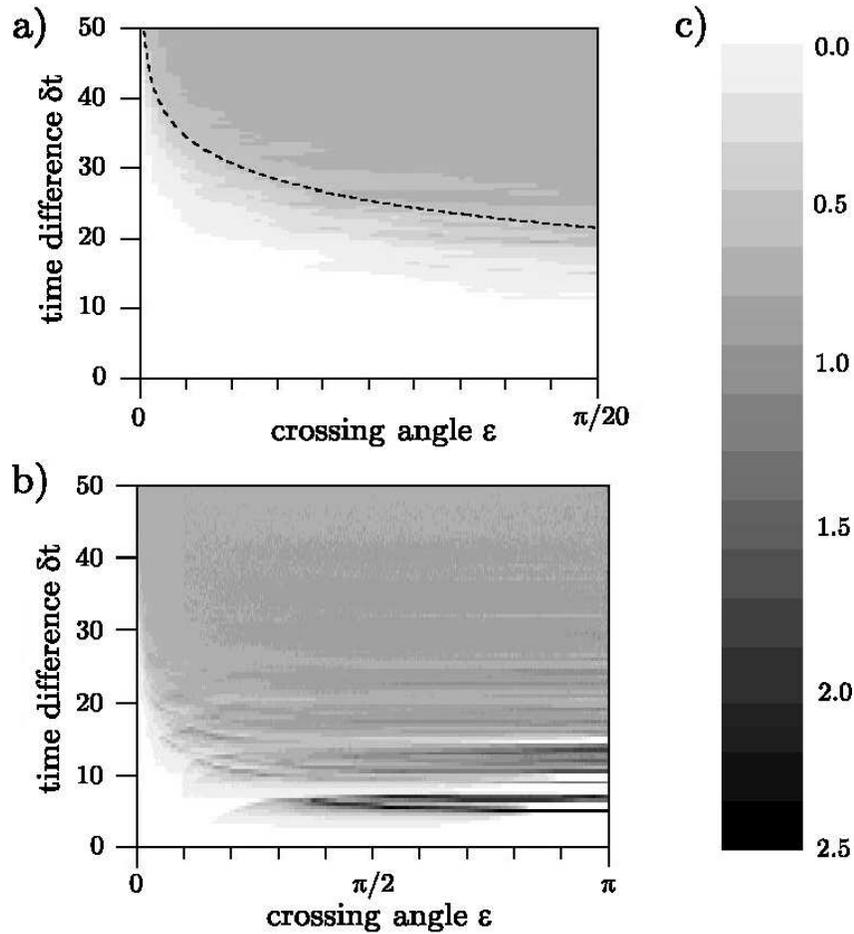}
\end{center}
\caption{
  Plot of the combined density of time differences $\delta t$ and crossing
  angles $\epsilon$ in the cardioid billiard: a) for $\epsilon<\frac{\pi}{20}$, b)
  for all angles. Normalization as explained in the text and used in
  Fig.  \ref{fig:diamond_results}; the resulting scale is shown in c).
  For small angles, we observe a threshold logarithmic in $\epsilon$, as
  indicated by a dashed line.}
\label{fig:cardioid_results}\end{figure}

The statistics of the remaining crossings confirms our predictions.
Again, the density of crossing angles and time differences, Fig.
\ref{fig:cardioid_results}, reveals a threshold for $\delta t$ depending
logarithmitically on the angle, and some system-specific
inhomogeneities close to that threshold.

Like in case of the desymmetrized diamond, these system-specific
features do not prevent universal behavior in the semiclassical limit,
since they are associated to classical time scales negligible compared
to $T_E$ and $T_H$.  A better understanding of these features could,
nevertheless, be helpful for mathematically rigorous work or for
studying deviations from universality outside the semiclassical limit.
Note that the Hadamard-Gutzwiller model also displays system-specific
structures, but of a different kind: In that model, the analog of
Figs. \ref{fig:diamond_results} and \ref{fig:cardioid_results} just
consists of dispersionless logarithmic curves corresponding each to
the family of orbit parts obtained by deformation of one periodic
orbit \cite{Sieber}.

\section{Sieber/Richter pairs give no additional contributions}

\label{sec:notau3}

We have seen that Sieber/Richter pairs contribute to the spectral form
factor only through a {\it correction} term inside $P_T(\x,\epsilon)$, of
subleading order in the orbit period.  Could a more careful analysis
of these orbit pairs reveal further corrections, also affecting the
form factor?

In the following, we shall identify several such corrections, but show
that their contributions to the form factor mutually cancel. We will
restrict ourselves to the Hadamard-Gutzwiller model, i.e., a surface
of constant negative curvature, tesselated into octagonic pieces
\cite{HGM,Gutzwiller,ThePaper,HeuslerPhD}.  This model is
homogeneously hyperbolic, i.e., the local stretching rates $\chi(\x)$ of
all phase-space points $\x$ and the Lyapunov exponents $\lambda_\gamma$ of all
periodic orbits coincide with the Lyapunov exponent of the system
$\lambda$.  As a consequence, we have $B_u(\x)= -B_s(\x)=m\lambda$, and may
neglect the $\x$ dependence of $C(\x)$.  Thus, when investigating the
statistics of encounters, we need not discriminate between different
points on the energy shell. To formulate that statistics, we work in
dimensionless units with $m=|{\bf p}|=\lambda=1$; in these units length,
period, and action of each orbit coincide, and will collectively be
denoted by $L$; likewise we will replace $t_{\rm enc}\to l_{\rm enc}$,
$T_H\to L_H$, and $\Delta S\to\Delta L$.  We then obtain a density of crossing
angles only,
\begin{equation}
\label{PLepsilon}
P_L(\epsilon)=\frac{\sin\epsilon}{\Omega}L(L-2l_{\rm enc}(\epsilon))\,,
\end{equation}
and Eq. (\ref{ksr_cr}) turns into
\begin{equation}
\label{good_sum}
K_{\rm SR}(\tau)=\frac{4}{L_H}\left\langle\sum_\gamma\int_{\epsilon>0} d\epsilon\,A_\gamma A_{\gamma'}\delta
\left(\tau L_H-\frac{L_\gamma+L_{\gamma'}}{2}\right)P_{L_\gamma}(\epsilon)\cos\frac{\Delta L}{\hbar}\right\rangle\,.
\end{equation}
Here, we deliberately avoided to replace $A_{\gamma'}\to A_\gamma$, or
$L_{\gamma'}\to L_\gamma$, and dropped the phases of $A_\gamma$ and $A_{\gamma'}$ since
the corresponding Maslov indices coincide. (The Maslov indices even
vanish, unless we divide the Hadamard-Gutzwiller model into pieces in
order to remove its symmetry.) Compared to Eqs.  (\ref{crossing_xe})
and (\ref{ksr_cr}), a factor 2 was shifted from the density
(\ref{PLepsilon}) to the prefactor in Eq. (\ref{good_sum}), since
$P_L(\epsilon)$ does not discriminate between the two points of traversal
and thus accounts for each crossing only once.

When evaluating (\ref{good_sum}), we want to avoid the potentially
dangerous approximations made in Chapter \ref{sec:tau2} and in
Appendix \ref{sec:crossings_overview}.  Three points require special
care: \newpage
\begin{itemize}
\item We need to take into account the difference between the
  stability amplitudes $A_\gamma$ and $A_{\gamma'}$. To do so, we write
\begin{equation}
A_\gamma=\frac{T_\gamma}{\sqrt{|\det(M_\gamma-1)|}}
=\frac{T_\gamma}{2\sinh\frac{\lambda_\gamma T_\gamma}{2}}
=\frac{L_\gamma}{2\sinh\frac{L_\gamma}{2}}
 \stackrel{L_\gamma\gg 1}{\approx} L_\gamma{\rm
  e}^{-L_\gamma/2}\,,
\end{equation}
see Eq. (\ref{amplitude_class}), and express $A_{\gamma'}$ as
\begin{equation}
\label{a_gamma_prime}
A_{\gamma'}\approx\frac{L_{\gamma'}}{L_\gamma}{\rm e}^{\Delta L/2}A_\gamma =\left(1-\frac{\Delta
    L}{L_\gamma}\right){\rm e}^{\Delta L/2}A_\gamma\,,
\end{equation}
depending on the length difference between $\gamma$ and $\gamma'$, $\Delta L=
L_\gamma-L_{\gamma'}$.
\item When treating the length (or action) difference, hyperbolic
  geometry allows to go beyond Sieber's and Richter's approximation
  $\Delta L\approx\frac{\epsilon^2}{2}$. As shown in \cite{ThePaper}, we have $\Delta
  L=-4\ln\cos\frac{\epsilon}{2}=\frac{\epsilon^2}{2}+\frac{\epsilon^4}{48}+\ldots$.
\item Finally, we work with the exact crossing density $P_L(\epsilon)$,
  without replacing $\sin\epsilon=\epsilon-\frac{\epsilon^3}{3}+\ldots$ by $\epsilon$.
\end{itemize}

The three corrections mentioned involve only
higher orders in $\epsilon$.  Upon applying %$\int d\epsilon\, {\rm e}^{{\rm i}\Delta S/\hbar}\ldots=
$\int d\epsilon\,{\rm e}^{{\rm i}\epsilon^2/(2\hbar)}\ldots$ the resulting contributions to
the form factor will be of higher order in $\hbar$, each factor $\epsilon^2$
being turned into a factor proportional to $\hbar$.  Therefore, the
changes arising will not affect the contribution of the correction
term inside $P_L(\epsilon)$, originating form the necessity of intervening
loops. The latter correction term still yields $K_{\rm SR,
  corr}(\tau)=-2\tau^2$.  We must, however, reexamine the contribution of
the leading term $\propto L^2$.  In Appendix \ref{sec:crossings_overview},
that term gave a contribution scaling like $\hbar^{-1}$, but vanishing
after averaging.  Additional corrections of relative order $\hbar$
therefore have a chance to survive in the semiclassical limit.  In
contrast, the length difference $\Delta L\propto \epsilon^2$, effectively of order
$\hbar$, can be safely neglected when directly compared to the orbit
period $L_{\gamma'}\propto\hbar^{-1}$.  We hence drop the summand $\frac{\Delta
  L}{L_\gamma}$ in (\ref{a_gamma_prime}) and replace
$\frac{L_{\gamma}+L_{\gamma'}}{2}\to L_\gamma$ in the $\delta$-function of Eq.
(\ref{good_sum}).  Altogether, we thus obtain
\begin{eqnarray}
\label{good_sum_lead}
K_{{\rm SR,lead}}(\tau)&=&\frac{4}{L_H}\left\langle\sum_\gamma A_\gamma^2\,\delta(\tau L_H-L_\gamma)
\int_{\epsilon>0} d\epsilon\, {\rm e}^{\Delta L/2}\left(\frac{\sin\epsilon}{\Omega}L_\gamma^2\right)\cos\frac{\Delta L}{\hbar}\right\rangle
\nonumber\\
&=&4\tau^3\frac{L_H^2}{\Omega}\left\langle\int_{\epsilon>0} d\epsilon\, {\rm e}^{\Delta L/2}\sin\epsilon
\,\cos\frac{\Delta L}{\hbar}\right\rangle,
\end{eqnarray}
where in the final step we have used the sum rule of Hannay and Ozorio
de Almeida.

The remaining integral can be evaluated by transforming to $\Delta
L=-4\ln\cos\frac{\epsilon}{2}$ as a new integration variable.  After
simplifying the integrand of Eq. (\ref{good_sum_lead}),
\begin{eqnarray}
d\epsilon\,\e^{\Delta L/2}\sin\epsilon&=&d\Delta L\left(\frac{d\Delta L}{d\epsilon}\right)^{-1}
\e^{\Delta L/2}\sin\epsilon \nonumber\\
&=&d\Delta L\left(2\tan\frac{\epsilon}{2}\right)^{-1}\left(\cos\frac{\epsilon}{2}\right)^{-2}\left(2\sin\frac{\epsilon}{2}\cos\frac{\epsilon}{2}\right) \nonumber\\
&=&d\Delta L\,,
\end{eqnarray}
we find
\begin{equation}
\label{srleadresult}
K_{{\rm SR,lead}}(\tau)=4\tau^3\frac{L_H^2}{\Omega}\left\langle\int_0^{\Delta L_{\rm
      max}}d\Delta L\cos\frac{\Delta L}{\hbar}\right\rangle = 4\tau^3\frac{L_H^2\hbar}{\Omega}\left\langle
  \sin\frac{\Delta L_{\rm max}}{\hbar}\right\rangle\,.
\end{equation}
Here $\Delta L_{\rm max}$ is the maximal length difference considered,
analogous to the maximal action difference $c^2$ following from
$|s|,|u|<c$ in Chapter \ref{sec:tau2}.  The resulting sine oscillates
rapidly in the semiclassical limit, and is annihilated by averaging.
We thus see that the additional corrections do not affect the spectral
form factor.

To achieve this result, it was of crucial importance to include {\it
  all three} corrections listed, of order $O(\epsilon^2)$ compared to the
terms considered in \cite{SR,Sieber}.  Each single correction would
yield a non-vanishing contribution of order $\tau^3$.  However, Eq.
(\ref{srleadresult}) implies that all three corrections taken together
mutually cancel.

\section{Summary}

We reviewed an alternative treatment of Sieber/Richter pairs,
determining the $\tau^2$ contribution to the form factor. This
alternative approach was based on self-crossings in two-dimensional
configuration space and the geometry of the stable and unstable
manifolds in phase space.  Each encounter contains one or more
self-crossings, or a narrowly avoided crossing.  The action difference
can be expressed as a function of the crossing angle and the stable
and unstable directions at the location of the crossing.  Numerical
investigations clearly demonstrate that the stretches of the relevant
encounters are separated by loops, and therefore the
traversals of the relevant crossings have minimal time differences
logarithmic in the crossing angle.  Moreover, our plots show some
system-specific structures related to very short loops.  We finally
presented a more careful analysis of Sieber/Richter pairs in the
Hadamard-Gutzwiller model, showing that three additional correction
terms give mutually canceling contributions to the form factor.

%% file: anhangb.tex
\chapter{Integrals involving $1/t_{\rm enc}$}

\stand

\label{sec:integral}

We want to evaluate the integral
\begin{equation}
\label{osc_integral}
\int_{-c}^cd^{l-1}sd^{l-1}u\frac{1}{t_{\rm
    enc}(s,u)}{\rm e}^{{{\rm i}}\Delta S/\hbar}
\end{equation}
over the $2(l-1)$ stable and unstable separations $s_{j}$, $u_{j}$
inside an $l$-encounter.  These variables determine both the duration
$t_{\rm enc}(s,u)$ of the encounter in question and its contribution
to the action difference $\Delta S=\sum_{j}s_{j}u_{j}$. We shall show that
the integral oscillates rapidly as $\hbar\to 0$ and thus may be neglected
in the semiclassical limit.  This was required in Section
\ref{sec:contribution} to show that certain terms in the multinomial
expansion of $w_T(s,u)$ do not contribute to the form factor; the
special case $l=2$ was treated in Section \ref{sec:contribution_sr}.

The key is the following change of picture: So far, all Poincar{\'e}
sections ${\cal P}$ inside a given encounter were integrated over; we
thus had to divide out the duration $t_{\rm enc}$.  Instead, we may
{\it single out a section ${\cal P}^{{\rm e}}$, fixed at the end of
  the encounter}, and only consider the stable and unstable
separations $s_{j}^{{\rm e}}$, $u_{j}^{{\rm e}}$ therein.  For
homogeneously hyperbolic dynamics, i.e., $\Lambda({\bf x},t)={\rm e}^{\lambda
  t}$ for all ${\bf x}$ and $t$, the separations inside ${\cal
  P}^{{\rm e}}$ are given by $s_j^{{\rm e}}= s_j{\rm e}^{-\lambda t_u}$,
$u_j^{{\rm e}}= u_j{\rm e}^{\lambda t_u}$ with $t_u$ denoting the time
difference between ${\cal P}$ and ${\cal P}^{{\rm e}}$.

We recall that the encounter ends when the largest of the unstable
components, say the $J$th one, reaches $\pm c$ such that $u_J^{{\rm
    e}}=u_J{\rm e}^{\lambda t_u}=\pm c$. All $l-1$ possibilities
$J=1,2,\ldots,l-1$ and the two possibilities for the sign $u_J^{{\rm
    e}}/c=\pm 1$ give additive contributions $I_J^\pm$ to the integral
(\ref{osc_integral}).  Each of these contributions is easily evaluated
after transforming the integration variables from $s_j$, $u_j$ to
$s_j^{{\rm e}}$, $u_j^{{\rm e}}$ (with $j\neq J$), $s_J^{{\rm e}}$, and
$t_u=\frac{1}{\lambda}\ln\frac{c}{|u_J|}$.  The Jacobian of that
transformation equals $\lambda c$.  The new coordinates determine the
action difference as $\Delta S=\sum_j s_j^{{\rm e}} u_j^{{\rm e}}=\sum_{j\neq J}
s_j^{{\rm e}} u_j^{{\rm e}}\pm s_J^{{\rm e}}c$.  The encounter duration
is given by $t_{\rm enc}(s,u)=t_{\rm enc}(s^{\rm e},u^{\rm e})
=\min_j\Big\{\frac{1}{\lambda}\ln\frac{c}{|s^{\rm e}_j|}\Big\}$, see Eq.
(\ref{tenc}).
Note that $t_{\rm enc}$ does not depend on the unstable coordinates
$u^{\rm e}$; the latter determine only the duration of the encounter head,
which is missing because the Poincar\'e section ${\cal P}^{\rm e}$
is placed in the end of the encounter.  Since our new coordinates are
restricted to the ranges $-c<s_j^{{\rm e}}<c$, $-c<u_j^{{\rm e}}<c$,
for $j\neq J$, $-c<s_J^{{\rm e}}<c$, and $0<t_u<t_{\rm enc}$, we obtain
\begin{eqnarray}
\label{osc_result}
I_J^\pm&=&\lambda c\int_{-c}^cd s_J^{{\rm e}}{\rm e}^{\pm {\rm i}s_J^{{\rm e}}c/\hbar}
\left(\prod_{j\neq J}\int_{-c}^cd s_j^{{\rm e}} du_j^{{\rm e}}{\rm e}
^{{\rm i}s_j^{{\rm e}}u_j^{{\rm e}}/\hbar}\right)\nonumber\\&&\times\underbrace{\frac{1}{t_{\rm enc}
(s^{{\rm e}},u^{{\rm e}})}
\int_0^{t_{\rm enc}(s^{{\rm e}},u^{{\rm e}})}d t_u}_{=1}  \nonumber\\
&\sim& \lambda (2\pi\hbar)^{l-2}2\hbar\sin\frac{c^2}{\hbar}\,.
\end{eqnarray}
Note that the divisor $t_{\rm enc}$ was canceled by the $t_u$
integral; moreover, the $2(l-2)$ integrals over $s_j^{{\rm e}}
,u_j^{{\rm e}}$, of the form already encountered in Eq.
(\ref{simple_integral}), gave the factor $(2\pi\hbar)^{l-2}$. Most
importantly, the factor $\sin\frac{c^2}{\hbar}$, provided by the integral
over $s_J^{{\rm e}}$, is a rapidly oscillating function of $c$ and
$\hbar$, annulled by averaging over these quantities; as shown in Section
\ref{sec:contribution_sr} averaging over $c$ is equivalent to
averaging over the energy $E$.  Thus, the integral
(\ref{osc_integral}), just the $2(l-1)$-fold of Eq.
(\ref{osc_result}), effectively vanishes as $\hbar\to 0$.  Note that
rapidly oscillating terms as in Eq. (\ref{osc_result}) are essentially
spurious and would not appear if smooth encounter cut-offs were used
(instead of our $|s|<c,|u|<c$).

%% file: anhangc.tex
\chapter{Extension to general hyperbolicity and $f>2$}

\label{sec:general}

\stand So far, some of our reasoning was restricted to two-dimensional
homogeneously hyperbolic systems.  Here, we extend our approach to
general fully chaotic dynamics with arbitrary numbers $f$ of degrees
of freedom.  The present results were published in \cite{Higherdim}
for the case of Sieber/Richter pairs, and in an Appendix of
\cite{LongPaper} for arbitrary orders in $\tau$.

\section{General hyperbolicity}

\label{sec:maths}

First, we drop the restriction to ``homogeneously hyperbolic''
dynamics, for which all phase space points ${\bf x}$ have the same
Lyapunov exponent $\lambda$ and the same stretching factor $\Lambda(t)={\rm
  e}^{\lambda t}$.  We extend our reasoning to general hyperbolic systems,
where the stretching factors $\Lambda({\bf x},t)$ may depend on ${\bf x}$.
In such systems the Lyapunov exponents of {\it almost all } points
still coincide with the ${\bf x}$ independent ``Lyapunov exponent of
the system'', whereas each periodic orbit may come with its own
Lyapunov exponent (see Section \ref{sec:classical}).

Most importantly, the divergence of the stretches involved in an
encounter depends on the local stretching factor of that encounter,
rather than the Lyapunov exponent of the system.  Consequently, our
formula (\ref{tenc}) for the encounter duration can only be read as an
approximation, and that approximation is now to be avoided.  We will
thus allow the duration $t_{\rm enc}^\alpha$ of the $\alpha$-th encounter to
depend not only on the stable and unstable separations $s_{\alpha j}$,
$u_{\alpha j}$, but also on the phase-space location of the piercing ${\bf
  x}_{\alpha 1}$ chosen as the origin of the corresponding Poincar{\'e}
section.  Together, the reference piercing $\x_{\alpha 1}$ and the
separations $s_{\alpha j}$, $u_{\alpha j}$ determine the positions of all
piercing points of the $\alpha$-th encounter and therefore clearly suffice
to determine its duration.  The changes arising will be important only
for showing that the contribution originating from the
$\frac{1}{t_{\rm enc}}$-integrals of Appendix \ref{sec:integral}
vanishes; recall that the reasoning in Appendix \ref{sec:integral}
explicitly required homogeneous hyperbolicity. In contrast, the terms
contributing to $K(\tau)$ remain unaffected, since for these terms all
occurrences of $t_{\rm enc}$ mutually cancel.

When generalizing the statistics of encounters of Section
\ref{sec:statistics}, we must extend $w_T(s,u)$ to a {\it density
  $w_T(\x,s,u)$ also depending on the points of the reference
  piercings} ${\bf x}=\{{\bf x}_{1 1},{\bf x}_{2 1},\ldots,{\bf x}_{V
  1}\}$.  We assume that all piercing points are statistically
uncorrelated; as we have seen, the only existing correlations are
related to loops shorter than the classical relaxation time $t_{\rm
  cl}\ll T_E$ and thus cannot affect the spectral form factor.  We
therefore expect a density of reference piercings and stable and
unstable coordinates
\begin{equation}
\label{wTx}
w_T(\x,s,u)=\frac{T(T-\sum_\alpha l_\alpha t_{\rm enc}^\alpha)^{L-1}}
{(L-1)!\Omega^{L}\prod_\alpha t_{\rm enc}^\alpha}\,,
\end{equation}
differing from the $w_T(s,u)$ of Eq. (\ref{density}) only by $t_{\rm
  enc}^\alpha=t_{\rm enc}^\alpha(\x_{\alpha 1},s_\alpha, u_\alpha)$ being a function of
${\bf x}_{\alpha 1}$, and by a factor $\frac{1}{\Omega^V}$ arising from the
Liouville density for each of the $V$ reference piercings. Our
expression for the form factor, Eq.  (\ref{ksum}), now turns into
\begin{eqnarray}
\label{ksumx}
K(\tau)=\kappa \tau+\kappa \tau\left\langle\sum_{\vec{v}} \frac{N(\vec{v})}{L}
\int d^V\mu(\x)\int d^{L-V}s\;d^{L-V}u \;
w_T(\x,s,u)\;{\rm e}^{\rm i\Delta S/\hbar}\right\rangle,
\end{eqnarray}
where the ${\bf x}$ integral refers to $V$ points ${\bf x}_{\alpha1}$ in
the energy shell, i.e., $d^{V}\mu({\bf x})=\prod_{\alpha=1}^V d^4 x_{\alpha
  1}\,\delta(H({\bf x}_{\alpha 1})-E)$.  Eqs. (\ref{wTx}) and (\ref{ksumx})
can easily be justified along the lines of Chapter \ref{sec:geometry}.
In the following we want, however, to give a mathematically more
careful derivation, showing explicitly how the equidistribution
theorem of \cite{Equidistribution} and the condition of mixing come
into play.

\subsubsection{Derivation of Eqs. (\ref{wTx}) and (\ref{ksumx})}

We first want to consider encounters of {\it one single periodic orbit
  $\gamma$}. To do so, we choose an arbitrary ``starting point'' ${\bf
  z}_0$ on $\gamma$, and let $\Phi_t({\bf z}_0)$ denote the image of ${\bf
  z}_0$ under evolution over the time $t$.  For the moment, we want to
restrict ourselves to systems without time-reversal invariance.

We now generalize the density $\rho(s,u,t)$ of Eq. (\ref{rho}) to a
density $\rho^\gamma({\bf x},s,u,t)$ of piercing times $t_{\alpha j}$, reference
piercings $\x_{\alpha 1}=\Phi_{t_{\alpha 1}}({\bf z}_0)$, and stable and
unstable coordinates $s$, $u$ associated to $\hat{s}$, $\hat{u}$ with
$\Phi_{t_{\alpha j}}({\bf z}_0)-\x_{\alpha 1}=\hat{s}_{\alpha j}\vs(\x_{\alpha 1})
+\hat{u}_{\alpha j}\vu(\x_{\alpha 1})$.  The density $\rho^\gamma({\bf x},s,u,t)$
shall refer to a fixed orbit $\gamma$, a fixed structure of the orbit pair
$(\gamma,\gamma')$ and fixed times $t_{\alpha 1}$ of the reference piercings; it
will be normalized such that integration over energy-shell regions for
the reference piercings $\x_{\alpha 1}$, and over intervals for the stable
and unstable coordinates $s_{\alpha j}$, $u_{\alpha j}$ ($j=1,\ldots,l_{\alpha}-1$)
and the remaining piercing times $t_{\alpha j}$ (now with $j$ running
$j=2,\ldots,l_{\alpha}$) yields the corresponding number of sets of piercings
inside $\gamma$.  For each orbit $\gamma$, the $\rho^\gamma(\x,s,u)$ thus defined
can be written as a product of $\delta$ functions,
\begin{eqnarray}
\label{rho_deltas}
&&
\rho^\gamma({\bf x},s,u,t)=\Theta_T(\x,s,u,t)
\prod_{\alpha=1}^V\delta(\Phi_{t_{\alpha 1}}({\bf z}_0)-{\bf x}_{\alpha 1})
\nonumber\\
&&\ \ \ \ \ \ \times
\prod_{j=2}^{l_\alpha}\delta\left(\Phi_{t_{\alpha j}}({\bf z}_0)-{\bf x}_{\alpha 1}-\hat{s}_{\alpha j}{\bf e}^s({\bf x}_{\alpha 1})
-\hat{u}_{\alpha j}{\bf e}^u({\bf x}_{\alpha 1})\right).
\end{eqnarray}
This product also involves a characteristic function $\Theta_T(\x,s,u,t)$
which returns 1 if the ordering of times corresponds to the structure
considered and the time differences suffice to have all encounter
stretches separated by loops; otherwise $\Theta_T(\x,s,u,t)$ is equal to
zero.  Eq. (\ref{rho_deltas}) can easily be generalized to
time-reversal invariant systems: If the $j$-th stretch of the $\alpha$-th
encounter is almost time-reversed with respect to the first one, we
simply have to replace $\Phi_{t_{\alpha j}}({\bf z}_0)\to{\cal T}\Phi_{t_{\alpha
    j}}({\bf z}_0)$.

In analogy to Section \ref{sec:statistics}, we integrate over the
piercing times and divide out the encounter durations, obtaining a
density of reference piercings and stable and unstable separations
only,
\begin{equation}
\label{wgammax}
w^\gamma({\bf x},s,u)=\frac{\int d^Lt\ \rho^\gamma({\bf
      x},s,u,t)} {\prod_\alpha t_{\rm enc}^\alpha({\bf x}_{\alpha 1},s_\alpha,u_\alpha)}.
\end{equation}

The periodic-orbit sum for the spectral form factor (\ref{ksum}) now
takes the form
\begin{eqnarray}
\label{kngamma}
 &&K(\tau)=\kappa\tau+\frac{\kappa}{T_H}\left\langle
\sum_{\vec{v}} \frac{N(\vec{v})}{L}
\int d^{V}\mu({\bf x})\int d^{L-V}s\;d^{L-V}u\;{\rm e}^{{\rm i}\Delta S/\hbar}
\right.\nonumber\\
&&\ \ \ \ \ \ \ \ \ \ \ \ \ \ \ \ \ \ \ \ \ \ \ \ \ \ \times\left.\left\{\sum_{\gamma}|A_\gamma|^2\delta(T-T_\gamma)
w^\gamma(\x,s,u)\;\right\}\right\rangle.
\end{eqnarray}
The form factor thus depends on the average
\begin{equation}
\frac{1}{T}\left\langle\sum_{\gamma}|A_\gamma|^2\delta(T-T_\gamma) w^\gamma(\x,s,u)\;\right\rangle_{\!\!\!\Delta T}
\end{equation}
of $w^\gamma(\x,s,u)$ over the ensemble of all periodic orbits with
periods inside a small window around $T$, weighted with the square of
their stability amplitudes.  We have to show that this average
coincides with the $w_T(\x,s,u)$ of Eq. (\ref{wTx}).

To proceed, we split the time integral of Eq.  (\ref{wgammax}) into an
integral over $0<t_{11}<T$, and further integrals over the differences
$t_{\alpha j}'=t_{\alpha j}-t_{11}$ of all other piercing times from the first
one.  Using that $\Phi_{t_{\alpha j}}({\bf z}_0)=\Phi_{t_{11}+t_{\alpha j}'}({\bf
  z}_0)=\Phi_{t_{11}}( \Phi_{t_{\alpha j}'}({\bf z}_0))$, we may thus
represent $w^\gamma$ as the average of an observable $f({\bf z})$ along
$\gamma$,
\begin{equation}
w^{\gamma}({\bf x},s,u)=\frac{1}{T}\int_0^T dt_{11}\, f(\Phi_{t_{11}}({\bf z}_0))\equiv
\big[f\big]_\gamma
\end{equation}
with
\begin{eqnarray}
\label{fauxiliary}
f({\bf z})&=&\frac{T}{\prod_\alpha t_{\rm enc}^\alpha({\bf x}_{\alpha 1},s_\alpha,u_\alpha)}
\int d^{L-1}t'\,\Theta_T(\x,s,u,t')\prod_{\alpha=1}^V\delta(\Phi_{t_{\alpha 1}'}({\bf z})-{\bf x}_{\alpha 1})
\nonumber\\&&\times
\prod_{j=2}^{l_\alpha}\delta\left(\Phi_{t_{\alpha j}'}({\bf z})-{\bf x}_{\alpha 1}-\hat{s}_{\alpha j}{\bf e}^s({\bf x}_{\alpha 1})
-\hat{u}_{\alpha j}{\bf e}^u({\bf x}_{\alpha 1})\right)\,.
\end{eqnarray}
Here were replaced $\Theta_T(\x,s,u,t)\to \Theta_T(\x,s,u,t')$, since the times
$t_{\alpha j}'=t_{\alpha j}-t_{11}$ have to obey the same ordering and the
same minimal distances as the times $t_{\alpha j}$.

Thus prepared, we can invoke the {\it equidistribution theorem} of
\cite{Equidistribution} (see Subsection \ref{sec:ergodicity}): If an
observable $f({\bf z})$ is averaged (i) along a periodic orbit $\gamma$
and (ii) over an ensemble of all $\gamma$ (in a small time window and
weighted with $|A_\gamma|^2$ as above), we obtain an energy-shell average
$\overline{f({\bf z})}$.  Hence, the periodic-orbit average of
$w^\gamma(\x,s,u)$ can be evaluated as
\begin{equation}
%\begin{eqnarray}
\label{equidistribution_used}
%&&
\frac{1}{T}\Big\langle\!\sum_\gamma |A_\gamma|^2\delta(T-T_\gamma)w^\gamma(\x,s,u)\Big
\rangle_{\!\!\!\Delta T}%\nonumber\\
%&&
\!\!\!\!=
\!\frac{1}{T}\Big\langle\!\sum_\gamma |A_\gamma|^2\delta(T-T_\gamma)\big[f\big]_\gamma\Big\rangle_{\!\!\!\Delta T}
\!\!\!\!=\!\!\textstyle{\int\!\frac{d\mu({\bf z})}{\Omega}f({\bf z})}\equiv\overline{f({\bf z})}\,.
%\end{eqnarray}
\end{equation}

For the observable given in Eq. (\ref{fauxiliary}), the energy-shell
average $\overline{f({\bf z})}$ can be calculated provided the
dynamics is {\it mixing} (see Subsection \ref{sec:ergodicity}), i.e.,
if for two observables $g({\bf z})$, $h({\bf z})$ we have
\begin{equation}
\label{mixing_repeat}
\overline{g({\bf z})h(\Phi_t({\bf z}))}
\xrightarrow[t\to\infty]{}\overline{g}\;\overline{h}\,.
\end{equation}
For finite $t$, we can then approximate
\begin{equation}
\label{mixing_approx}
\overline{g({\bf z})h(\Phi_t({\bf z}))}
\approx\overline{g}\;\overline{h}\,,
\end{equation}
if $t$ is large enough to treat ${\bf z}$ and $\Phi_t({\bf z})$ as
uncorrelated. Neglecting correlations between subsequent piercings, we
repeatedly invoke Eq.  (\ref{mixing_approx}) for the average of the
product of $\delta$ functions in Eq.  (\ref{fauxiliary}).  We thus end up
with a product of averages of the individual $\delta$ functions with
$\Phi_{t_{\alpha 1}'}({\bf z})$ and $\Phi_{t_{\alpha j}'}({\bf z})$ replaced by
${\bf z}$.  These averages are easily calculated, yielding $V$ factors
$\int\frac{d\mu({\bf z})}{\Omega}\delta({\bf z}-{\bf x}_{\al 1})=\frac{1}{\Omega}$
and $L-V$ factors $\int \frac{d\mu({\bf z})}{\Omega}\delta({\bf z}-{\bf x}_{\al
  1}-\hs_{\al j}\vs(\x_{\al 1}) -\hu_{\al j}\vu(\x_{\al
  1}))=\frac{1}{\Omega}$.  The periodic-orbit average of $w^\gamma(\x,s,u)$,
and thus the energy-shell average of $f({\bf z})$, now turns into
\begin{eqnarray}
\label{wfancy}
\overline{f({\bf z})}
= \frac{T}{\prod t_{\rm enc}^\alpha}\int d^{L-1}t'\frac{\Theta_T(\x,s,u,t')}{\Omega^L}
=\frac{T(T-\sum_\alpha l_\alpha
  t_{\rm enc}^\alpha)^{L-1}} {(L-1)!\Omega^{L}\prod_\alpha t_{\rm
    enc}^\alpha}\,.
\end{eqnarray}
In the final step, the $t'$ integral was evaluated in the same way as
the $t$ integral of Section \ref{sec:statistics}.  Indeed, the
periodic-orbit average of $w^\gamma(\x,s,u)$ yields the same $w_T(\x,s,u)$
as predicted in Eq. (\ref{wTx}) and the expression (\ref{kngamma}) for
the spectral form factor is brought to the same form as in Eq.
(\ref{ksumx}).

\subsubsection{Contributions to the form factor}

The $t_{\rm enc}$-independent terms in the multinomial expansion of
$w_T(\x,s,u)$ yield the same contributions to the form factor as in
Section \ref{sec:contribution}, since the additional divisor $\Omega^V$ is
canceled by integration over ${\bf x}$.  Summation thus gives a
$K(\tau)$ faithful to the predictions of random-matrix theory.  All
other contributions can be neglected in the semiclassical limit,
because they are either of a too low order in $T$ or proportional to
integrals involving $\frac{1}{t_{\rm enc}(\x,s,u)}$ (where we dropped
the encounter label $\alpha$).

\subsubsection{Integrals involving $1/t_{\rm enc}(\x,s,u)$}

\enlargethispage{0.5mm}

To show that integrals of the form
\begin{equation}
\label{osc_integral_fancy}
\int \frac{d\mu({\bf x})}{\Omega}\int_{-c}^cd^{l-1}sd^{l-1}u\frac{1}{t_{\rm
    enc}({\bf x},s,u)}{\rm e}^{{{\rm i}}\Delta S/\hbar}\,
\end{equation}
vanish after averaging, we reason similarly as in Appendix
\ref{sec:integral}.  For each contribution $I_J^\pm$, we transform from
${\bf x},s,u$ to phase-space points ${\bf x}^{{\rm e}}$ and
separations $s^{{\rm e}}_j,u^{{\rm e}}_j$ ($u^{{\rm e}}_J=\pm c$ fixed)
inside a Poincar{\'e} section ${\cal P}^{{\rm e}}$ in the encounter end,
and the separation $t_u$ between ${\cal P}$ and ${\cal P}^{{\rm e}}$.
For general hyperbolic dynamics, the stable and unstable coordinates
are related by $s_j^{{\rm e}}=\Lambda({\bf x},t_u)^{-1}s_j$ and $u_j^{{\rm
    e}}=\Lambda({\bf x},t_u)u_j$; see Eq. (\ref{linearize}).  The
Jacobian\footnote{ In particular, we have $\frac{du_J}{dt_u}
  =\frac{d\,\Lambda(\x,t_u)^{-1}}{dt_u}u_J^{\rm e} =-\Lambda({\bf
    x},t_u)^{-1}\frac{d\ln|\Lambda({\bf x},t_u)|}{dt_u}u_J^{{\rm e}}
  =-\Lambda({\bf x},t_u)^{-1}\chi({\bf x}^{{\rm e}})u_J^{{\rm e}}$ with
  $u_J^{{\rm e}}=\pm c$, where we used that $\Phi_{t_u}({\bf x})={\bf
    x}^{{\rm e}}$.  The factor $\Lambda({\bf x},t_u)^{-1}$ is compensated
  by the remaining transformations $u_j\to u_j^{{\rm e}} (j\neq J)$ and
  $s_j\to s_j^{{\rm e}}$.  For $I_J^+$ (that is, positive $u_J$) the
  minus sign in $\frac{d u_J}{d t_u}$ is compensated because the upper
  bound $c$ for $u_J$ corresponds to the lower bound $0$ for $t_u$.
  For $I_J^-$ the minus sign is canceled by the minus in $u_J^{\rm
    e}=-c$.  } of this transformation now reads $\chi({\bf x}^{{\rm
    e}})c$ with the local stretching rate defined as $\chi(\Phi_t({\bf
  x}))=\frac{d\ln|\Lambda({\bf x},t)|}{dt}$ (see Subsection
\ref{sec:hyperbolicity}).  We thus obtain
\begin{eqnarray}
\label{osc_result_fancy}
 I_J^\pm&=&\int \frac{d\mu({\bf x}^{\rm e})}{\Omega}\chi({\bf x}^{{\rm e}})c
\int_{-c}^cd s_J^{{\rm e}}{\rm e}^{\pm {\rm i}s_J^{{\rm e}}c/\hbar}
\left(\prod_{j\neq J}\int_{-c}^cd s_j^{{\rm e}} du_j^{{\rm e}}
{\rm e}^{{\rm i}s_j^{{\rm e}}u_j^{{\rm e}}/\hbar}\right)   \nonumber\\
&&\times\underbrace{\frac{1}{t_{\rm enc}({\bf x}^{\rm e},s^{{\rm e}},u^{{\rm e}})}
\int_0^{t_{\rm enc}({\bf x}^{\rm e},s^{{\rm e}},u^{{\rm e}})}d t_u}_{=1}
\;,
\end{eqnarray}
coinciding with Eq. (\ref{osc_result}) since the energy-shell average
of the local stretching rate yields the Lyapunov exponent of the
system $\lambda$.  Thus, all $I_J^\pm$ vanish after averaging.  Incidentally,
we could have dropped the argument $u^{\rm e}$ of $t_{\rm enc}(\x^{\rm
  e},s^{\rm e},u^{\rm e})$ because the unstable coordinates only
determine the duration of the encounter head; since ${\cal P}^{\rm e}$
is placed in the end of the encounter, this duration must be zero.

\enlargethispage{0.5mm}

We have to check that Eq. (\ref{osc_result_fancy}) remains valid if
the local stretching rate becomes negative for some regions inside the
energy shell.  In this case, the unstable coordinates can temporarily
shrink rather than grow. In the end of the encounter, the unstable
coordinates may therefore oscillate between values larger and smaller
than $\pm c$.  The Poincar{\'e} section ${\cal P}^{\rm e}$ may then be
placed at any position where $u_J$ reaches $\pm c$.  Consequently, the
mapping $\x,s,u\to \x^{\rm e},s^{\rm e},u^{\rm e},t_u$ becomes {\it
  multi-valued}.\footnote{The encounter duration $t_{\rm enc}$, too,
  becomes non-unique.  However, the differences between the possible
  $t_{\rm enc}$'s are negligible compared to the overall duration of
  order Ehrenfest time.}  When transforming from Eq.
(\ref{osc_integral_fancy}) to (\ref{osc_result_fancy}), each possible
set of coordinates $\{\x^{\rm e},s^{\rm e}, u^{\rm e},t_u\}$ arising
from the same $\{\x,s,u\}$ is taken into account separately.  The
corresponding contributions come, however, with different signs: All
sections ${\cal P}^{\rm e}$ traversed with growing $|u_J|$ and thus
$\chi(\x^{\rm e})>0$ contribute with a positive sign, whereas all ${\cal
  P}^{\rm e}$ traversed with shrinking $|u_J|$ and thus $\chi(\x^{\rm
  e})<0$ contribute with a negative sign.  (If we wanted to count all
${\cal P}^{\rm e}$ positively, we would have to use the absolute value
$|\chi(\x^{\rm e})|c$ of the Jacobian.)  Due to the asymptotic growth of
$|u_J|$, each $\{\x,s,u\}$ leads to one more section ${\cal P}^{\rm
  e}$ traversed with growing $|u_J|$ than with shrinking $|u_J|$.
Summing over all ${\cal P}^{\rm e}$, we see that only one contribution
to Eq.  (\ref{osc_result_fancy}) remains effective.  Hence, our
coordinate transformation remains valid even in case of negative local
stretching rates.

\subsubsection{Physical interpretation}

\begin{figure}[t]
\begin{center}
\includegraphics[scale=0.24]{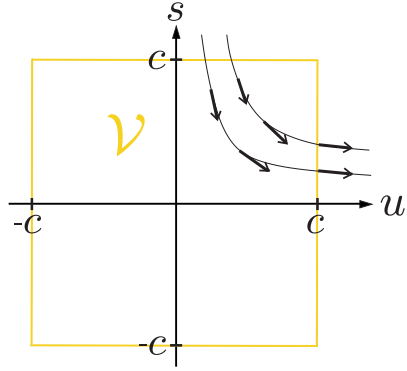} %0.26
\end{center}
\vspace{-0.5cm}

\caption{
  Motion of piercing points through a Poincar{\'e} section ${\cal P}$
  inside a 2-encounter.  The box contains stable and unstable
  coordinates within the ranges $|s|,|u|<c$ corresponding to the
  volume $\V$ defined in the text.  As ${\cal P}$ is shifted, the
  unstable components grow and the stable ones shrink, traveling on a
  hyperbola $\Delta S=s u$; arrows denote the direction of motion.  At end
  of the encounter, the piercing points traverse the line $u=c$.  Of
  course, there are also trajectories cutting through the remaining
  four quadrants of $\V$.  }
\label{fig:travel}
\end{figure}

The transformation leading from Eq. (\ref{osc_integral_fancy}) to Eq.
(\ref{osc_result_fancy}), and its ``homogeneous'' counterpart in
Appendix \ref{sec:integral}, have an interesting physical
interpretation. The piercing points of each $l$-encounter are
described by coordinates $\x,s,u$ restricted to the volume
$\V=\Big\{(\x,s,u)\;\Big|\;|s_j|,|u_j|<c\ {\rm for }\ 
j=1,\ldots,l-1\Big\}$.  As our Poincar{\'e} section ${\cal P}$ is shifted, the
piercing points travel through $\V$: $\x$ follows the corresponding
orbit, the stable coordinates shrink and the unstable coordinates
grow.  Each encounter thus corresponds to one ``trajectory'' cutting
through the volume $\V$. For the example of a 2-encounter, we depicted
in Fig. \ref{fig:travel} the range of stable and unstable coordinates
$|s|,|u|<c$ belonging to $\V$, and two trajectories corresponding to
the piercing points of two different encounters.

We now have to sum over encounters or, equivalently, over trajectories
cutting through $\V$.  There are essentially two ways to perform such
sums: On the one hand, we may integrate over $\V$ and over time (that
is, over the time of the reference piercing). Each trajectory is thus
counted for the time $t_{\rm enc}$ it spends inside $\V$, and we
subsequently have to {\it divide by $t_{\rm enc}$}. This approach was
taken when deriving the $w_T(\x,s,u)$ of Eq. (\ref{wTx}), or the
$w_T(s,u)$ of Eq.  (\ref{wT}). Note that in these cases, the integrand
depended on the probability density for $\x$, $s$, and $u$. In Eq.
(\ref{osc_integral_fancy}) we dropped proportionality factors arising
from the time integration and the probability, and only integrated
over $\e^{{\rm i}\Delta S/\hbar}$.

%\begin{floatingfigure}{10cm}%[t]
%\begin{wrapfigure}{r}{8cm}
%\end{floatingfigure}

On the other hand, we can count each trajectory {\it when it leaves
  the volume $\V$}.  Each trajectory finally leaves $\V$ through one
of the $2(l-1)$ faces defined by $u_J=\pm c$, $J=1,\ldots,l-1$, with the
remaining stable and unstable coordinates restricted to $(-c,c)$.  (In
contrast trajectories first enter through the faces defined by $s_J=\pm
c$.)  To count trajectories leaving $\V$ through one of these faces,
we consider the current density leading out of $\V$. We thus multiply
the integrand, i.e., the ``density'' associated to $\x$, $s$, and $u$,
with the velocity of leaving $\V$.  The unstable coordinates change
with the velocity $\frac{du_j}{dt}=\chi(\x)u_j$, see Eq.
(\ref{stretching}).  The velocity component perpendicular to the face
with $u_J=\pm c$ is therefore given by $\chi(\x^{\rm e})c$; here we
denoted the corresponding phase-space point by $\x^{\rm e}$ and fixed
the sign such that the velocity of trajectories leaving $\V$ is taken
as positive.  The resulting current density has to be integrated over
the face of $\V$ considered, and over time.  Dropping proportionality
factors arising from the time integral and from probability
considerations, we are thus led to Eq.  (\ref{osc_result_fancy}), with
the Jacobian $\chi(\x^{\rm e})c$ interpreted as a velocity.  If the
local stretching rate becomes negative, the trajectories can reenter
into $\V$ through the same face, giving a negative contribution to the
flux in Eq.  (\ref{osc_result_fancy}).  However, since each trajectory
finally leaves $\V$, all entries and exits ultimately sum up to yield
one positive contribution.

Eqs. (\ref{osc_integral_fancy}) and (\ref{osc_result_fancy}) hence
represent two equivalent ways of summing over encounters, or
trajectories crossing $\V$. The approach of Eq.
(\ref{osc_integral_fancy}) is better suited for terms contributing to
$K(\tau)$, whereas the approach of Eq.  (\ref{osc_result_fancy}), first
employed in \cite{Turek} for Sieber/Richter pairs in two-dimensional
systems, yields an easier treatment of the vanishing terms.

\section{More than two degrees of freedom}

\label{sec:multi}

Our results can easily be extended to dynamics with any number $f$ of
degrees of freedom. The applicability of the Gutzwiller trace formula
to such systems has been extensively studied in
\cite{Primack,SmilanskyFreedoms}.

For dynamics with arbitrary $f\geq 2$, the Poincar{\'e} section ${\cal P}$
at point ${\bf x}$ is {\it spanned by $f-1$ pairs of stable and
  unstable directions} ${\bf e}^s_m({\bf x})$, ${\bf e}^u_m({\bf x})$
($m=1,2,\ldots,f-1$), as in Eq.  (\ref{decompose_multi}).  The mutual
normalization of these directions was fixed in Eq. (\ref{norm_multi}).
Each pair of directions comes with separate stretching factors
$\Lambda_m({\bf x},t)$ and stretching rates $\chi_m(\x)$, and a separate
Lyapunov exponent $\lambda_m$.

Our results carry over if we write out the additional index $m$: The
piercing points of a given encounter are described by components
$\hs_{jm}$, $\hu_{jm}$ ($j=2,\ldots ,l$; $m=1,\ldots,f-1)$.  Still defining
the encounter as the region where all these components are inside
$(-c,c)$, the durations of heads and tails, see Eqs.
(\ref{tu}-\ref{ts}), generalize to
\begin{equation}
  t_u=\min_{j,m}\left\{\frac{1}{\lambda_m}\ln\frac{c}{|\hu_{jm}|}\right\},
  \ \ \ \;
  t_s=\min_{j,m}\left\{\frac{1}{\lambda_m}\ln\frac{c}{|\hs_{jm}|}\right\}.
\end{equation}
The action difference related to one encounter is now given by $\Delta
S=\sum_{j,m} s_{jm}u_{jm}$, with $s_{jm},u_{jm}$ defined by the same
coordinate transformation as in Subsection
\ref{sec:action_difference}.  The integral over $(L-V)(f-1)$ pairs of
stable and unstable components $s_{\alpha jm}$, $u_{\alpha jm}$ in the second
line of Eq. (\ref{contribution}) yields $(2\pi\hbar)^{(L-V)(f-1)}$, which
is just what we need since the Heisenberg time now reads
$T_H=\frac{\Omega}{(2\pi\hbar)^{f-1}}$.

Given that the encounter ends as soon as one unstable component, say
$u_{JM}$, reaches $\pm c$, the $\frac{1}{t_{\rm enc}}$-integral of
Appendices \ref{sec:integral} and \ref{sec:maths} is split into
components $I_{JM}^\pm$, with $\lambda$ replaced by $\lambda_M$, and $\chi({\bf x})$
by $\chi_M({\bf x})$.  These components can be evaluated as in case of
two degrees of freedom.

\section{Summary}

We extended our results to general fully chaotic systems with
arbitrary number $f$ of degrees of freedom. For inhomogeneously
hyperbolic systems, the duration of each encounter not only depends on
the differences between piercings, but also on the phase-space
location of the first piercing. This dependence had to be taken into
account when showing that integrals involving $\frac{1}{t_{\rm enc}}$
effectively vanish.  Moreover, we showed explicitly how to implement
the necessary average over periodic orbits.  Our results carry over to
systems with $f>2$ if we write out the additional index $m$, numbering
the $f-1$ pairs of stable and unstable directions.

%% file: anhangd.tex
\chapter{Encounter overlap}

\stand

\label{sec:overlap_appendix}

So far, we have confined ourselves to encounters whose {\it stretches
  are separated by intervening loops, i.e., do not overlap}.  To
justify this, we now want to show that encounters with overlapping
stretches do not give rise to additional orbit pairs, and therefore do
not contribute to the form factor.  We must distinguish between three
special cases.  As already mentioned in Section \ref{sec:loops}, if
stretches of two {\it different} encounters overlap, these encounters
have to be regarded as a {\it single encounter}.  Generalizing the
results of Section \ref{sec:min_dist}, we will show that two {\it
  antiparallel} stretches without an intervening loop have to be
treated as {\it one single stretch}.  Finally, we shall see that
encounters with overlapping {\it parallel} stretches need not be
considered because the related partners can also be accessed by
reconnections inside non-overlapping encounters.

\section{Antiparallel encounter stretches}

\label{sec:antiparallel_overlap}

\begin{figure}
\begin{center}
  \includegraphics[scale=0.6]{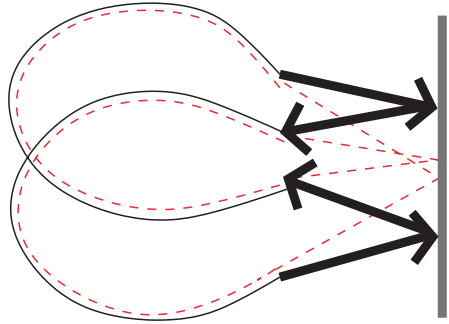}
\end{center}
\caption{Full line: Sketch of an orbit undergoing two
  almost self-retracing reflections from a hard wall. The encounter
  depicted by thick full lines could be regarded as a 3-encounter or a
  4-encounter with overlapping stretches, or as a 2-encounter with
  both stretches separated by intervening loops. Dashed line: Partner
  orbit obtained by reconnecting the four ports of the above
  encounter.}
\label{fig:retracer_clover}
\end{figure}

Two antiparallel encounter stretches can follow each other without an
intervening loop only if the encounter involves an almost
self-retracing reflection from a hard wall, as in Fig.
\ref{fig:retracer_examples}. If the encounter contains more than two
stretches, the remaining stretches must be reflected from the same
wall; an example, with two almost self-retracing reflections, is
depicted in Fig. \ref{fig:retracer_clover}.  Note, however, that the
almost self-retracing piece in Fig.  \ref{fig:retracer_examples}, and
each of the two such pieces in Fig.  \ref{fig:retracer_clover}, has
only two ports. Therefore, the depicted encounters should rather be
interpreted in a different way, with each of these pieces viewed as
{\it a single stretch}, folded back onto itself.  The stretches thus
defined are separated by loops.  Hence, there is no need to consider
antiparallel encounters without intervening loops.

Critical readers may wish to check that with this interpretation we do
not loose any orbit pairs.  For the encounter in Fig.
\ref{fig:retracer_examples}, this was already shown in Section
\ref{sec:min_dist}. Here we want to briefly repeat the main ideas. The
encounter in question has only two ports.  Since there is no way to
reshuffle connections between just two ports, we expect no partner
orbit.  To check this expectation, we attempt to construct a partner
orbit as follows: We place a Poincar{\'e} section inside the encounter.
The encounter pierces through this section in two phase-space points.
Now, we imagine that the orbit is ``cut open'' in both points and thus
split into two parts.  These parts have four loose ends.  Generalizing
the idea of ``reconnecting ports'', we can try to reconnect these
loose ends.  We then obtain an -- unphysical -- periodic trajectory
following one part of the orbit $\gamma$, and the time-reversed of the
other part.  One might assume that a periodic orbit $\gamma'$ can be found
by slightly deforming this trajectory.  However, by linearizing the
equations of motion around the above trajectory, we showed that this
suspected partner $\gamma'$ coincides with the original orbit $\gamma$.  Hence
there is no (off-diagonal) partner orbit.

Let us now turn to the encounter in Fig. \ref{fig:retracer_clover},
involving two almost self-retracing reflections.  In this example, we
can obtain a partner orbit by changing connections between the four
ports, like for any 2-encounter; this partner is drawn as a dashed
line.  We will show that no further partners can be found if we
artificially try to interpret the depicted encounter as, say, a
4-encounter.  Let us place a Poincar{\'e} section ${\cal P}$ somewhere
inside the encounter, and cut the orbit open in the four piercing
points.  A partner orbit $\gamma'$ can be found if we reconnect the
resulting eight loose ends, and subsequently look for a nearby
classical periodic orbit.  Using the ideas of Subsection
\ref{sec:action_difference}, the necessary reconnections can be
performed in three successive steps.  Each step affects only two
piercings, and changes the connections of the corresponding loose
ends.  We can show that two of these steps effectively do not change
the orbit.  Let us identify the loose end preceding the upper
self-retracing reflection in Fig. \ref{fig:retracer_clover} with the
upper right ``port'' in Fig. \ref{fig:steps}.  Then, this end is
engaged in reconnection steps involving all other loose ends, or
``ports'', on the same side. Among these, one step also involves the
loose end following the same reflection.  This step does not change
the orbit, just like reconnections inside the 2-encounter of Fig.
\ref{fig:retracer_examples}.  The remaining two steps thus suffice to
obtain the partner orbit $\gamma'$.  Repeating the same reasoning for the
second reflection, we can show that one further step may be dropped.
Hence, one step suffices to obtain the partner orbit, which therefore
can be seen as originating from reconnections in a 2-encounter rather
than a 4-encounter.  We thus see that it suffices to consider the
encounter of Fig.  \ref{fig:retracer_clover} as a 2-encounter with
both stretches separated by intervening loops, rather than a
4-encounter without intervening loops. Similar arguments apply to
encounters involving arbitrarily many almost self-retracing
reflections.\footnote{ Note that the 2-encounter in Fig.
  \ref{fig:retracer_clover} could be viewed either as parallel or as
  antiparallel, but only the antiparallel variant yields a partner
  orbit.  The situation becomes more complicated for larger
  encounters.  For example an encounter of three almost self-retracing
  stretches can alternatively be viewed as $\triplearrow[0]$,
  $\triplearrow[9]$, $\triplearrow[10]$, or $\triplearrow[6]$; the
  first stretch by definition points from left to right.  Each of
  these possibilities leads to a different division of ports into
  ``left'' and ``right'', and to different partner orbits, and thus
  has to be considered as a different encounter.  All these encounters
  lead to different structures, and to different choices of piercing points,
  and are therefore taken into account
  separately.}

\section{Antiparallel fringes}

\label{sec:fringes}

\begin{figure}
\begin{center}
  \includegraphics[scale=0.3]{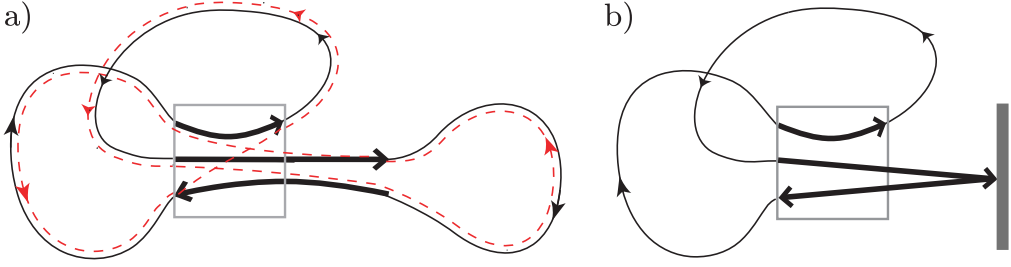}
\end{center}
\caption{
  3-encounter (in the box) with a ``fringe'' where only {\it two}
  antiparallel stretches remain close. Stretches are separated by a
  loop in a) but not in b); an almost self-retracing reflection arises
  in the latter case. A partner orbit reconnecting the ports of all
  three stretches (dashed line) is obtained only in case a).}
\label{fig:fringe}
\end{figure}

Recall that we define an $l$-encounter as a region inside a periodic
orbit in which $l$ orbit stretches come close up to time reversal.
Attached to the sides of the encounter are ``fringes'' in which only
some of the $l$ stretches remain close while others have already gone
astray, as shown in Fig. \ref{fig:fringe}a.  We shall now demonstrate
that these fringes do not affect the spectral form factor.

As a first example, let us assume that two {\it antiparallel}
stretches remain close after the end of the encounter, as in Fig.
\ref{fig:fringe}a. If there is no intervening loop, the orbit has to
undergo an almost self-retracing reflection (see Fig.
\ref{fig:fringe}b).  No connections can be switched between the two
stretches involved.  Thus, we cannot build a partner orbit in which
all three stretches of the encounter are changed.  Of course, we can
reshuffle connections between, e.g., the upper and lower stretches in
Fig.  \ref{fig:fringe}b, without changing the connections of the
middle stretch.  The arising partner is, however, associated to a
2-encounter rather than a 3-encounter.  Hence, the 3-encounter in Fig.
\ref{fig:fringe}b has to be disregarded.  The same arguments apply to
any encounter in which two antiparallel ``fringe'' stretches follow
each other without an intervening loop.  (Note that the results in the
end of Appendix \ref{sec:antiparallel_overlap} carry over to fringes
as well.)

On the other hand, if the ``fringe'' stretches are separated by an
external loop, a partner orbit can be obtained as usual, by
reshuffling connections between all stretches of the encounter, see
Fig. \ref{fig:fringe}a.  To make sure there is an intervening loop, we
impose minimal separations between piercing points, similar to those
used for excluding overlap between {\it encounters} in Section
\ref{sec:statistics}.  Consider two subsequent antiparallel stretches,
say, a stretch $j$ leading from left to right, and a stretch $j+1$
leading from right to left.  After these stretches have pierced
through a Poincar{\'e} section ${\cal P}$ with unstable coordinates
$\hat{u}_{\alpha j}$ and $\hat{u}_{\alpha,j+1}$, they will remain close for a
time $\frac{1}{\lambda}\ln\frac{c}{|\hat{u}_{\alpha j}-\hat{u}_{\alpha,j+1}|}$.
This duration contains both the time $t_u$ till the end of the $\alpha$-th
encounter, see Eq. (\ref{tu}), and an additional time span {\it after}
the end of the encounter, which gives the duration of the fringe.
Using the unstable coordinates $\hat{u}_{\alpha j}^{\rm e}$ of piercing
through a section in the end of the encounter (depending on
$\hat{u}_{\alpha j}$ via $\hat{u}_{\alpha j}^{\rm e}=\hat{u}_{\alpha j}{\rm e}^{\lambda
  t_u}$), this additional time may be written as
$\frac{1}{\lambda}\ln\frac{c}{|\hat{u}_{\alpha j}^{\rm e}-\hat{u}_{\alpha,j+1}^{\rm
    e}|}$.  Hence, the minimal time difference $2t_u$ between
piercings demanded in Section \ref{sec:statistics} has to be
incremented by a time $\frac{2}{\lambda}\ln\frac{c}{|\hat{u}_{\alpha j}^{\rm
    e}-\hat{u}_{\alpha, j+1}^{\rm e}|}$, depending on the unstable
coordinates only via $\hat{u}^{\rm e}$ or, equivalently, via
coordinates $u^{\rm e}$ defined through a coordinate transformation as
in Subsection \ref{sec:action_difference}.

Similarly, if the $j$-th stretch points from right to left and the
$(j+1)$-st stretch points from left to right, the corresponding
minimal distance has to be incremented by an amount purely depending
on the stable coordinates $s^{\rm b}$ in the beginning of the
encounter.  The contribution of each encounter to the total sum of
minimal separations can therefore be written in the form
\begin{equation}
\label{texcl}
t_{\rm excl}^\alpha=l_\alpha t_{\rm enc}^\alpha+\Delta
  t_s^\alpha(s^{{\rm b}}) +\Delta t_u^\alpha(u^{{\rm e}}) \end{equation}
with $\Delta t_s$ and $\Delta t_u$ functions of $s^{\rm b}$ and $u^{\rm e}$
only; the precise form of these functions will not be needed in the
following.  The numerator in the density of phase-space separations
$w_T(s,u)$, Eq.  (\ref{density}), has to be modified accordingly; each
$l_\alpha t_{\rm enc}^\alpha$ is replaced by $t_{\rm excl}^\alpha$.

We proceed to show that the summands $\Delta t_s^\alpha$ and $\Delta t_u^\alpha$ do
not affect the spectral form factor.  By reasoning as in Subsection
\ref{sec:contribution}, we see that only those terms of the
multinomial expansion of the numerator in $w_T(s,u)$ contribute to
$K(\tau)$ which involve a product of all $t_{\rm excl}^\alpha$.  Due to
$t_{\rm excl}^\alpha\neq l_\alpha t_{\rm enc}^\alpha$, these terms now have to be
written as (compare Eq. (\ref{wcontr}))
\begin{eqnarray}
  \frac{w_T^{\rm contr}(s,u)}{L}&=&h(\vec{v})
  \left(\frac{T}{\Omega}\right)^{L-V} \prod_\alpha\frac{t_{\rm excl}^\alpha}{l_\alpha
    t_{\rm enc}^\alpha}\nonumber \\ &=&
    h(\vec{v})
  \left(\frac{T}{\Omega}\right)^{L-V}\prod_\alpha\left[1+\frac{\Delta t_s^\alpha+\Delta
      t_u^\alpha}{l_\alpha t_{\rm enc}^\alpha}\right],%\nonumber
\end{eqnarray}
and contribute to the form factor as (compare Eq.
(\ref{contribution}))
\begin{equation}
\label{contribution_fringes}
%&&
\kappa \tau h(\vec{v}) \left(\frac{T}{\Omega}\right)^{L-V}\prod_\alpha\left[\int
  d^{L-V}s\,d^{L-V}u
  \left(1+\frac{\Delta t_s^\alpha+\Delta
      t_u^\alpha}{l_\alpha t_{\rm enc}^\alpha}\right){\rm
    e}^{\frac{\rm i}{\hbar}\sum_{j}s_{\alpha j}u_{\alpha j}}\right].
\end{equation}
The integral in Eq. (\ref{contribution_fringes}) coincides with the
one in Eq. (\ref{contribution}) up to the fringe corrections
$\propto\frac{\Delta t_s^\alpha}{t_{\rm enc}^\alpha}$ and $\propto\frac{\Delta t_u^\alpha}{t_{\rm
    enc}^\alpha}$.  By reasoning similarly as in Appendix
\ref{sec:integral}, one easily shows that the resulting integrals
vanish in the semiclassical limit: When integrating over $\frac{\Delta
  t_u}{t_{\rm enc}^\alpha}$, the additional factor $\Delta t_u^\alpha(u^{\rm e})$
affects only the integral over $u^{\rm e}$ inside Eq.
(\ref{osc_result}), whereas the rapidly oscillating integral over
$s_J^{\rm e}$ remains unchanged. For $\frac{\Delta t_s^\alpha}{t_{\rm
    enc}^\alpha}$ we have to consider a surface of section in the
beginning rather than in the end of the encounter.  Then, the integral
over the unstable coordinate $u_{J'}^{\rm b}$ associated to the
largest stable coordinate $s_{J'}^{\rm b}=\pm c$ oscillates rapidly as
$\hbar\to 0$; this integral remains unaffected by the factor $\Delta
t_s(s^{\rm b})$.  For antiparallel orbit stretches, fringe corrections
to the form factor are hence revealed as negligible in the
semiclassical limit.

The present treatment can immediately be generalized to $f>2$, if we
write out the additional index $m$, and to inhomogeneously hyperbolic
systems, if we allow $t_{\rm enc}$ to depend on $\x$. In the latter
case, $\Delta t_s$ will be a function of both $s^{\rm b}$ and the
phase-space points ${\bf x}^{\rm b}$ in the beginning of the
encounter, and $\Delta t_u$ will depend on $u^{\rm e}$ and ${\bf x}^{\rm
  e}$ in the end of the encounter.

\section{Parallel encounter stretches}

\label{sec:parallel_overlap}

We now turn to {\it parallel} encounter stretches.  We shall see that
if two subsequent parallel stretches of an encounter overlap or follow
each other after a relatively short loop, the encounter will have a
very peculiar structure.

\begin{figure}
\begin{center}
  \includegraphics[scale=0.21]{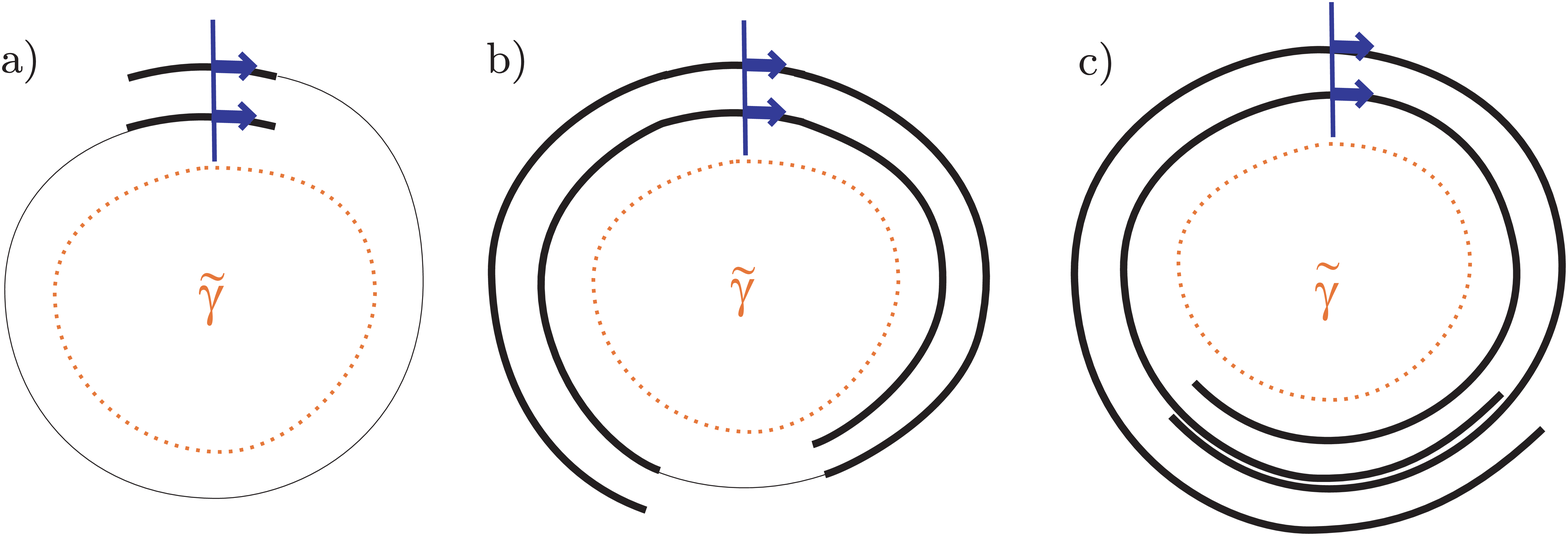}
\end{center}
\caption{
  Two parallel encounter stretches (thick lines) a) separated by a
  comparatively long loop (thin line), b) separated by a short loop,
  c) overlapping in a region depicted by two very close thick lines.
  The long orbit respectively follows a) slightly more than one, b)
  slightly less than two, c) more than two periods of the shorter
  orbit $\tilde{\gamma}$ (dotted line). Also depicted are Poincar{\'e}
  sections approximately in the center of the encounters, each with
  two piercing points.  }
\label{fig:parloop}
\end{figure}

Whenever two points of a periodic orbit $\gamma$ are close in phase space,
the orbit part between these points must be almost periodic -- and
hence in the vicinity of a shorter periodic orbit $\tilde{\gamma}$.  The
two phase-space points belong to mutually close encounter stretches,
which thus are near to $\tilde{\gamma}$ as well. Typically, these
stretches are short compared to the intervening loop. Hence $\gamma$
follows only slightly more than one revolution of $\tilde{\gamma}$ (Fig.
\ref{fig:parloop}a).

In contrast, if the stretches are only separated by a short loop (see
Fig. \ref{fig:parloop}b), the orbit $\gamma$ will remain close to
$\tilde{\gamma}$ for almost two periods: Each of the two stretches follows
nearly one period of $\tilde{\gamma}$, and the short intervening loop,
too, remains in the vicinity of $\tilde{\gamma}$.

Finally, if the stretches overlap, $\gamma$ will contain two or more
repetitions of $\tilde{\gamma}$; see Fig. \ref{fig:parloop}c for an
example of $\gamma$ following slightly more than two periods of
$\tilde{\gamma}$.

All further stretches of the encounters considered also have to come
close to $\tilde{\gamma}$. The orbit $\gamma$ will thus {\it approach
  $\tilde{\gamma}$ several times}, and in case of overlap or near-overlap
of stretches {\it at least one approach lasts significantly longer
  than one revolution of $\tilde{\gamma}$}.

We must now look for partner orbits related to such encounters.  To do
so, we shall follow the procedure outlined in Appendix
\ref{sec:antiparallel_overlap}: We place a Poincar{\'e} section ${\cal P}$
somewhere inside the encounter, and imagine the orbit $\gamma$ cut into
parts in its points of piercing through ${\cal P}$. We then change
connections between the loose ends of these parts, and slightly deform
the resulting closed trajectory to obtain a classical periodic orbit.
\footnote{ To avoid confusion, one must clearly distinguish between:
  {\it encounter stretches}, where the orbit $\gamma$ comes close to
  itself; {\it loops} between these stretches; {\it orbit parts}
  between two piercing points, containing parts of the corresponding
  encounter stretches and the intervening loop; and {\it regions where
    $\gamma$ approaches a shorter orbit $\tilde{\gamma}$}, possibly containing
  several encounter stretches and loops. }

The crucial point is now the following: If $\gamma$ follows nearly two, or
even several, revolutions close to $\tilde{\gamma}$, we have large freedom
in selecting piercing points.  For example, the encounter of Fig.
\ref{fig:par}c contains a region with three mutually close lines,
corresponding to only {\it two encounter stretches}. Placing a
Poincar{\'e} section inside this region, we find {\it three piercing
  points}. We need to select two of them.  If the encounter contains
further stretches, these stretches will also pierce through our
section, possibly several times, and we have to select one piercing
for each stretch. Some of the possible choices of piercings formally
correspond to {\it different encounters}, even though they yield {\it
  the same partner orbit}.  Therefore blindly considering all
encounters and associating with each of them a partner orbit we would
count certain orbit pairs several times.

In the following, we shall identify the conditions under which
piercing points of different encounters yield the same partner.  We
will demonstrate that within each family of encounters related to the
same partner, only for one member the stretches are separated by loops
exceeding certain minimal durations.  When evaluating the form factor,
we will include only this representative encounter and disregard all
other members of the family, to avoid overcounting of orbit pairs. The
resulting contribution to the form factor will turn out the same as if
we only demand the intervening loops to be positive; this is why the
latter condition was used in the main part.

\subsection{Parallel 3-encounters}

\label{sec:parallel3}

\begin{figure}[t]
\begin{center}
  \includegraphics[scale=0.22]{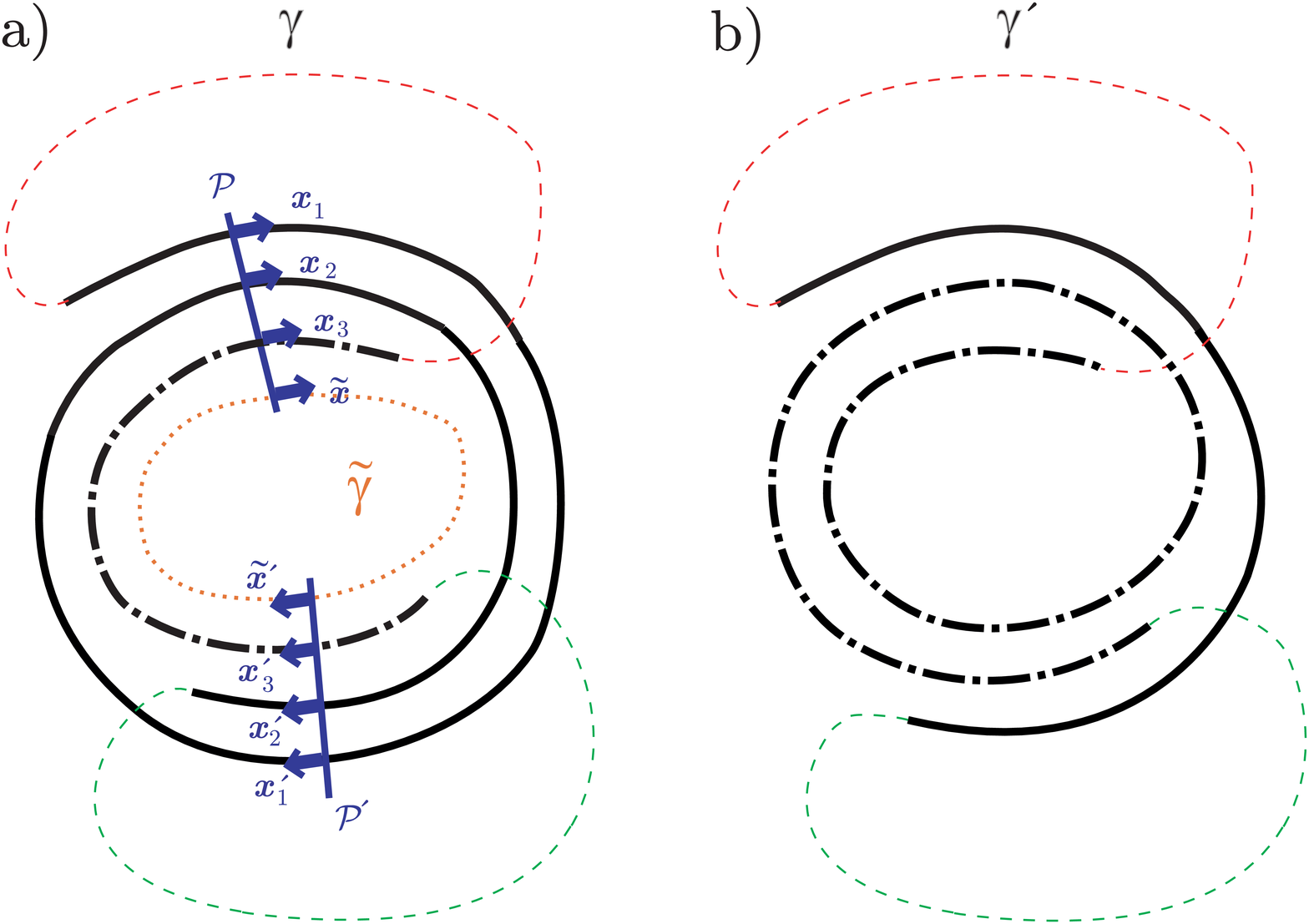}
\end{center}
\caption{
  a) An orbit $\gamma$ approaches a shorter orbit $\tilde{\gamma}$ (dotted
  line) in two regions (marked by thick full and dash-dotted lines).
  It pierces three times through each of the Poincar{\'e} sections ${\cal
    P}$ and ${\cal P}'$; the two sets of piercings belong to two
  different encounters.  b) Reconnections in either encounter lead to
  the same partner $\gamma'$, with one revolution of $\tilde{\gamma}$
  transposed between the full and dash-dotted regions.}
\label{fig:par}
\end{figure}

We first focus on the example of a parallel 3-encounter inside an
orbit $\gamma$.  The stretches of this encounter pierce through a Poincar{\'e}
section ${\cal P}$ in phase-space points ${\bf x}_1$, ${\bf x}_2$, and
${\bf x}_3$, see Fig. \ref{fig:par}a.  We shall derive a restriction
on the time difference $t_{12}$ between $\x_1$ and $\x_2$. Since ${\bf
  x}_1$ and ${\bf x}_2$ are close in phase space, the part of $\gamma$
between these points approximately follows a periodic orbit
$\tilde{\gamma}$ with period close to $t_{12}$.  The orbit $\gamma$ will also
stay close to $\tilde{\gamma}$ for some time before ${\bf x}_1$ and after
${\bf x}_2$.  The whole region close to $\tilde{\gamma}$ is drawn as a
thick full line in Fig.  \ref{fig:par}a.  In the vicinity of ${\bf
  x}_3$, the orbit $\gamma$ has to approach $\tilde{\gamma}$ for a second
time. This second approach is represented by a dash-dotted line.

The piercings ${\bf x}_1$, ${\bf x}_2$, and ${\bf x}_3$ cut $\gamma$ into
three parts. When reconnecting the ``loose ends'' of these parts to
form a partner orbit, we change the ordering of orbit parts. This
reordering may be interpreted as cutting out the orbit part leading
from ${\bf x}_1$ to ${\bf x}_2$ (close to $\tilde{\gamma}$) and
reinserting it between the two other parts, i.e., when $\gamma$ traverses
${\bf x}_3$.  In other words, one revolution of $\tilde{\gamma}$ is
transposed between the two regions approaching $\tilde{\gamma}$.  The
resulting partner is shown in Fig.  \ref{fig:par}b, with one
revolution of $\tilde{\gamma}$ taken out from the region depicted by the
thick full line and reinserted inside the dash-dotted region.  (Recall
that after transposing orbit parts the resulting trajectory must be
slightly deformed to yield a classical periodic orbit.)

Crucially, {\it different sets of piercing points may yield the same
  partner}.  We have already seen that the Poincar{\'e} section may be
moved continuously through the encounter.  The piercing points will
then all be shifted by the same amount of time. All sets of piercings
thus obtained belong to the same encounter and yield the same partner.

In the present scenario, the freedom in choosing piercing points is
even larger.  To show this, let us compare two sets of piercings, the
above points $\x_1$, $\x_2$, $\x_3$ in a Poincar{\'e} section ${\cal P}$,
and piercings $\x_1'$, $\x_2'$, $\x_3'$ in a (possibly) different
Poincar{\'e} section ${\cal P}'$.  We assume that the new
piercings are also close to $\tilde{\gamma}$, with ${\bf x}_1'$
and ${\bf x}_2'$ inside the same region of approach to $\tilde{\gamma}$ as
${\bf x}_1$ and ${\bf x}_2$, and ${\bf x}_3'$ in the same region as
${\bf x}_3$, as in Fig. \ref{fig:par}a.  Furthermore, we demand that the points
$\x_1'$ and $\x_2'$ enclose one revolution of $\tilde{\gamma}$, just
like $\x_1$ and $\x_2$.  Then, upon reconnection again one revolution
of $\tilde{\gamma}$ (the piece leading from ${\bf x}_1'$ to ${\bf x}_2'$)
is cut out from the first region approaching $\tilde{\gamma}$ and
reinserted inside the second region of approach (at position ${\bf
  x}_3'$).  Hence we obtain the same partner as for the ``old''
piercings.

The phase-space points $\x_1'$, $\x_2'$, $\x_3'$ will typically be
shifted along the orbit compared to their counterparts $\x_1$, $\x_2$,
$\x_3$.  The first two points must both be shifted by the same amount
of time $t_A$, to guarantee that they remain separated by one period
of $\tilde{\gamma}$.  The third one may be shifted by a different time
$t_B$.  After these shifts, all phase-space points must meet inside
the same Poincar{\'e} section ${\cal P}'$, such that the orbit $\gamma$
pierces through ${\cal P}'$ in $\x_1'$, $\x_2'$, and $\x_3'$.

We have seen that for $t_A=t_B$, i.e., if all points are shifted by
the same amount of time, the piercings ${\bf x}_1'$, ${\bf x}_2'$, and
${\bf x}_3'$ form part of the same encounter as ${\bf x}_1$, ${\bf
  x}_2$, and ${\bf x}_3$, crossed by a different Poincar{\'e} section.
However, if $t_A\neq t_B$ our ``new'' piercings belong to a different
encounter than the ``old'' ones, with both encounters yielding the
same partner orbit.  In the example of Fig. \ref{fig:par}a, the points
${\bf x}_1'$ and ${\bf x}_2'$ are shifted to the future ($t_A>0$)
compared to their counterparts ${\bf x}_1$ and ${\bf x}_2$, whereas
${\bf x}_3'$ is located to the past of ${\bf x}_3$ ($t_B<0$).
Nevertheless, both sets of piercing points give rise to the partner
orbit depicted in Fig.  \ref{fig:par}b, with one revolution of
$\tilde{\gamma}$ transposed between the two regions approaching
$\tilde{\gamma}$.

Not all encounters allow for such alternative piercings.  Roughly
speaking, to find alternative piercings with $t_A>0$ and $t_B<0$, the
orbit $\gamma$ must follow $\tilde{\gamma}$ long enough such that points
shifted in two opposite directions can meet again in the same Poincar{\'e}
section ${\cal P}'$.  There is a chance for such piercing points to exist if
the encounter stretches overlap or are separated by short loops, and
therefore $\gamma$ follows several periods of $\tilde{\gamma}$.  We proceed to
determine the precise conditions.

First of all, {\it how far can the points be shifted without leaving
  the vicinity of $\tilde{\gamma}$?} To answer this question, we need to
determine the stable and unstable coordinates of the phase-space point
$\tilde{{\bf x}}$ in which the orbit $\tilde{\gamma}$ intersects the
section ${\cal P}$. The trajectories passing through $\tilde{{\bf x}}$
and ${\bf x}_1$ remain close at least for one period of $\tilde{\gamma}$
after which they are carried to $\tilde{{\bf x}}$ and ${\bf x}_2$,
respectively. Thus by reasoning similarly as in Subsection
\ref{sec:manifolds} we see that $\tilde{{\bf x}}$ and ${\bf x}_1$ have
approximately the same unstable component
$\hat{\tilde{u}}\approx\hat{u}_1=0$.\footnote {The error introduced by this
  approximation is negligible compared to the other unstable
  differences. Linearizing the equations of motion, we obtain the
  unstable component $\hat{\tilde{u}}$ of $\tilde{\x}$ as
  $\hat{u}_2-\hat{\tilde{u}}={\rm e}^{\lambda
    t_{12}}(\hat{u}_1-\hat{\tilde{u}})$ with $\hat{u}_1=0$ and thus
  $\hat{\tilde{u}}-\hat{u}_1=\hat{\tilde{u}}=(1-{\rm e}^{\lambda
    t_{12}})^{-1}\hat{u}_2$.  The difference between $\hat{\tilde{u}}$
  and $\hat{u}_1$ vanishes like ${\cal O}({\rm e}^{-\lambda t_{12}})$
  compared to $\hat{u}_2$.}

Likewise the trajectories passing through $\tilde{{\bf x}}$ and ${\bf
  x}_2$ remain close for large negative times, such that the stable
component of $\tilde{{\bf x}}$ may be approximated by
$\hat{\tilde{s}}\approx\hat{s}_2$.

The first two phase-space points may be shifted to the future as long
as both remain close to $\tilde{\gamma}$.  The second point starts to
deviate significantly from $\tilde{\gamma}$ earlier than the first one.
Therefore, we can shift both phase-space points until the unstable
separation between $\x_2$ and $\tilde{\x}$, i.e.,
$\hat{u}_2-\hat{\tilde{u}}\approx\hat{u}_2-\hat{u}_1$, grows beyond our
bound $c$.  This will happen after a time
$\frac{1}{\lambda}\ln\frac{c}{|\hat{u}_2-\hat{u}_1|}$.  We will thus stay
in the vicinity of $\tilde{\gamma}$ while
\begin{equation}
\label{condition1}
t_A<\frac{1}{\lambda}\ln\frac{c}{|\hat{u}_2-\hat{u}_1|}.
\end{equation}

The third point may be shifted to the past as long as the stable
component of its separation from $\tilde{\gamma}$, i.e.,
$\hat{s}_3-\hat{\tilde{s}}\approx\hat{s}_3-\hat{s}_2$ remains below $c$;
this leads to the restriction
\begin{equation}
\label{condition2}
-t_B<\frac{1}{\lambda}\ln\frac{c}{|\hat{s}_3-\hat{s}_2|}.
\end{equation}

The shifted points $\x_1'$, $\x_2'$, and $\x_3'$ allow for
reconnections only if they {\it meet inside the same Poincar{\'e} section
  ${\cal P}'$}.  This restriction is trivially fulfilled if all
phase-space points are shifted by the same amount of time $t_A=t_B$
from one section ${\cal P}$ to a different section ${\cal P}'$.
Likewise, the points will meet inside the same section if, say, $\x_1$
and $\x_2$ effectively perform one more rotation around $\tilde{\gamma}$
than $\x_3$. After one such rotation, $\x_1$ and $\x_2$ end up in the
same Poincar{\'e} section as they started from.  More generally, the
shifted points will meet in the same Poincar{\'e} section if $\x_1$ and
$\x_2$ perform an integer number of rotations more, or less, than
$\x_3$.  This means that the time difference $t_A-t_B$ must involve an
integer number of periods of $\tilde{\gamma}$, i.e., an integer multiple
of $t_{12}$.  The restriction of having ${\bf x}_1'$, ${\bf x}_2'$,
and ${\bf x}_3'$ located inside the same Poincar{\'e} section thus boils
down to
\begin{equation}
\label{condition3}
t_A-t_B=n\,t_{12}
\end{equation}
with $n=-2,-1,0,1,2,\ldots$.  If we restrict ourselves to $t_A>0$,
$t_B<0$, we must have positive $n=1,2,\ldots$.  Combining Eqs.
(\ref{condition1}-\ref{condition3}), we are led to
\begin{equation}
\label{nrestriction}
n\,t_{12}<\frac{1}{\lambda}\ln\frac{c}{|\hat{u}_2-\hat{u}_1|}
+\frac{1}{\lambda}\ln\frac{c}{|\hat{s}_3-\hat{s}_2|}\,.
\end{equation}
If the separation $t_{12}$ is sufficiently large, i.e., if
\begin{equation}
\label{parallel_min_raw}
t_{12}>\frac{1}{\lambda}\ln\frac{c}{|\hat{u}_2-\hat{u}_1|}
+\frac{1}{\lambda}\ln\frac{c}{|\hat{s}_3-\hat{s}_2|}\,,
\end{equation}
Eq. (\ref{nrestriction}) has no positive solution $n$.  In this case,
no alternative encounter can be obtained by shifting the first two
phase-space points to the future and the third one to the past;
otherwise such encounters can be found, one for each possible $n\geq 1$.
For the case $t_A<0$, $t_B>0$, analogous reasoning yields a similar
condition.

We are now prepared to {\it single out one representative encounter
  for each orbit pair}. Inside each family of equivalent encounters we
take the member for which the time between piercings ${\bf x}_2$ and
${\bf x}_3$ is smallest.  This time difference is decreased by all
shifts with $t_A>0$, $t_B<0$, and increased if $t_A<0$, $t_B>0$.
($t_A$ and $t_B$ with equal sign need not be considered since we may
shift ${\cal P}'$ until the signs are opposite.)  Hence from the
chosen encounter no alternative one may be accessible through a shift
with $t_A>0$, $t_B<0$. Consequently our representative encounter has
to satisfy Eq.  (\ref{parallel_min_raw}). Using the stable coordinates
$\hat{s}_j^{\rm b}=\hat{s}_j{\rm e}^{\lambda t_s}$ in the beginning of the
encounter and the unstable coordinates $\hat{u}_j^{\rm
  e}=\hat{u}_j{\rm e}^{\lambda t_u}$ in the end, and recalling that $t_{\rm
  enc}=t_s+t_u$, we may rewrite this condition as\footnote {
  Alternatively, we could have selected the encounter with the largest
  rather than the smallest time difference between $\x_2$ and $\x_3$.
  This would yield a condition symmetric to (\ref{min_parallel}), but
  with stable and unstable coordinates interchanged.  This condition
  would yield the same result for the spectral form factor.}
\begin{equation}
\label{min_parallel}
t_{12}>t_{\rm enc}+\frac{1}{\lambda}\ln\frac{c}{|\hat{u}_2^{\rm e}-\hat{u}_1^{\rm e}|}
+\frac{1}{\lambda}\ln\frac{c}{|\hat{s}_3^{\rm b}-\hat{s}_2^{\rm b}|}\,.
\end{equation}
Minimal separations between, say, the second and third piercing, are
obtained from Eq. (\ref{min_parallel}) by cyclic permutation.

Eq. (\ref{min_parallel}) can be {\it interpreted} as follows:
Demanding $t_{12}>t_{\rm enc}$ implies that we consider only
encounters whose stretches do not overlap, as postulated in the main
part.  The additional increment in Eq. (\ref{min_parallel}) can be
regarded as a minimal loop duration.  If the loop in between the
encounter stretches is very large, $\gamma$ will follow $\tilde{\gamma}$ for
only slightly more than one period and there will be no alternative
encounter.  If the loop slightly exceeds our threshold, there
typically are alternative encounters, but the one considered is the
encounter representative for our orbit pair.  The boundary between
both scenarios is irrelevant for our considerations.

The minimal loop duration depends purely on $\hat{s}^{\rm b}$ and
$\hat{u}^{\rm e}$.  In the language of Appendix \ref{sec:fringes}, the
second summand in Eq. (\ref{min_parallel}) gives the duration of the
``fringe'' where the first two stretches remain close after the end of
the encounter; likewise the third summand represents the duration of
the fringe where the second and third stretch come close before the
beginning of the encounter.

Due to the exclusive dependence on $\hat{s}^{\rm b}$ and $\hat{u}^{\rm
  e}$, respectively, the additional summands in Eq.
(\ref{min_parallel}) {\it do not affect the form factor} in the
semiclassical limit.  Again the sum of minimal time differences is of
the same form, Eq.  (\ref{texcl}), as met in case of antiparallel
fringes, and the reasoning of Appendix \ref{sec:fringes} carries over
accordingly.

Focusing on the example of a parallel 3-encounter, we have thus
justified the treatment in the main part: We have shown that
encounters with overlapping stretches must be excluded, and that
corrections due to the exclusion of encounter stretches separated by
short loops do not affect $K(\tau)$ in the semiclassical limit.

\subsection{General $l$-encounters}

The previous line of reasoning may be generalized to $l$-encounters
with arbitrary $l$. For simplicity, we first consider orbit pairs
$(\gamma,\gamma')$
differing in one encounter with all $l$ stretches almost parallel.
This encounter pierces $l$ times through a Poincar{\'e}
section ${\cal P}$, cutting the orbit $\gamma$ into $l$ parts.
We label each of the $l$ orbit parts by the same number
as the following piercing point; this label coincides with the number
of the loop and the entrance port included in the corresponding part.
As before, the second orbit part, between the first two piercing points,
is close to a shorter periodic orbit $\tilde{\gamma}$.
We shall see that the partner $\gamma'$ can be obtained by
performing reconnections inside two separate sets of piercing
points, and afterwards cutting out the part close to $\tilde{\gamma}$
and reinserting it in a different place.
In analogy to the preceding Subsection,
the two sets of
points may be shifted independently, without changing the partner orbit
$\gamma'$.

We will again employ
methods of permutation theory.
The order of orbit parts in $\gamma$ is given by the permutation
$P^\gamma=P_{\rm loop}=(1,2,\ldots,l)$.  In the partner orbit $\gamma'$, the $l$
parts are ordered differently.  This ordering is described by a
permutation $P\neq P^\gamma$.  The permutation $P$ also determines the
encounter permutation $P_{\rm enc}=P_{\rm loop}^{-1}P$, giving the
connections between ports, or loose ends, inside $\gamma'$.  For orbit
pairs differing in one $l$-encounter, $P_{\rm enc}$ has a single cycle
of length $l$.

We can proceed from $\gamma$ to $\gamma'$ in two steps. In the first step, we
go from $\gamma$ to an intermediate orbit $\gamma_{\rm int}$. This orbit has
the parts $1,3,4,\ldots,l$ ordered as in $\gamma'$, but the second orbit part
still follows the first. When reordering the orbit parts, the first
piercing point, on the border between the first and the second orbit
part, remains practically unaffected.  All other piercing points are
changed.  To describe $\gamma_{\rm int}$ in terms of permutations, it is
technically easier to consider the first two orbit parts as one single
part, labelled by the index 1.
(Treating the first two orbit parts as one means that the first
piercing point, unchanged in $\gamma_{\rm int}$, is disregarded, and that
the second piercing point, following the two initial orbit parts, is
renamed into the first.)  The ordering of orbit parts in $\gamma_{\rm
  int}$ is then given by a permutation $Q$ obtained from the
permutation $P$, describing $\gamma'$, by leaving out the label 2.
Likewise, we have to eliminate the element 2 from the permutation
$P_{\rm loop}$, leading to a new permutation $Q_{\rm loop}$.  The
intra-encounter connections of $\gamma_{\rm int}$ are thus described by
the permutation $Q_{\rm enc}=Q_{\rm loop}^{-1}Q$.  The cycle structure
of $Q_{\rm enc}$ was already investigated in Subsection
\ref{sec:unitary_recursion}, where for formal reasons we eliminated
$L=l$ rather than $2$; all our results carry over if we eliminate 2
instead.  We have seen that inside $Q_\Enc$, the $l$-cycle of $P_{\rm
  enc}$ is split in two cycles, given in Eq.  (\ref{breakup}).  Hence,
$\gamma_{\rm int}$ may be obtained from $\gamma$ by independent
reconnections inside two
separate sets of encounter stretches.

In the second step, we go from $\gamma_{\rm int}$ to the partner $\gamma'$.
We thus have to cut out the second orbit part,
close to $\tilde{\gamma}$, and reinsert it into
its destined place.

We are now prepared to generalize the arguments of the previous
Subsection. The $l$ piercing points can be divided in two sets
$A$ and $B$, corresponding to the two cycles of $Q_{\rm enc}$;
the first piercing point, not taking part in reconnections,
must be included in the same set $A$ as the second one.
We can now show that the same partner $\gamma'$ arises if the points
of $A$ and $B$ are shifted independently by respective times $t_A$ and $t_B$
(e.g., $t_A>0$ and $t_B<0$),
provided that all piercings remain in the vicinity of $\tilde{\gamma}$
and end up inside the same Poincar\'e section
${\cal P}'$.
First, note that reconnections inside $A$ yield the same
result as before,
since the phase-space points of $A$ were all shifted by an equal amount of time
and stayed mutually close.
Similar arguments apply to the reconnections inside $B$.
Next, we transpose the orbit part enclosed
between the first two piercings.
This part still contains one revolution of $\tilde{\gamma}$,
given that
the first and second point were
shifted by the same time $t_A$
and remained in the vicinity of $\tilde{\gamma}$.
Hence, for both the ``old'' and the ``new'' piercings,
one revolution of $\tilde{\gamma}$
is cut out from the same region approaching $\tilde{\gamma}$,
and reinserted inside a different region.
Consequently, in both cases we obtain the same $\gamma'$.

To guarantee that all points of
$A$ remain near $\tilde{\gamma}$, Eq.  (\ref{condition1}) must be
replaced by
\begin{equation}
t_A<\frac{1}{\lambda}\ln\frac{c}{\max_{j\in A}|\hat{u}_1-\hat{u}_j|}\,.
\end{equation}
Similarly, the points included in $B$ will stay in the vicinity of $\tilde{\gamma}$ if
\begin{equation}
-t_B<\frac{1}{\lambda}\ln\frac{c}{\max_{j\in B}|\hat{s}_2-\hat{s}_j|}\,.
\end{equation}
As above, we choose a representative encounter from which no other
encounter is accessible through a shift with $t_A>0$ and $t_B<0$.  The
piercing points of this representative encounter now have to satisfy
the restriction
\begin{equation}
t_{12}>t_{\rm enc}+\frac{1}{\lambda}\ln\frac{c}{\max_{j\in A}|\hat{u}^{\rm e}_1-\hat{u}^{\rm e}_j|}
+\frac{1}{\lambda}\ln\frac{c}{\max_{j\in B}|\hat{s}^{\rm b}_2-\hat{s}^{\rm b}_j|}\,,
\end{equation}
generalizing Eq. (\ref{min_parallel}).  Again, the form factor remains
unaffected by the increment depending purely on $\hat{s}^{\rm e}$ and
$\hat{u}^{\rm e}$.

Our results directly carry over to orbit pairs differing in several
parallel encounters, and can be further generalized to time-reversal
invariant systems. In the latter case, some of the remaining piercing
points may be approximately time-reversed with respect to $\x_1$ and
$\x_2$.  Such time-reversed points need to be shifted by an amount of
time $-t_A$ or $-t_B$ rather than $t_A$ or $t_B$, to make sure that
all points are moved into the same direction.  Moreover, our reasoning
extends to $f>2$ and inhomogeneous hyperbolicity as outlined in the
antiparallel case.

\subsection{Symbolic dynamics}

\label{sec:por}

We would like to illustrate the above results by a simple example,
elucidating their relation to symbolic dynamics.  We consider a
parallel 3-encounter, with two overlapping stretches and -- in
contrast to the preceding Subsections -- no fringes attached.  In
symbolic dynamics, all three encounter stretches will then be
characterized by the same sequence $\ES$.  Whenever two subsequent
stretches overlap, so do the corresponding symbol sequences. This is
only possible if $\ES$ has a special structure.  The simplest symbol
sequences allowing for such overlap are of the form $\ES=\XS\YS\XS$
with $\XS$ and $\YS$ arbitrary subsequences.  The latter structure
permits two subsequent strings $\ES$ to overlap in $\XS$ and merge to
a symbol sequence $\XS\YS\XS\YS\XS$, the middle $\XS$ denoting the
overlapping piece.  The form $\ES=\XS\YS\XS$ obviously limits the
length of the overlap ${\cal X}$ to less than half the length of
${\cal E}$.\footnote{Otherwise, if the ``overlap'' extends over more
  than half the length of $\mathcal E$ we have two pieces ${\mathcal
    E}=({\cal A}{\cal B})^{n+1}{\cal A}$ overlapping in $({\cal
    A}{\cal B})^{n}{\cal A}$ and merging to $({\cal A}{\cal
    B})^{n+2}{\cal A}$, now $n>0$. In the following, we will restrict
  ourselves to $n=0$.}

\begin{figure}%[t!]
\begin{center}
  \includegraphics[scale=0.24]{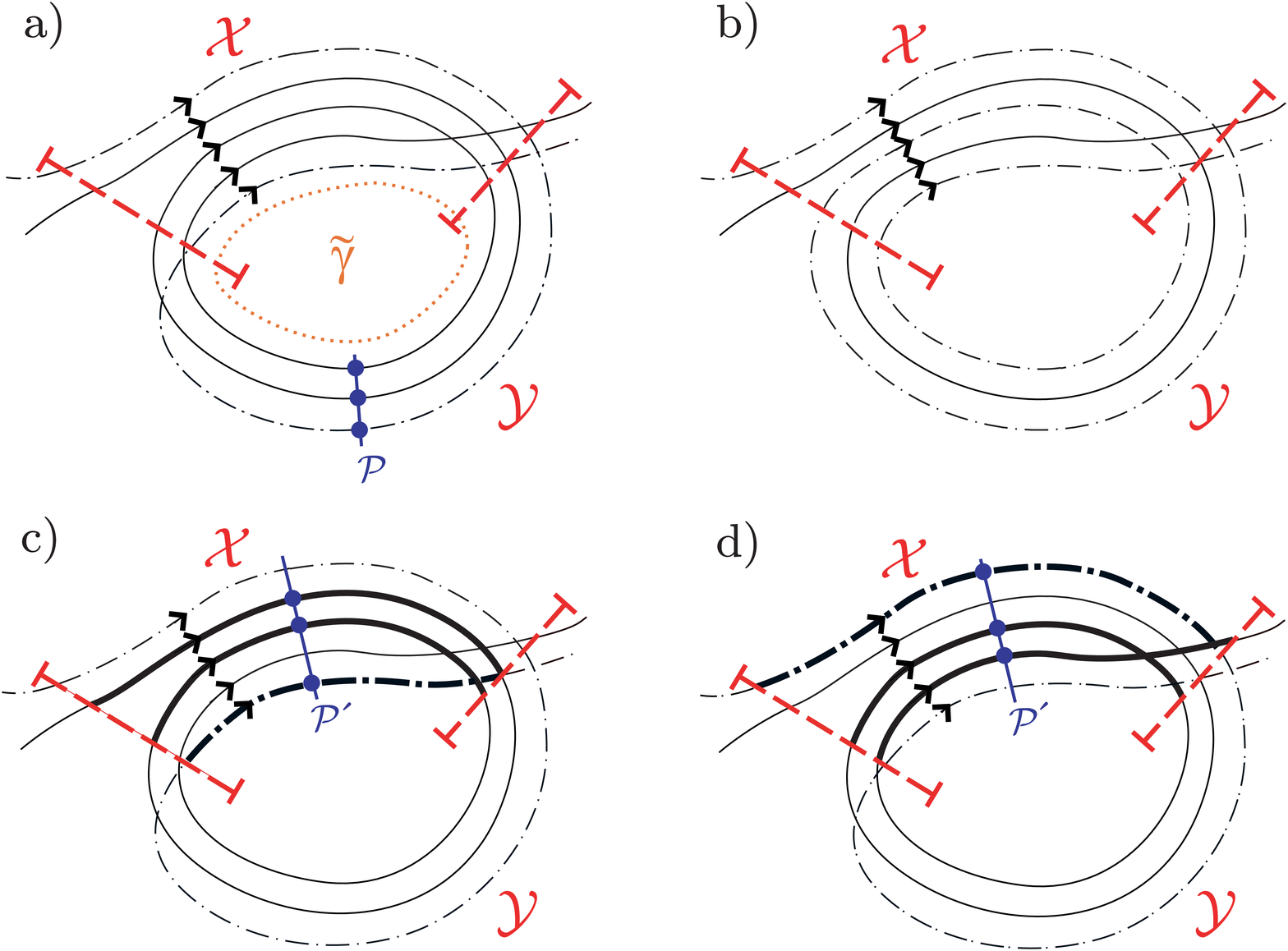}
\end{center}
\caption{
  Overlap of parallel encounter stretches: a) An orbit $\gamma$ twice
  approaches a shorter orbit $\tilde{\gamma}=\XS\YS$ (dotted line); the
  approaches, depicted by thick full and dash-dotted lines, have
  symbol sequences $\XS\YS\XS\YS\XS$ and $\XS\YS\XS$.  Also depicted
  is a Poincar{\'e} section ${\cal P}$ inside the region ${\cal Y}$, with
  three piercings.  b) The partner $\gamma'$ has one repetition of
  $\tilde{\gamma}$ shifted between the two regions.  c), d) Two shorter
  3-encounters (named (i) and (ii) in the text) can be found in the
  region ${\cal X}$.  They are depicted by thick lines and pierce
  three times through a Poincar{\'e} section ${\cal P'}$ inside ${\cal
    X}$.  The three encounters depicted in a), c), and d) all yield
  the same partner b).  The encounter in d) is chosen as a
  representative encounter: Starting from this representative, no
  alternative encounter can be found if we shift the piercing points
  of the thick full stretches to the future, and the one of the thick
  dash-dotted stretch to the past.  }
\label{fig:por}
\end{figure}

The orbit $\gamma$ now contains two regions approaching a shorter orbit
$\tilde{\gamma}$ with symbol sequence $\XS\YS$: one region with the
sequence $\XS\YS\XS\YS\XS$ formed out of the merger of two stretches,
and one region with a sequence $\XS\YS\XS$ consisting of the remaining
third stretch. If we denote the two intervening loops by $\LS_1$ and
$\LS_2$, $\gamma$ will have the symbol string
\begin{equation}
\gamma=(\XS\YS\XS\YS\XS)\LS_1(\XS\YS\XS)\LS_2\,.
\end{equation}
This scenario is illustrated in Fig. \ref{fig:por}a. Two Poincar{\'e}
sections, represented by dashed lines, divide the shorter orbit
$\tilde{\gamma}$ in two pieces, with corresponding symbol sequences $\XS$
and $\YS$.  The longer orbit $\gamma$ approaches $\tilde{\gamma}$ twice. The
first approach, depicted by a full line, contains two full revolutions
of $\tilde{\gamma}=\XS\YS$, and one additional piece close to the segment
with symbol sequence $\XS$ (not shown in the picture).  After a loop
with sequence $\LS_1$, $\gamma$ remains close to $\tilde{\gamma}$ for one full
revolution of $\tilde{\gamma}$, and one additional piece with the symbol
sequence $\XS$.  This second approach is represented by a dash-dotted
line.  The orbit $\tilde{\gamma}$ finally closes itself after an
additional loop with the sequence $\LS_2$ (again not depicted).

As shown previously, the partner orbit $\gamma'$ has one revolution of
$\tilde{\gamma}=\XS\YS$ transposed between the two regions close to
$\tilde{\gamma}$; see Fig. \ref{fig:por}b.  In symbolic dynamics, it is
therefore characterized by
\begin{equation}
\gamma'= (\XS\YS\XS)\LS_1(\XS\YS\XS\YS\XS)\LS_2\,.
\end{equation}

If we place a Poincar{\'e} section ${\cal P}$ inside the region with
symbol sequence $\YS$, the orbit $\gamma$ will pierce through this section
in three phase-space points. The latter section may be shifted freely
through the encounter.  The piercing points are then all shifted by
the same amount of time, either to the preceding or to the following
pieces with symbol sequences $\XS$.  All these sets of piercings
correspond to the same encounter and lead to the same partner orbit.

We have already seen that there is even more freedom in choosing
piercing points: The two points inside the region $\XS\YS\XS\YS\XS$
and the point inside $\XS\YS\XS$ may be shifted independently, as long
as they remain inside these regions and end up inside the same
Poincar{\'e} section ${\cal P}'$; recall that the two points inside
$\XS\YS\XS\YS\XS$ must both be shifted by the same amount of time.
This gives two additional possibilities:
\begin{enumerate}
\item[(i)] The piercing points of $\XS\YS\XS\YS\XS$ may be shifted to
  the past, whereas the piercing of $\XS\YS\XS$ is shifted to the
  future.  The shifted points are placed inside the first and second
  $\XS$ of $\XS\YS\XS\YS\XS$, and inside the final $\XS$ in
  $\XS\YS\XS$.
\item[(ii)] Alternatively, the points of $\XS\YS\XS\YS\XS$ may be
  shifted to the future, i.e., to the second and third $\XS$, whereas
  the point inside $\XS\YS\XS$ is shifted to the past, i.e., to the
  first $\XS$.
\end{enumerate}
In both cases, the phase-space points have to be shifted until they
meet in the same Poincar{\'e} section ${\cal P}'$.  The two additional
possibilities are depicted in Figs.  \ref{fig:por}c and d, where the
pieces $\XS$ containing the piercings of (i) and (ii) are highlighted
by thick full and dash-dotted lines.  The corresponding encounters
just consist of these pieces $\XS$.  For instance in case (i), see
Fig.  \ref{fig:por}c, the upper thick full stretch has a large
distance from the other stretches before the beginning of the sequence
$\XS$, whereas the thick dashed stretch departs from the others
immediately after the end of $\XS$.  Hence, the two alternative
encounters involve shorter stretches than the original encounter, and
all these stretches are separated by intervening loops.

Altogether, we have now found three encounters yielding the same
partner, depicted in Fig. \ref{fig:por}b.  Among these, the encounter
corresponding to (ii) (see Fig. \ref{fig:por}d), containing the last
two $\XS$ in $\XS\YS\XS\YS\XS$ and the first $\XS$ in $\XS\YS\XS$, is
chosen as a representative encounter; from this representative no
further encounter is accessible through a shift to the future in
$\XS\YS\XS\YS\XS$ and to the past in $\XS\YS\XS$.  The condition of
Eq.  (\ref{min_parallel}) implies that the other two encounters are
ignored: The encounter of Fig. \ref{fig:por}a contains overlapping
stretches, whereas the encounter of Fig. \ref{fig:por}c involves one
loop shorter than required by Eq. (\ref{min_parallel}).  The
one-to-one correspondence between orbit pairs and encounters is thus
restored.

\section{Summary}

We have shown that the spectral form factor is determined by
encounters whose stretches are separated by intervening loops, and
that encounters with overlapping stretches must be disregarded.  If
two {\it antiparallel} stretches follow each other without an
intervening loop they have to be treated as a single stretch, folded
back into itself after an almost self-retracing reflection.  This
self-retracing reflection may also occur inside the ``fringes'' of the
encounter, where some encounter stretches have already departed from
the rest.  To treat such fringes correctly, we have to slightly
increase the minimal distance between piercing points, but this
increase does not affect the spectral form factor in the semiclassical
limit.  If {\it parallel} stretches overlap or nearly overlap, the
orbit has to follow multiple
revolutions of a shorter orbit $\tilde{\gamma}$.  In this situation, {\it
  several different encounters} lead to the {\it same partner orbit}.
To select one of these encounters, we must again impose a minimal
distance between piercing points slightly larger than in the main
part, but find the same value for the form factor as $\hbar\to 0$.